\documentclass[10pt,twoside]{book}

\usepackage[
    paperwidth=156mm,
    paperheight=234mm,
    bindingoffset=0.25in,
    centering,
    marginparwidth=0in,
    textwidth=4.25in,
    marginparsep=0em,
    top=1.25in,
    bottom=1.25in,
]{geometry}

\usepackage{graphicx}
\usepackage[table,dvipsnames]{xcolor}

\usepackage{algorithm}
\usepackage{algpseudocode}
\usepackage[tbtags]{amsmath}
\usepackage{amsbsy,amssymb,amsfonts,amsthm}
\usepackage{cancel}
\usepackage{caption}
\usepackage{environ}
\usepackage{enumitem}
\usepackage{epsdice}
\usepackage{fancyhdr}
\usepackage{fancyvrb}
\usepackage{hyperref}
\usepackage{ifthen}
\usepackage{latexsym}
\usepackage{listings}
\usepackage{makecell}
\usepackage{makeidx}
\usepackage[framemethod=TikZ]{mdframed}
\usepackage[novbox]{pdfsync}
\usepackage{physics}
\usepackage{sectsty}
\usepackage{tabularx}
\usepackage{tikz}
\usepackage{titlesec,titletoc}
\usepackage{url}
\usepackage{multirow}

\definecolor{lightgray}{RGB}{220,220,220}
\definecolor{myblue}{cmyk}{1 0.35 0 0.5}
\definecolor{myyellow}{cmyk}{0.5 0.1 0.9 0.1}

\sectionfont{\color{myblue}\sffamily}
\chapterfont{\color{myblue}\sffamily}
\subsectionfont{\color{myblue}\sffamily}

\newtheorem{theorem}{Theorem}[chapter]

\newtheorem{lemma}{Lemma}[chapter]

\newtheorem{definition}{Definition}[chapter]

\newtheorem{assumption}{Assumption}[chapter]

{\begin{list}               
    {$\bullet$ \hfill}{
        \setlength{\leftmargin}{\parindent}
        \setlength{\parsep}{0.04\baselineskip}
        \setlength{\itemsep}{0.5\parsep}
        \setlength{\labelwidth}{\leftmargin}
        \setlength{\labelsep}{0em}}
    }
{\end{list}}

\newcommand{\boxedthm}[1]{
\begin{mdframed}[roundcorner=5pt,middlelinewidth=2pt,backgroundcolor=myblue!10]
\vspace{-1ex}
#1
\end{mdframed}
}

\newcommand{\boxedmsg}[1]{
\begin{mdframed}[roundcorner=5pt,middlelinewidth=2pt,backgroundcolor=blue!10]
\vspace{-1ex}
#1
\end{mdframed}
}

\newcommand{\boxedeg}[1]{
\begin{mdframed}[roundcorner=5pt,middlelinewidth=2pt,backgroundcolor=myyellow!10]
\vspace{-1ex}
#1
\end{mdframed}
}

\newcommand{\keyword}[1]{\textcolor{myblue}{\textsf{\textbf{#1}}}}
\newsavebox{\mybox}

\newcounter{example}[chapter]
\renewcommand{\theexample}{\arabic{example}}

\providecommand{\eref}[1]{Equation~(\ref{#1})}
\providecommand{\cref}[1]{Chapter~\ref{#1}}

\providecommand{\fref}[1]{Figure~\ref{#1}}

\providecommand{\R}{\ensuremath{\mathbb{R}}}

\providecommand{\E}{\ensuremath{\mathbb{E}}}

\providecommand{\Pb}{\ensuremath{\mathbb{P}}}

\providecommand{\Var}{\mathrm{Var}}

\providecommand{\abs}[1]{\lvert#1\rvert}

\providecommand{\bydef}{\overset{\text{def}}{=}}

\renewcommand{\vec}[1]{\ensuremath{\boldsymbol{#1}}}
\providecommand{\mat}[1]{\ensuremath{\boldsymbol{#1}}}

\providecommand{\calA}{\mathcal{A}}
\providecommand{\calB}{\mathcal{B}}

\providecommand{\calD}{\mathcal{D}}

\providecommand{\calF}{\mathcal{F}}
\providecommand{\calG}{\mathcal{G}}
\providecommand{\calH}{\mathcal{H}}
\providecommand{\calI}{\mathcal{I}}
\providecommand{\calJ}{\mathcal{J}}

\providecommand{\calL}{\mathcal{L}}

\providecommand{\calN}{\mathcal{N}}

\providecommand{\calR}{\mathcal{R}}
\providecommand{\calS}{\mathcal{S}}
\providecommand{\calT}{\mathcal{T}}

\providecommand{\mA}{\mathbf{A}}
\providecommand{\mB}{\mathbf{B}}
\providecommand{\mC}{\mathbf{C}}
\providecommand{\mD}{\mathbf{D}}
\providecommand{\mE}{\mathbf{E}}
\providecommand{\mF}{\mathbf{F}}
\providecommand{\mG}{\mathbf{G}}
\providecommand{\mH}{\mathbf{H}}
\providecommand{\mI}{\mathbf{I}}
\providecommand{\mJ}{\mathbf{J}}

\providecommand{\mL}{\mathbf{L}}

\providecommand{\mS}{\mathbf{S}}
\providecommand{\mT}{\mathbf{T}}
\providecommand{\mU}{\mathbf{U}}

\providecommand{\ve}{\mathbf{e}}
\providecommand{\vf}{\mathbf{f}}
\providecommand{\vg}{\mathbf{g}}
\providecommand{\vh}{\mathbf{h}}

\providecommand{\vk}{\mathbf{k}}

\providecommand{\vn}{\mathbf{n}}

\providecommand{\vp}{\mathbf{p}}

\providecommand{\vr}{\mathbf{r}}
\providecommand{\vs}{\mathbf{s}}
\providecommand{\vt}{\mathbf{t}}
\providecommand{\vu}{\mathbf{u}}
\providecommand{\vv}{\mathbf{v}}
\providecommand{\vw}{\mathbf{w}}
\providecommand{\vx}{\mathbf{x}}
\providecommand{\vy}{\mathbf{y}}

\providecommand{\mGamma}{\mat{\Gamma}}

\providecommand{\mPhi}{\mat{\Phi}}

\providecommand{\mSigma}{\mat{\Sigma}}

\providecommand{\valpha}{\vec{\alpha}}
\providecommand{\vbeta}{\vec{\beta}}

\providecommand{\vtheta}{\vec{\theta}}

\providecommand{\vxi}{\vec{\xi}}

\providecommand{\vrho}{\vec{\rho}}
\providecommand{\vvarrho}{\vec{\varrho}}

\providecommand{\vvarphi}{\vec{\varphi}}

\providecommand{\vpsi}{\vec{\psi}}

\providecommand{\vutilde}{\boldsymbol{\widetilde{u}}}

\providecommand{\vxhat}{\boldsymbol{\widehat{x}}}

\providecommand{\Var}{\mathrm{Var}}

\providecommand{\to}{\rightarrow}

\newcommand{\subjectto}{\mathop{\mathrm{subject\, to}}}
\newcommand{\argmin}[1]{\mathop{\underset{#1}{\mbox{argmin}}}}

\newcommand{\minimize}[1]{\mathop{\underset{#1}{\mathrm{minimize}}}}

\usepackage{filemod}
\usepackage{subfig}
\usepackage{pdfpages}
\renewcommand{\vu}{\mathbf{u}}
\newcommand{\comment}[1]{}

\usepackage[english]{babel}
\addto\captionsenglish{}
\graphicspath{{pix/}}
\makeindex

\makeatletter

\def\@makechapterhead#1{%
  \vspace*{50\p@}%
  {\parindent \z@ \raggedright \color{myblue}\sffamily
    \ifnum \c@secnumdepth >\m@ne
      \if@mainmatter
        \Huge\bfseries \thechapter.\space%
      \fi
    \fi
    \interlinepenalty\@M
    \Huge \bfseries #1\par\nobreak
    \vskip 40\p@
  }
}

\begin{document}

\setcounter{page}{1}
\thispagestyle{empty}

\frontmatter

\begin{titlepage}
    \vspace*{1cm}

    \begin{center}
    {\Huge

    \noindent\textbf{Computational Imaging \\ through \\Atmospheric Turbulence}}

            \vspace{0.5cm}

    \end{center}

    \vspace{1.5cm}
    \begin{flushright}
    \noindent{\large \textbf{Stanley H. Chan \\ Nicholas Chimitt}}
    \end{flushright}

    \vfill


\end{titlepage} 
\chapter*{}
This book is based upon work supported in part by the Intelligence Advanced Research Projects Activity (IARPA) under Contract No. 2022‐21102100004, and in part by the National Science Foundation under the grants CCSS-2030570 and IIS-2133032. The views and conclusions contained herein are those of the authors and should not be interpreted as necessarily representing the official policies, either expressed or implied, of IARPA, or the U.S. Government. The U.S. Government is authorized to reproduce and distribute reprints for governmental purposes notwithstanding any copyright annotation therein.

\let\cleardoublepage\clearpage
\chapter*{Preface}

The physics of imaging through atmospheric turbulence has been around for more than 80 years, and over this time it has generated a rich collection of texts and papers. We came to the field with an image processing background without having even read an optics textbook. Upon reading some of the imaging through turbulence literature, we very quickly realized the depth and breadth of the subject, but more strikingly the lack of an easy-to-read reference for people with a background like us.

The goal of this book is to provide a short introduction to the subject from the perspective of an image processing person. By image processing, we are thinking of scientists and engineers working on inverse problems in imaging systems with the goal of recovering signals from corrupted measurements. To this end, we are targeting readers who would like to know the physics of atmospheric turbulence so that they can improve their algorithms. Because of the specific perspective we take here and the targeted audience group, we shall not take a very rigorous physics-based approach. Unless the reader is already familiar with wave optics, the learning barrier will be so high that an average person would not be able to master the concepts quickly. Democratizing the ideas and educating the image processing community is an important mission of this book.

As we write this book, we aim in delivering the ``big pictures'' of the subject. Whenever needed, we streamline background materials including probability, optics, and optimization. Some sacrifices in the material are made to balance precision and clarity. Therefore, we do not regard this book as any substitution of the great optics books of our time. Whenever possible, we will connect the technical details back to our theme of computational imaging.

We would like to thank a lot of people who offered generous feedback to us:
Jeremy Bos,
Chris Dainty,
Russell Hardie,
Dan LeMaster,
Kevin Miller,
Casey Pellizzari,
Michael Roggemann,
Mike Rucci,
Jason Schmidt, and
Mark Spencer.
We also like to thank our fellow colleagues and students at Purdue: Charlie Bouman, Mark Bell, Mary Comer, and Amy Reibman, who examined several Ph.D. dissertations containing materials used in this text. Two members of our group are particularly instrumental to our turbulence project: Xingguang Zhang and Zhiyuan Mao. We thank the continuous support of IARPA and the Michigan State University team in the BRIAR program, especially Xiaoming Liu, Arun Ross, Anil Jain, Atlas Wang, Humphrey Shi, and their students.

We also wish to thank Mark de Jongh with Now Publisher in reaching out to us and helping to make this book possible. In addition, a big part of the text is presented and recorded at the 2022 IEEE Conference on Computer Vision and Pattern Recognition (CVPR), and the 2022 IEEE International Conference on Image Processing. Readers interested in watching the video recordings can go to \url{https://www.youtube.com/watch?v=g_VY0KToV_s&t=2s}.

\vspace{8ex}
\begin{flushright}
Nicholas Chimitt and Stanley H. Chan\\
West Lafayette, IN, United States\\
June 2023
\end{flushright}

\tableofcontents

\mainmatter
\chapter{Introduction}
\vspace{-6ex}
\noindent\textcolor{myblue}{\rule{\textwidth}{4pt}}
\vspace{1ex}

\section{Scope of This Book}

\subsection{What is Atmospheric Turbulence?}
\begin{figure}[h]
\centering
\includegraphics[width=\linewidth]{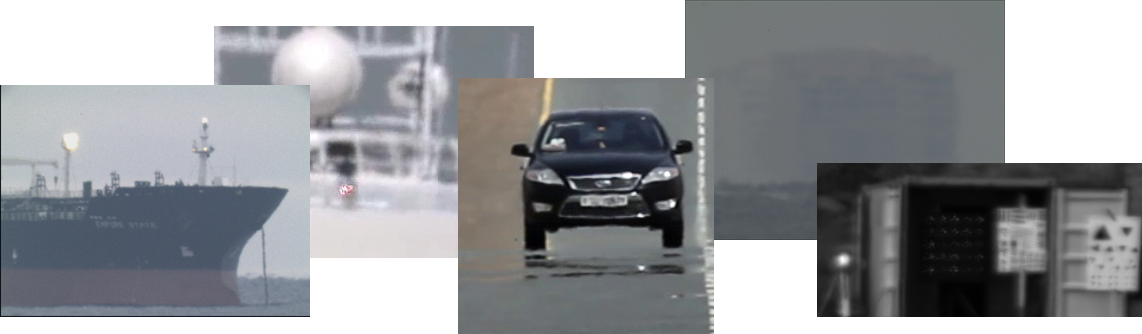}
\caption{Examples of imaging through atmospheric turbulence. Notice that the captured images look wavy and blurry.}
\label{fig: Ch0 example ground to ground turbulence}
\end{figure}

Our Earth's atmosphere is a beautiful creation. Not only does it provide necessities for our lives, but it also makes the way we \emph{see} things interesting. The following phenomena should be familiar to many of you: Suppose that you are standing in a desert. There is a car located at some distance from your camera. You take a picture of the car, and you see that the image of the car is wavy and blurry. To give an idea of what the images would look like, we invite you to take a look at the examples in \fref{fig: Ch0 example ground to ground turbulence}. A natural question, of course, is whether it is possible to recover the original image from these corrupted measurements. The subject of capturing and recovering images associated with atmospheric turbulence is broadly known as \keyword{imaging through atmospheric turbulence}.

But exactly what is atmospheric turbulence? When we talk about ``atmospheric turbulence'', we have the tendency to associate the word with roars of gusty winds and unstable air. The word ``turbulence'' carries a lot of assumptions that may vary from person to person. To begin to describe what turbulence is, we refer to the famous Vincent van Gogh painting, \emph{The Starry Night} (1889). The painting shows a few scientific phenomena that may correlate well with what you think. Firstly, there is a swirling air flowing in the middle of the sky. This kind of unstable air flow with a directional movement creates a subjective feeling of ``turbulent''. This painting does not convey a sense of chaos, but instead a calm, peaceful night. The same is true for turbulence; turbulence is not necessarily something that is aggressive and forceful, but instead may be calm and slow-moving. Furthermore, if we look at the sky, the stars do not appear as points but as big blobs. The dispersion of the light from a point to a blob is due to a \keyword{point spread function}. In turbulence, we call it the \keyword{blur} effect. The blur effect reduces the spatial resolution of the object.

\begin{figure}[h]
\centering
\includegraphics[width=0.75\linewidth]{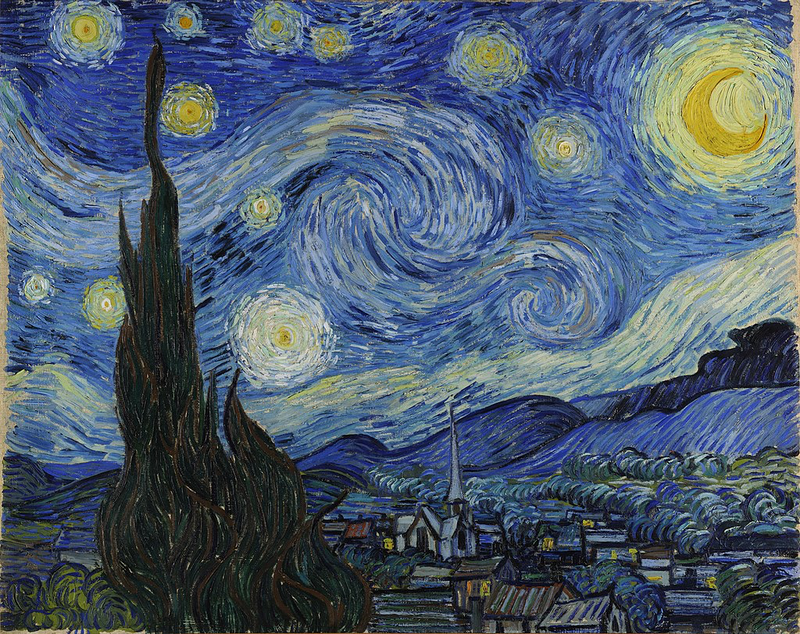}
\caption{A photograph of Vincent van Gogh’s The Starry Night (1889), which currently hangs in the Museum of Modern Art in New York.
}
\label{fig: Ch0 Van Gogh}
\end{figure}

While the above Van Gogh painting definitely captures some essence of turbulence that correlates with how we interpret the word, there is an interesting binary definition of the word ``turbulent''. In fluid dynamics, flow is either laminar or turbulent. There is nothing in between. Laminar flow occurs when all elements of fluid move in the same direction smoothly, everything else is turbulent. The steam pouring out of an engine exhibits turbulence, similar to cream as it spreads throughout a cup of coffee. So, if you agree with this definition, then turbulence is everywhere in our lives. In fact, right under your nose, the mixing molecules are experiencing turbulence!

\begin{figure}[h]
\centering
\includegraphics[width=0.7\linewidth]{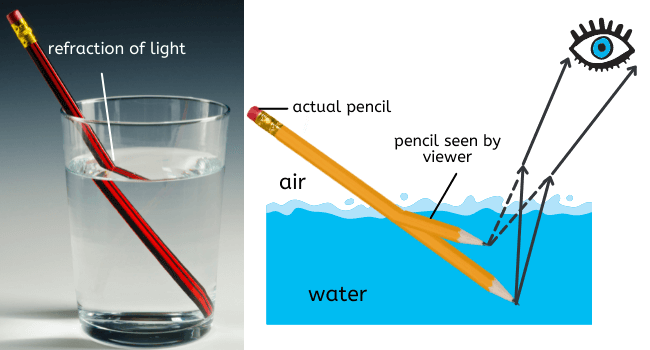}
\caption{Physics of refraction. \url{https://smartclass4kids.com/refraction-of-light/}}
\label{fig: Ch0 Water}
\end{figure}

So, when will we encounter atmospheric turbulence in images? If air is flowing under your nose, why don't you see the effects of turbulence when reading this book? The answer has to do with how light propagates across different layers of medium with a changing \keyword{index of refraction}\index{refraction! index of}. \fref{fig: Ch0 Water} shows a common experiment we all can do at home. Suppose we put a pencil in a cup of water, the pencil will appear as bent. The bending of the pencil is caused by the difference in the index of refraction of air and water. Since water molecules are more densely packed, it reduces the speed of light. In optics, we say that the water introduces a phase delay. If we replace water with another transparent liquid, we will change the index of refraction and so the bending will appear differently.

This same phenomenon happens in the air but in a more nuanced fashion. We have linked density and air previously, taking this a step further, we can imagine that as the air undergoes turbulent motion some areas will have collisions while others will have particles moving away from each other. The collisions will have more particles per unit area (thus higher density) than the other areas. This will cause microscopic fluctuations in the density of the medium, and thus the refractive index will also have these fluctuations. The temperature, humidity, wind velocity, altitude, etc. will all affect this behavior.

As light propagates through this medium, it will experience phase delays just as in the case of the pencil in water. For the pencil, there is a significant shifting of its observed location, but we can still easily recognize it as a pencil. In the atmosphere, the light will propagate through the medium and become dented up along the way to the observer, and the subsequent phase delay will have structured randomness. This can be imagined to some degree as looking down at the pencil in the glass of water while you're shaking the glass slightly. This will not produce such a sharp shift but instead smaller random shifts, and therefore it may be difficult to recognize what we are looking at until the water becomes still once again. This random phase-induced pixel shifting in the turbulence literature is known as \keyword{tilt}. For the atmosphere, this motion will be constant and inescapable, but there will be times when it is less aggressive than others.

The visibility of the turbulent effect depends on many factors. However, the two most important ones are distance and heat.
\begin{itemize}
\setlength\itemsep{0ex}
\item \textbf{Distance}. The atmospheric turbulence effect would not be visible if the distance between the object and the camera is close. The reason is that the propagation of light requires a path. If the index of refraction is changing slightly in a local segment, then we need to accumulate many of these segments in order to create the turbulence effects. If the object is placed very close to the camera, there is simply not enough distance to accumulate the turbulence effects.
\item \textbf{Heat}. The index of refraction is mainly determined by the density of air. If the temperature is high, air tends to be less densely located. But more critically is the instability of the temperature. Because the atmosphere exists in the presence of wind, external heat sources, and so on, the mixing of warm and cold air is nearly ever-present. This mixing occurs in a turbulent fashion and will contribute to the effects that we see in images.
\end{itemize}

It is easy to imagine that besides distance and heat, any environmental factor could contribute to the amount of turbulence:
\begin{itemize}
\item \textbf{Altitude}. The density of air is different at different altitudes. When it is near the ground, the air is denser than that in the stratosphere. So, if we take a ground-to-sky picture (e.g., for astrophotography), we will experience atmospheric turbulence. Most of the distortions would occur near the ground, and less so high up in the sky. This is mostly caused by the heat stored on the ground and sea.
\item \textbf{Humidity}. Humidity affects the temperature and amount of heat. A locally more humid region will have a different heating and cooling mechanism than a dry region. Therefore, the turbulence effect will also be different.
\item \textbf{Time of day}. Turbulence is stronger during the day and weaker during the night. This is again due to instability in temperature.
\item \textbf{Wind velocity}. The wind will act as a mechanism to influence the mixing of hot and cool air, which may help to speed up the turbulent mixing. The wind will most notably influence how fast a turbulent image changes in time.
\end{itemize}

\begin{figure}[h]
\centering
\includegraphics[width=0.8\linewidth]{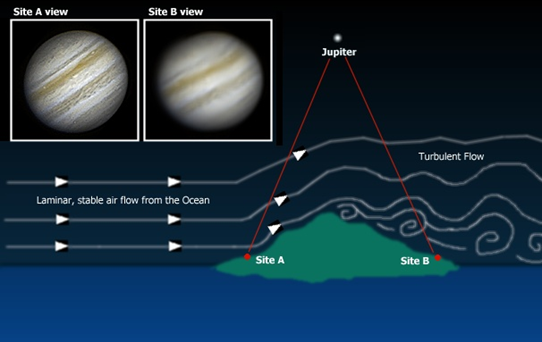}
\caption{Examples of imaging through atmospheric turbulence. Notice that the changing atmosphere makes the captured images look wavy and blurry. Image source: \url{https://www.findlight.net/blog/deformable-mirrors-in-astronomy/}}
\label{fig: Ch0 example ground to sky turbulence}
\end{figure}

Because of the locality of the weather conditions (temperature, humidity, etc.), the optical path from the object to the camera could vary greatly between the two locations. For example, imagine that you want to take a picture of Jupiter. If you stand at one site and the air above you is rapidly flowing, you will see a severely distorted image. However, if we go to another site where the air is stable, the image will have a little atmospheric effect. As we can see from the illustration in \fref{fig: Ch0 example ground to sky turbulence}, the amount of turbulence along the optical path between the camera and the object will determine the quality of the image.

One of the most difficult things in imaging through atmospheric turbulence is the \emph{randomness} of the phenomenon. A turbulence-distorted image is a \emph{realization} of a random process. Therefore, if you take a picture now and another picture one second later, the two pictures will appear different even if the object is stationary. Various mathematical models can tell us the structure of the randomness, e.g., the covariance matrix or the autocorrelation function, but we will never know the exact realization. Therefore, unlike the bending of a pencil in water which we can predict exactly what will happen, turbulence can vary significantly even if the turbulence parameters remain the same. Such randomness does not only make the parameter estimation task extremely difficult, but it also makes image recovery difficult. Moreover, data collection is difficult too because there is no way we can capture the turbulence field. The best we can capture would be the distorted images. The lack of ground truth, model, precise measurements, etc. makes the subject significantly more challenging than many mainstream image restoration tasks such as deblurring.

\subsection{Why Study This Problem?}
If imaging is so difficult, why are we still interested in the subject? For years, imaging through turbulence is a priority for space-related missions. Some defense-related industries are interested in the subject for ground-to-ground applications. More recently, there is an increasing interest in long-range biometric applications. In what follows, we highlight a few reported applications in the literature.

\begin{figure}[h]
\centering
\includegraphics[width=\linewidth]{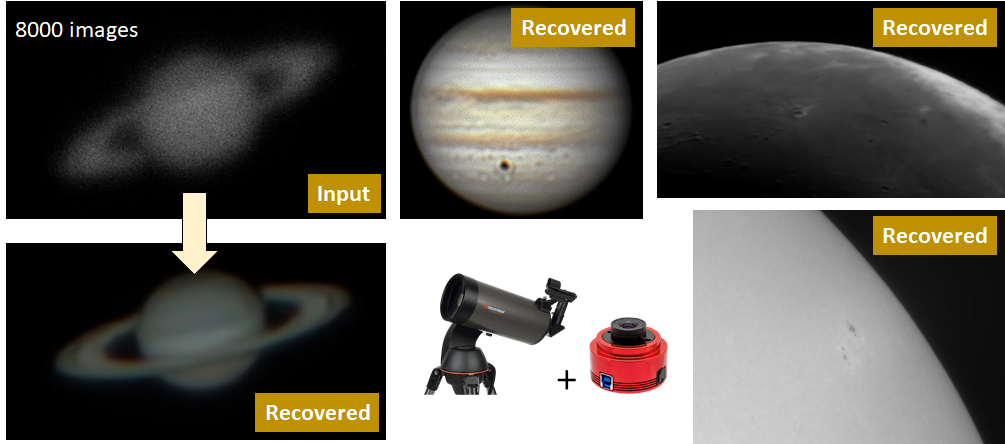}
\caption{Imaging through atmospheric turbulence has an important usage in astronomy. Shown in this figure are some of the real captures we did in July 2022, in West Lafayette, Indiana. By developing image reconstruction algorithms, we can recover various planetary objects.}
\label{fig: Ch0 recovery 2}
\vspace{-2ex}
\end{figure}

\textbf{Astronomy}. One of the earliest motivations for developing imaging through turbulence is astrophotography. As one can imagine, this has connections with space missions including observing planets, the moon, stars, or galaxies. At least for the past century, physicists and engineers have spent a tremendous amount of effort trying to understand the birth of the universe, the behavior of black holes \cite{akiyama_2019_a}, and seeking a habitable planet other than Earth. However, seeing through a ground telescope is fundamentally limited by the atmosphere. Even if we can identify and track the target using mechanical instruments, the images are still severely distorted by atmospheric turbulence. \fref{fig: Ch0 recovery 2} shows a raw Saturn image we captured in West Lafayette in July 2022. The telescope used was a fairly low-end one that we could purchase from the internet. The captured images are processed carefully using many advanced deep-learning algorithms. As we can observe, the image resolution has been significantly improved. From there, physicists can perform various analyses to answer their scientific questions.

\begin{figure}[h]
\centering
\includegraphics[width=\linewidth]{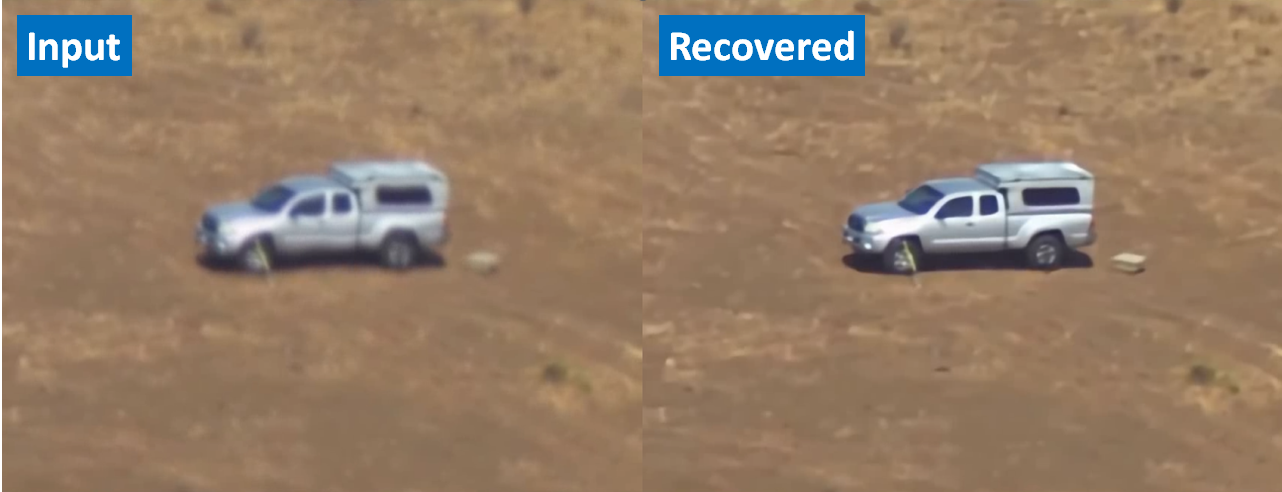}
\caption{A monitoring application of techniques developed for imaging through turbulence. Shown in this figure is a reconstruction result for an airborne image.}
\label{fig: Ch0 recovery}
\end{figure}

\textbf{Defense Missions}. The development of imaging through turbulence is never separable from defense missions. Unmanned aerial vehicles (UAVs) constantly carry out remote sensing tasks. However, due to the atmosphere, the images being captured would be severely distorted as shown in \fref{fig: Ch0 recovery}. If we want to perform any downstream computer vision tasks such as identifying an object, understanding the action, or tracking the object, we need methods to compensate for the turbulence effect.

Some readers at this point may ask: both the current example and the previous example assume a conventional camera. Why not use a more sophisticated adaptive optics system for the astronomical application, and perhaps use a camera with a different wavelength or even use radar / LiDAR for the monitoring application? The answer is typically associated with the constraint of the cameras we have. This could include the cost, size, weight, and power. Conventional cameras are easily accessible. They are also significantly less expensive than specialized cameras. As such, it is easier to deploy conventional cameras at a large volume. For some types of missions, the added functionality of specialized cameras may not justify the cost.

\begin{figure}[h]
\centering
\includegraphics[width=\linewidth]{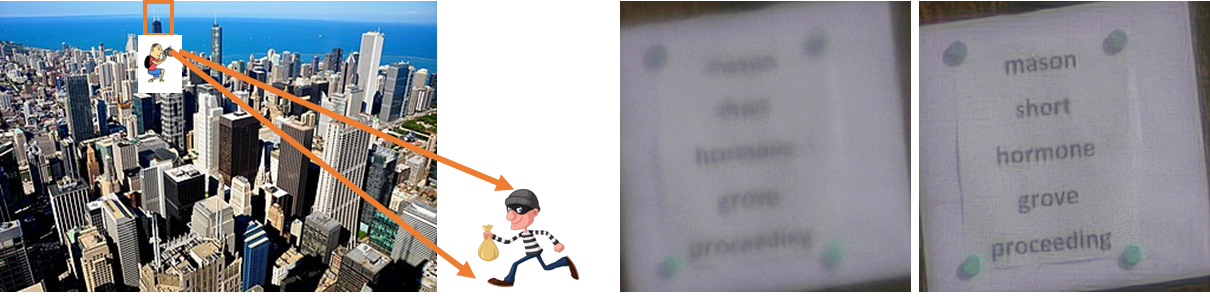}
\caption{A monitoring application of techniques developed for imaging through turbulence. Shown in this figure is a reconstruction result for an airborne image.}
\label{fig: Ch0 recovery 3}
\end{figure}

\textbf{Biometrics}. Another increasingly popular application of imaging through atmospheric turbulence is human and object recognition. The idea is to place long-range cameras in strategically chosen locations for monitoring purposes. With the distance being long and ground-to-ground, the turbulence effect becomes unavoidable. \fref{fig: Ch0 recovery 3} shows a cartoon illustration of the problem. In the same figure, we also show a realistic image recovery of a long-distance object. In this example, we use optical character recognition as the task. Without appropriately processing the image, it would be extremely difficult to recognize the pattern in the image.

In computer vision and image processing, imaging through atmospheric turbulence is often an upstream task. It means that we are interested in \emph{modeling} the physical phenomenon of the turbulence, and \emph{processing} of the images using algorithms. Some prior work develops \emph{hardware} to acquire images in a different way, such as the speckle imaging or adaptive optics. Although any of these can be the terminal task, they can also be fed into other downstream tasks such as detection and tracking. A frequently asked question is whether it is necessary to recover the images before running any of these recognition algorithms. We do not think there is a definitive answer at this moment, because some people argue that no matter what restoration we do, the features we use to perform the restoration are likely the same as the features we use to perform recognition. So why not just do the recognition directly? Others argue that the prior information and signal structure being explored by restoration are different from recognition. For example, in restoration, we typically use a large spatial-temporal window to align the features with the help of some specially chosen loss function (loss functions in the sense of deep learning). Integrating these techniques in recognition would basically require us to perform the restoration.

Another important aspect of the problem is the end goal. An astrophysicist probably has a different expectation from law enforcement. For astronomy, we care about a \emph{small} \keyword{field of view} because the stars are so far away. We want very high resolution and we can afford to buy big telescopes. For law enforcement, it is more likely that we will use standard cameras to see a scene with a \emph{large} field of view. The difference between a small field of view image and a large field of view image is the turbulence type. If the field of view is small, the turbulence will most likely be correlated, sometimes perfectly correlated. This is an easier problem in the sense of deblurring because the blur will be spatially invariant. However, the object will usually be small and so we will have a weak signal. If the field of view is large, the turbulence will a lot harder to handle because one point in the image may not correlate strongly with another point in the image. Therefore, algorithms developed for astrophysics may not apply to a ground-to-ground problem. The simulation parameters will be different, and the restoration techniques will also be different.

As far as this book is concerned, we are mostly interested in ground-to-ground imaging. This means that we are more focused on computer vision and image processing instead of physics and astronomy. Under this context, it makes sense for us to aim for speed, scale, and image appearance. If we want to train a neural network to reconstruct the image, we expect to have a real-time simulator that can be integrated into the restoration algorithm to predict the distortion. In physics applications, we will likely have less urgency for speed as we will be more interested in resolution and potentially new discoveries. Therefore, as you read this book, we hope that you can appreciate the signal processing aspect of the problem.

\section{Computational Imaging}
One may look at the history of imaging through atmospheric turbulence, which has been studied in its modern form for 80 years since Kolmogorov, and wonder: What new angles can we take? With almost sure probability, every aspect of the problem would have some prior work, from hardware to algorithms. In parallel, signal processing and computer vision have experienced a paradigm shift in recent years with the advent of deep learning methodologies. The broader community has dedicated a considerable amount of time to these methods which are constantly advancing and replacing older methodologies. Of particular interest to us for this book is deep learning for \keyword{image reconstruction}.

The angle we take in this book is what we called \keyword{computational imaging} through turbulence, contrasting the classical imaging through turbulence. The word ``computational'' highlights the unique conceptual difference that we are \emph{not} solving the inverse problem in a post-processing manner. Instead, we emphasize the significance of having a \keyword{co-design} of the computational forward model and the computational algorithm. This co-design philosophy is particularly important when we start to build neural networks to recover images. We want the network to acknowledge the physics of turbulence. Conversely, we want the forward image formation model to be compatible with the neural network, e.g., being differentiable, scalable, and fast.

\subsection{Computational Camera}
The spirit of computational imaging is to co-design the camera's optics and the image processing algorithm. Perhaps the most common view of computational imaging is to co-design the acquisition unit (typically a camera) and the reconstruction algorithm. For the co-design to make sense, the camera should have a configurable component. For example, in coded aperture  \cite{Levin_2007_a}, the configurable component is the spatial coding of the mask. In lensless imaging \cite{Antipa_2018_a}, the configurable component is the diffuser. The configuration component can also be a light source as in the case of structured illumination where the configurable component is the illumination pattern. Magnetic resonance and computed tomography imaging also share a similar spirit to these methods as the acquisition occurs in a different domain. Other examples of these co-design concepts include light-field imaging \cite{Veeraraghavan_2007_a}, holography, plenoptic camera, and ghost imaging or on the image sensor level, cameras such as the event camera, 1-bit quanta image sensors, time-of-flight detectors, and focal plane arrays. We refer to this co-design of hardware and algorithms as \keyword{computational camera} models. These models are at the intersection of computer vision, optics, and signal processing.

In a conventional camera, we can think of the camera as being a passive device. If we use $\vx \in \R^d$ to denote the ground truth image in the object plane, and $\vy \in \R^d$ to denote the observed image in the image plane, the camera can be mathematically described as the mapping $\calG$ from $\vx$ to $\vy$:
\begin{equation}
\underset{\text{observed image}}{\underbrace{\vy}} = \underset{\text{camera}}{\underbrace{\calG}} \;\;\; (\underset{\text{true image}}{\underbrace{\vx}}).
\end{equation}
In a conventional (non-computational) camera, the camera model $\calG$ is often fixed and known.

A post-processing image reconstruction algorithm for the conventional camera is to find the best estimate $\widehat{\vx}$ by minimizing a certain loss function:
\begin{equation}
\widehat{\vx} = \argmin{\vx} \;\; \text{Loss}(\calG(\vx), \vy).
\label{eq: Ch0 inverse problem 1}
\end{equation}
For example, if we prefer the sum square loss, then we have
\begin{equation}
\text{Loss}(\calG(\vx), \vy) = \|\calG(\vx)-\vy\|^2.
\end{equation}
Other loss functions may be used or a \keyword{regularization function}\index{regularization function} for $\vx$ to improve the optimization search space.

The key observation of the above equations is that in a conventional capture-then-reconstruct camera setting, the camera is \emph{pre-defined}. While we use the model $\calG$ during reconstruction, the design of $\calG$ is completely separated from the reconstruction. We almost always want $\calG$ to produce the ideal image and for good reasons.

In a computational camera, the above isolation of the camera and the algorithm is replaced by a co-design philosophy. The first thing we do is to \emph{parameterize} the camera with a finite set of tunable knobs $\vtheta$. This $\vtheta$ can contain any attributes of the camera. For example, we can input the exposure time, exposure pattern \cite{Raskar_2006_a}, coded aperture pattern, etc. By changing the parameter $\vtheta$, we effectively change how the image is acquired.

The presence of the tunable parameter $\vtheta$ makes the overall design interesting. Instead of solving a one-variable optimization in \eref{eq: Ch0 inverse problem 1}, we are conceptually solving the joint optimization
\begin{equation}
\widehat{\vx},\;\widehat{\vtheta} = \argmin{\vx, \, \vtheta} \;\; \text{Loss}(\calG_{\vtheta}(\vx), \vy).
\label{eq: Ch0 inverse problem 2}
\end{equation}
This optimization says that while we are looking for the best estimate $\vx$, we are also asking the camera $\calG_{\vtheta}$ to be simultaneously configured. If we change $\vx$, $\calG_{\vtheta}(\vx)$ will change and so the loss will change. Similarly, if we change $\vtheta$, $\calG_{\vtheta}$ will change and so the loss will also change. The conventional camera philosophy tells us to fix a $\vtheta$ by standard metrics (aberrations, field of view, etc.) and estimate the best $\hat{\vx}$. The computational camera says to design $\calG_{\vtheta}$ in such a way that aids in our estimation of $\hat{\vx}$.

So are we done yet? No. There are a few issues regarding the optimization defined in \eref{eq: Ch0 inverse problem 2}. Firstly, it does not capture the ``design'' of a reconstruction algorithm. The reconstruction algorithm is not about finding one particular $\vx$, but a mapping that takes any measurement $\vy$ and return us an estimate. Secondly, when we talk about camera design, we are not interested in one particular ground truth image $\vx$. We are generally interested in all images. Finally, $\calG_{\vtheta}$ is typically realized through hardware. This will complicate our numerical optimization, $\calG_{\vtheta}$ doesn't necessarily have a gradient, thus \emph{true} joint optimization will be challenging.

To respond to the first two requests, we realize that in any image reconstruction literature, the reconstruction algorithm is always a mapping that takes $\vy$ and gives us an estimate $\widehat{\vx}$. Sometimes, perhaps most of the time, the reconstruction algorithm will require side information such as knowledge about $\calG_{\vtheta}$. In this way, we can write
\begin{equation*}
\widehat{\vx} = \underset{\text{reconstruction}}{\underbrace{\calR(\vy, \calG_{\vtheta})}} ,
\end{equation*}
where $\calR$ is a function that takes the measurement $\vy$ and return us the estimate $\widehat{\vx}$.

The reconstruction algorithm can take a variety of forms. For example, in the pre-deep-learning era, the reconstruction is often a regularized minimization, e.g.,
\begin{equation*}
\calR_{\lambda}(\vy, \calG_{\vtheta}) = \argmin{\vx} \;\; \|\vy -\calG_{\vtheta}(\vx)\|^2 + \lambda \; g(\vx),
\end{equation*}
where $g(\vx)$ is a regularization function and $\lambda$ is the corresponding regularization parameter. If we choose $g(\vx) = \|\vx\|_{\text{TV}}$, then we have a total-variation minimization. We put a subscript $\lambda$ underneath $\calR$ to emphasize that the reconstruction mapping is dependent on the regularization parameter $\lambda$.

In the deep-learning era, we often use a deep neural network as the reconstruction. In this case, we parameterize $\calR$ as $\calR_{\vpsi}$ so that
\begin{equation}
\widehat{\vx} = \underset{\text{neural network}}{\underbrace{\calR_{\vpsi}( \vy, \; \calG_{\vtheta} )}}.
\end{equation}
The way to think about $\vpsi$ is that it represents the weights of the neural network. When the reconstruction neural network is trained, we will send the measurement $\vy$ together with the camera model $\calG_{\vtheta}$ to the network. The network will then return us $\widehat{\vx}$.

For learning-based approaches, the reconstruction can be formulated as
\begin{align}
\widehat{\vpsi}
&= \argmin{\vpsi} \;\; \E_{\vx} \Big[ \text{Loss}(\widehat{\vx}, \vx) \Big] \notag \\
&= \argmin{\vpsi} \;\; \E_{\vx} \Big[ \text{Loss}\Big(\calR_{\vpsi}( \vy, \; \calG_{\vtheta} ), \vx\Big) \Big].
\end{align}
The symbol $\E[\cdot]$ in the equation denotes the statistical expectation. We introduce the expectation in the equation because $\text{Loss}(\widehat{\vx}, \vx)$ is only for one particular image $\vx$. If we want to co-design the system for \emph{all} images, then we need to take the average over all possible images we want to study. For example, if we are building a computational camera for human faces, then the expectation is taken over all the face images. In deep learning, we can consider $\vx$ as a training sample. By taking the statistical expectation over $\vx$, we ask the minimization to be taken with respect to the entire training set.

Regarding the issue of the gradient of $\calG_{\vtheta}$, if the camera $\calG_{\vtheta}$ is simplistic enough that a gradient calculation is possible, this can be formulated as a joint optimization problem:
\begin{align}
(\widehat{\vpsi}, \widehat{\vtheta})
&= \argmin{\vpsi, \vtheta} \;\; \E_{\vx} \Big[ \text{Loss}(\widehat{\vx}, \vx) \Big] \notag \\
&= \argmin{\vpsi, \vtheta} \;\; \E_{\vx} \Big[ \text{Loss}\Big(\calR_{\vpsi}( \vy, \; \calG_{\vtheta} ), \vx\Big) \Big]. \label{eq: Ch0 inverse problem 3}
\end{align}
This typically requires some potentially limiting assumptions. One potential assumption is that \keyword{geometric optics}\index{geometric optics} is sufficient in describing the behavior of the system. We present the following example of a simple $\calG_{\vtheta}$ which may allow for such gradient computation.

\boxedeg{
\vspace{2ex}
\textbf{Example}. Consider a compressed sensing camera where we are interested in capturing a low-dimensional (compressed) signal instead of the full signal $\vx$. The compression scheme we use is a wide matrix $\mG \in \R^{m \times d}$ where $m \ll d$ so that the measured signal
\begin{equation*}
\vy  = \calG_{\vtheta}(\vx)\bydef  \underset{\text{encoded signal}}{\underbrace{\mG \vx}} \quad + \underset{\text{noise term}}{\underbrace{\vn}}
\end{equation*}
has a much lower dimension than the true signal $\vx$. Notice that we introduce a noise term to make the measurement more realistic. To reconstruct the signal we create a reconstruction algorithm (could be a neural network) that does
\begin{equation*}
\vxhat = \calR_{\vpsi}( \vy ).
\end{equation*}
Here, we can think of $\calR_{\vpsi}$ as a neural network parameterized by the weight $\vpsi$.

To train the neural network and simultaneously find the optimal sensing matrix $\mG$, we perform the joint optimization
\begin{equation*}
\widehat{\mG}, \widehat{\vpsi} = \argmin{\mG, \vpsi} \;\; \E_{\vx, \vn} \Big[ \text{Loss}( \calR_{\vpsi}( \mG \vx + \vn), \vx ) \Big].
\end{equation*}
The expectation is taken with respect to both the signal $\vx$ and noise $\vn$ because they are random.

If we consider the hardware feasibility, we can further pose constraints on $\mG$. For example, we can require $\mG$ to be binary so that it can be implemented through digital micro-mirror devices (DMD).
}

\begin{figure}[h]
\centering
\includegraphics[height=3.5cm]{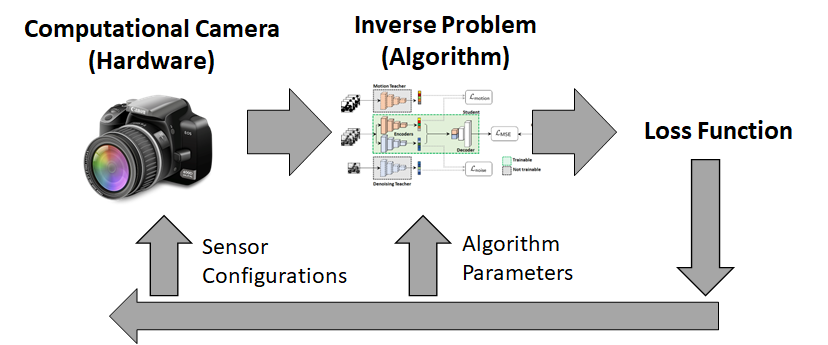}
\caption{Computational camera aims at co-designing the acquisition unit (often a camera) and the image reconstruction algorithm.}
\label{fig: Ch0 CompImagA}
\end{figure}

\fref{fig: Ch0 CompImagA} illustrates the conceptual diagram of a typical computational camera setup. There is a camera $\calG_{\vtheta}$ with a few configurable knobs $\vtheta$, e.g., exposure pattern, gain pattern, aperture mask, etc. Using the compressed sensing example above, $\vtheta$ would mean the sensing matrix $\mG$. As the ground truth image $\vx$ is passed to the camera, we obtain a measurement $\vy$. The measurement is passed to a reconstruction method $\calR_{\vpsi}$ to obtain an estimate $\vxhat$. The estimate is compared with the ground truth $\vx$ to generate a loss. The loss $\text{Loss}(\vxhat,\vx)$ is fed back to a training algorithm to update the configurable knobs $\vtheta$ and the reconstruction algorithm $\vpsi$. The process continues until the expected loss is minimized.

The iterative procedure we described above is for \emph{training}. During inference (i.e., deployed to the field), we fix $\vtheta$ to take measurements. For every captured signal $\vy$, we send it to the reconstruction algorithm to produce an estimate $\vxhat$. Since the whole computational camera is trained to minimize the \emph{average} loss over a \emph{large} collection of typical scenarios, it will work reasonably well on similar scenarios. Some reconstruction algorithms do not require training, e.g., total variation minimization. Some camera configurations would prefer on-the-fly characteristics (i.e., they can change from one configuration to another as the scene changes). So, there is a wide range of variety here.

\subsection{Computational Image Formation}
A particular challenge for the computational camera line of work is that the gradient of $\calG_{\vtheta}$ may be difficult or impossible to derive for complex systems. This leads us to draw a distinction that is separate from the computational camera. One essential requirement of the camera and algorithm co-optimization is that the reconstruction algorithm $\calR_{\vpsi}$ can ``understand'' what the camera $\calG_{\vtheta}$ is doing, and vice versa. In particular, the optimization we need to solve is a \emph{joint} optimization
\begin{equation*}
\widehat{\vpsi}, \widehat{\vtheta} = \argmin{\vpsi, \vtheta} \;\; \E_{\vx} \Big[ \text{Loss}\Big(\calR_{\vpsi}( \vy, \; \calG_{\vtheta} ), \vx\Big) \Big].
\end{equation*}

To explain the issue of ``mutual understanding'', we need a neural network concept known as \keyword{back propagation}\index{back propagation}. To train a neural network $\calR_{\vpsi}$, we need to back propagate the gradient of the loss $\nabla_{\vpsi} \text{Loss}(\widehat{\vx},\vx)$ so that we can make an update of the network weights through a stochastic gradient descent algorithm. Similarly, to update the camera model $\calG_{\vtheta}$, we also need some kind of optimization algorithm to inform us how to update the camera parameter $\vtheta$.

For generality, let's suppose that we use a gradient descent algorithm to update both $\vtheta$ and $\vpsi$, denoting the overall parameter as $\vvarphi = [\vpsi, \vtheta]$. 
Back propagation creates some difficulty. By chain rule, finding $\nabla_{\vvarphi} \text{Loss}$ requires $\nabla_{\vvarphi} \calR_{\vpsi}(\vy,\calG_{\vtheta})$. This will in turn require us to take the gradient of the camera $\nabla_{\vtheta}\calG_{\vtheta}$. However, $\nabla_{\vtheta}\calG_{\vtheta}$ is impossible to obtain unless $\calG_{\vtheta}$ takes a certain mathematical form that allows differentiation. A purely physical device would not allow it to happen, because there is nothing called the ``gradient of Canon 5D $f$/2.4 camera, with respect to the exposure time''. Because of this physical-to-algorithmic gap, we almost always use mathematical equations $\calG_{\vtheta}$ to represent the actual hardware.

\begin{figure}[h]
\centering
\includegraphics[height=3.5cm]{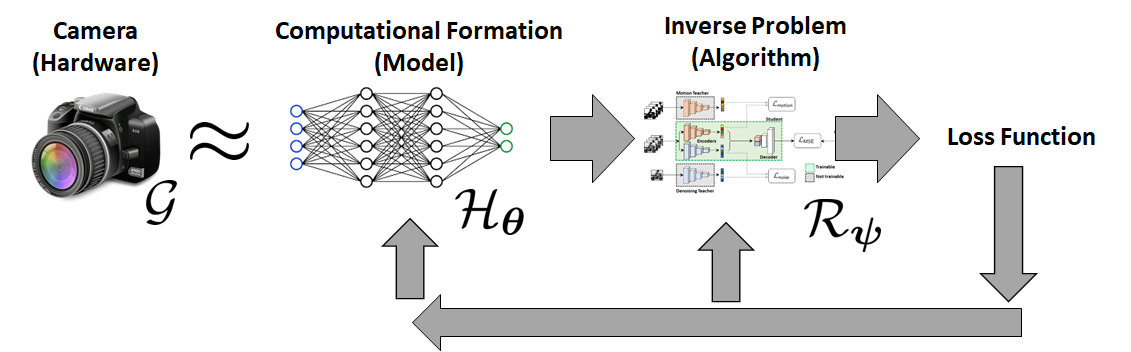}
\caption{Computational camera aims to co-design the acquisition unit (often a camera) and the image reconstruction algorithm.}
\label{fig: Ch0 CompImagC}
\end{figure}

When we use a mathematical model to substitute the true camera $\calG$, we are no longer working with a computational \emph{hardware}. To make things clear, we need to separate the true camera and a model to approximate the camera. Let's denote the true camera as $\calG$. The true camera is something we can operate, but we cannot compute the gradient. One workaround solution is to use a neural network $\calH_{\vtheta}$ to approximate $\calG$. Neural rendering, for example, is exactly following this line of thought. Instead of trying to directly handle the camera (which does not really have a ``gradient''), we train a neural network $\calH_{\vtheta}$ such that $\calH_{\vtheta} \approx \calG$. The training of such a neural network (if we can ignore the reconstruction part) can be done using
\begin{equation}
\widehat{\vtheta} = \argmin{\vtheta} \E_{\vx} \Big[ \text{Loss}(\calH_{\vtheta}(\vx), \calG(\vx))\Big].
\end{equation}
Basically, we are looking for a set of network weights such that the overall network $\calH_{\vtheta}$ can mimic the true camera. Substituting this idea into the computational camera pipeline, we can draw \fref{fig: Ch0 CompImagC}.

The co-optimization of the image formation model (as opposed to the camera) and the reconstruction algorithm can be formulated via a joint optimization
\begin{align}
\widehat{\vtheta}, \widehat{\vpsi} =
\argmin{\vtheta, \vpsi} \;\;\; &
\underset{\text{learn the camera model}}{\underbrace{\E_{\vx} \Big[ \text{Loss}(\calH_{\vtheta}(\vx), \calG(\vx))\Big]}} \notag \\
&\qquad +
\underset{\text{reconstruct the signal}}{\underbrace{\E_{\vx} \Big[\text{Loss}\Big(\calR_{\vpsi}( \vy, \; \calH_{\vtheta} ), \vx\Big) \Big]}}.
\end{align}
This is a harder problem, but the two terms are intuitive. The first term is the \emph{computational image formation} problem where we try to find a model that describes the camera. The second term is the reconstruction problem where we construct an algorithm to recover the image.

Using a neural network as $\calH_{\vtheta}$ is one of many options. Some problems, such as developing new optical designs through nano-photonics \keyword{meta-materials} \cite{Khorasaninejad_2016_a}, would fall under the same umbrella. A piece of meta-material can be considered as an engineered glass with some desired phase properties. To design the phase profile, we have to use numerical wave approximations to mimic how light passes through the meta-material. If the true meta material is denoted as $\calG$, then any numerical wave approximation is $\calH_{\vtheta}$. As we try to find the best approximation $\calH_{\vtheta}$, we also need to bare in mind that the reconstruction $\calR_{\vpsi}$ can support. Therefore, even though the joint optimization looks abstract and complicated, it does capture the essence of the co-design of optics and reconstruction.

As we use a numerical model to approximate the hardware, we are no longer working with a computational camera as we are not configuring the camera directly. What we are instead doing is working with a computational model which is summarized as an equation, algorithm, or neural network. This progression from a camera to a computational camera model creates a sub-field in computational imaging, which we call \keyword{computational image formation}\index{computational image formation}.

\subsection{Connecting with Imaging Through Turbulence}
At this point, we can go back to atmospheric turbulence. Readers who are familiar with a computational camera might think that for atmospheric turbulence, we can probably build some kind of computational camera to offset the reconstruction difficulty. In fact, the subject of \keyword{adaptive optics} is developed for that purpose. However, adaptive optics has both merits (such as a much better phase compensation) and limitations (such as big, expensive, limited resolution, and requiring a reference star). We shall not discuss it in detail here.

In imaging through turbulence, the degradation process is governed by \emph{nature}. Only God knows exactly how each point on the object plane is mapped to a digital value on the image plane. Using the terminology we defined previously, we must acknowledge that there is an \emph{unknown} target function $\calG$ that takes the input and gives us the output\footnote{The terminology ``target function'' we use here comes from the literature of machine learning. A target function in machine learning refers to the ground truth mapping that we aim to learn from data. A target function is never known in machine learning because it is the subject of interest. In learning, our goal is to find a meaningful approximation to the target function by minimizing some kind of training loss.}:
\begin{equation}
\vy = \calG(\vx).
\end{equation}
We stress that the function $\calG$ is unknown because it represents nature.

The complexity of turbulence is not only that we do not know the equation of the distortion. Since each turbulent instant is a realization of a random event, there is literally no way we will know exactly how light is being distorted. This is a very different (and harder) scenario compared to classical problems such as non-blind image deconvolution where we \emph{know} the blur kernel and we also \emph{know} the spatial invariance of the convolution. Even in the case of blind deconvolution, we know the structure of the problem.

But how about the models from the work of Kolmogorov and others created many decades ago? Aren't they experimentally verified to be accurate? Yes, but what they have proposed are models, not the target function. The models are what we called the \keyword{hypothesis functions}\index{hypothesis function}. Some hypothesis functions are more accurate whereas some are easier for computation. Mathematically, if we denote these hypotheses as $\calH_{\vtheta}$, then we have
\begin{equation}
\widehat{\vy} = \calH_{\vtheta}(\vx),
\end{equation}
where $\widehat{\vy} \in \R^d$ denotes the prediction made by the hypothesis function $\calH_{\vtheta}$. Simply put, we can call $\widehat{\vy}$ as the numerically \emph{simulated} image, and $\calH_{\vtheta}$ as a \keyword{simulator}\index{simulator}. The hypothesis function $\calH_{\vtheta}$ is parameterized by the parameter $\vtheta \in \R^p$. For example, the vector $\vtheta$ can contain the turbulence strength, temperature, etc.

The essence of $\calH_{\vtheta}$ is the \keyword{parameterization} of nature. We are not saying that $\calH_{\vtheta}$ \emph{is} nature, but $\calH_{\vtheta}$ is a good \emph{approximation} of nature. Being an approximation means that there is approximation error. It also means that we need criteria to choose a better approximation from a worse approximation. The criteria are often multifaceted. Some may have a stronger emphasis on accuracy over a narrow field of view, while some may want an extremely fast model. The trade-off is often determined \emph{before} we even build the simulator. This means that the choice of $\calH_{\vtheta}$ is isolated from the choice of any downstream image processing tasks including image restoration.

\begin{figure}[h]
\centering
\includegraphics[width=\linewidth]{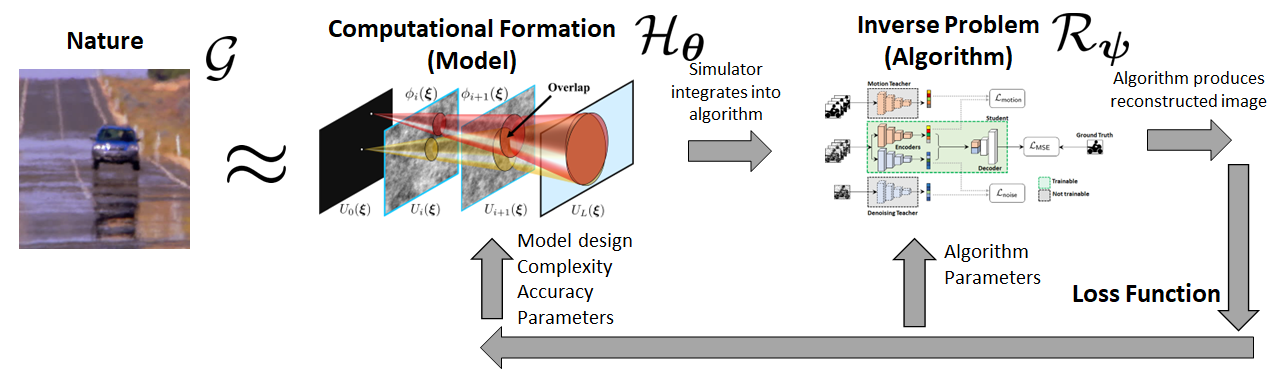}
\caption{Computational imaging through turbulence focuses on co-designing the approximation to nature (which is a computational building block) with the image reconstruction algorithm. }
\label{fig: Ch0 computational imaging concept 2}
\end{figure}

At this point, we should be able to draw the connection between computational imaging and imaging through turbulence. Instead of having hardware (e.g. a camera) $\calH_{\vtheta}$ to perform the image formation, we use a computational model $\calH_{\vtheta}$ to perform the light propagation (see \fref{fig: Ch0 computational imaging concept 2}). Although $\calH_{\vtheta}$ is not the same as nature, it creates an artificial process that mimics it. This artificial process adds a tremendous amount of degrees of freedom, giving us an optimization problem.
Just like the camera configurations in computational cameras which should be co-optimized with the downstream reconstruction, in imaging through turbulence, the co-optimization is about the model $\calH_{\vtheta}$ and the reconstruction algorithm $\calR_{\vpsi}$. Therefore, this concept falls squarely within the framework of computational image formation along with differentiable light transport and differentiable optics.

\section{Outline of This Book}
After describing computational imaging, we can now elaborate on the approaches we take in this book. There are several components:

\textbf{Chapter 2. Image Formation Model}. In this Chapter, we present the foundations of wave propagation physics. To this end, we need to discuss the concept of Fourier Optics. The way we view Fourier Optics is that it is a \emph{signals and systems} perspective for wave optics. It provides us with the necessary mathematical tools such as the Huygens-Fresnel principle, Fresnel diffraction, Fraunhofer diffraction, spatial coherence, point spread function, optical transfer function, and the image formation equation.

To readers who have prior knowledge about this subject, we have a tendency to treat Fourier Optics as the ground truth $\calG$. We argue that this is not the case. Under all the sufficient conditions laid out by Fourier Optics, we can build a highly reliable model $\calH_{\vtheta}$ that can approximate $\calG$ up to the assumptions we make. For example, if we assume that the object is sufficiently far away from us, then Fraunhofer diffraction theory would be enough to describe how wave diffracts and propagates. However, as we step immediately outside the assumption, our model $\calH_{\vtheta}$ will fail. Therefore, one important emphasis as we discuss the concepts is the conditions under which our derivations are correct.

\textbf{Chapter 3. Modeling Turbulence}. The objective of this Chapter is to provide an overview of the other half of the foundation. Since we are interested in atmospheric turbulence, we need to understand its origin and its impact to the image formation process. The approach we take here is to first describe the index of refraction and its statistical properties under the Kolmogorov model. From there, we can discuss how light propagates through a stack of phase screens determined by the index of refraction statistics. The wave propagation model can be implemented numerically through a technique known as split-step propagation.

The model we conclude this Chapter with, namely the split-step propagation, is another model $\calH_{\vtheta}$. Compared to the model we derive in Chapter 2, the new model is customized for atmospheric turbulence. However, the design of $\calH_{\vtheta}$ in this Chapter is solely aiming to match $\calG$. We care little about the speed, and we almost never care about what image reconstruction algorithm we are going to use to recover the image. In fact, historically speaking, turbulence simulators are purely evaluated based on the statistics of the simulated data compared to the theoretical statistics. The expectation is that if the simulator is faithful, we can use it to \emph{evaluate} any reconstruction algorithm but not \emph{co-optimize} with the reconstruction algorithm.

We should also clarify a misconception about the known theoretical turbulence model. Under various assumptions, the theoretical model people developed over the past century is at best a \emph{model}. Therefore, it is not the true $\calG$ but only a very good $\calH_{\text{\tiny theoretical}}$. The numerical simulator we use in practice, including the split-step propagation, is another level of approximation to the theoretical model. Therefore, the relationship is best described as
\begin{equation}
\underset{\text{nature}}{\underbrace{\calG}}
\overset{\text{\tiny unknown}}{\approx}
\underset{\text{theoretical model}}{\underbrace{\calH_{\text{\tiny theoretical}}}}
\overset{\text{\tiny varies}}{\approx}
\underset{\text{numerical model}}{\underbrace{\calH_{\text{\tiny split step}}}}
\end{equation}

When we say ``our statistics match with the theoretical prediction'', as many papers in the literature declare, we mean that $\calH_{\text{\tiny theoretical}} \approx \calH_{\text{\tiny split step}}$. The implicit assumption/hope here is that the approximation error in $\calH_{\text{\tiny theoretical}} \approx \calG$ is sufficiently small. In many situations, we know that this assumption is valid. But in some cases, for example, when we start to consider amplitude attenuation on top of the typical phase distortion, we need to resort to a more advanced theory to reduce the gap between  $\calH_{\text{\tiny theoretical}}$ and $\calG$.

\textbf{Chapter 4. Propagation-Free Modeling and Simulation}. This Chapter is one of the two core Chapters of our book. Taking into consideration of the reconstruction algorithm, existing turbulence simulators cannot serve the reconstruction purpose for many reasons:
\begin{itemize}
\setlength\itemsep{0ex}
\item \textbf{Slow}. Methods such as split-step is remarkably computationally expensive. We need to loop through all the pixels of the image one by one and propagate them from one phase screen to another. The speed is too low to be even meaningful for the co-design with reconstruction algorithms.
\item \textbf{Non-differentiable}. For an image formation model to be integrable as part of the reconstruction, a fundamental criterion is being \emph{differentiable} in the sense of back propagation. Models such as the split-step propagation do not satisfy this criterion because the simulation steps are sequentially executed. Therefore, the gradient $\nabla_{\vtheta}\calH_{\vtheta}$ is very hard to compute.
\item \textbf{Inaccurate}. Some methods such as geometric deformation + Gaussian blur are fast and differentiable, but they barely mimic nature. As a result, they provide a poor approximation of the equation $\calH_{\vtheta} \approx \calG$.
\end{itemize}

With these deficiencies in mind, we present a new line of work in the recent literature that aims to bring the model fast, differentiable, and accurate. The idea is based on a propagation-free method where the distortion is implemented through a random \emph{sampling} process rather than propagation. For the sampling to make sense, we need to derive a collection of statistical results. Some of the results are based on known literature, whereas some are new. More interestingly, some implementations require (shallow) neural networks to bridge the numerical gap.

The conclusion of this Chapter is a series of approximations that maintains the majority of the accuracy of the split-step method while offering a significant speed up and differentiability. This will allow us to integrate the image formation model with the reconstruction.

\textbf{Chapter 5. Image Restoration}. In this final Chapter, we discuss how the computational image formation model $\calH_{\vtheta}$ can be used together with the image restoration algorithm $\calR_{\vpsi}$. We will review the classical results in lucky imaging, frame alignment, and image deconvolution. These results will be used to guide the design of many of the latest deep neural network designs. In particular, we show how $\calH_{\vtheta}$ can be used to synthesize training data. This data synthesis process is not only done prior to the training process but it can be integrated \emph{during} the training by providing multi-scale distortions. Furthermore, building blocks of $\calH_{\vtheta}$ such as geometric warping and spatially varying blur can be decoupled during the training process. Since we have a powerful $\calH_{\vtheta}$, we can synthesize partially degraded images to guide the reconstruction at different stages. The simulator $\calH_{\vtheta}$ can also be used during inference as a mechanism to estimate the turbulence strength. Some recent methods send feedback signals from the reconstruction algorithm $\calR_{\vpsi}$ to the simulator $\calH_{\vtheta}$ by asking $\calH_{\vtheta}$ to generate a ``re-distorted'' image so that we know how well our turbulence estimation is. As you can see here, a plethora of approaches can be developed when $\calH_{\vtheta}$ is jointly optimized with $\calR_{\vpsi}$.

\chapter{Image Formation Model}
\vspace{-6ex}
\noindent\textcolor{myblue}{\rule{\textwidth}{4pt}}
\vspace{1ex}

\label{sec: sec1}
The subject of this book is imaging through atmospheric turbulence. To describe such a phenomenon, we must understand both the process by which an image is formed and the way the atmosphere interferes with this process. This Chapter details image formation without consideration of any degradation by the atmosphere. Our approach here is to describe the signals and systems concepts that are familiar to readers coming from an image processing background. We will begin with a high-level discussion of optics and image formation.

\section{Ray Optics or Wave Optics?}
\label{sec: sec1_1}

\subsection{From Ray Optics to Wave Optics}
When modeling a physical phenomenon, one must carefully select a mathematical model which serves the problem one seeks to analyze. For the purpose of modeling light, there are two commonly utilized models in the image processing and computer vision communities. They are the \keyword{ray tracing model}\index{ray tracing} and the \keyword{wave model}. Ray tracing is useful in various forms of computer graphics, especially for lens design and rendering. In ray tracing, we ``trace'' light rays through space or materials, with the end result being the distribution of rays in a scene. The wave model arises from \keyword{Maxwell's equations} \cite{Born_1999_a, Jackson_1999_a}. The wave model is incredibly accurate, but it is also analytically more difficult compared to ray tracing.

Readers with a computer graphics background may say: While it is hard to deny that the wave model is more carefully aligned with nature, it is arguably overkill for computer graphics applications. Modern GPUs are capable of performing ray tracing in real time, however, the same can not be said about evaluating the wave equations. Therefore, given its appropriateness, the ray model is far preferable, allowing us to form strikingly realistic images for various graphics applications \cite{Ramamoorthi_2009_a, Christensen_2016_a}. We completely agree with this argument. In fact, we believe that ray optics is simple, elegant, and appropriate for applications exactly in the space such as rendering. However, for imaging through atmospheric turbulence, the random fluctuation of the medium is not as easy as one would imagine. If we choose ray tracing, we may miss critical subtleties in the nature of light. \index{ray tracing! rendering}

Readers with the optics background may then say: If ray tracing has deficiencies in modeling nature, we should then pursue the wave model. As long as we are not working with quantum matters, the wave model is sufficient to describe essentially all phenomena we would possibly need for engineering purposes. Specifically, we can describe the diffraction effects, phase delays, amplitude attenuation, etc, to a high degree of accuracy. We completely agree on this point, too. In fact, for almost one century since Kolmogorov, we have seen the remarkable power of wave optics theory in describing imaging through atmospheric turbulence. However, the price we need to pay is the level of mathematical rigor and computation. Wave models can be extremely complicated and sometimes intractable.

So, what should we do then? In this book, we will take an \emph{approximate} form of the wave model. Our model is \keyword{Fourier Optics}\index{Fourier optics} combined with the \keyword{thin lens approximation}\index{thin lens! model} as described by Goodman \cite{Goodman_2005_a}. The motivation for this model arises when one attempts to describe a system with a lens using the wave model. Before we discuss Fourier Optics and the lens models, we shall first highlight the key aspects at a high level.

\subsection{This Book: Fourier Optics}
\begin{figure}[t]
    \centering
    \begin{tabular}{cc}
    \includegraphics[width=0.4\linewidth]{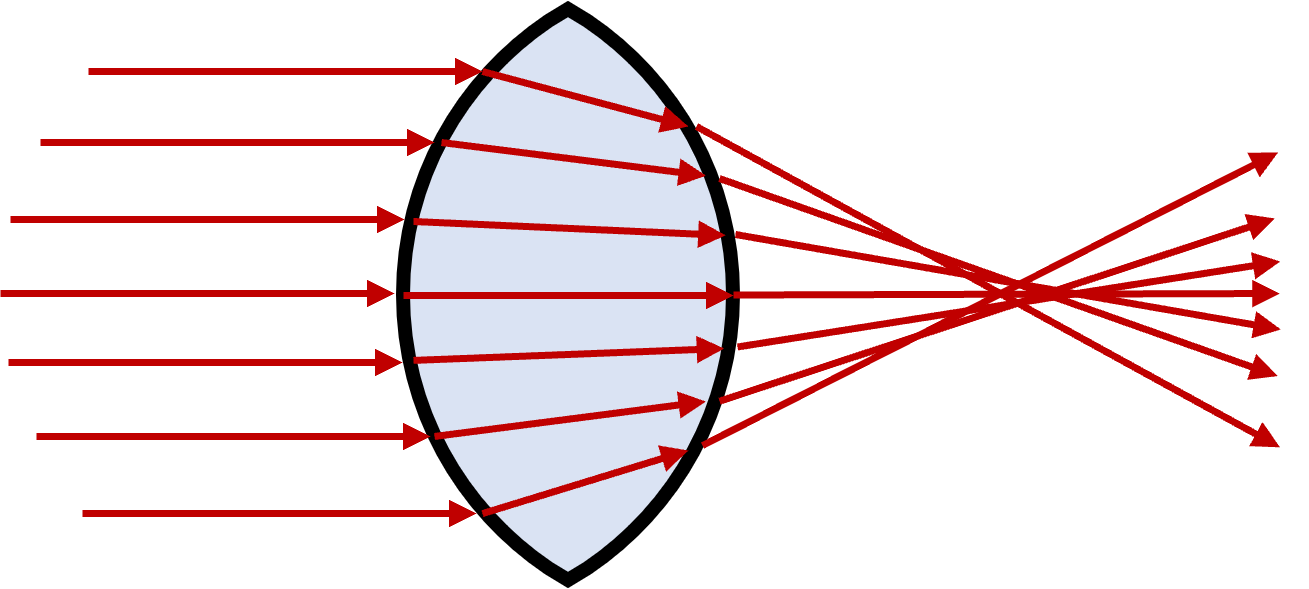}&
    \includegraphics[width=0.4\linewidth]{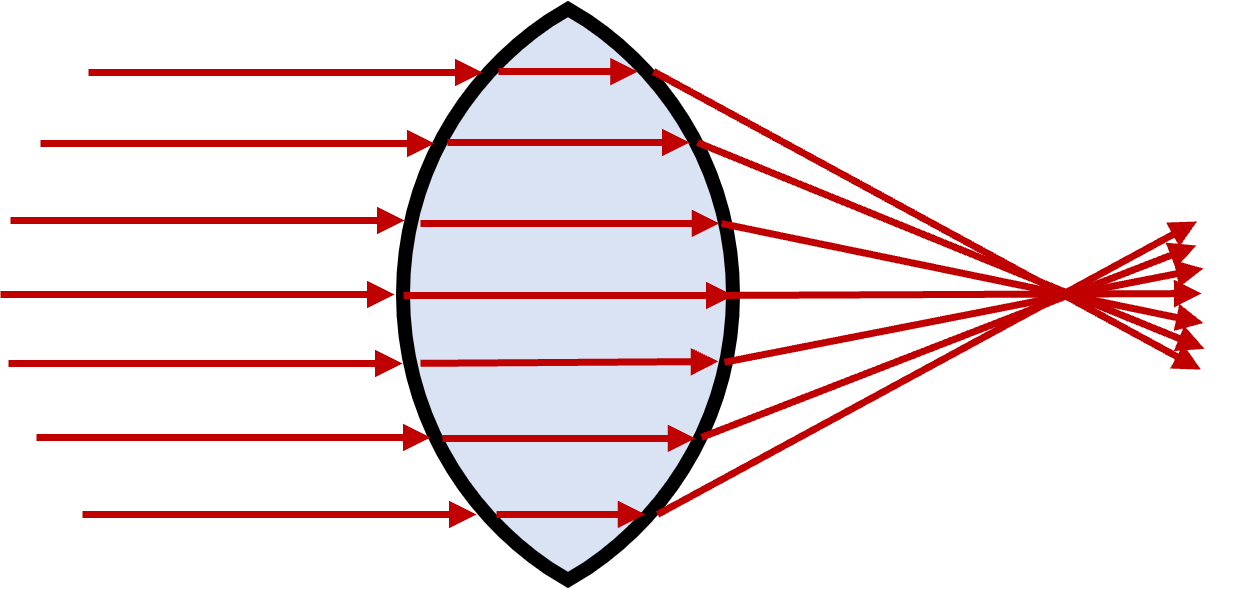}\\
    (a) Standard refraction & (b) Thin lens approximation
    \end{tabular}
    \caption{Visualizing the difference between (a) ray tracing and (b) the thin lens approximation. The thin lens model will not account for refractive issues such as the circle of confusion which arises in (a).}
    \label{fig: ch1_three_models}
\end{figure}

To distinguish between the ray tracing and the model taken in this book, we visually present how these two models describe light's interaction with a lens in \fref{fig: ch1_three_models}. Beginning with ray tracing, this model describes light as rays that travel in straight lines and bend when passing through media such as a lens via \keyword{Snell's law}\index{refraction! Snell's law},
\begin{equation*}
    \frac{\sin \theta_1}{\sin \theta_2} = \frac{n_2}{n_1}.
\end{equation*}
This bending of light is known as \keyword{refraction}\index{refraction}, a key feature of ray tracing. If we know the angle which the incident ray forms with the surface normal, $\theta_1$, and the \keyword{indices of refraction}\index{refraction! index of} $n_1$ and $n_2$ for both regions of space, we can compute the direction of the ray as it travels through the lens. This is shown in \fref{fig: ch1_three_models} by the change in direction of the rays as they pass through the entrance and exit of the lens.

Instead of rays of light, the wave model describes a wave as it passes through the lens. This is given as the solution to a differential equation with the boundary conditions as determined by the lens. The wave model captures both refractive and the wave-like \keyword{diffractive}\index{diffraction} effects of an imaging system. The trade-off is the difficulty in finding general solutions to systems \cite{Goodman_2005_a}, even in the case of the simplistic system shown in \fref{fig: ch1_three_models}. Therefore, the full wave model will not be suitable for our more application-driven goals.

The Fourier Optics model combined with the thin lens approximation does something a bit different. Instead of needing to solve a differential equation, a simple rule is proposed: the thicker the lens, the more the wave is delayed in phase -- a point we will revisit more appropriately later in this Chapter. In \fref{fig: ch1_three_models}, the thickest part of the lens corresponds to the wave moving slower for longer, hence being delayed relative to the wave that traversed through thinner sections of the lens. Roughly speaking, the less the wave is delayed, the less it will bend. We visualize just the effect of the thin lens approximation in \fref{fig: ch1_three_models}, though when combined with the concept of Fourier Optics, a proper depiction would include waves. For simplicity, we opt for rays in this case.

Although refraction is approximated for a simpler rule (i.e. by the thickness of the lens and not Snell's law), the model successfully captures the diffractive effects of the wave model. For brevity we will often refer to the Fourier Optics model with the thin lens approximation as simply ``Fourier Optics''.

We summarize the key features of ray tracing, Fourier Optics, and the full wave model as follows:
\begin{enumerate}
    \item \textbf{Ray tracing:} Ray tracing models the light as rays, which change direction upon interaction with a surface boundary. The change in direction is dictated by Snell's law. Ray tracing describes the refractive effects of light.
    \item \textbf{Wave model:} The wave model requires analysis of a differential equation which often \emph{cannot} be solved in closed form. The wave model incorporates \emph{both} refractive and diffractive effects.
    \item \textbf{Fourier Optics:} Fourier Optics models the light as a superposition of planar waves. Additionally, it is often paired with the thin lens approximation giving us a simple analytic tool that results in a \keyword{linear space-invariant model} of an imaging system. Fourier Optics can describe the diffractive effects seen in imaging.
\end{enumerate}
The advantage of Fourier Optics is exactly the fact that an imaging system with a lens can be written in terms of a Fourier transform. Additionally, the entire imaging system's response can be written in a traditional linear systems framework, which leads us to our next introductory topic.

\subsection{Linear Space-Invariant Optical Systems}
A concept typical to an image processing person is that of an input signal, $J(\vx)$, passed through a system, $h(\vx)$, to produce $I(\vx)$, the output signal. A general system of this sort is one of the core focuses of the classical book from Oppenheim and Willsky \cite{Oppenheim_1996_a}. Assuming the system is linear and space-invariant (LSI)\index{linear space-invariant systems}, we may write the input-output relationship via \keyword{convolution}\index{convolution} as
\begin{equation}
I(\vx) = h(\vx)  \circledast J(\vx),
\label{eq: ch1 lti}
\end{equation}
where $\circledast$ denotes the convolution operator. Of course, it is also known that this may be equivalently written in the Fourier domain as
\begin{equation*}
\mathfrak{Fourier}\{I(\vx)\} = \mathfrak{Fourier}\{h(\vx)\}\mathfrak{Fourier}\{J(\vx)\},
\end{equation*}
where $\mathfrak{Fourier}\{I(\vx)\}$ denotes the Fourier transform of the input signal. The goal of this book is to impart the knowledge of turbulent imaging and embed it within such a simple linear systems relationship.

It may be natural at this point to wonder what the system $h$ would entail in an optical system. We will denote $I(\vx)$ to represent the observed optical signal at a location $\vx$. For the sake of modeling, we are more interested in the terms on the right-hand side of \eref{eq: ch1 lti}. In the case of $J(\vx)$, this will represent the ``true'' object, to be defined more precisely within our context in due time. As a result, $h(\vx)$ must represent everything else; it must represent how the imaging system forms an image, including effects such as diffraction, any errors in our focusing, perturbations by the medium, and so on. For the purposes of this book, $h(\vx)$ is where the majority of the optics exist.

The key result of this Chapter will be the equations that describe image formation for two different types of light. We will take the following imaging system to be our example:
\begin{equation*}
I(\vx) = |h(\vx)|^2 \circledast J (\vx).
\end{equation*}
While this expression is slightly different from our first example, the idea remains the same. In fact, we will repeatedly return to this equation in a slightly more complicated form. More specifically, the expression written in terms of more physical entities will prove to be the more useful:
\begin{equation}
I(\vx) = |\mathfrak{Fourier} \{ P(\vu) e^{j \phi(\vu)} \}|^2 \circledast  J(\vx).
\label{eq: ch1 big idea 1}
\end{equation}
This equation is a direct result of the Fourier Optics model. Here we define $P(\vu)$ to be an indicator function describing the \keyword{aperture} of the imaging system and $\phi(\vu)$ as the \keyword{phase error} of the imaging system. Regardless of your familiarity with these concepts at this point, the purpose of presenting \eref{eq: ch1 big idea 1} is to demonstrate a simple fact: we can think of our optical system using typical Fourier-based analysis and reasoning if we know what to ``plug in''.

This, to many engineers, should come as a bit of a relief; the end goal is to arrive at the familiar destination of linear systems theory. There will be many details within this Chapter that are critical to arriving at and understanding \eref{eq: ch1 big idea 1}, though once the assumptions and limitations are understood, some of these details can be disregarded and we can continue onward from this more comfortable perspective.

\newpage
\section{Waves}

\subsection{The Scalar Wave Equation}
\label{sec: sec1_2}
\begin{figure}[ht]
    \centering
    \includegraphics[width=0.6\linewidth]{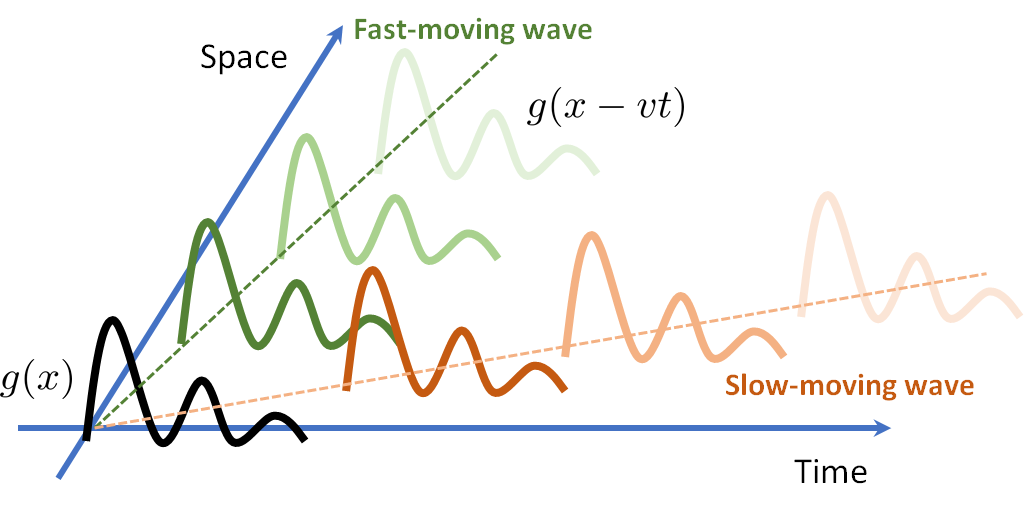}
    \caption{A visualization of two waves moving at two different speeds through a space-time domain $S$. The fast-moving wave takes a shorter time to reach a new position compared to a slow-moving wave.}
    \label{fig: Ch1_wave_movement}
\end{figure}

We begin by considering the one-dimensional wave equation\index{waves}. Let us define a \keyword{template function} $g(x)$ over a space-time domain $S = \{(x,t) \;|\; x \in \R, t \in \R\}$ \cite{Hecht_2015_a}. Using $g(x)$, we construct another function $f: S \rightarrow \R$ which is a shifted version of $g(x)$:
\begin{equation}
f(x,t) = g(x - vt),
\label{eq: ch1 scalar f(x,t)}
\end{equation}
where $v \in \R$ is the speed of propagation along a particular direction [m/s]. A different magnitude $v$ will lead to a different trajectory of the template function through $S$, as shown in \fref{fig: Ch1_wave_movement}. When light propagates through a vacuum, its speed is given as $v = c$ where $c = 3 \times 10^8$ [m/s] is the speed of light.\index{speed of light}

Our goal is to derive an equation to summarize the behavior of $f(x, t)$ over $S$. To this end, let us take the first-order derivative in space:
\begin{equation}
\pdv{f(x,t)}{x} = \pdv{g(x-vt)}{x} = \pdv{g(x - vt)}{(x-vt)}\cdot \pdv{(x-vt)}{x} = \pdv{g(r)}{r},
\label{eq: ch1 partial f partial x}
\end{equation}
where we defined $r = x - vt$. Similarly, the first-order derivative in time is
\begin{equation}
\pdv{f(x,t)}{t} = \pdv{g(x - vt)}{t} = \pdv{g(x-vt)}{(x-vt)} \cdot \pdv{(x-vt)}{t} = -v \pdv{g(r)}{r},
\label{eq: ch1 partial f partial t}
\end{equation}
Taking the derivatives of \eref{eq: ch1 partial f partial x} and \eref{eq: ch1 partial f partial t} again, we obtain a pair of second-order derivatives:
\comment{
\begin{align}
    \pdv[2]{f(x,t)}{x} &= \frac{\partial}{\partial x} \pdv{f(x,t)}{x} &= \frac{\partial}{\partial x} \frac{\partial g(r)}{\partial r}  &= \frac{\partial^2g(r)}{\partial r^2},&
    \label{eq: ch1 partial f partial x2}\\
    \pdv[2]{f(x,t)}{t} &= \frac{\partial}{\partial t} \pdv{f(x,t)}{t} &=
    \frac{\partial}{\partial t} \left(-v\frac{\partial g(r)}{\partial r}\right) &=
    v^2 \pdv[2]{g(r)}{r}.&
    \label{eq: ch1 partial f partial t2}
\end{align}
}
\begin{align}
    \pdv[2]{f(x,t)}{x}
    = \frac{\partial}{\partial x} \pdv{f(x,t)}{x}
    = \frac{\partial^2g(r)}{\partial r^2},
    \label{eq: ch1 partial f partial x2}\\
    \pdv[2]{f(x,t)}{t}
    = \frac{\partial}{\partial t} \pdv{f(x,t)}{t}
    = v^2 \pdv[2]{g(r)}{r}.
    \label{eq: ch1 partial f partial t2}
\end{align}
Combining \eref{eq: ch1 partial f partial x2} and \eref{eq: ch1 partial f partial t2}, we obtain a partial differential equation that connects space and time,
\begin{equation}
    \pdv[2]{f(x,t)}{x} = \frac{1}{v^2}\pdv[2]{f(x,t)}{t}.
    \label{eq: ch1 1d wave equation 1}
\end{equation}
\eref{eq: ch1 1d wave equation 1} is known as the \keyword{one-dimensional wave equation}\index{wave equation! one dimensional}. If we were to instead write the velocity as $v = c/n$, then \eref{eq: ch1 1d wave equation 1} will become
\begin{equation}
    \pdv[2]{f(x,t)}{x} = \frac{n^2}{c^2}\pdv[2]{f(x,t)}{t}.
    \label{eq: ch1 1d wave equation 2}
\end{equation}
Writing the velocity in this way motivates us to consider $c$ to be the maximum speed and $n$ to be the attenuation. In the case of optics, $n$ is known as the index of refraction.\index{refraction! index of}

We may also extend the wave equation to higher dimensions. This will give us the high-dimensional \keyword{scalar wave equation}\index{wave equation! high dimensional (scalar)}:
\begin{equation}
    \laplacian f(\vx, t) - \frac{n^2}{c^2} \pdv[2]{f(\vx, t)}{t} = 0,
    \label{eq: ch1_general_wave_eq}
\end{equation}
where $\vx = [x, y, z] \in \R^3$ are the spatial coordinates of the wave, and
\begin{equation*}
\laplacian = \pdv[2]{}{x} + \pdv[2]{}{y} + \pdv[2]{}{z}
\end{equation*}
is the Laplacian operator. If polarization is involved, we can extend the scalar wave equation to a \keyword{vector field} $\vf(\vx,t) = (f_1(\vx,t), \ldots, f_p(\vx,t))$, representing the different polarizations of the wave.

\subsection{Helmholtz Equation}
\label{sec: sec1_3}
We now seek to utilize \eref{eq: ch1_general_wave_eq} using a set of elementary solutions. It is likely not a surprise an elementary solution would be a sinusoidal \emph{wave},
\begin{equation}
    f(\vx,t) = A(\vx) \cos( 2\pi \nu t + \theta(\vx)),
\end{equation}
with spatially varying amplitude $A(\vx)$ and phase $\theta(\vx)$. While this form is certainly valid, it is often mathematically convenient to instead use a complex representation. Accordingly, we will typically consider a \keyword{complex wave function}:\index{complex wave function}
\begin{equation}
    u(\vx,t) = U(\vx)e^{-j2\pi \nu t}.
    \label{eq: Ch1 wave equation general form}
\end{equation}
The function $U(\vx)$ represents the complex envelope, or \keyword{phasor}\index{phasor} \cite{Hecht_2015_a, Goodman_2005_a}, which is then modulated by the time-evolving term $e^{-j2\pi \nu t}$.

This particular form of elementary solution presented in \eref{eq: Ch1 wave equation general form} is useful as it separates space and time into two distinct terms. In most situations we will be interested in finding $U(\vx)$; if we further wish to describe the time behavior, this can be achieved by simply appending the temporal modulation term. We note the convention of a negative sign in $e^{-j2\pi \nu t}$ results in a clockwise rotation for an increase in radians. In order to relate the two forms of solutions, we can look at the real component of $u(\vx,t)$, writing \eref{eq: Ch1 wave equation general form} as
\begin{align*}
    \mathfrak{Real}\{u(\vx,t)\} &= A(\vx) \cos( 2\pi \nu t + \theta(\vx)), \\
    &= f(\vx, t),
\end{align*}
where $A(\vx)$ is the scalar amplitude and $\theta(\vx)$ is the phase, and further defining
\begin{equation}
    U(\vx) = A(\vx)\exp\{-j\theta(\vx)\}.
\end{equation}

Let us investigate a few physical properties of our complex envelope. Assuming a non-trivial phase function $\theta(\vx)$, the form of $U(\vx)$ suggests the cyclical nature of the envelope. For a constant value $\varphi \in [0, 2\pi)$, the set $\{ \vx \;|\; \angle U (\vx) = \varphi \}$ defines a set of \keyword{wavefronts}\index{wavefront}. The distance between neighboring wavefronts is then said to be the \keyword{wavelength}\index{wavelength} of the wave, denoted as $\lambda$ [m]. We also define the \keyword{optical frequency}\index{optical frequency} $\nu$ (pronounced as ``nu'') [s$^{-1}$] to be
\begin{equation}
    \nu = \frac{v}{\lambda},
\end{equation}
where $v$ is the velocity of the wave [m/s]. The optical frequency describes the rate at which the wave oscillates in time as it arrives.

\begin{figure}
    \centering
    \begin{tabular}{ccc}
    \includegraphics[width=0.3\linewidth]{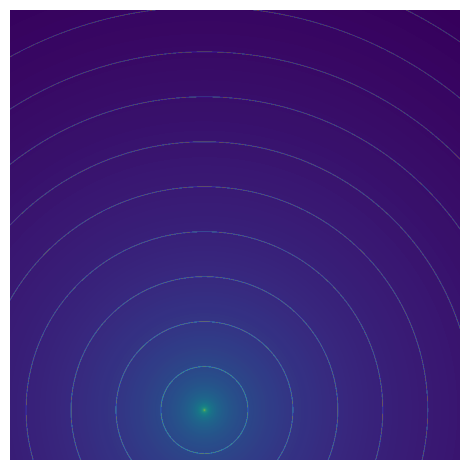}&
    \includegraphics[width=0.3\linewidth]{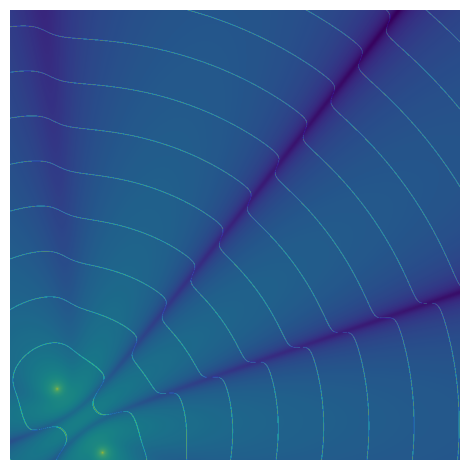}&
    \includegraphics[width=0.3\linewidth]{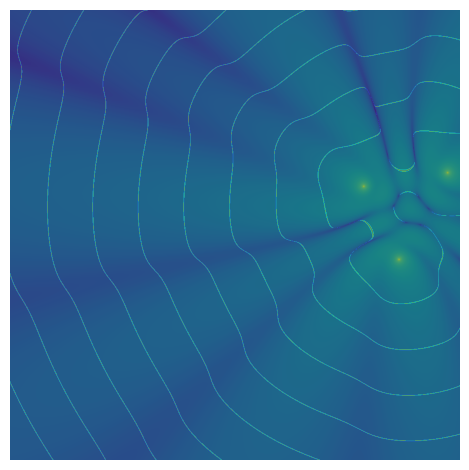} \\
    (a) One point & (b) Two points & (c) Three points
    \end{tabular}
    \caption{Here we show the magnitude of the field created by one, two, and three points. A bright spot corresponds to a large magnitude while a dark region indicates a smaller magnitude. In addition, we also show the wavefronts which are approximately 0 radians.}
    \label{fig: ch1_wavefronts}
\end{figure}

This concept of wavefronts and wavelengths may benefit from being presented visually. To this end, we provide \fref{fig: ch1_wavefronts}, where we show the following consideration:
\begin{equation*}
    V(\vx) = \abs{\sum_{i=1}^{N} U(\vx; \vx_i)} = \abs{\sum_{i=1}^{N} \frac{\exp{-j (2\pi / \lambda) \abs{\vx - \vx_i}}}{\abs{\vx - \vx_i}}},
\end{equation*}
where we have chosen the amplitude function $A(\vx; \vx_i) = 1/\abs{\vx - \vx_i}$ and phase function $\phi(\vx; \vx_i) = (2\pi / \lambda) \abs{\vx - \vx_i}$. Simply put, there are sources of waves located at various positions $\vx_i$, known as \keyword{point sources}\index{point source}, whose amplitude and phase vary as a function of distance from the point $\vx_i$. In addition to the magnitude of this summation, we also show the wavefronts $\{  \vx \;|\; \angle V (\vx) = 0\}$. We show examples of one, two, and three sources of waves. In this figure, the background shading indicates the magnitude of the complex wave, which varies in a somewhat complicated fashion due to the interaction of the waves in the complex space, along with the wavefronts indicated as solid lines.

With the intent of using our form of the complex wave function $u(\vx,t)$ within the scalar wave equation of \eref{eq: ch1_general_wave_eq}, we present the following pair of derivatives:
\begin{align}
  \laplacian u(\vx, t) &=
  \left\{\laplacian U(\vx)\right\}e^{-j2\pi\nu t},
  \label{eq: Ch1 partial u1} \\
  \pdv[2]{u(\vx,t)}{t} &= (j2\pi \nu)^2 U(\vx)e^{-j2\pi\nu t}.
  \label{eq: Ch1 partial u2}
\end{align}
Substituting \eref{eq: Ch1 partial u1} and \eref{eq: Ch1 partial u2} into \eref{eq: ch1_general_wave_eq}, we can write the scalar wave equation in terms of the phasor,
\begin{equation*}
\laplacian U(\vx) = -\left(2 \pi \nu\right)^2 \frac{n^2}{c^2} U(\vx).
\end{equation*}
Letting $k = 2\pi \nu n /c = 2\pi/\lambda_\nu$ be the \keyword{wave number} (where $\lambda_\nu = c/(\nu n)$ is the wavelength), we arrive at the \keyword{Helmholtz equation}.

\boxedthm{
\begin{definition}[\keyword{Helmholtz Equation}\index{Helmholtz equation}]
\label{thm: Helmholtz}
Let the complex wave equation be $u(\vx,t) = U(\vx)e^{-j 2 \pi \nu t}$ where $\vx = [\vx,z]^T \in \R^3$ and $\nu$ is optical frequency. The Helmholtz equation of the scalar field $U(\vx)$ is given by
\begin{equation}
\laplacian U(\vx) + k^2 U(\vx) = 0,
\label{eq: ch1 Helmholtz equation 2}
\end{equation}
where $\laplacian = \pdv[2]{}{x} + \pdv[2]{}{y} + \pdv[2]{z}$ is the Laplacian operator, and $k = 2\pi/\lambda$ is the wave number.
\end{definition}
}

\subsection{Types of Waves}
All complex wave functions taking the form of $u(\vx,t) = U(\vx)e^{-j2\pi \nu t}$ where $U(\vx)$ is twice differentiable will satisfy the Helmholtz equation. Three forms that will be particularly useful to us are planar, spherical, and parabolic waves:\index{waves! common types of}
\boxedthm{
\vspace{-1ex}
\begin{definition}[\keyword{Planar wave}]\index{planar wave}
The scalar field of the \keyword{planar wave} is
\begin{equation}
    U(\vx) = A_0 e^{-j \vk^T\vx} = A_0 e^{-j (k_x x + k_y y + k_z z)},
\end{equation}
where $\vk = [k_x, k_y, k_z]^T$ is the \keyword{wave vector}\index{wave vector} such that $k_x^2+ k_y^2 + k_z^2 = k^2$ and $k = 2\pi/\lambda$ is the wave number. The constant $A_0$ represents the amplitude.
\end{definition}
}
\boxedthm{
\vspace{-1ex}
\begin{definition}[\keyword{Spherical wave}]\index{spherical wave}
The scalar field $U(r)$ of the \keyword{spherical wave} is
\begin{equation}
    U(r) = \frac{A_0}{r} e^{-j k r},
    \label{eq: ch1 spherical wave}
\end{equation}
where $r = |\vx| = \sqrt{x^2 + y^2 + z^2}$ is the radius between $\vx$ and the origin.
\end{definition}
}
\boxedthm{
\vspace{-1ex}
\begin{definition}[\keyword{Paraboloidal wave}]\index{paraboloidal wave}
The scalar field of a\newline\keyword{paraboloidal wave} is given by
\begin{equation}
    U(\vx) = \frac{A_0 e^{-jkz} }{z} e^{-jk \left( \frac{|\vx|^2}{2z}\right)}.
    \label{eq: ch1 parabola wave 1}
\end{equation}
\end{definition}
}

\begin{figure}[ht]
\centering
\includegraphics[width=0.95\linewidth]{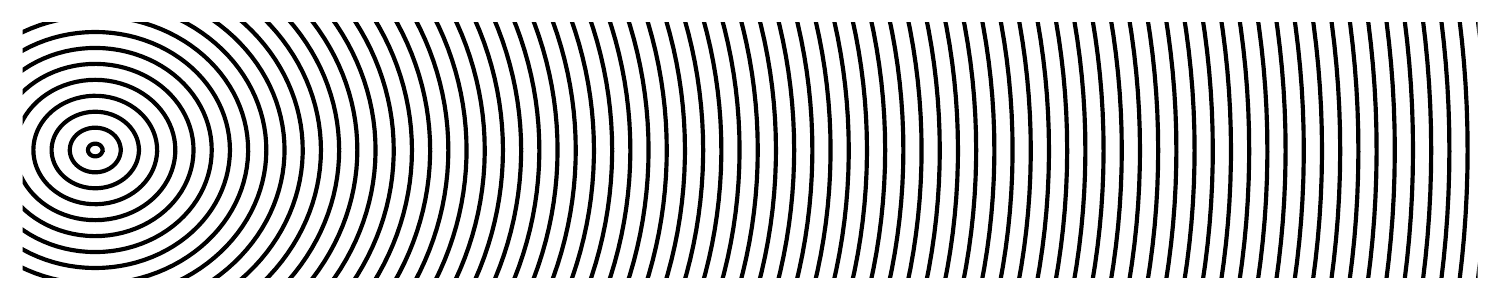}
\vspace{-2ex}
\caption{As a spherical wave propagates outwards, at a sufficient distance it becomes suitably paraboloidal (or parabolic), and further yet planar. We note that one can approximate a planar wave with a parabolic one. These ranges are not mutually exclusive, but rather the most appropriate approximation (in most cases) for the associated wave at a distance.}
\label{fig: Ch1 spherical plane wave}
\vspace{-2ex}
\end{figure}

Suppose we have the case of a spherical wave emitting energy as shown in \fref{fig: Ch1 spherical plane wave}. As we consider the wavefronts of the phasor increasingly further from the origin of the wave, we notice that the spherical wave begins to look like a parabolic wave. Increasing our distance further, we note the wave's tendency towards appearing as a plane wave. This approximate nature of a spherical wave turning into a parabolic wave and then a planar wave will be a useful conceptual and analytic tool for the purposes of this Chapter.

\subsection{Why Waves?}
Thus far we have dedicated our energy towards describing the mathematics behind wave motion in a general sense. There is nothing tying it specifically to optics at this point! There are two ways in which one may consider this reality, one more formally from Maxwell's equations and the other more intuitively through experiment.

Maxwell's equations\index{wave equation! electromagnetic} in differential form can be written as \index{Maxwell's equations}
\begin{align}
    \begin{aligned}
        \vnabla \times \mE &= -\pdv{\mB}{t},
        \hspace{0.75in} &\vnabla \times \mB &= \mu_0\left(\epsilon_0\pdv{\mE}{t} + \mJ\right), \\
        \vnabla \cdot \mE &= \rho/\epsilon_0, &\vnabla \cdot \mB &= 0.
    \end{aligned}
    \label{eq: 6_Maxwell}
\end{align}
As a first step towards developing the wave equation, we restrict our analysis to a vacuum, thus finding the ``pure" form of electromagnetic wave propagation undisturbed by outside forces. If we further define $\epsilon_0\mu_0 = 1/c^2$, this allows our equations \eqref{eq: 6_Maxwell} to take the form
\begin{align}
    \begin{aligned}
        \curl{\mE} &= -\pdv{\mB}{t},
        \hspace{0.75in} &\curl{\mB} &= \frac{1}{c^2}\pdv{\mE}{t}, \\
        \div{\mE} &= 0, &\div{\mB} &= 0,
    \end{aligned}
    \label{eq: 6_Maxwell_vacuum}
\end{align}
where $c$ is the speed of light.

Through the use of the triple scalar product and $BAC-CAB$ \cite{Feynman_1965_a} rule, we can obtain the wave equation. Starting with taking the curl of the first equation in \eqref{eq: 6_Maxwell_vacuum},
\begin{align*}
    \curl{\left(\curl{\mE}\right)} &= \curl\left({-\pdv{B}{t}}\right), \\
    \grad{\left(\div{\mE}\right)} - \div{\grad{}}\mE &= -\pdv{}{t}\left(\curl{\mB}\right), \\
    \laplacian{\mE} &\overset{(a)}{=} \frac{1}{c^2} \pdv[2]{\mE}{t},
\end{align*}
where arriving at the LHS of (a) utilized the divergence property of the field $\mE$ in free space while the RHS used the definition of the curl of the magnetic field.
By symmetry, we can state the same equation for the magnetic field $\mB$. We therefore present the following definition:
\boxedthm{
\begin{definition}[\keyword{Electromagnetic Wave Equation}]
    The homogeneous, 3-dimensional form of the electromagnetic wave equation is given by
    \begin{equation}
        \laplacian{\mE} - \frac{1}{c^2} \pdv[2]{\mE}{t} = 0,
        \label{eq: ch1_hom_EM_WE}
    \end{equation}
    and predicts that an electromagnetic wave travels at the speed of light -- a constant $c$.
\end{definition}
}
This suggests that electromagnetic radiation travels at a constant speed $c$, the very same speed of light we are familiar with. Maxwell himself suggested that light must be an electromagnetic wave \cite{Goodman_2005_a}, with experimental evidence supporting this assertion in the proceeding years. Therefore, if we accept that light is electromagnetic radiation then it may be modeled as a wave.

This represents a mathematical reason why light should be treated as a wave. There are, however, more experimental reasons as to why we may choose to do so. One of the most convincing experimental reasons to model light as a wave is the phenomenon of diffraction.

\section{The Diffraction Problem}
\label{sec: sec1_4}
\begin{figure}
    \centering
    \includegraphics[width=0.65\linewidth]{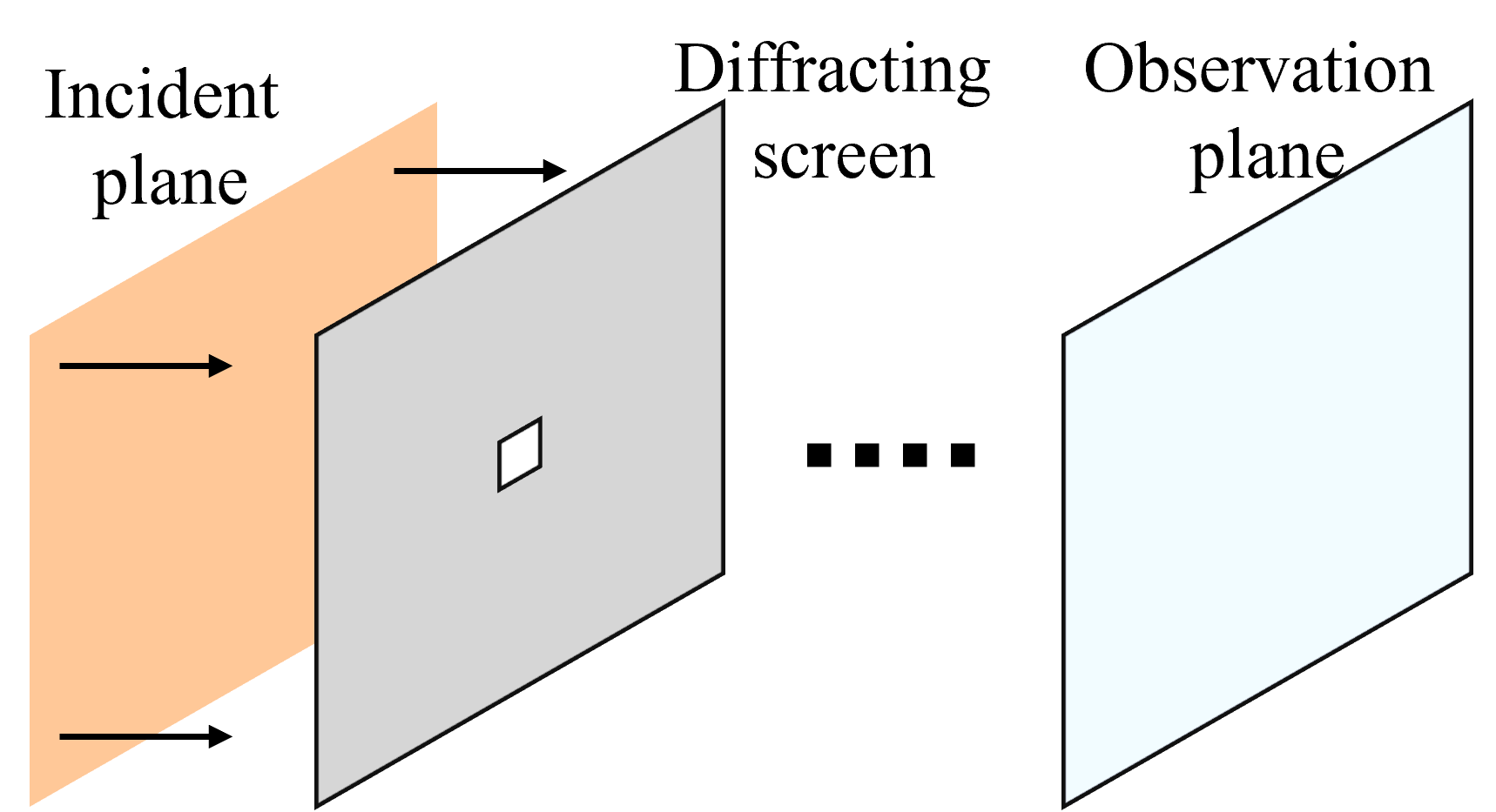}
    \caption{(a) The physical system we will be analyzing which gives rise to diffraction. An incident wave propagates toward the diffracting screen. We are interested in describing the wave at the observation plane. (b) Two possible diffraction patterns from [Top] a square aperture and [Bottom] a circular aperture.}
    \label{fig: ch1_diff_problem}
\end{figure}

The primary concern of \cref{sec: sec1_4} is the phenomena of diffraction\index{diffraction}\index{diffraction}. Diffraction is a clear case for the wave model and is seen most obviously in the case of a wave that is incident upon a large, non-transmitting screen with a small hole. The wave on the side opposite to that of the incidence is said to demonstrate diffractive effects. We show the system we will analyze to arrive at the mathematical description of diffraction in \fref{fig: ch1_diff_problem}.

Due to the wave-like behavior, we will require the Helmholtz equation (\eref{eq: ch1 Helmholtz equation 2}) to describe it mathematically. Problems within optics that arise from the Helmholtz equation are typically stated as boundary problems: if we know a subset of values for $U(\vx)$, we are then interested in knowing the distribution of the field at \emph{another} set of locations. While the nature of the boundaries and the sufficient conditions for problems to be well-posed (see \cite{Born_1999_a}) are beyond the scope of this book, they underlie the problems in the proceeding discussions. We instead opt for a simpler introduction first and present the main mathematical result afterward.

\subsection{The Huygens-Fresnel Principle}
\begin{figure}[ht]
\centering
\includegraphics[width = 0.45\linewidth]{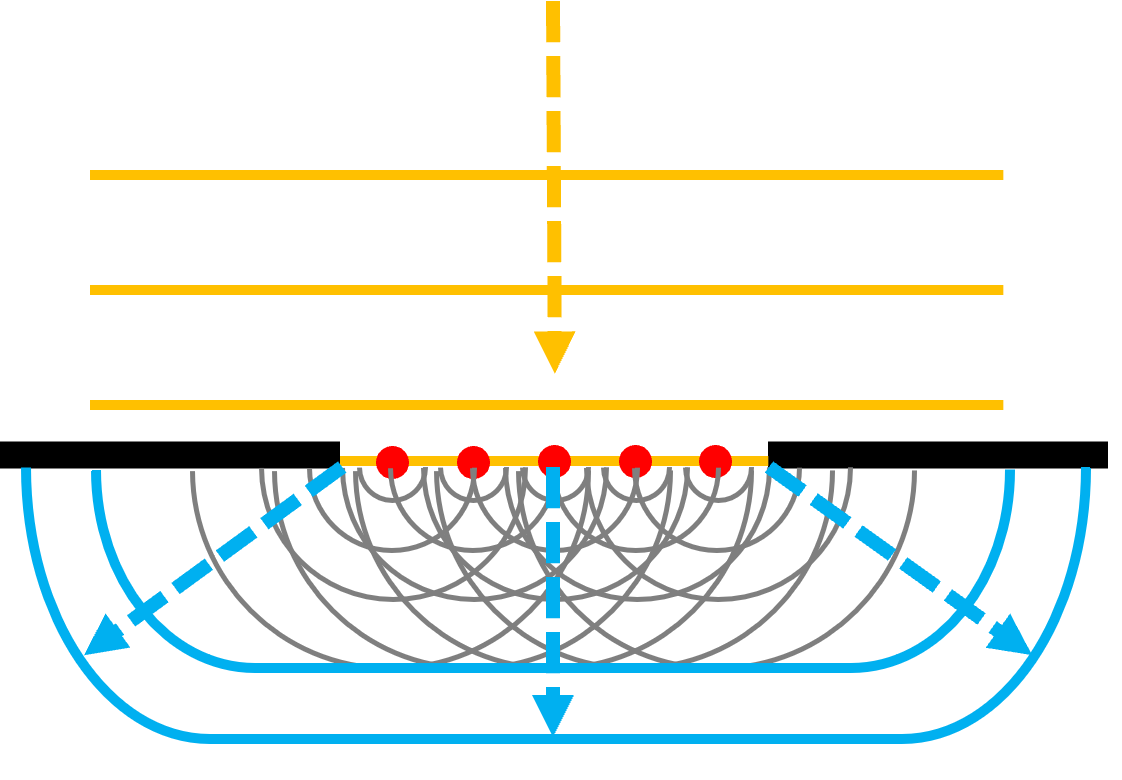}
\caption{The Huygens-Fresnel Principle states that the point source radiates outwards spherically, and every point along the wavefront can be considered as a secondary source.}
\label{fig: Ch1_Huygens-Fresnel Principle}
\end{figure}

The \keyword{Huygens-Fresnel principle (HFP)}\index{Huygens-Fresnel principle} is a natural choice in presenting the intuition behind how light is modeled as a wave. We begin with the definition:
\boxedthm{\begin{definition}[\keyword{Huygens-Fresnel principle}]
The propagation of the wave follows two principles:
\begin{enumerate}
\setlength\itemsep{0ex}
    \item Sources of waves radiate outwards spherically;
    \item Every point along any wavefront can be considered a secondary source, referred to as \emph{wavelets}, which interact.
\end{enumerate}
\end{definition}}
The Huygens-Fresnel principle tells us in simple terms how to begin with a particular wave distribution on a surface and end with the distribution on a different surface. We visualize what happens to a plane wave arriving at a diffracting screen in \fref{fig: Ch1_Huygens-Fresnel Principle}. Here we can see the secondary wavefronts cause the wave to ``leak'' beyond the shadow of the screen.

\begin{figure}
    \centering
    \begin{tabular}{cccc}
	Ray optics & Wave optics & Ray optics & Wave optics \\
	 \includegraphics[width=0.2\linewidth]{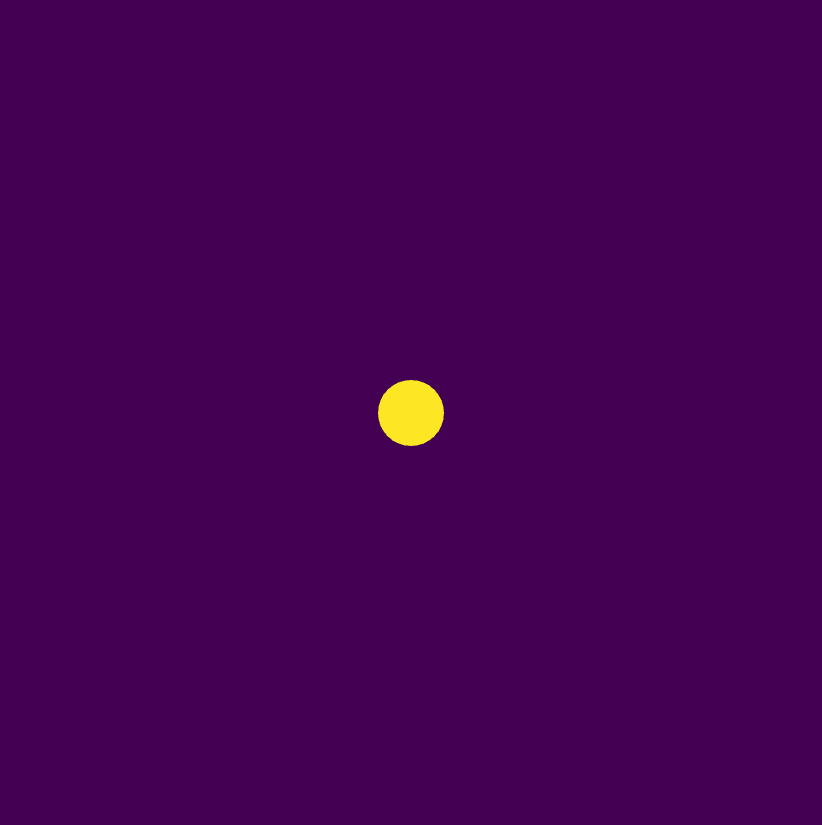} &
     \includegraphics[width=0.2\linewidth]{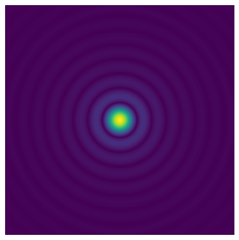} &
	 \includegraphics[width=0.2\linewidth]{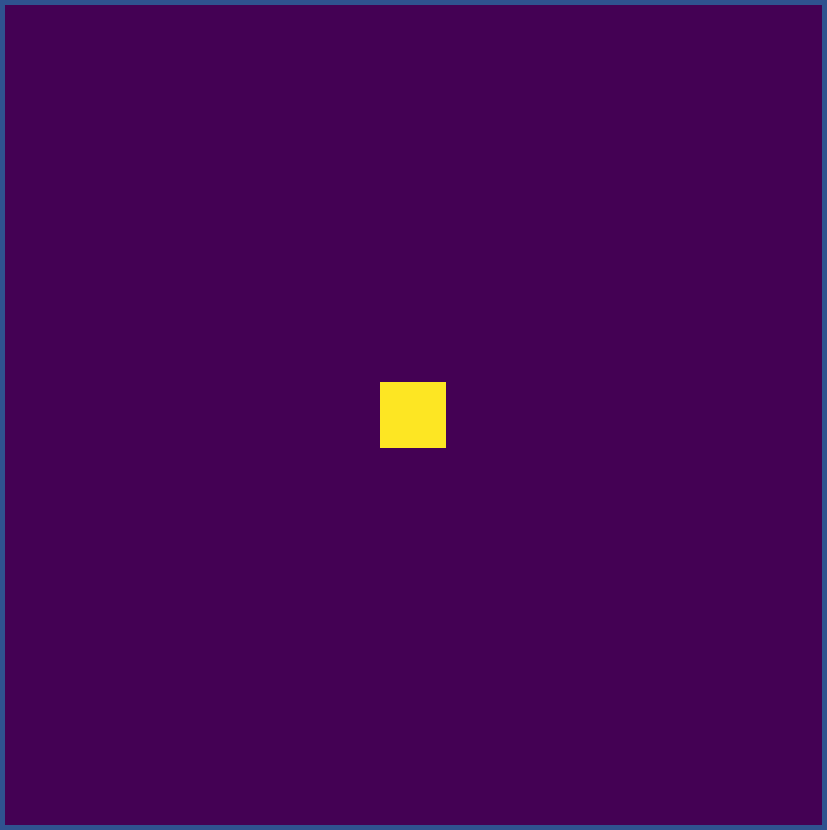} &
     \includegraphics[width=0.2\linewidth]{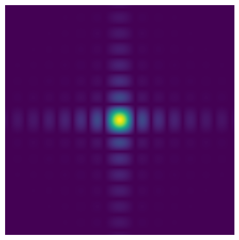} \\
   \end{tabular}
   \begin{tabular}{ccc}
	(a) Circular aperture & \hspace{9ex} & (b) Square aperture
   \end{tabular}
    \caption{A representation of the field on the observation plane as predicted by ray tracing and the wave model. The ray optics solution does not explain how the energy leaks out beyond the shadow of the aperture, which is a mismatch from physical reality.}
    \label{fig: ch1_rays_vs_hfp}
\end{figure}

Let us contrast the two forms of propagation models as shown in \fref{fig: ch1_rays_vs_hfp}: ray optics and wave optics as described by the HFP. Ray tracing would predict the pattern an observer would see to be the same shape as the aperture, only rays that are propagating through the gap will arrive on the observation plane. Furthermore, utilizing the fact that rays travel in a straight line, we should expect to see a copy of the aperture as an observer. The HFP instead predicts that some light will leak out of the shadow of the aperture and cause the pattern to be distributed over a larger area than that predicted by ray optics.

One may notice the rippling effects in the diffraction pattern. This can be explained by the second rule of the HFP which states that the waves interact. This means that the waves may either constructively or destructively interfere with one another. Using the concept of  phasors as before, we can understand this interference to arise from the real and imaginary components, thus contributing interference. Furthermore, the distance between the peaks in the diffraction pattern will be directly related to the wavelength of the incident light, a point we will present later in this Chapter.

The preceding discussion, of course, contains no mathematical concepts though it is often useful to have the Huygens-Fresnel principle in mind when solving problems of this nature. It is a helpful guiding tool that can aid in the visualization of propagation through various objects, and can even be used to explain refraction, phase perturbations due to atmospheric turbulence, and so on.

\subsection{Rayleigh-Sommerfeld Diffraction}
\index{Rayleigh-Sommerfeld diffraction}One of the mathematical approaches which describe diffractive effects is known as the Rayleigh-Sommerfeld diffraction integral, with a closely related alternative being the Fresnel-Kirchhoff diffraction integral \cite{Goodman_2005_a}. The Rayleigh-Sommerfeld approach is often regarded as the most rigorous approach to formulating the solution to the diffraction effect that is in agreement with Maxwell's equations. The formulation seeks to describe a wave after incidence upon a diffracting screen. The problem is easily stated, yet the details involved are beyond the scope of this book. We would suggest the interested reader to Goodman \cite{Goodman_2005_a} for an easy-to-follow treatment, and those who are interested in the rigorous development to Born and Wolf \cite{Born_1999_a}.

\begin{figure}[ht]
\centering
\includegraphics[width=0.7\linewidth]{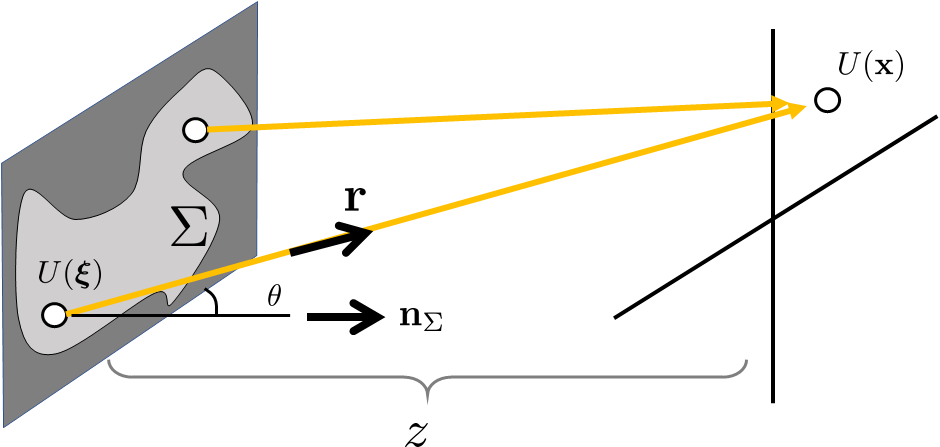}
\caption{When there are more than two points in the object plane, the observed field at position $\vx$ will be the superposition of the fields emitting from the source plane $\Sigma$.}
\label{fig: Ch1_Huygens_two_point}
\end{figure}

We depict the geometry of the problem in \fref{fig: Ch1_Huygens_two_point}. As in the case of the HFP, we wish to write the wave in the observation plane as a superposition of the secondary wavelets. The Rayleigh-Sommerfeld diffraction integral does just this, it describes the phasor at point $\vx$ due to the field at the aperture is given by the integration over the aperture $\Sigma$, which is stated in Theorem~\ref{thm: Ch1 Rayleigh Sommerfeld}.

\boxedthm{
\begin{theorem}[\keyword{Rayleigh-Sommerfeld diffraction}]\index{Rayleigh-Sommerfeld diffraction! definition}
\label{thm: Ch1 Rayleigh Sommerfeld}
The wave observed at a point $\vx$ after diffracting through a surface $\Sigma$ is described by
\begin{equation}
U(\vx) = \frac{1}{j\lambda} \iint_{\Sigma} U(\vxi) \frac{\exp \left\{jkr\right\}}{r} \cos\theta \; d\vxi,
\label{eq: ch1_HFP}
\end{equation}
where $\Sigma$ is the source surface composed of many individual point sources, $\theta$ as the angle between the surface normal and the vector $\vr$ pointing from $\vxi$ to $\vx$ (with $r = |\vr|$), and $k=2\pi/\lambda$ as the wavenumber.
\end{theorem}
}

While \eref{eq: ch1_HFP} is provided without proof, it can be understood somewhat intuitively. In fact, it resembles a mathematical form of the Huygens-Fresnel principle! The Rayleigh-Sommerfeld integral describes the wave at a point $\vx$ as a superposition of spherical waves which are weighted by their source phasor amplitudes, similar to the case of secondary wavelets in the HFP. Due to the fact that the spherical wave is modeled as a complex exponential, we will have our interaction as required. There is additionally a directivity constant which can be understood to account for the direction of propagation of the original wave. The scaling constant $1/(j\lambda)$ is a technical point that can be explained rigorously via the Rayleigh-Sommerfeld theory \cite{Goodman_2005_a}. For the purpose of this book, we can treat the term as something that comes out of the derivative of the exponential $e^{jkr}$.

\subsection{Fresnel Diffraction}
\index{Fresnel diffraction}The Rayleigh-Sommerfeld diffraction in Theorem~\ref{thm: Ch1 Rayleigh Sommerfeld} is a useful result for our specific configuration of \fref{fig: ch1_diff_problem}. This result can describe the wave along any parallel plane on the right side of the diffracting screen. However, this ability comes at a cost: the complicated behavior of the wave near the screen will make finding general closed forms solutions challenging. As a result, it will prove to be analytically intractable in most situations of interest. To make sense of the result and to derive something we can apply to image formation we need a set of approximations. The first approximation we will introduce is known as the \keyword{Fresnel approximation}\index{Fresnel approximation}. The Fresnel approximation takes inspiration from the fact that at a sufficient distance, a spherical wave can be approximated as a parabolic wave. Therefore, if we are only interested in describing the plane at a distance sufficiently far away from the screen, we may be able to obtain a more useful result via an approximation of the general Rayleigh-Sommerfeld result.

To begin our approximate analysis, we first note that $\cos\theta = z/r$ in \eref{eq: ch1_HFP}, thus it may instead be written as
\begin{equation}
U(\vx) = \frac{z}{j\lambda} \iint_{\Sigma} U(\vxi) \frac{\exp\{jkr\}}{r^2} d\vxi.
\label{eq: ch1 Fresnel Diffraction 1}
\end{equation}
We first consider the approximation for $r^2$. By using Taylor's expansion, $\sqrt{1+b} \approx 1 + \frac{1}{2}b - \frac{1}{8}b^2 + \ldots$. Substituting this into the distance $r$, we can show that
\begin{align}
r
&= \sqrt{z^2 + |\vx-\vxi|^2} = z \sqrt{1+\left|\frac{\vx-\vxi}{z}\right|^2} \approx z \left[1+ \frac{1}{2}\left|\frac{\vx-\vxi}{z}\right|^2 \right] \approx z,
\label{eq: ch1 Fresnel Diffraction 3}
\end{align}
if $|\vx-\vxi|^2 \ll z^2$.

With additional calculations by substituting \eref{eq: ch1 Fresnel Diffraction 3} into \eref{eq: ch1_HFP}, we can show that
\begin{align}
U(\vx)
&= \frac{z}{j\lambda} \iint_{\Sigma} U(\vxi) \frac{\exp \left\{jk z \left[1+ \frac{1}{2}\left|\frac{\vx-\vxi}{z}\right|^2 \right] \right\}}{z^2} d\vxi \notag \\
&= \frac{e^{jkz}}{j\lambda z} \iint_{-\infty}^{\infty} U_{\Sigma}(\vxi) \exp \left\{\frac{jk}{2z} |\vx-\vxi|^2  \right\} d\vxi,
\label{eq: ch1 Fresnel Diffraction 4}
\end{align}
where $U_{\Sigma}(\vxi)$ is the incident field restricted to the aperture, defined as
\begin{equation*}
U_{\Sigma}(\vxi)
=
\begin{cases}
U(\vxi), &\qquad \vxi \in \Sigma,\\
0, &\qquad \text{otherwise}.
\end{cases}
\end{equation*}
Note that we didn't utilize the approximation of \eref{eq: ch1 Fresnel Diffraction 3} in the exponential term in \eqref{eq: ch1 Fresnel Diffraction 4}. This is for two reasons: (1) it is multiplied by the wavenumber $k$ which is typically a large value, certainly in the case of the visible spectrum; (2) a small change in phase will cause a potentially significant change in the exponential. Thus, it is hard to justify the approximation in this particular case \cite{Goodman_2005_a}.

Expanding the quadratic term inside the exponent of \eref{eq: ch1 Fresnel Diffraction 4} allows us to write
\begin{equation}
|\vx-\vxi|^2 = |\vx|^2 - 2 \vx^T \vxi + |\vxi|^2.
\label{eq: ch1 Fresnel Diffraction 5}
\end{equation}
Substituting \eref{eq: ch1 Fresnel Diffraction 5} into \eref{eq: ch1 Fresnel Diffraction 4}, we can show that the observed field is
\begin{align}
U(\vx) = \frac{e^{jkz}}{j\lambda z} e^{j \frac{k}{2z}|\vx|^2} \iint_{-\infty}^{\infty} \left\{U_\Sigma(\vxi) e^{j\frac{k}{2z}|\vxi|^2}\right\} e^{-j \frac{2\pi}{\lambda z} \vx^T\vxi} d\vxi.
\end{align}
An important observation here is that the integration is the Fourier transform of the field $U_\Sigma(\vxi) e^{j\frac{k}{2z}|\vxi|^2}$. Splitting the terms in $U_\Sigma(\vxi) e^{j\frac{k}{2z}|\vxi|^2}$, it can be seen that the field is composed of three parts: (1) the incident field $U(\vxi)$, (2) the aperture $\Sigma$, and a quadratic phase distortion $e^{j\frac{k}{2z}|\vxi|^2}$. This result is known as the \keyword{Fresnel diffraction integral}.

\boxedthm{\begin{theorem}
The \keyword{Fresnel diffraction integral}\index{Fresnel diffraction! definition} is defined as
\begin{equation}
U(\vx) = \frac{e^{jkz}}{j\lambda z} e^{j \frac{k}{2z}|\vx|^2} \iint_{-\infty}^{\infty} \left\{U_\Sigma(\vxi) e^{j\frac{k}{2z}|\vxi|^2}\right\} e^{-j \frac{2\pi}{\lambda z} \vx^T\vxi} d\vxi,
\label{eq: ch1 Fresnel diffraction main}
\end{equation}
where $U_{\Sigma}(\vxi)$ is the incident field passing through a finite aperture $\Sigma$. Equivalently, \eref{eq: ch1 Fresnel diffraction main} can be written as
\begin{equation}
U(\vx) = \frac{e^{jkz}}{j\lambda z} e^{j \frac{k}{2z} |\vx|^2} \mathfrak{Fourier}
\left\{U_\Sigma(\vxi) e^{j\frac{k}{2z}|\vxi|^2}\right\}\bigg|_{\vf = \frac{\vx}{\lambda z}},
\end{equation}
where the Fourier frequencies are evaluated at $\vf = \vx/(\lambda z)$.
\end{theorem}}

\subsection{Fraunhofer Approximation}
\index{Fraunhofer diffraction}
The Fresnel approximation is valid when we may suitably approximate the emergent waves as parabolic. For propagation distances even further away we may instead choose to use planar waves, allowing us to further simplify the result of Fresnel diffraction. With the approximation of planar waves, the quadratic phase function in \eref{eq: ch1 Fresnel diffraction main} can be approximated as\index{Fraunhofer approximation}
\begin{equation*}
e^{j\frac{k}{2z}|\vxi|^2} \approx 1.
\end{equation*}
When such an approximation is valid, the Fresnel diffraction integral is simplified to
\begin{equation}
    U(\vx) = \frac{e^{jkz}e^{j \frac{k}{2z} |\vx|^2} }{j\lambda z} \iint_{-\infty}^{\infty}  U_\Sigma(\vxi)  e^{-j \frac{2\pi}{\lambda z} \vx^T\vxi} \; d\vxi,
\end{equation}
which we recognize as the Fourier transform of the incident field $U_\Sigma(\vx)$. This integral is known as the \keyword{Fraunhofer diffraction integral}.

\boxedthm{
\begin{theorem}
The \keyword{Fraunhofer diffraction integral} is defined as
\begin{equation}
    U(\vx) = \frac{e^{jkz}e^{j \frac{k}{2z} |\vx|^2} }{j\lambda z} \iint_{-\infty}^{\infty}  U_\Sigma(\vxi)  e^{-j \frac{2\pi}{\lambda z} \vx^T\vxi} d\vxi,
    \label{eq: ch1 Fraunhofer definition}
\end{equation}
or equivalently, in terms of Fourier transform:
\begin{equation}
U(\vx) = \frac{e^{jkz} e^{j \frac{k}{2z} |\vx|^2} }{j\lambda z}  \mathfrak{Fourier}
\big\{ U_\Sigma(\vxi) \big\}\bigg|_{\vf = \frac{\vx}{\lambda z}}.
\label{eq: ch1 Fraunhofer via Fourier}
\end{equation}
\end{theorem}
}

The Fraunhofer diffraction integral presents an important physical result. The behavior of a wave incident upon a diffracting screen can be modeled as a device that is closely related to the Fourier transform of the input wave (specifically, the part which passes through the screen). Furthermore, since the Fourier transform is well-studied, we may apply standard Fourier results and intuitions to these systems. We present two examples of Fraunhofer patterns below.

\boxedeg{
\vspace{2ex}
\textbf{Example}. (\keyword{Square Aperture}) Suppose that the incident field is given by the square aperture
\begin{equation}
U_{\Sigma}(\vxi) = \text{rect}\left(\frac{\xi}{2w_X}\right)\text{rect}\left(\frac{\eta}{2w_Y}\right),
\end{equation}
where $w_X$ and $w_Y$ are the half-widths of the aperture. Using \eref{eq: ch1 Fraunhofer via Fourier}, the Fraunhofer diffraction pattern is given by
\begin{align*}
U(\vx) = \frac{e^{jkz} e^{j\frac{k}{2z}|\vx|^2}}{j\lambda z} \mathfrak{Fourier}\Big\{ U_{\Sigma}(\vxi) \Big\}\bigg|_{f_x = \frac{x}{\lambda z}, f_y = \frac{y}{\lambda z}}.
\end{align*}
The Fourier transform of $U_{\Sigma}(\vxi)$ is
\begin{equation*}
\mathfrak{Fourier}\Bigg\{ \text{rect}\left(\frac{\xi}{2w_X}\right)\text{rect}\left(\frac{\eta}{2w_Y}\right) \Bigg\} = A \;  \text{sinc}(2w_X f_X) \text{sinc}(2w_Y f_Y),
\end{equation*}
where $A = 4w_Xw_Y$. Thus, the Fraunhofer diffraction pattern is
\begin{equation}
U(\vx) = \frac{e^{jkz} e^{j\frac{k}{2z}|\vx|^2}}{j\lambda z} A \; \text{sinc}\left(\frac{2w_X x}{\lambda z}\right) \text{sinc}\left(\frac{2w_Y y}{\lambda z}\right),
\label{eq: Ch1 Fraunhofer square}
\end{equation}
where we substituted $f_X = x/(\lambda z)$ and $f_Y = y/(\lambda z)$. \hfill $\square$
}

\begin{figure}[ht]
\centering
\includegraphics[width=0.95\linewidth]{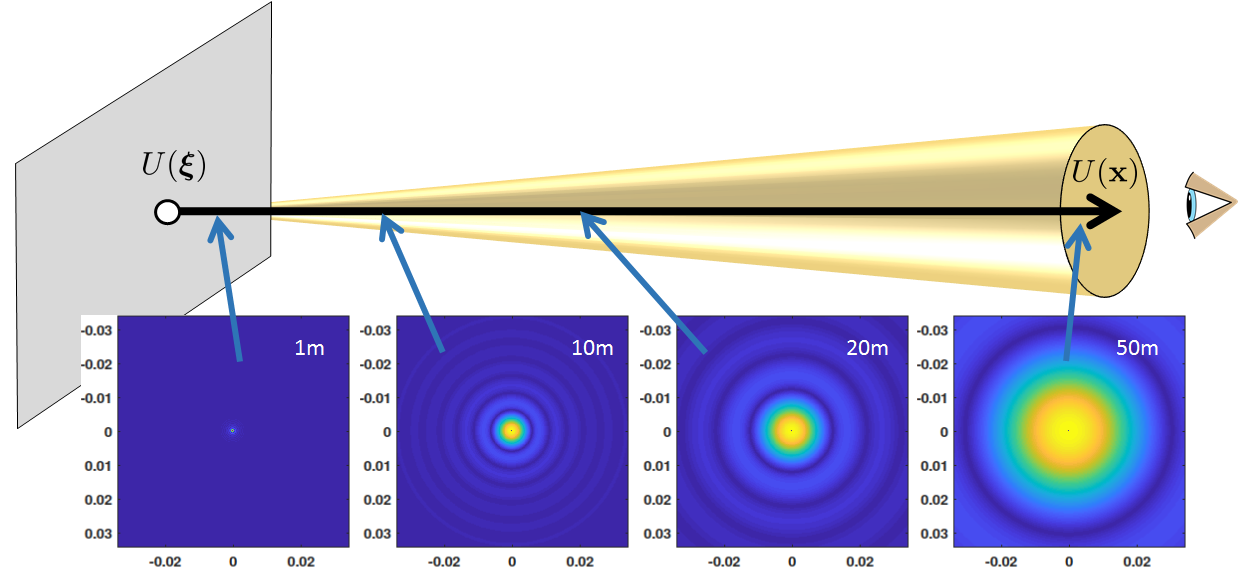}
\caption{Fraunhofer diffraction pattern of a circular aperture. Notice that the spread of the diffraction pattern grows as the wave propagates.}
\label{fig: Ch1 Fraunhofer circular}
\end{figure}

\boxedeg{
\vspace{2ex}
\textbf{Example}. (\keyword{Circular Aperture}) Consider the circular aperture function of aperture radius $D$:
\begin{equation}
    U_{\Sigma}(\rho) = \text{circ}\left( \frac{\rho}{D}\right) =
        \begin{cases}
    1, & \rho<D,\\
    \frac{1}{2}, & \rho = D,\\
    0, & \text{otherwise}.
    \end{cases}
\end{equation}
where $\rho = |\vxi| = \sqrt{\xi^2+\eta^2}$ is the radius in the object plane. Since the circular aperture is circularly symmetric, we perform the calculation in the polar coordinate. The Fraunhofer diffraction pattern is then given by the Fourier-Bessel transform $\mathfrak{Bessel}$:
\begin{align*}
U(r) = \frac{e^{jkz}}{j\lambda z} \exp\left\{ j \frac{kr^2}{2z}\right\} \mathfrak{Bessel}\Big\{ \text{circ}\left( \frac{\rho}{D} \right)\Big\}\Bigg|_{f = r/(\lambda z)},
\end{align*}
where $r = |\vx| = \sqrt{|\vx|^2}$ is the radius of the coordinate in the image plane. The Fourier-Bessel transform of the circular aperture is
\begin{equation}
\mathfrak{Bessel}\Big\{ \text{circ}\left( \frac{\rho}{D} \right)\Big\} = D^2 \frac{J_1(2\pi D f)}{D f} = 2 \cdot \pi D^2 \cdot \frac{J_1(2\pi D f)}{2 \pi D f}.
\label{eq: Ch1 Fraunhofer circular 1}
\end{equation}
Substituting $f = r/(\lambda z)$ into \eref{eq: Ch1 Fraunhofer circular 1} yields
\begin{equation}
U(r) = \frac{e^{jkz}}{j\lambda z} \exp\left\{ j \frac{kr^2}{2z}\right\} \cdot (\pi D^2) \cdot
\underset{\text{jinc}(r)}{ \underbrace{\left[ 2 \frac{J_1(2\pi D r / (\lambda z))}{ 2\pi D r / (\lambda z)}\right]}},
\label{eq: ch1_ex_jinc}
\end{equation}
where $J_1$ is the Bessel function of the first kind (first order). The Fraunhofer diffraction pattern is shown in \fref{fig: Ch1 Fraunhofer circular}. \hfill $\square$
}

\subsection{Diffraction Patterns}
Thus far, we have developed a way of describing the field as a result of incidence with a diffracting screen. If we were to place a sensor to detect the diffracted wave, we would not observe the complex wave directly. Instead, we would see the power per unit area.\index{diffraction! patterns}

The notion of power in our system is a straightforward consideration by the fact that our waves are modeled as sinusoidal. Therefore, the power at a position $\vx$ can be written as
\begin{align*}
P(\vx) &\propto \E[\vert A(\vx) \cos(2 \pi \nu t + \theta(\vx)) \vert^2 ], \\
&\propto |A(\vx)|^2 / 2,
\end{align*}
where the proportionality is a result of the quantum efficiency and other physical properties of the system and $\E[\cdot]$ denotes the statistical \keyword{expectation}. We will return to the reason for this expectation in a later subsection, for now we can just consider it as a time averaging operation. Ignoring the physical quantities defining the power, we define a function known as the \keyword{intensity}\index{intensity} of a wave to be proportional to the power \cite{Goodman_2005_a}. The intensity is given as
\begin{align}
I(\vx) &= \E[ | U(\vx) \exp\{-j2\pi\nu t\} |^2 ], \nonumber\notag \\
&= |U(\vx)|^2. \label{eq: Ch1 intensity definition basic}
\end{align}
Instead of referring to the power directly, we typically prefer intensity for its simplicity. Furthermore, we drop the expectation in \eqref{eq: Ch1 intensity definition basic} due to the time invariance of the phasor.

This result implies that the observed pattern on a screen or sensor will simply be proportional to the magnitude square of the phasor field. Therefore, our discussions resulting in Fresnel and Fraunhofer diffraction can be applied to imaging scenarios through the application of \eref{eq: Ch1 intensity definition basic}.

It is useful to note that in the above discussion, we have used the term intensity instead of \keyword{irradiance}\index{irradiance}, with the former term being the one used in Goodman's book. The units that we wish to describe are [W/m$^2$], whereas sometimes the term intensity corresponds to units [W/m$^2$/sr] where sr is known as a steradian. In this book we will use the term intensity to refer to power per unit area, thus being identical to the irradiance. This is done to be in coordination with Goodman and to adopt the terminology most often used by the computer vision and image processing communities.

\subsection{Diffraction Limit}
\begin{figure}[ht]
\centering
\includegraphics[width=0.5\linewidth]{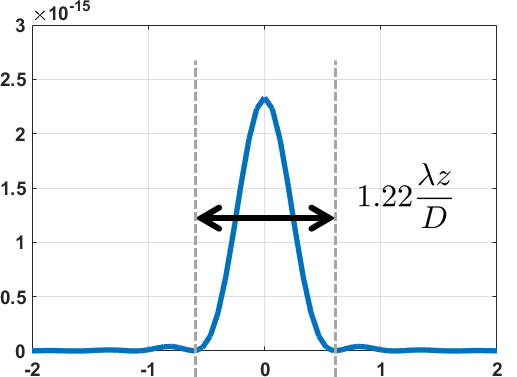}
\caption{The width of the central lobe of the Airy disc is given by $d = 1.22\lambda z / D$.}
\label{fig: Ch1 Airy disc}
\end{figure}

\index{diffraction! limit}
The Fraunhofer diffraction pattern derived for a circular aperture \eref{eq: ch1_ex_jinc} plays a significant role in imaging. If we consider the intensity of the field, we can show that
\begin{equation}
I(r) = |U(r)|^2 = \left(\frac{2\pi D^2}{\lambda z}\right)^2 \left[2 \frac{J_1\left(\frac{2\pi D r}{\lambda z}\right)}{\frac{2\pi D r}{\lambda z}}\right]^2.
\label{eq: Ch1 Airy disc}
\end{equation}
The intensity distribution $I(r)$ is called the \keyword{Airy disc}\index{Airy disc}. The shape of the Airy disc is shown in \fref{fig: Ch1 Airy disc}. We marked the width of the central lobe of the Airy disc, which is measured between the two zero-crossings:
\begin{equation}
d = 1.22 \frac{\lambda z}{D}.
\label{eq: Ch1 Rayleigh criteria}
\end{equation}

\begin{figure}[ht]
\centering
\begin{tabular}{ccc}
\hspace{-1ex}\includegraphics[width=0.3\linewidth]{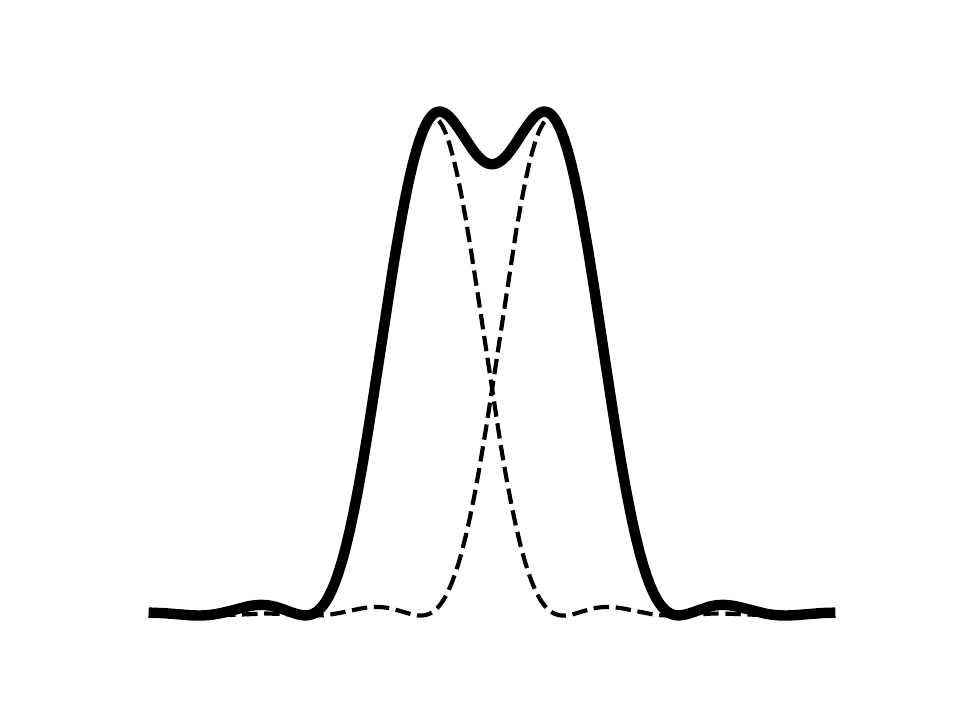}&
\hspace{-2ex}\includegraphics[width=0.3\linewidth]{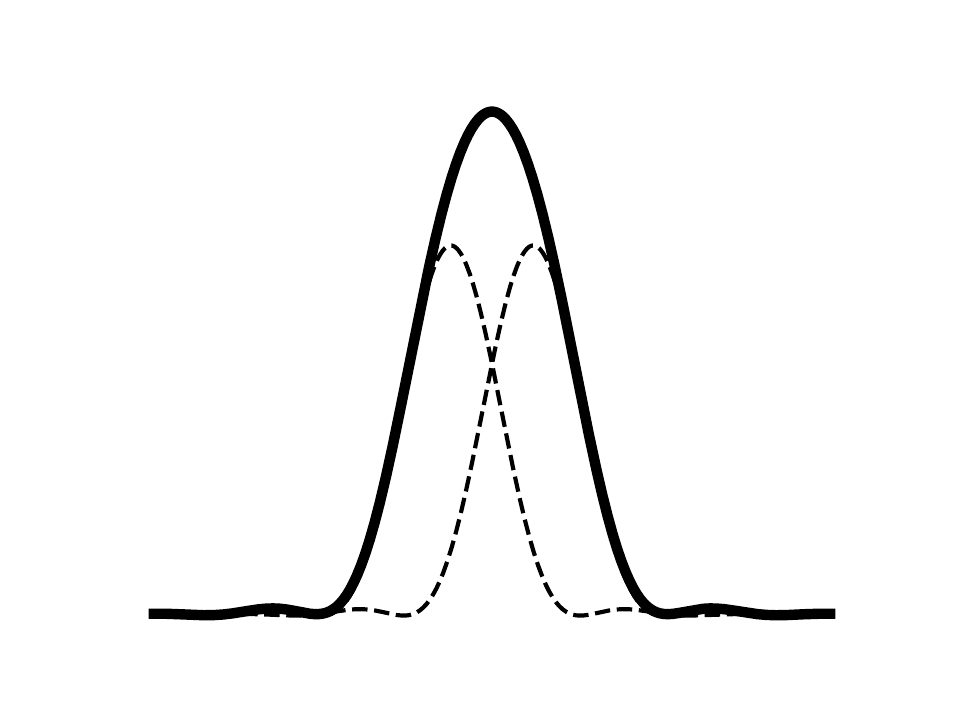}&
\hspace{-2ex}\includegraphics[width=0.3\linewidth]{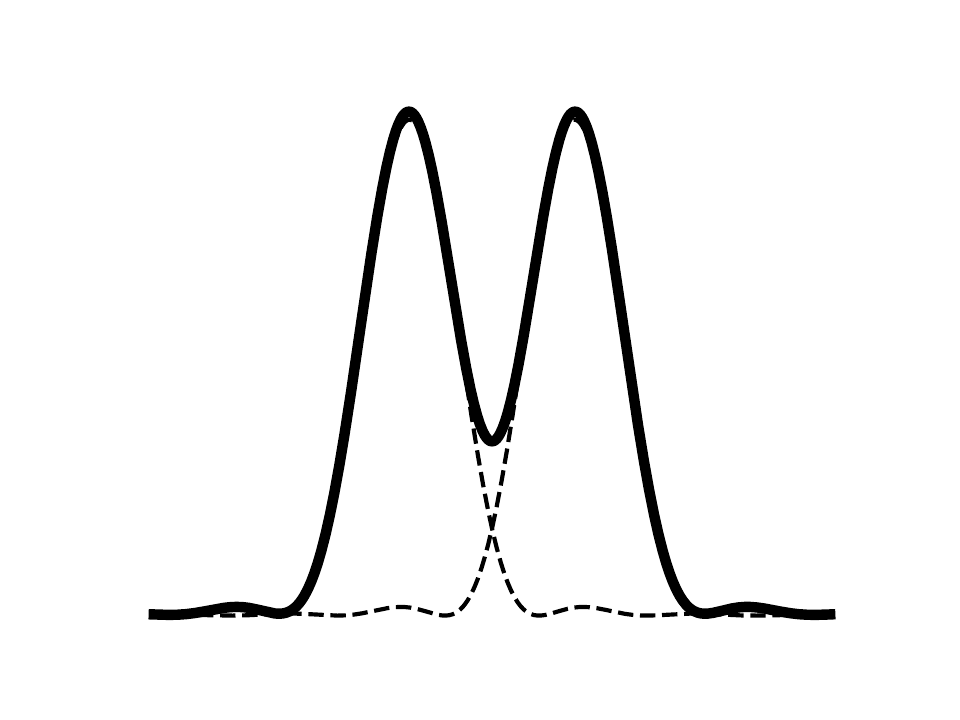}\\
(a) Rayleigh criteria & (b) unresolved & (c) resolved
\end{tabular}
\caption{As the wave leaves from a screen consisting of two pinholes, the superposition of the diffracted waves will make the two points unresolvable. The critical separation of the two pinholes is called the Rayleigh criteria.}
\vspace{-2ex}
\label{fig: Ch1 Rayleigh}
\end{figure}

Suppose that there is an aperture containing two pinholes where the centers are separated by a distance $\delta$. As the wave leaves these two pinholes, two diffraction patterns are generated. If $\delta$ is large, we expect that the two diffraction patterns barely overlap, and so we can resolve the two points. However, as $\delta$ reduces, we will eventually reach the point where the two diffraction patterns blend into each other to the point of being indistinguishable. Between these two extremes, there exists a critical state which corresponds to the minimum separation where the two patterns are still able to be resolved. At this threshold, we say that we have reached the \keyword{Rayleigh criteria}\index{Rayleigh criteria}. The situation is illustrated in \fref{fig: Ch1 Rayleigh}.

Mathematically, the minimum resolvable separation is defined at the point of half of the main lobe width,
\begin{equation}
\delta = 0.61 \frac{\lambda z}{D}.
\label{eq: Ch1 Rayleigh half}
\end{equation}
The factors that define \eref{eq: Ch1 Rayleigh half} are the wavelength $\lambda$, the propagation path $z$, and the aperture radius $D$. A short wavelength $\lambda$ causes the spread of the diffraction to be small. Thus, we can resolve the two points with a small $\delta$. A long propagation path $z$ increases the spread of the diffraction pattern. Thus, the two points are better resolved if $z$ is small. Due to the Fourier relationship of diffraction, a large aperture will shrink the size of the diffraction pattern, which will decrease $\delta$. The opposite case similarly increases $\delta$ through the same Fourier scaling relationship.

This leads us to introduce the \keyword{numerical aperture}\index{numerical aperture} (NA) as $\text{NA} = D/z$. The numerical aperture allows us to write the Rayleigh criteria as
\begin{equation}
\delta = 0.61 \frac{\lambda}{\text{NA}}.
\label{eq: Ch1 Rayleigh half 2}
\end{equation}
Specifically, the numerical aperture measures the half-angle subtended by the exit pupil when viewed from the image plane. NA is large when $z \ll D$.

\section{Lenses}
\label{sec: sec1_5}
An imaging system almost always contains a lens, and with good reason. Previously, our discussion of waves has taken place with diffracting screens and observations at \emph{long} distances. This feels different from the case of an imaging system where we are interested in relatively short distances. Fortunately, the way a thin lens affects propagation will allow us to use our previous results.

\subsection{Phase Distortion by a Lens}
\begin{figure}[ht]
\centering
\includegraphics[width=0.75\linewidth]{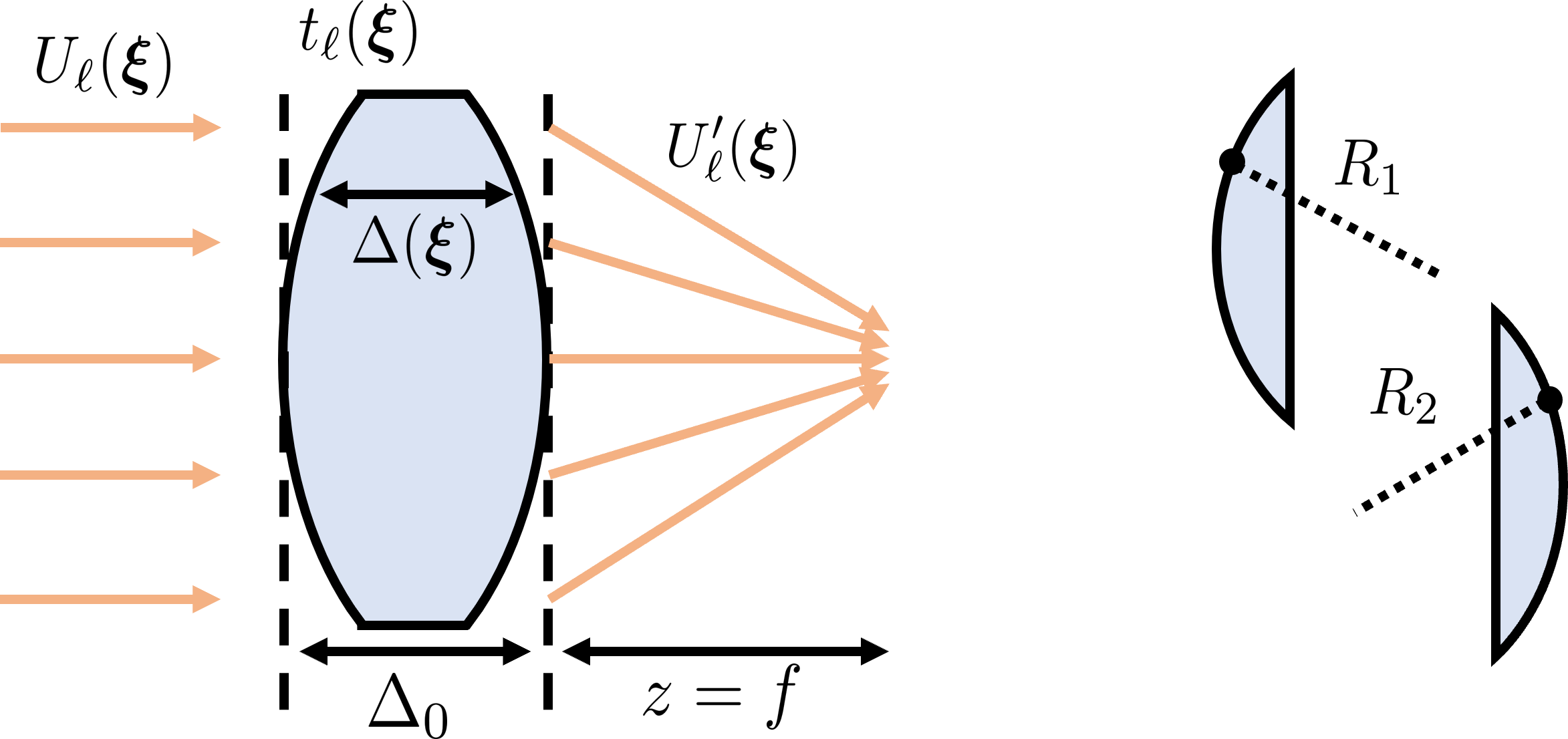}
\caption{The geometry of a lens. As a ray travels from the left to the right, the radius of the left surface is denoted as $R_1$, whereas the radius of the right surface is denoted as $R_2$.}
\label{fig: ch1 lens}
\end{figure}

A lens is typically composed of glass, thus having a higher index of refraction relative to free space. As a result of this relative increase, the speed of light is attenuated according to the index of refraction. Consider a lens with a shape outlined in \fref{fig: ch1 lens}. We denote the thickness of the lens at coordinate $\vxi$ as $\Delta(\vxi)$, and the maximum thickness as $\Delta_0$. The \keyword{phase delay} $\phi(\vxi)$ caused by the lens consists of two parts: the high refractive index region provided by the lens and the free space region which is anywhere outside the lens. The sum of the two terms gives the phase delay $\phi(\vxi)$
\begin{equation}
\phi(\vxi) = \frac{2\pi}{\lambda}\underset{\text{lens}}{\underbrace{n\Delta(\vxi)}} + \frac{2\pi}{\lambda} \underset{\text{free space}}{\underbrace{[\Delta_0 - \Delta(\vxi)]}},
\label{eq: Ch1 lens phase delay 1}
\end{equation}
where $n$ is the refractive index of the lens.

At this point, we have defined the function $\phi(\vxi)$, which we intend to apply to the wave. If we wish to apply it to the input field, i.e. the wave incident upon the lens surface, we face the problem of bending through the lens. However, if we consider the ``rays'' to enter and leave without deviating from their original path of propagation, then we can simply write the delay in the wave to be a function of \emph{only} the phase delay at a point $\vxi$.

This is the model of a \keyword{thin lens}\index{thin lens! model}\index{thin lens}, which is an idealized scenario in which the lens is thin enough such that any refractive effects can be approximated by the thickness. Continuing with our usage of the complex representation, we take the complex exponential of $\phi(\vxi)$ in \eref{eq: Ch1 lens phase delay 1}, giving us the \keyword{lens transformation}:\index{thin lens! transformation}
\begin{align}
t_\ell(\vxi)
&= \exp\left\{j\phi(\vxi) \right\} = \exp\left\{jk \Delta_0\right\}  \exp\left\{jk(n-1)\Delta(\vxi)\right\}.
\label{eq: Ch1 lens phase delay 2}
\end{align}
When an incident field $U_{\ell}(\vxi)$ reaches the lens, the lens changes its phase by multiplying $U_{\ell}(\vxi)$ with $t_{\ell}(\vxi)$:
\begin{equation}
U_{\ell}'(\vxi) = U_\ell(\vxi) P(\vxi) t_\ell(\vxi),
\label{eq: Ch1 lens phase delay 3}
\end{equation}
where $P(\vxi)$ is the \keyword{pupil function} such that
\begin{equation}
P(\vxi) =
\begin{cases}
1, &\qquad \vxi  \text{ is inside the aperture},\\
0, &\qquad \text{otherwise}.
\end{cases}
\label{eq: ch1 lens fresne3}
\end{equation}
The resulting field $U_{\ell}'(\vxi)$ is located immediately behind the lens, as shown in \fref{fig: ch1 lens}.

The lens in \fref{fig: ch1 lens} can be modeled as an intersection of two circles plus a constant, with radii $R_1$ and $R_2$ for left and right circles accordingly. Assuming $\vxi$ is sufficiently smaller than the two radii, the \keyword{paraxial approximation}\index{paraxial approximation} allows us to write the thickness of the lens as the parabolic function \cite{Goodman_2005_a}:
\begin{equation}
\Delta(\vxi) \approx \Delta_0 - \frac{|\vxi|^2}{2}\left(\frac{1}{R_1} - \frac{1}{R_2}\right).
\label{eq: ch1 lens thickness function}
\end{equation}
Combined with the previous definitions, if we define the \keyword{focal length}\index{focal length} $f$ of a lens to be
\begin{equation}
\frac{1}{f} = (n-1)\left(\frac{1}{R_1} - \frac{1}{R_2}\right),
\label{eq: ch1 focal length}
\end{equation}
we can write \eref{eq: Ch1 lens phase delay 3} as
\begin{equation}
U_\ell'(\vxi) = U_{\ell}(\vxi) P(\vxi) \exp\left\{-j\frac{k}{2f}|\vxi|^2\right\},
\label{eq: ch1 lens fresne2}
\end{equation}
where we have removed any constant phase terms.

As in the case of our analysis of diffraction, we are additionally interested in how the wave behaves as it moves away from the lens. For this particular problem, we are concerned with the distribution of the field at the plane of distance $z = f$. Application of Fresnel integral results in
\begin{align}
U_f(\mathbf{u})
&= \frac{e^{jkf}}{j\lambda f} e^{j\frac{k}{2f}|\mathbf{u}|^2} \iint_{-\infty}^{\infty} \left\{U_\ell'(\vxi)e^{j\frac{k}{2f}|\vxi|^2} \right\} e^{-j\frac{2\pi}{\lambda f} \vxi^T\mathbf{u}} \; d\vxi \notag\nonumber\\
&= \frac{e^{jkf}}{j\lambda f} e^{j\frac{k}{2f}|\mathbf{u}|^2} \iint_{-\infty}^{\infty} U_\ell(\vxi)P(\vxi) e^{-j\frac{2\pi}{\lambda f} \vxi^T\mathbf{u}} \; d\vxi.
\label{eq: ch1 lens fresne4}
\end{align}
Comparing with \eref{eq: ch1 Fraunhofer definition}, we realize that \eref{eq: ch1 lens fresne4} is the Fraunhofer diffraction pattern of the windowed field incident on the lens.

This is a somewhat surprising result. In our previous discussion of diffraction, the Fraunhofer pattern without a lens occurred a very far distance away! Now, it occurs at the \keyword{focal plane}\index{focal plane}. This highlights the key mathematical simplification by a lens, specifically an idealistic thin lens.

\subsection{Impulse Response of a Lens}
\label{sec: ch1 impulse response lens}
\begin{figure}[ht]
\centering
\includegraphics[width=0.8\linewidth]{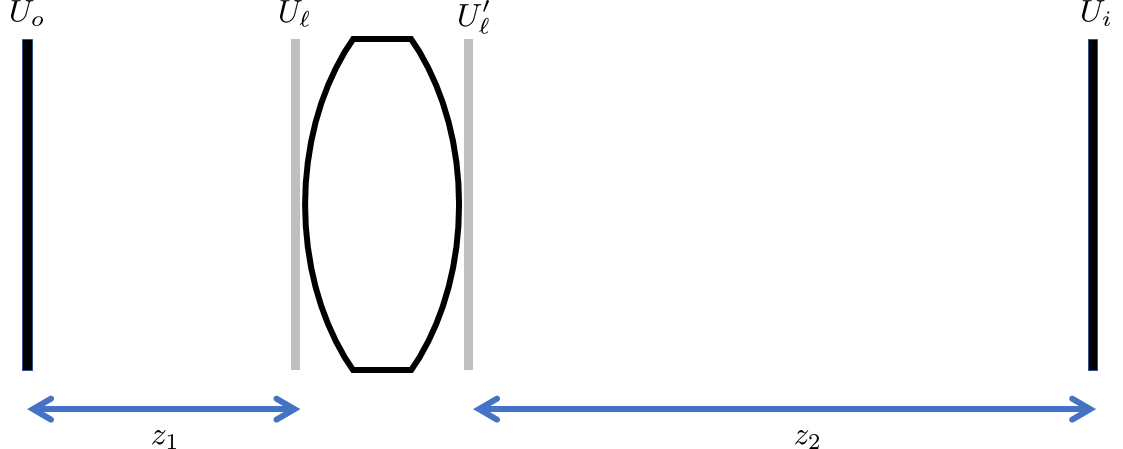}
\caption{Configuration for studying the impulse response of a lens, with the ``impulse'' being a field we specify at a distance.}
\label{fig: Ch1 lens_2}
\end{figure}

In the previous subsection, we discussed how the phase of an incident wave is transformed by a lens. In this subsection, we take a step further to study how the wave emitted from a point source at a specified distance is transformed by a lens. To analyze such an impulse response, one way is to derive the field observed at various stages as in \fref{fig: Ch1 lens_2}.

Starting from the left-hand side of \fref{fig: Ch1 lens_2}, we consider a \keyword{point source}\index{point source} as the input field $U_o$. A point source is described mathematically by a delta function:
\begin{equation}
U_o(\vx) = \delta(\vx-\vu),
\label{eq: Ch1 lens image 1}
\end{equation}
where $\vu$ specifies the center of the delta function $\delta(\cdot)$. Through operator notation, we wish to present the result of
\begin{equation}
U_i(\vx) = \mathfrak{Fresnel}[z_2] \left\{ t_\ell(\vxi) P(\vxi) \mathfrak{Fresnel}[z_1] \{ \delta(\vx - \vu) \} \right\},
\label{eq: Ch1 lens image 2}
\end{equation}
with $\mathfrak{Fresnel}[z]\{\cdot\}$ denoting Fresnel diffraction of $\{\cdot\}$ evaluated at a distance $z$.
For a detailed derivation, we would point the reader towards Goodman's Fourier Optics \cite{Goodman_2005_a}. We do, however, highlight the fact that within this derivation, the assumption most relevant to us is the assumption that the \keyword{lens law}\index{lens law}
\begin{equation}
\frac{1}{z_1} + \frac{1}{z_2} = \frac{1}{f},
\label{eq: Ch1 lens simplification 1}
\end{equation}
is satisfied. This effectively assumes that the object is being focused properly.

With simplifications as fully elaborated by \cite{Goodman_2005_a}, \eref{eq: Ch1 lens image 2} can be simplified to
\begin{align}
U_i(\vx)
&\approx \frac{1}{\lambda^2 z_1z_2} \iint_{-\infty}^{\infty} P(\vxi) \notag\\
&\qquad \times \exp\left\{ -jk \left[\left(\frac{u}{z_1} + \frac{x}{z_2}\right)\xi + \left(\frac{v}{z_1} + \frac{y}{z_2}\right)\eta \right]\right\} \; d\vxi.
\label{eq: Ch1 lens image 6a}
\end{align}
Defining the \keyword{magnification factor}\index{magnification} (where the minus sign accounts for the image inversion)
\begin{equation}
M = -z_2/z_1,
\end{equation}
it follows that
\begin{align}
\underset{h(\vx,\vu)}{\underbrace{U_i(\vx)}}
&\approx \frac{1}{\lambda^2 z_1z_2} \iint_{-\infty}^{\infty} P(\vxi) \exp\left\{ -j\frac{2\pi}{\lambda z_2} (\vx-M\vu)^T\vxi \right\} \; d\vxi.
\label{eq: Ch1 lens image 6}
\end{align}
This result leads us to define the \keyword{impulse response of a lens}.
\boxedthm{
\begin{theorem}[\keyword{Impulse Response of a Lens}]
Consider the optical setup shown in \fref{fig: Ch1 lens_2}. If the input is $U_o(\vx) = \delta(\vx-\vu)$, the output is the impulse response, defined as
\begin{align}
h(\vx,\vu)
&\approx \frac{1}{\lambda^2 z_1z_2} \iint_{-\infty}^{\infty} P(\vxi) \exp\left\{ -j\frac{2\pi}{\lambda z_2} (\vx-M\vu)^T\vu \right\} \; d\vxi.
\label{eq: Ch1 lens image 7}
\end{align}
where $M = -z_2/z_1$ and $P(\vu)$ is the aperture.
\end{theorem}
}

\subsection{The Amplitude Spread Function}
The impulse response in \eref{eq: Ch1 lens image 7} is so named because $h(\vx,\vu)$ is the outcome of the optical system with a lens when the input $U_o(\vu)$ is the delta function. For general input $U_o(\vu)$, the output is given by
\begin{equation}
U_i(\vx) = \iint_{-\infty}^{\infty} h(\vx,\vu) U_o(\vu) \; d\vu.
\label{eq: Ch1 psf 1}
\end{equation}
We can think of \eref{eq: Ch1 psf 1} as the superposition of the incident field $U_o$ at different locations $\vu$ weighted by $h(\vx,\vu)$. To make the equation useful, we present a sequence of normalization so that \eref{eq: Ch1 psf 1} becomes a convolution.

Inspecting \eref{eq: Ch1 lens image 7}, we consider a change of variables $\widetilde{\vu} = M\vu$, also noting $h(\vx, \vu)$ may be written as $h(\vx - \vutilde)$. Without loss of generality, we may set $\vutilde=0$, resulting in the definition of the \keyword{amplitude spread function (ASF)}\index{amplitude spread function}:
\boxedthm{
\begin{theorem}[\keyword{Amplitude spread function}]
\label{thm: ch1 lens convolution}
Consider the optical setup shown in \fref{fig: Ch1 lens_2}. The amplitude spread function of the system is
\begin{equation}
h(\vx)
= \frac{A}{\lambda z_2} \iint_{-\infty}^{\infty} P(\vxi) \exp\left\{ -j\frac{2\pi}{\lambda z_2} \vx^T\vxi \right\} \; d\vxi,
\label{eq: Ch1 theorem lens convolution 2}
\end{equation}
which is the Fourier transform of $P(\vxi)$ evaluated at frequencies $(\tfrac{x}{\lambda z_2}, \tfrac{y}{\lambda z_2})$ (multiplied with a constant) and with $A = 1/(\lambda z_1)$, though often for notational convenience, we can set the constant $\tfrac{A}{\lambda z_2}$ to unity.
The resulting field formed in the image plane is
\begin{align}
U_i(\vx) = \widetilde{h}(\vx) \circledast U_g(\vx),
\label{eq: Ch1 theorem lens convolution 3}
\end{align}
where $\circledast$ denotes the 2D convolution, with $\widetilde{h}(\vx) = \tfrac{1}{M}h(\vx)$ and $U_g(\vutilde) = \frac{1}{|M|}  U_o\left(\frac{\vutilde}{M}\right)$.
\end{theorem}
}
The result in \eref{eq: Ch1 theorem lens convolution 3} is significant for a few reasons:
\begin{itemize}
\setlength\itemsep{0ex}
\item The incident field $U_g(\vutilde)$ is related to what we will soon introduce as the \keyword{ideal image} produced by \keyword{geometrical optics}\index{geometrical optics}. Except for the magnification factor $|M|$, there are no other distortions caused by the optics.
\item In the presence of a finite-aperture lens, \eref{eq: Ch1 theorem lens convolution 3} implies that diffraction will cause a \keyword{diffraction-limited blur} via the convolution with $h(\vx)$.
\item To eliminate the effect of the diffraction-limited blur, one can make the aperture much bigger than the image sensor. According to \eref{eq: Ch1 theorem lens convolution 2}, a pupil function with an infinitely large radius will turn $h(\vx)$ to the delta function. In this case, the observed field $U_i$ in \eref{eq: Ch1 theorem lens convolution 3} is the geometrical optics result $U_g$.
\item The ASF is an intrinsic property of an optical system. It is determined by the wavelength $\lambda$, propagation distance $z_2$, and the geometry of the aperture $P(\vx)$. For example, a square aperture has a different ASF compared to a circular aperture.
\end{itemize}

\section{Image Formation}
\label{sec: sec1_6}
Thus far discussions have primarily been about the electromagnetic field. In \cref{sec: sec1_6}, we will further discuss how \emph{images} are formed. Towards the end of \cref{sec: sec1_6}, we will introduce the concepts of \keyword{amplitude transfer function} and \keyword{optical transfer function}, two major concepts within Fourier optics.

\subsection{Coherent and Incoherent Imaging}
The images we observe (by our eyes or by image sensors) are the intensities of the electromagnetic fields. Using the definitions of $U_i(\vx)$ in \eref{eq: Ch1 theorem lens convolution 3} of Theorem~\ref{thm: ch1 lens convolution} and intensity in \eref{eq: Ch1 intensity definition basic}, we consider the image intensity written in terms of the source phasor:
\begin{align}
I_i(\vx)
&= \E\left[|U_i(\vx)|^2\right] = \E\left[|\widetilde{h}(\vx) \circledast U_g(\vx)|^2\right] = \E\left[ \left|\int \widetilde{h}(\vx-\vutilde) U_g(\vxi) \; d\vutilde \right|^2 \right] \notag\nonumber \\
&= \iint \widetilde{h}(\vx-\vutilde_1) \widetilde{h}^*(\vx-\vutilde_2) \E\left[U_g(\vutilde_1)U_g^*(\vutilde_2)\right] d\vutilde_1 d\vutilde_2 ,
\label{eq: ch1 coherent incoherent 0}
\end{align}
where $\E[\cdot]$ denotes the expectation and $(\cdot)^*$ denotes the complex conjugate. For the proceeding discussions, we will no longer consider the waves to be fully deterministic; the waves we will now consider will be stochastic in nature. Let us ask: where does the randomness come from?

\begin{figure}[ht]
\centering
\includegraphics[width=\linewidth]{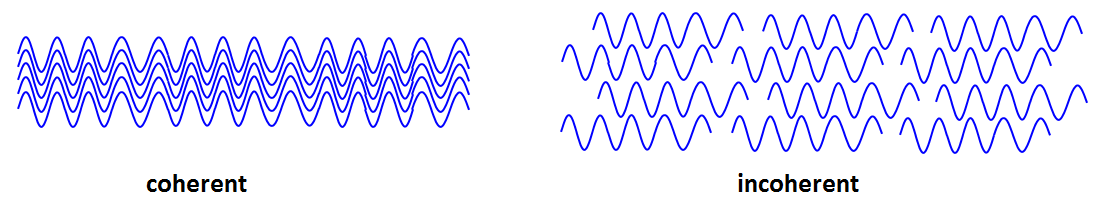}
\caption{A coherent source generates trains of continuous waves where the phases are completely correlated. Note, however, that two correlated waves do not necessarily have the same initial values. They can be $180^\circ$ out of phase but correlated. An incoherent source generates waves where the instantaneous phases are uncorrelated. One way to understand them is to visualize the waves as functions with discontinuous phases. Source: [\url{https://www.schoolphysics.co.uk/age16-19/Wave\%20properties/Wave\%20properties/text/Coherent\%20and\%20incoherent/index.html}}
\label{fig: ch1 coherent incoherent}
\end{figure}

A distributed light source generates many waves simultaneously. Depending on the type of the light source, the \keyword{instantaneous phases} of the waves are \keyword{correlated}\index{correlation} to different extents. The correlation of the phases is described through the concept of \keyword{coherence}\index{coherence}. Coherence can be measured in time, space, and spectrum. For the purposes of this book, we shall focus on \keyword{spatial coherence}\index{coherence! spatial}. We are mostly interested in two types of light sources: coherent sources and incoherent sources. \fref{fig: ch1 coherent incoherent} shows an example of both.

The core difference between a coherent source (and thus a coherent illumination) and an incoherent source needs to be traced back to how the waves are generated. If we have a laser exciting electrons to emit energy synchronously, the resulting electromagnetic fields will be trains of continuous waves with a constant phase profile as shown in \fref{fig: ch1 coherent incoherent}. (Note, however, that the initial values of the phases do not need to be identical. For example, one wave can be $180^\circ$ out of phase with another wave but they propagate at the same frequency and direction.) We say that such illumination is \keyword{coherent} because the phases of the two waves are completely correlated --- if we know how the first wave propagates, we also know how the other waves propagate (assuming that we know the initial phases.) A coherent illumination is obtained when the waves are originated from the same source.

An \keyword{incoherent}\index{incoherent} illumination has the opposite behavior. Oftentimes, an incoherent illumination is obtained when the source produces waves randomly. One way to think of these randomly generated waves is that they are ``broken trains of waves'' instead of a continuous wave. Many light sources such as a tungsten bulb generate waves in such a way because electrons at different parts of the tungsten emit energy differently. When waves are emitted randomly in an independent fashion, the instantaneous phases are uncorrelated. Most of our everyday experience is illuminated by such incoherent sources.

Returning to \eref{eq: ch1 coherent incoherent 0}, we say the spatial coherence of two waves is measured using the \keyword{mutual intensity}\index{mutual intensity}:
\begin{equation}
J_g(\vutilde_1,\vutilde_2) = \E\left[ U_g(\vutilde_1) U_g^*(\vutilde_2) \right].
\label{eq: Ch1 mutual intensity}
\end{equation}
The way to understand \eref{eq: Ch1 mutual intensity} is to treat the two incident fields $U_g(\vutilde_1)$ and $U_g^*(\vutilde_2)$ as \keyword{random variables}\index{random variable} drawn spatially from the same random process $U_g$. The expectation in \eref{eq: Ch1 mutual intensity}, which applies to the product of the two terms, is the correlation of the two samples.\footnote{For readers who are less familiar with random processes, we recall that the autocorrelation function $R_{X}(\xi_1,\xi_2)$ of a random process $X(\xi)$ is defined as $R_X(\xi_1,\xi_2) = \E[X(\xi_1)X(\xi_2)]$. The two instants $X(\xi_1)$ and $X(\xi_2)$ are two random variables; they may or may not be correlated. Nevertheless, the joint expectation $\E[X(\xi_1)X(\xi_2)]$ is well-defined. Mapping the 1D random process $X(\xi)$ to a 2D random process $U_g(\vxi)$, $J_g$ is therefore the autocorrelation function of $U_g$.}

As we have stated, coherent illumination exhibits perfect correlation in the waves. Given that the complex envelope of a wave may be written $U_g(\vx) = A_g(\vx) \exp\{ -j\phi(\vx)\}$, we may say that for a coherent wave the following is true:
\begin{equation}
\phi(\vutilde_1) - \phi(\vutilde_2) = \phi(\vutilde_2 - \vutilde_1).
\end{equation}
In this case, the phase difference is a \keyword{deterministic} function of its separation. The expectation accordingly is dropped, with mutual intensity function
\begin{align}
J_g(\vutilde_1,\vutilde_2) &= A_g(\vutilde_1) A_g(\vutilde_2) \exp\{ j\phi(\vutilde_2 - \vutilde_1) \}, \notag\nonumber\\
&= U_g(\vutilde_1) U_g^*(\vutilde_2).
\label{eq: ch1 coherent Jg}
\end{align}
Substituting \eref{eq: ch1 coherent Jg} in \eref{eq: ch1 coherent incoherent 0}, $I_i(\vx)$ becomes
\begin{align}
I_i(\vx)
&= \iint \widetilde{h}(\vx-\vutilde_1) \widetilde{h}^*(\vx-\vutilde_2) U_g(\vutilde_1)U_g^*(\vutilde_2) d\vutilde_1 d\vutilde_2 \notag \\
&= \left|\iint \widetilde{h}(\vx-\vutilde_1) U_g(\vutilde_1) \; d\vutilde_1\right|^2 = \left| \widetilde{h}(\vx) \circledast U_g(\vx) \right|^2.
\label{eq: ch1 I_i coherent}
\end{align}
Therefore, for coherent light sources, the observed image $I_i(\vx)$ is the magnitude square of the convolved signal $\widetilde{h}(\vx) \circledast U_g(\vx)$. Compared with \eref{eq: ch1 coherent incoherent 0}, we notice that the only difference is the removal of the expectation operator $\E[\cdot]$.

\boxedthm{
\begin{theorem}[\keyword{Coherent Image Formation}]\index{image formation! coherent}
\label{thm: ch1 coherent}
If the light source is \keyword{coherent}, then the observed image $I_i(\vx)$ is
\begin{equation}
I_i(\vx)
= \left| \widetilde{h}(\vx) \circledast U_g(\vx) \right|^2.
\label{eq: ch1 I_i coherent main}
\end{equation}
\end{theorem}
}

When the source is incoherent, $U_g(\vx)$ will be an independent random process. Therefore, two spatial samples $U_g(\vutilde_1)$ and $U_g^*(\vutilde_2)$ are two independent random variables. The mutual intensity function is then given by
\begin{equation}
J_g(\vutilde_1,\vutilde_2) = A_g(\vutilde_1) A_g(\vutilde_2)
\E[\exp\{ j(\phi(\vutilde_2) - \phi(\vutilde_1)) \} ].
\label{eq: ch1 incoherent Jg 0}
\end{equation}
Further assuming that the phase function is uniformly distributed as $\phi(\vx) \sim [0, 2\pi)$, we can show that
\begin{align}
J_g(\vutilde_1,\vutilde_2)
&=
\begin{cases}
I_g(\vutilde_1) &\qquad \mbox{if} \;\; \vutilde_1 = \vutilde_2, \\
0, &\qquad \mbox{otherwise},
\end{cases}\notag\\
&= I_g(\vutilde_1) \delta(\vutilde_1-\vutilde_2),
\label{eq: ch1 incoherent Jg}
\end{align}
where we defined
\begin{equation}
I_g(\vutilde_1) \; \bydef \; \E\left[ |U_g(\vutilde_1)|^2 \right].
\label{eq: ch1 I_g}
\end{equation}
In other words, $J_g(\vutilde_1,\vutilde_2)$ is either $I_g(\vutilde_1)$ (when the coordinates $\vutilde_1$ and $\vutilde_2$ coincide) or zero (when the coordinates are different). Substituting \eref{eq: ch1 incoherent Jg} in \eref{eq: ch1 coherent incoherent 0}, we show that
\begin{align}
I_i(\vx)
&= \iint \widetilde{h}(\vx-\vutilde_1) \widetilde{h}^*(\vx-\vutilde_2)  \Big[ I_g(\vutilde_1) \delta(\vutilde_1-\vutilde_2) \Big] d\vutilde_1 d\vutilde_2 \notag \\
&= \iint  |\widetilde{h}(\vx-\vutilde_1)|^2 I_g(\vutilde_1) d\vutilde_1 = \left|  \widetilde{h}(\vx) \right|^2 \circledast I_g(\vx).
\label{eq: ch1 I_i incoherent}
\end{align}
Therefore, the observed image is the convolution of $\left|  \widetilde{h}(\vx) \right|^2$ and $\left| U_g(\vx) \right|^2$.

\boxedthm{
\begin{theorem}[\keyword{Incoherent Image Formation}]\index{image formation! incoherent}
\label{thm: ch1 incoherent}
If the light source is incoherent, then the observed image $I_i(\vx)$ is
\begin{equation}
I_i(\vx)
= \left| \widetilde{h}(\vx) \right|^2 \circledast I_g(\vx).
\label{eq: ch1 I_i incoherent main}
\end{equation}
\end{theorem}
}
\eref{eq: ch1 I_i incoherent main} describes a fundamental relationship we will consider throughout the course of this book. This equation describes the relationship between the input field's intensity and the output fields intensity through the quantity $\left|  \widetilde{h}(\vx) \right|^2$, which we refer to as the \keyword{point spread function (PSF)}\index{point spread function}.

\subsection{Imaging as a Space-Invariant System}
The PSF brings us back to our intent which was stated towards the beginning of this Chapter, to write an imaging system with a lens in a standard LSI fashion. So much of computer vision and image processing takes place using incoherent imaging, thus the PSF represents the quantity that will describe a majority of imaging systems of interest to the average computer vision researcher. Furthermore, the goal of Chapter 2 will be to develop the proper background to properly describe a turbulent PSF.

We now wish to interpret \eref{eq: ch1 I_i incoherent main} beginning with recalling \eref{eq: Ch1 theorem lens convolution 2}:
\begin{equation*}
h(\vx)
= \frac{A}{\lambda z_2} \iint_{-\infty}^{\infty} P(\vxi) \exp\left\{ -j\frac{2\pi}{\lambda z_2} \vx^T\vxi \right\} \; d\vxi,
\end{equation*}
and $h(\vx) = M\widetilde{h}(\vx)$. We can observe this result is merely the Fraunhofer pattern  \eref{eq: Ch1 theorem lens convolution 2} evaluated at distance at $z_2$ and with leading complex terms removed. Equivalently, we may write this using the Fourier transform as
\begin{equation}
h(\vx)
= \frac{A}{\lambda z_2} \mathfrak{Fourier}
\big\{ P(\vxi) \big\}\bigg|_{\vf = \frac{\vx}{\lambda z_2}}.
\end{equation}
Examining \eref{eq: ch1 I_i incoherent main}, we may then think of incoherent image formation in the following way. First, we form the geometric optics predicted image. This is simply ray tracing; each point will map to a single point on the imaging plane with no effects by diffraction. Second, we take this geometrically predicted image and blur it according to the magnitude-squared diffraction pattern.

\begin{figure}
\centering
\begin{tabular}{ccc}
\includegraphics[width=0.4\linewidth]{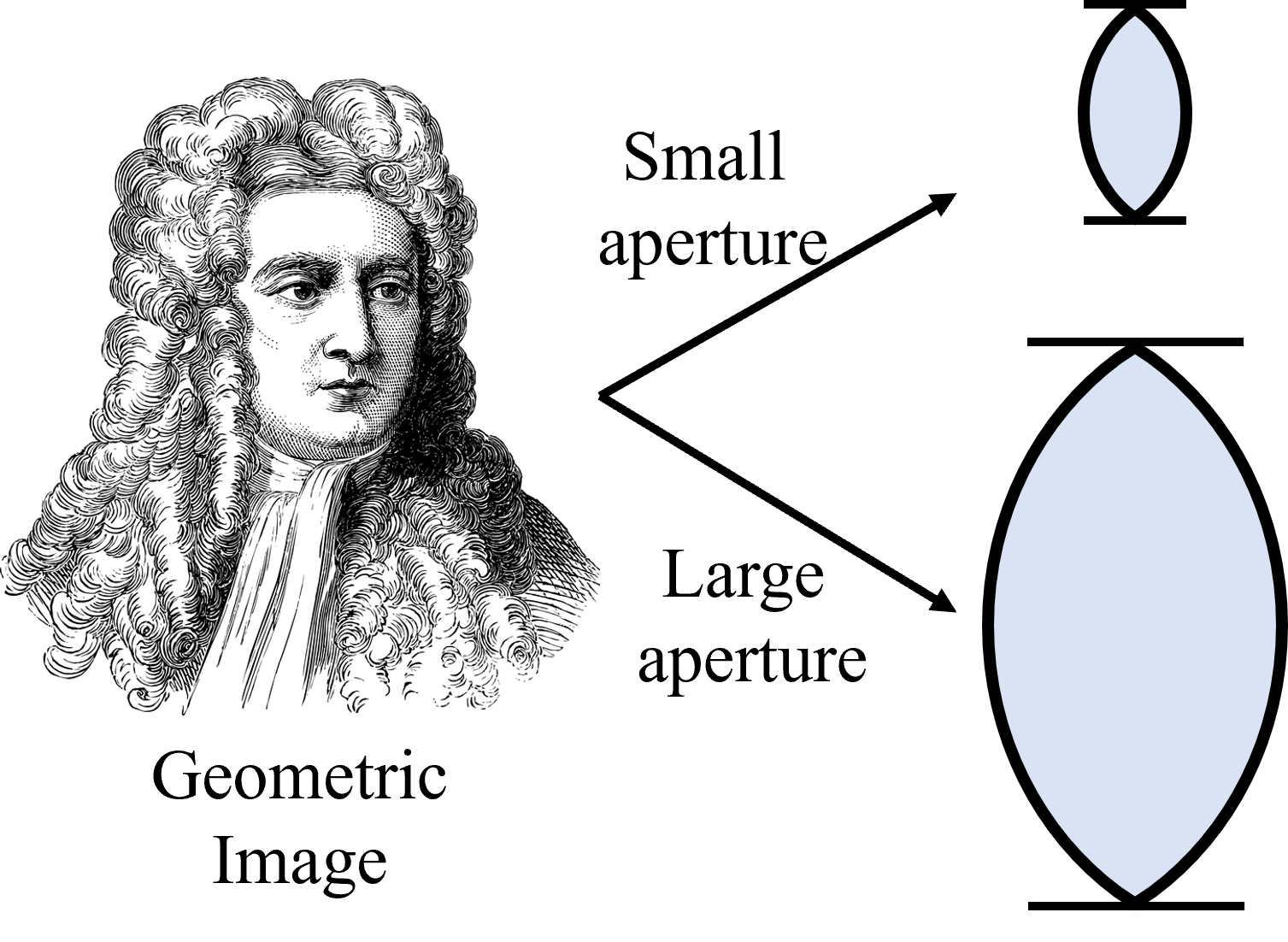} &
\includegraphics[width=0.25\linewidth]{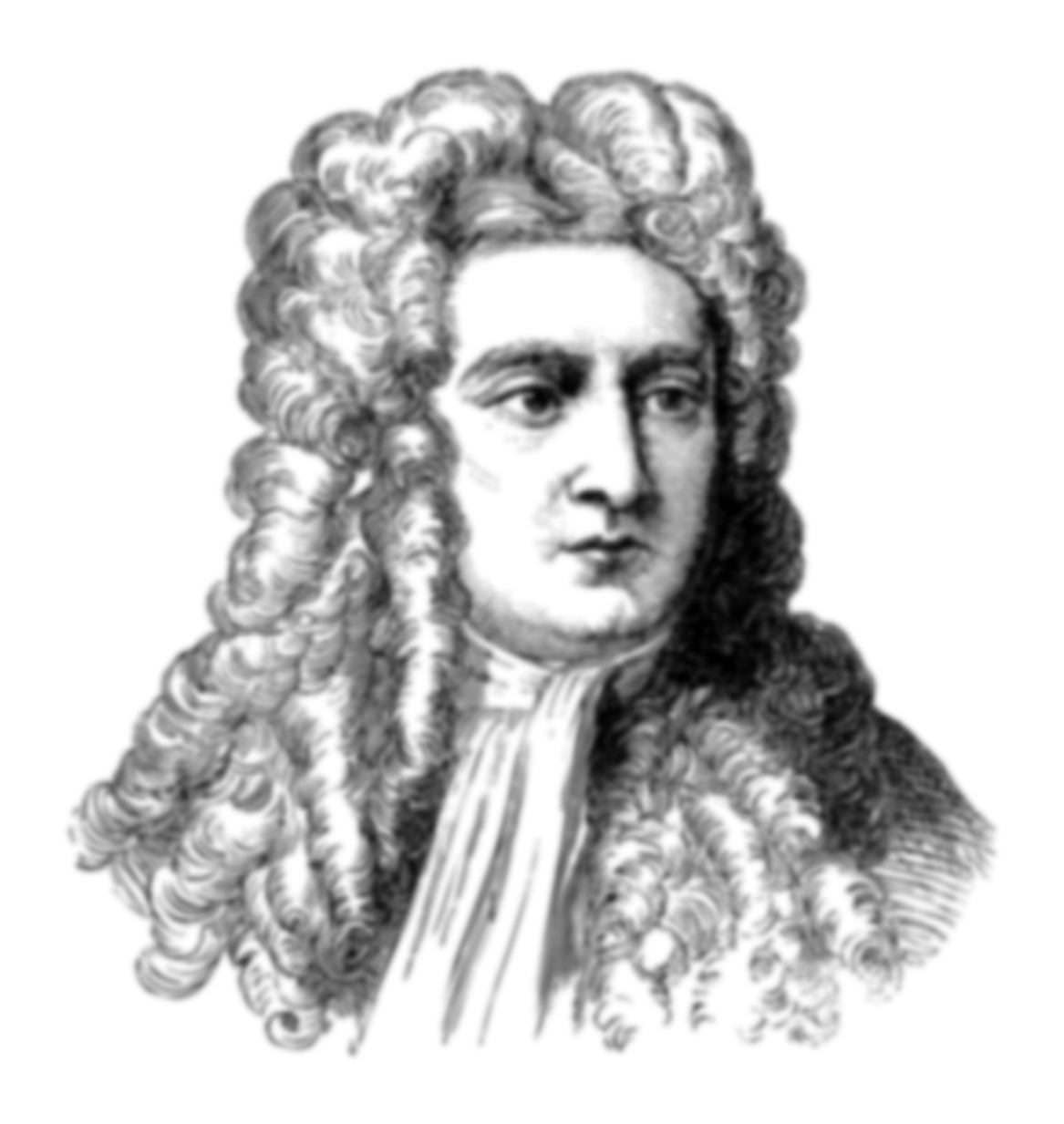} &
\includegraphics[width=0.25\linewidth]{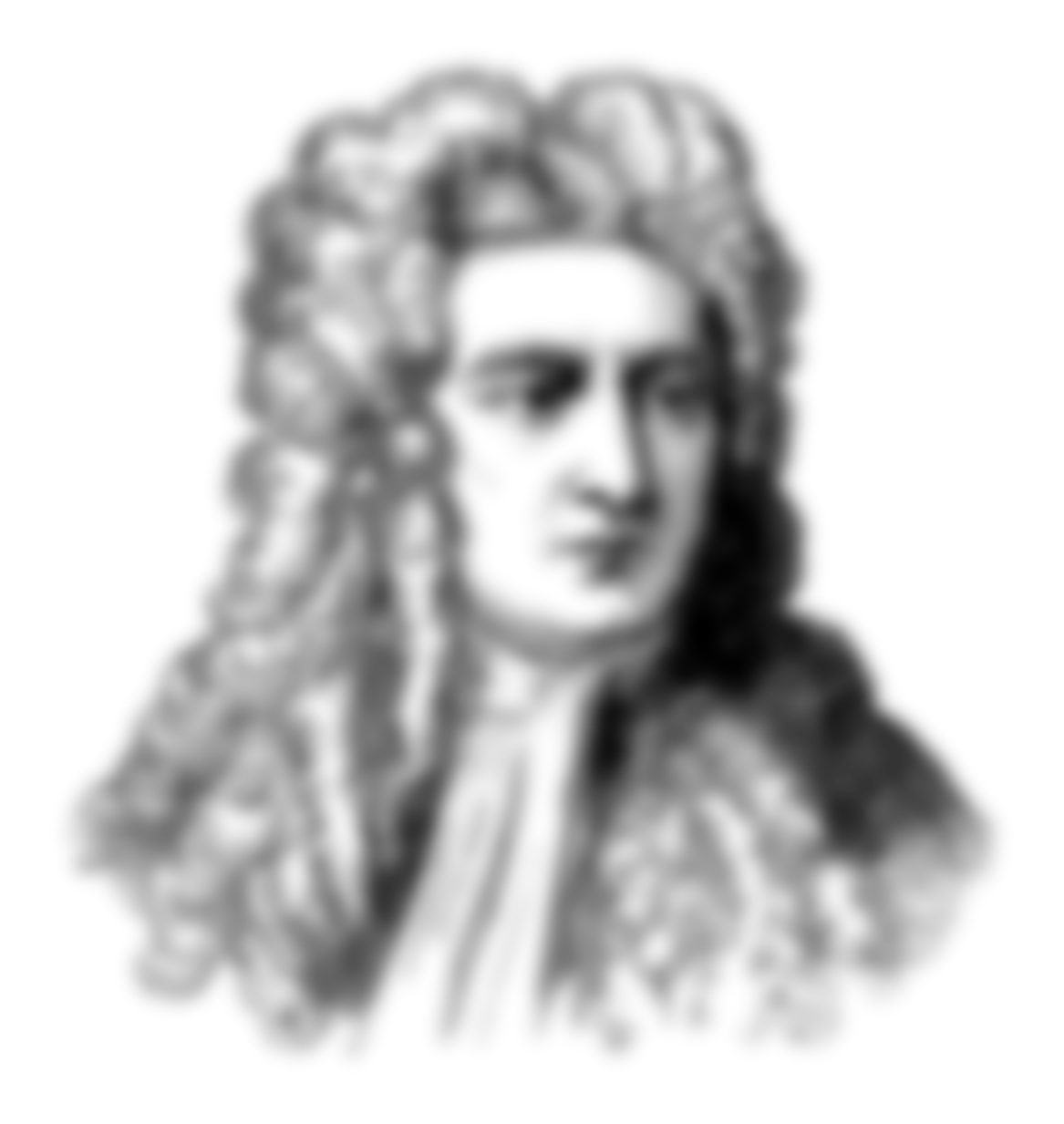} \\
(a) Two imaging systems & (b) Large aperture & (c) Small aperture
\end{tabular}
\caption{If we have a geometric image such as a perfect intensity representation of Newton (a), two imaging systems will view it differently. (b) A larger aperture will produce a smaller PSF while (c) a small aperture produces a larger PSF.}
\label{fig: newton}
\end{figure}

A typical lens-aperture system blurs the ``ideal'' image by a diffraction kernel. Mathematically, we may write the image formation by a perfectly in-focus system as
\begin{equation*}
I_i(\vx) = |\mathfrak{Fourier} \{ P(\vxi) \}|^2 \circledast I_g(\vx).
\end{equation*}
We will typically leave off such terms which correspond to the resizing of the Fourier transform for brevity as this convolution should be understood to be done according to the proper system parameters.
This accounts for the diffraction limit by the blur acting as a low pass filter. Due to the scaling property of the Fourier transform, a larger aperture corresponds to a smaller blur, which we present a visualization of this in \fref{fig: newton}.

This leads us to consider the case in which the imaging system is not perfectly in-focus. An assumption within our development of \eref{eq: ch1 I_i incoherent main} is that our system is focused, assuming that lens law (see \eref{eq: Ch1 lens simplification 1}) was satisfied. Let us remove this assumption, instead assuming there is a slight focusing error such that\index{defocus}
\begin{equation}
\frac{1}{z_1} + \frac{1}{z_2} - \frac{1}{f} = \epsilon.
\end{equation}
The resulting impulse response will then be given by \cite{Goodman_2005_a}
\begin{equation}
h(\vx)
= \frac{A}{\lambda z_2} \iint_{-\infty}^{\infty} P(\vxi)
\exp \left\{j\frac{k\epsilon}{2}|\vxi|^2 \right\}
\exp\left\{ -j\frac{2\pi}{\lambda z_2} \vx^T\vxi \right\} \; d\vxi.
\label{eq: Ch1 defocus}
\end{equation}
Using this in place of our derivation for incoherent imaging, the image will then be formed according to
\begin{equation*}
I_i(\vx) = |\mathfrak{Fourier} \{ P(\vxi) e^{j k \epsilon |\vxi|^2 / 2} \}|^2 \circledast I_g(\vx).
\end{equation*}
More generally, we may write\index{image formation! with phase errors}
\begin{equation}
I_i(\vx) = |\mathfrak{Fourier} \{ P(\vxi) e^{j k \phi(\vxi)} \} |^2 \circledast I_g(\vx),
\end{equation}
where $\phi(\vxi)$ gives the path-error in meters. We will return to this concept in more detail for the purposes of imaging through atmospheric turbulence.

\subsection{Amplitude Transfer Function}
The image formation equations for a coherent system in Theorem~\ref{thm: ch1 coherent} and an incoherent system in Theorem~\ref{thm: ch1 incoherent} have one thing in common: they are both expressed using a convolution. The underlying mathematical assumption we made is that the image formation process is spatially invariant and so the system is determined by the impulse response. For such a system, it is also common to describe image formation in the Fourier domain. To see how this can be made possible, we define
\begin{align*}
G_g(\vf) \bydef \mathfrak{Fourier}\{U_g(\vx)\} = \int_{-\infty}^{\infty} U_g(\vx) \exp\left\{-j2\pi\vf^T\vx\right\} \; d\vx\\
G_i(\vf) \bydef \mathfrak{Fourier}\{U_i(\vx)\} = \int_{-\infty}^{\infty} U_i(\vx) \exp\left\{-j2\pi\vf^T\vx\right\} \; d\vx,
\end{align*}
where $\vf = [f_X,f_Y]$ is the frequency. In addition, we define the amplitude transfer function to be the Fourier transfer of the ASF:
\boxedthm{
\begin{definition}[\keyword{Amplitude transfer function}\index{amplitude transfer function}]
The amplitude transfer function is
\begin{align*}
H(\vf) \bydef \mathfrak{Fourier}\{h(\vx)\} = \int_{-\infty}^{\infty} h(\vx) \exp\left\{-j2\pi\vf^T\vx\right\} \; d\vx.
\end{align*}
\end{definition}
}

For \keyword{coherent imaging systems}\index{imaging system! coherent}, the observed field is given by (see \eref{eq: Ch1 theorem lens convolution 3})
\begin{equation}
U_i(\vx) = h(\vx) \circledast U_g(\vx),
\label{eq: Ch1 coherent field}
\end{equation}
where we have dropped the constant phase terms. Taking the magnitude squares on both sides, we obtain
\begin{equation}
I_i(\vx) = |h(\vx) \circledast U_g(\vx)|^2.
\label{eq: Ch1 coherent intensity}
\end{equation}
Therefore, for coherent illumination, the observed intensity is the magnitude square of the convolution.

Going back to \eref{eq: Ch1 coherent field}, the standard property of the Fourier analysis tells us that
\begin{equation}
G_i(\vf) = H(\vf) G_g(\vf).
\end{equation}
The amplitude transfer function, according to \eref{eq: Ch1 theorem lens convolution 2} is therefore
\begin{align*}
H(\vf)
&\bydef \mathfrak{Fourier}\left\{ \frac{A}{\lambda z_2} \int_{-\infty}^{\infty} P(\vxi) \exp\left\{ -j\frac{2\pi}{\lambda z_2} \vx^T\vxi \right\} \; d\vxi  \right\}\\
&= \int_{-\infty}^{\infty} \left\{ \frac{A}{\lambda z_2} \int_{-\infty}^{\infty} P(\vxi) e^{ -j\frac{2\pi}{\lambda z_2} \vx^T\vxi } \; d\vxi \right\} e^{-j2\pi\vf^T\vx} \; d\vx
\end{align*}
which is nothing but taking Fourier transform twice. It is not difficult to show that a function $f(\vx)$ transformed twice by the Fourier transform is the flipped version $f(-\vx)$. Therefore,
\begin{align*}
H(\vf) = P(-\lambda z_2 \vf) = P(\lambda z_2 \vf),
\end{align*}
where we ignored the constant $A/(\lambda z_2)$ and used the property of a symmetry circular pupil function that $P(\vx) = P(-\vx)$. Therefore, the amplitude transfer function is essentially the pupil function with a scaled coordinate.

\boxedeg{
\vspace{2ex}
\textbf{Example}. Suppose that the pupil function is
\begin{equation}
P(\vxi) = \text{rect}\left(\frac{\xi}{2w}\right) \text{rect}\left(\frac{\eta}{2w}\right).
\end{equation}
The amplitude transfer function is just the pupil function with a scaled coordinate
\begin{align*}
H(\vf)
&= P(\lambda z_2 \vf) \\
&= \text{rect}\left(\frac{\lambda z_2 f_\xi}{2w}\right) \text{rect}\left(\frac{\lambda z_2 f_\eta}{2w}\right).
\end{align*}
}

\subsection{Optical Transfer Function}
For \keyword{incoherent imaging systems}\index{imaging system! incoherent}, the observed intensity $I_i(\vx)$ follows from \eref{eq: ch1 I_i incoherent main} that
\begin{align}
I_i(\vx)
&= \left| h(\vx) \right|^2 \circledast I_g(\vx) \notag\\
&= \kappa \int_{-\infty}^{\infty} \left| h(\vx-\vu) \right|^2 I_g(\vu) \; d\vu,
\end{align}
where $\kappa$ is the constant $1/|M|^2$ resulting from Theorem~\ref{thm: ch1 lens convolution}. The convolution here is defined for $|h|^2$ and $I_g$. Thus, we consider the following terms:
\begin{align*}
\calG_g(\vf) &= \mathfrak{Fourier}\{I_g(\vx)\} = \frac{\int_{-\infty}^{\infty} I_g(\vx)\exp\{-j2\pi\vf^T\vx\} d\vx}{ \int_{-\infty}^{\infty} I_g(\vx) d\vx},\\
\calG_i(\vf) &= \mathfrak{Fourier}\{I_i(\vx)\} = \frac{\int_{-\infty}^{\infty} I_i(\vx)\exp\{-j2\pi\vf^T\vx\} d\vx}{ \int_{-\infty}^{\infty} I_i(\vx) d\vx},
\end{align*}
where we normalize the Fourier transforms by the signal powers so that the maximum intensities of $\calG_g$ and $\calG_i$ are the unity. We then define the optical transfer function:
\boxedthm{
\begin{definition}[\keyword{Optical transfer function}]\index{optical transfer function}
The optical transfer function is
\begin{align*}
\calH(\vf) &= \mathfrak{Fourier}\{|h(\vx)|^2\} = \frac{\int_{-\infty}^{\infty} |h(\vx)|^2\exp\{-j2\pi\vf^T\vx\} d\vx}{ \int_{-\infty}^{\infty} |h(\vx)|^2 d\vx}.
\end{align*}
\end{definition}
}

In Fourier domain \eref{eq: ch1 I_i incoherent main} becomes
\begin{equation}
\calG_i(\vf) = \calH(\vf)\calG_g(\vf).
\label{eq: Ch1 incoherent intensity}
\end{equation}
Comparing \eref{eq: Ch1 coherent intensity} with \eref{eq: Ch1 incoherent intensity}, we observe that the intensity of the image in an incoherent illumination is the convolution of the magnitude squares.

\begin{figure}[ht]
\centering
\includegraphics[width=\linewidth]{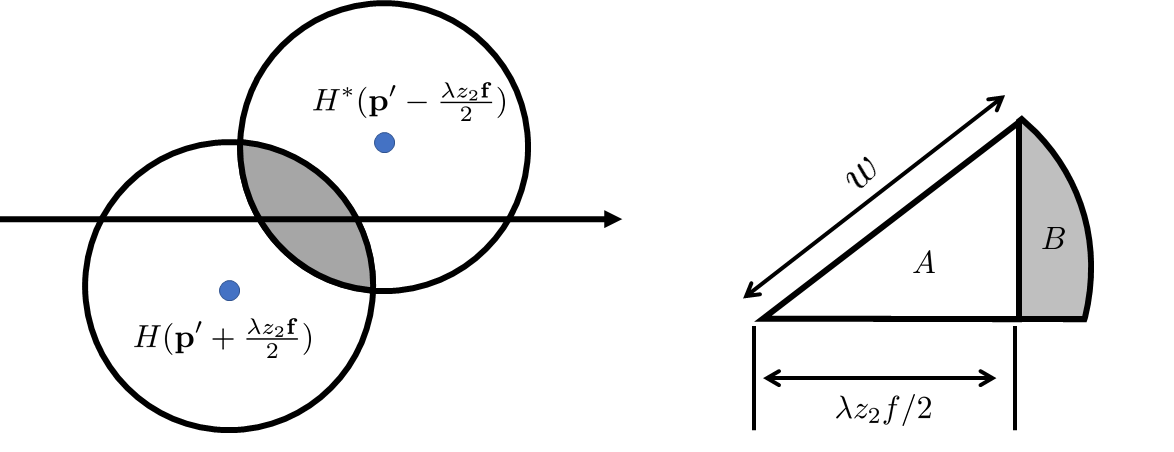}
\caption{The optical transfer function $\calH(\vf)$ is the autocorrelation function of $H(\vf)$ evaluated at position given by the two circles. This translates to computing the ratio between the overlapping area and the total area.}
\label{fig: Ch1 OTF}
\end{figure}

With the ATF and OTF defined as
\begin{align*}
H(\vf) &= \mathfrak{Fourier}\{h(\vx)\},\\
\calH(\vf) &= \mathfrak{Fourier}\{\vert h(\vx)\vert^2\},
\end{align*}
we can write the relationship between the optical transfer function $\calH$ and the amplitude transfer function $H$ is given by\index{amplitude transfer function! relation to OTF}\index{optical transfer function! relation to ATF}
\begin{align}
\calH(\vf)
&= \frac{ \int_{-\infty}^{\infty} H(\vp') H^*(\vp'-\vf) \; d\vp' }{ \int_{-\infty}^{\infty} |H(\vp')|^2 \; d\vp' },
\label{eq: Ch1 OTF and ATF}
\end{align}
where we note the term $|h(\vx)|^2$ turned into convolution in the Fourier domain.
Therefore, $\calH(\vf)$ is the auto-correlation of the $H(\vx)$. The geometry of $\calH(\vf)$ can be seen by doing a simple change of variable. Letting $\vp+\tfrac{\vf}{2}$, \eref{eq: Ch1 OTF and ATF} can be re-centered as
\begin{align}
\calH(\vf) = \frac{ \int_{-\infty}^{\infty} H(\vp'+\tfrac{\vf}{2})H^*(\vp'-\tfrac{\vf}{2}) \; d\vp' }{ \int_{-\infty}^{\infty} |H(\vp')|^2 \; d\vp' }.
\label{eq: Ch1 OTF and ATF 2}
\end{align}
Referring to \fref{fig: Ch1 OTF}, $\calH(\vf)$ is the ratio
\begin{equation*}
\calH(\vf) = \frac{\text{area of overlap}}{\text{total area}}.
\end{equation*}

For a circular aperture, the amplitude transfer function is
\begin{equation}
H(\vf) = P(\lambda z_2 \vf).
\end{equation}
Substituting into \eref{eq: Ch1 OTF and ATF 2}, we have
\begin{equation}
\calH(\vf) = \frac{P(\lambda z_2 \vf) \circledast P(\lambda z_2 \vf)}{P(0,0) \circledast P(0,0)},
\end{equation}
which we can recognize as simply an autocorrelation of the scaled pupil.
The overlapping region can be computed by referring to \fref{fig: Ch1 OTF}. The details for the evaluation of the circular OTF may be found in Goodman \cite{Goodman_2005_a}. Using geometry, it can be shown that\index{optical transfer function! of circular aperture}
\begin{align}
\calH(f) &= \begin{cases}
\frac{2}{\pi}\left[\arccos\left(\frac{f}{2f_0}\right) - \frac{f}{2f_0}\sqrt{1-\left(\frac{f}{2f_0}\right)^2}\right], &\qquad f \le 2f_0,\\
0, &\qquad \text{otherwise},
\end{cases}
\label{eq: Ch1 circular OTF}
\end{align}
where we defined the cutoff frequency
\begin{equation}
f_0 = \frac{w}{\lambda z_i},
\end{equation}
where here $f_0$ corresponds to the cutoff frequency of the coherent system of the same aperture, meaning the incoherent version of the same system will have a frequency cutoff that is twice the size.
The OTF obtained in \eref{eq: Ch1 circular OTF} is important, as it is the OTF of an incoherent \keyword{diffraction-limited} system. Assuming a perfect lens without distortions such as defocus, the system has a fundamental limit due to diffraction.

\section{Space-Variant Systems}
Thus far we have discussed optical systems with a single invariant point spread function $\abs{h(\vx)}^2$. This, of course, is in line with the standard signals and systems view of LSI systems. As we progress towards imaging through turbulence, we will no longer be able to justify this assumption of spatial invariance; the situations we will encounter may even have a different PSF (alternatively, an impulse response) for virtually \emph{every location in the image}. These differences will be correlated in space but may be wildly varying across the entire field of view. Therefore, we must re-examine our view of convolution to include spatially varying systems correctly. 

\subsection{Two Models for Convolution}
Convolution between an input $J(\vx)$ and an impulse response $h(\vx)$ is often written as\index{convolution}\index{convolution}
\begin{equation}
    I(\vx) = J(\vx) \circledast h(\vx) = \int J(\vu) h(\vx - \vu) d\vu.
    \label{eq: ch1_conv_og}
\end{equation}
This is a simplification of a more general equation. The convolution equation is a special case of the \keyword{superposition integral}\index{superposition}
\begin{equation}
    I(\vx) = \int h(\vx, \vu) J(\vu) d\vu.
    \label{eq: ch1_svc_temp}
\end{equation}
If the system is space-invariant, the impulse response can be simplified to $h(\vx,\vu) = h(\vx - \vu)$.

The superposition integral represents a more general framework for describing linear systems, of which the LSI systems are a subset. The LSI system particularly lends itself to the interpretation offered by Oppenheim and Willsky \cite{Oppenheim_1996_a} in which the impulse response is flipped, shifted, then multiplied and integrated. This is a literal interpretation of \eqref{eq: ch1_conv_og}. The alternative to this is to interpret convolution through \eqref{eq: ch1_svc_temp} which suggests the response is a weighted summation of impulse responses centered at each $\vu$.

We now introduce the possibility for the impulse response to vary as a function of position. However, we will choose two specific parameterizations, one in which the impulse response is parameterized by $\vx$, and the other where it is parameterized by $\vu$. We first consider the impulse response to vary as a function of $\vu$, leading us to write \eref{eq: ch1_svc_temp} as\index{convolution! scattering}
\begin{equation}
    I(\vx) = (J \circledast h_{\vu}) (\vx) = \int J(\vu) h_{\vu}(\vx - \vu) d\vu,
    \label{eq: ch1_scat_conv}
\end{equation}
This implies $h(\vx, \vu) = h_{\vu}(\vx - \vu)$. The alternative form of convolution rests upon the shift-and-integrate algorithm towards convolution often presented in a signals and systems course. The output at the coordinate $\vx$ will be dictated by the impulse response at $\vx$. Following the flip and shift principle, this leads us to write\index{convolution! gathering}
\begin{equation}
    I(\vx) = (J \circledast h_{\vx}) (\vx) = \int J(\vu) h_{\vx}(\vx - \vu) d\vu
    \label{eq: ch1_gath_conv}
\end{equation}
This implies $h(\vx, \vu) = h_{\vx}(\vx - \vu)$.

\begin{figure}
\centering
\begin{tabular}{cc}
\includegraphics[width=0.45\linewidth]{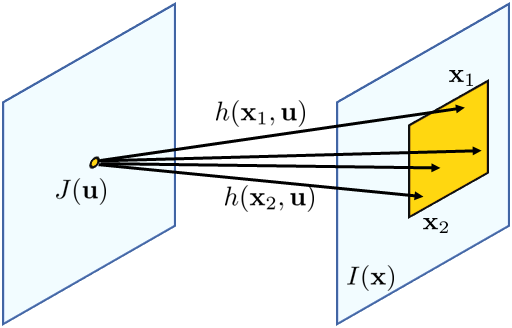} &
\includegraphics[width=0.45\linewidth]{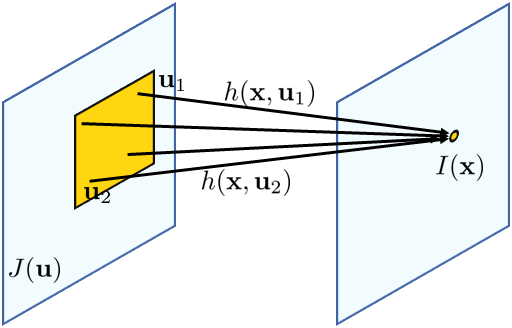}\\
(a) Scattering convolution & (b) Gathering convolution
\end{tabular}
\caption{A visualization of the difference between scattering and gathering convolutions. Source: \cite{Chimitt_2023_a}.}
\label{fig: scat_gath}
\end{figure}

Note that \eqref{eq: ch1_scat_conv} writes the impulse response as a function of the \emph{source} location. That is, the location of an impulse response in the input signal will dictate the form of the impulse response as suggested by $h_{\vu}$. \eref{eq: ch1_gath_conv} instead parameterizes the impulse response by the \emph{receiver} location $h_{\vx}$. This leads us to refer to the two forms of convolution in the following way: we will refer to \eqref{eq: ch1_scat_conv} as \keyword{scattering convolution} and \eqref{eq: ch1_gath_conv} as \keyword{gathering convolution}. We visualize this difference in interpretation in \fref{fig: scat_gath}.

To provide some insight into the terminology used for these two forms of convolution, first consider a signal which is comprised of a single impulse response located at $\vu$. As $\vx$ is varied, because of the fact that $I(\vx) = h_{\vu}(\vx - \vu)$, the output is simply the impulse response $h_{\vu}(\vx)$ centered at $\vu$. Thus, it tells us how the impulse response located at $\vu$ \emph{spread} to the neighboring points. Thus, the energy has been scattered from one point to many.

The alternative is to suppose we are interested in a receiver location $\vx$. This fixes our impulse response as $h_{\vx}$. However, note that this is already what \eqref{eq: ch1_gath_conv} is doing; $h_{\vx}$ is fixed in this integral. To compute the output, the impulse response, parameterized by source location $\vx$ is flipped and shifted $\vx$, then multiplied and integrated by the signal $J(\vu)$. Furthermore, through integration, the convolution takes the energy that exists within the input signal, multiplies and integrates over it, and places it in one location at $\vx$. Because of the fact the energy goes from many locations to one, it is referred to as gathering convolution.

To contrast the two forms in a different way, consider the input signal to be an impulse response $J(\vx) = \delta(\vx - \vu)$. The output by scattering would be
\begin{equation*}
    I(\vx) = h_{\vu}(\vx - \vu),
\end{equation*}
whereas the output by gathering would be
\begin{equation*}
    I(\vx) = h_{\vx}(\vx - \vu).
\end{equation*}
Due to the fact that $\vx$ need not equal $\vu$, the two responses may be different. Furthermore, the gathering result is harder to interpret conceptually as the impulse response changes with each $\vx$, where in the case of scattering the impulse response is fixed as $h_{\vu}$. With this book focusing on imaging through turbulence, a system that varies spatially, we need to determine which one is appropriate for our purposes.

\subsection{Which Model is Correct for Imaging?}
The short answer is that for modeling the wave propagation phenomenon, the scattering equation is the correct model. Here are some elaborated answers.

As we showed, the incoherent imaging system is linear in intensity, and thus far we have further assumed it to be spatially invariant. If we relax this assumption of spatial invariance, we must rely on the superposition integral to relate the input and output of a general system,
\begin{equation*}
    I(\vx) = \int h(\vx, \vu) J(\vu) d\vu.
\end{equation*}
This leaves us with a choice in determining how to choose $h(\vx, \vu)$. Specifically, should $h(\vx, \vu)$ be parameterized by the \emph{source} or \emph{receiver} variable? This choice is analogous to the case of deciding whether or not to perform scattering or gathering convolution \cite{Chimitt_2023_a}, with parameterization by source location $\vu$ corresponding to scattering and receiver location $\vx$ to gathering.

One way this question can be answered is through the terminology used by the optics community. Instead of using the term impulse response, filter, or kernel when discussing Fourier Optics, we have opted to use the term ``point spread function''. This term emphasizes the scattering model. The point spread function determines how a \emph{point} (or impulse) will \emph{spread} across the sensor plane (or, more generally, the surface of interest). The point spread function may change as a function of point location, but it will still describe how a particular point will spread across the sensor plane. It is also suggested through the visualization of the two in \fref{fig: scat_gath} that this matches our intuition with image formation; a point source located at a position will spread energy across the focal plane via propagation.

These concepts can be demonstrated more mathematically and through numerical experiments. With the current background, the introduction of these concepts would be somewhat challenging. Therefore, we would suggest the interested reader to a paper by Chimitt et al. \cite{Chimitt_2023_a} which provides the proof and helpful visualizations comparing the two forms of convolution and showcases why scattering is the proper choice in modeling imaging.

To tie the end of this Chapter with the beginning, the response of a linear, space-invariant imaging system is given as
\begin{equation}
I_i(\vx) = |\mathfrak{Fourier} \{ P(\vxi) e^{j \phi(\vxi)} \}|^2 \circledast  I_g(\vx).
\end{equation}
For the rest of this book, we will often allow the PSF to vary spatially through the following equation,
\begin{equation}
I_i(\vx) = \left(|\mathfrak{Fourier} \{ P(\vxi) e^{j \phi_\vu(\vxi)} \}|^2 \overset{\vu}{\circledast}  I_g\right) (\vx).
\end{equation}
This is a special class of PSFs that are parameterized by their phase error $\phi_\vu$, which we allow to be spatially varying as a function of point source location $\vu$. Our notation here is used to make the operation clear. First, the phase is defined by $\vxi$ which spans the aperture plane. From this, the PSF is formed and, after being parameterized by coordinate $\vx / (\lambda z)$, applied to $I_g$. This scaling of coordinate $\vx$ will typically be left off for simplicity in notation. We additionally emphasize the fact that the convolution is spatially varying with dependence on the \emph{source} location by the notation above.

\section{Summary}
\label{sec: sec1_7}
This Chapter presented Fourier optics combined with the thin lens model as well as the image formation process and spatially varying convolution. We summarize these as:

\keyword{Part 1 Wave equation}: We started by showing a class of waves that satisfy the scalar wave equation $\nabla^2 u(\vx,t) = \frac{n^2}{c^2} \frac{\partial^2 u(\vx,t)}{\partial t^2}$. If we further assume that $u(\vx,t) = U(\vx)\exp\{-j2\pi\nu t\}$, then the waves should satisfy the Helmholtz equation: $\nabla^2 U(\vx) + k^2 U(\vx) = 0$. All our subsequent discussions are about waves that satisfy the Helmholtz equation.

\keyword{Part 2 Diffraction}: We used the Huygens-Fresnel principle to model how waves move from one point to another point. Based on HFP, we then show the Rayleigh-Sommerfeld diffraction integral equation: $U(\vx) = \frac{1}{j\lambda} \int_{\Sigma} U(\vxi) \frac{\exp\{jkr\}}{r} \cos \theta d\vxi$. The Rayleigh-Sommerfeld equation is general, and so we introduce two approximations. The first approximation leads to the Fresnel diffraction. With a second approximation, the Fresnel diffraction is simplified to the Fraunhofer diffraction.

\keyword{Part 3 Lens}: All imaging systems use a lens with a finite aperture. The lens is equivalent to applying a convolutional kernel $\vert h(\vx) \vert^2$ to the field $U(\vx)$. The convolutional kernel $\vert h(\vx) \vert^2$ is the Fourier transform of the pupil function. For example, if the lens is circular, then $\vert h(\vx) \vert^2$ is the Bessel function (or the Airy disc). The kernel $h(\vx)$ is also known as the point spread function. In the absence of any distortion, the resulting field will have an intrinsic resolution limit determined by the width of the $\vert h(\vx) \vert^2$.

\keyword{Part 4 Image Formation}: We are mostly interested in two types of sources: coherent and incoherent. For coherent light, the image (not the field) is $I_i(\vx) = |h(\vx) \circledast U_g(\vx)|^2$. For incoherent light, the image is $I_i(\vx) = |h(\vx)|^2 \circledast I_g(\vx)$. Unless we use a laser source, typical long-range imaging through turbulence using a conventional camera is incoherent. The amplitude transfer function is $H(\vf) = \mathfrak{Fourier}\{h(\vx)\}$. The optical transfer function is $\calH(\vf) = \mathfrak{Fourier}\{\vert h(\vx)\vert^2\}$.

\keyword{Part 5 Spatially Varying Convolution}: A set of tools for representing spatially varying convolution were presented along with the notation $h_{\vu}(\vx) \overset{\vu}{\circledast} J(\vx)$. This is based on the perspective of superposition, with the chosen model for imaging as scattering convolution. 
\chapter{Modeling Turbulence}
\vspace{-6ex}
\noindent\textcolor{myblue}{\rule{\textwidth}{4pt}}
\vspace{1ex}

\label{sec: sec2}

The aim of this Chapter is to present the statistical model for imaging through the atmosphere. This begins with first understanding the statistical model for the atmosphere's index of refraction. The model adopted here is that the atmosphere's index of refraction is a random, \keyword{turbulent} process that changes according to altitude, temperature, pressure, wind, humidity, and other factors. This Chapter will introduce one of the more prevalent models, how it fits into our understanding of image formation, and how we may simulate turbulent effects on an image.
\index{atmospheric turbulence}

\begin{figure}[h]
\centering
\includegraphics[width=\linewidth]{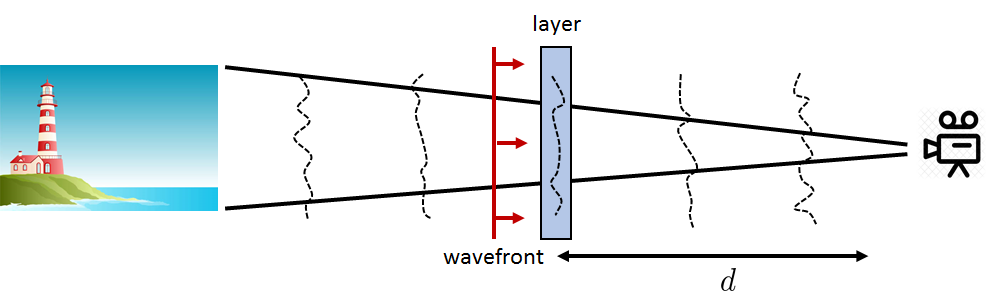}
\caption{A schematic diagram illustrating the partition of a horizontal optical path into many small segments. In this Chapter, we will analyze the statistical properties of the layers and their impact on waves propagating through them.}
\label{fig: ch2 layer}
\end{figure}

Consider the imaging situation presented in \fref{fig: ch2 layer}, in which a camera is placed a long distance away from an object (for our purposes, 500 meters or more will be suitable). In this situation, a wave that would have propagated as a planar wave in a vacuum is instead distorted by the atmosphere. These distortions are analogous to the case of a thin lens, however, instead of the thickness of the lens being carefully chosen to produce a converging, focused wave, the atmosphere generates a low quality ``dented'' wavefront. This denting of the wavefront will affect the image formation process by injecting a phase error into our system. We start with an overview to give the reader a sense of what these effects look like on an image and follow this with a brief summary of the rich history of the subject.

\section{Overview and History}

\subsection{The Image Formation Model}
In Chapter 2, we introduced the spatially invariant optical system's image formation equation,
\begin{equation}
I_i(\vx) = \abs{h(\vx)}^2 \circledast  I_g(\vx),
\end{equation}
which relates the observed image $I_i(\vx)$ to the geometric image $I_g(\vx)$ through the application of a spatially invariant PSF $\abs{h(\vx)}^2$,
\begin{equation}
\abs{h(\vx)}^2 = \abs{\mathfrak{Fourier} \{ P(\vxi) e^{j \phi(\vxi)} \}}^2_{\vf = \vx/(\lambda z)}.
\end{equation}
This particular PSF is described both by the aperture $P(\vxi)$ and phase error $\phi(\vxi)$. In Chapter 2, we presented an example where $\phi(\vxi)$ described the focusing error by an imaging system. In this Chapter, we inject the turbulence model into $\phi(\vxi)$. Therefore, if we have a model for $\phi(\vxi)$, our efforts from the previous Chapter will be put to good use.

\begin{figure}
    \centering
     \begin{tabular}{ccc}
	\includegraphics[width=0.3\linewidth]{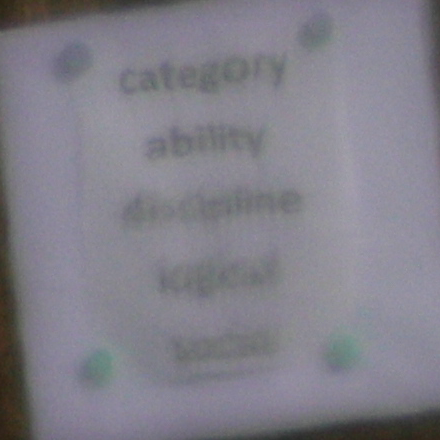} &
	\includegraphics[width=0.3\linewidth]{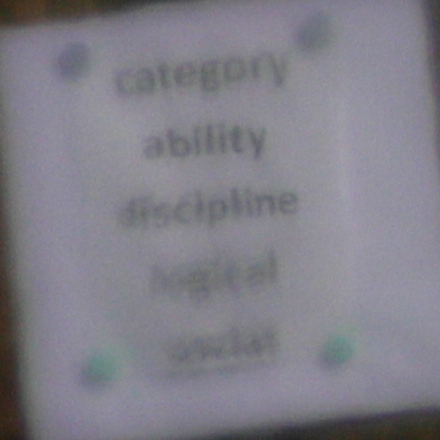} &
	\includegraphics[width=0.3\linewidth]{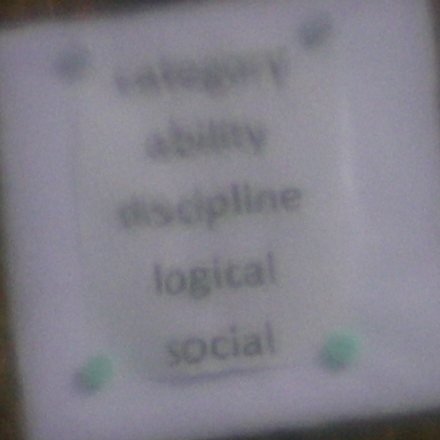}
\end{tabular}
    \caption{Three images of a board with printed text taken over a long distance (approximately 500 meters). Note the spatially varying effects, some words are blurred while others are legible.}
    \label{fig: turb_images}
\end{figure}

This leads us to encounter one of our first problems: the assumption of spatial invariance does \emph{not} hold true in the case of observations through the atmosphere. \fref{fig: turb_images} illustrates this effect, where the degradations appear to vary from point to point. This suggests a \keyword{spatially variant} imaging model, a point we will visit more appropriately in later portions of the Chapter. With what we have described so far, $\phi$ is where the turbulence will be inserted. Therefore, we must allow $\phi$ to vary as a function of position $\vu$ in the object plane. This leads us to write the PSF equation as
\begin{equation}
\abs{h_{\vu}(\vx)}^2 = \abs{\mathfrak{Fourier} \{ P(\vxi) e^{j \phi_{\vu}(\vxi)} \}}^2_{\vf = \vx/(\lambda z)}.
\label{eq: ch2_turb_psf}
\end{equation}
along with the image formation equation for a spatially variant system to be
\begin{equation}
I_i(\vx) = \left(\abs{h_{\vu} }^2  \overset{\vx}{\circledast} I_g \right) (\vx).
\end{equation}
Here, we denote $\vu$ as the input or source location and $\vx$ as the output or receiver location.

It is important to note that \eref{eq: ch2_turb_psf} is an approximation. Throughout this Chapter, and in fact the entire book, we will rely upon the assumption as follows:
\boxedmsg{
\vspace{2ex}
\textbf{Assumption}. We assume that the turbulent distortions can be adequately summarized by the phase distortions.
\vspace{1ex}
}
The assumption means that the amplitude distortions are negligible. If we take into account the amplitude distortions, a more general PSF equation would be
\begin{equation}
\abs{h_{\vu}(\vx)}^2 = \abs{\mathfrak{Fourier} \{ A_\vu(\vxi) P(\vxi) e^{j \phi_{\vu}(\vxi)} \}}^2_{\vf = \vx / (\lambda z)},
\end{equation}
where $A_\vu(\vxi)$ is a spatially varying amplitude component which also varies per location in the object plane. Our assumption enables us to describe levels of turbulence that fall in the weak to moderate levels of distortions. We can justify this in two ways: (1) The assumption of phase-exclusive distortions is common to many varieties of simulation and turbulent imaging literature; (2) the reconstruction of wide FOV incoherent images through strong turbulence is ill-posed. We view (2) as the proper justification for the purposes of this book; the problems we want to solve will be challenging even in this weak to moderate range. Therefore, the lack of generality will not cost us too greatly. We will comment more directly on the limitations of this model when the proper background has been introduced.

\subsection{Characteristics of Turbulence-Distorted Images}
At this point, a reader who is familiar with the image processing literature may see the images in \fref{fig: turb_images} and feel the need to jump to the conclusion that this is a standard deblurring problem. We emphasize that this is not the case. While Chapter 5 will provide a thorough discussion of the algorithms for these types of distortions, we wish to describe a few key characteristics of the problem here. This should hopefully convey the difficulties of reconstruction as well as give a sense as to why such a careful description of the atmospheric model is needed.

\begin{figure}[h]
    \centering
    \includegraphics[width=0.95\linewidth]{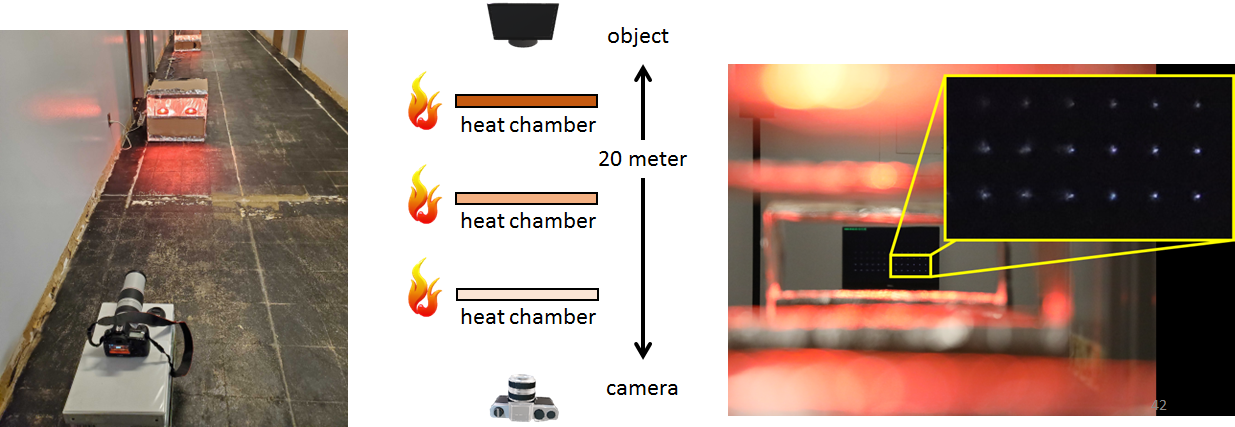}
    \caption{A physical setup of observing the PSF through a series of heat chambers. On the monitor, we display a grid of points. As light passes through the heat chambers, the image captured by the camera will suffer from turbulence.}
    \label{fig: ch2_visual_turb_point_source}
\end{figure}

Turbulent images are comprised of two key effects: pixel-shifting and blur. Both of these effects will vary per pixel. We provide a visualization of this in \fref{fig: ch2_visual_turb_point_source}. We further note that a blur kernel $\abs{h_\vu}^2$ has no closed form expression in the \keyword{image space}. So, we will need to describe it exclusively by $\phi_\vu$. Commonly used parametric models such as the Gaussian blurs are not appropriate to be the \keyword{prior model} because the turbulence blur kernels are not even symmetric. Note that the point sources on the screen are very small near the bottom right of the zoomed in portion while the points near the upper left are wider and more complicated, once again highlighting the spatially varying nature of the problem.

One must also consider the fact that the atmosphere changes over time. This means that the degradation will fluctuate between multiple frames. This has a significant impact on reconstruction approaches, influencing the notion of \keyword{lucky imaging}. Lucky imaging rests upon the fact that due to the atmosphere changing in time, every once in a while one will observe a frame \emph{or region} which has minimal degradation. Furthermore, pixel-shifting can be so strong that for a video with a moving object, it can be difficult to algorithmically determine if a motion is due to object motion or turbulence. All of these degradations compound to create a problem that requires the incorporation of temporal information, lucky imaging methodologies, deblurring, and un-warping.

\subsection{Historical Developments}
\index{atmospheric turbulence! history}\index{Kolmogorov}
The theory of atmospheric modeling that we will be discussing in this book arises from statistical models of turbulence. The modern form of statistics and probability can be traced back to Kolmogorov's axiomatic approach to probability \cite{Kolmogorov_1956_a} (originally published in 1933). Amazingly, a core insight giving rise to the dominant form of modern turbulence theory is \emph{also} a result of Kolmogorov's work \cite{Kolmogorov_1941_a, Frisch_1995_a} in which he proposed a statistical model which describes how turbulence transfers energy down from the large fluctuations to the small fluctuations of the medium. Kolmogorov and a student of his, Obukhov, published additional papers on this subject (such as \cite{Obukhov_1941_a, Obukhov_1941_b, Kolmogorov_1941_b}), further elaborating on various elements of the model.

\index{Tatarskii}These insights alone are not enough to bring us to where the field stands today. There is a big gap missing -- the works of Obukhov and Kolmogorov just model the turbulence. The question as to what happens to an optical wave propagating through the medium is not addressed. Continuing from the previous lineage, a student of Obukhov by the name of Tatarskii (alternatively spelled Tatarski in some sources) did exactly this in the form of an excellent manuscript \cite{Tatarski_1967_a} (the manuscript being a converted version of his Ph.D. thesis!). In this book, Tatarskii carefully applied the Kolmogorov theory within Maxwell's equations, resulting in the foundation that so many works derive from in the field of optical turbulence. There is little denying that the work of Tatarskii is likely \emph{the} foundation of the field.

\index{Fried}From here, a wide range of applications emerged\footnote{As we write this book, we recognize our limitations of not being trained in a traditional physics manner. It is therefore sometimes hard for us to appreciate the raw physics that has emerged from Tatarskii's work. However, on the applications side, especially towards generating large-scale datasets, we hope to share our experience with the readers.}. One of the more prominent researchers from this era (1960s - 1980s) is Fried. Fried published a vast amount of papers over a long career, where we highlight the following papers which were particularly helpful when we started to work on this topic: \cite{Fried_1965_a, Fried_1966_a, Fried_1975_a, Fried_1976_a, Fried_1978_a}. Fried proposed the \keyword{coherence diameter} $r_0$, which will be of significant importance for the latter majority of this book and is often referred to as the \keyword{Fried parameter}. Of course, we will return to its definition once the proper preliminaries have been presented.

Aside from Fried, there are others such as Ishimaru \cite{Ishimaru_1978_a, Ishimaru_1977_a, Tatarskii_1993_a} who worked with Tatarskii and wrote his own book on the subject. Greenwood contributed to adaptive optics in the presence of turbulence, with the \keyword{Greenwood frequency} attributed to his name which quantifies aspects of the temporal evolution of the atmosphere within the context of adaptive correction \cite{Greenwood_1977_a}. Hufnagel \cite{Hufnagel_1964_a, Hufnagel_1978_a} contributed to what is now regarded as the Hufnagel-Valley turbulence path model, a model we will introduce later in this Chapter. Roggemann and Welsh \cite{Roggemann_1996_a, Bos_2015_a, Welsh_1997_a} have both published numerous works within the field and an influential book that offers a thorough explanation of the necessary turbulent optics which is arguably more accessible to the average reader than previous texts. Additionally, there are those such as Tyson \cite{Tyson_1996_a, Tyson_2010_a} particularly oriented in the direction of astronomical imaging and adaptive optics.

With some of our admitted subjectivity, there is one paper that stands out among the rest. This is the paper by Noll \cite{Noll_1976_a} which may very well be our favorite paper on the topic. Noll proposes a basis expansion for the phase distortion (Fried also did this almost 10 years earlier \cite{Fried_1965_a}, however, Noll's has emerged as the more dominant approach). We would also pair this endorsement of Noll's paper with Roddier \cite{Roddier_1990_a} which offers a slight correction to Noll's paper and additional exposition, among other contributions. This basis expansion will solve the issue regarding the phase distortion $\phi$'s description in a simplistic manner. Noll's paper is so influential to the authors, in fact, that virtually the entire simulation approach proposed by us and collaborators (\cite{Chimitt_2020_a, Mao_2021_a, Chimitt_2022_a}) is inspired by Noll's work. This paper, and our various modifications and interpretation, will be the core of the discussion in Chapter 4.

This Chapter will follow the same chronology as the historical development described here. We will first introduce the Kolmogorov model and related models. From here, we will apply it to optical propagation through the atmosphere (following the more simplistic approach along the lines of Roggemann and Welsh \cite{Roggemann_1996_a}) following various applications. Finally, we will close the Chapter with the classical approach to simulating these effects in an imaging system known as \keyword{split-step propagation}, often referred to as simply ``split-step".

\section{The Atmosphere's Index of Refraction}
\label{sec: sec2_1}
\index{turbulence}
A key component to the theory we are to discuss is that the atmospheric index of refraction (which we will typically refer to as simply the atmosphere) is distributed randomly in a turbulent fashion. A reasonable question may then be: why is the atmosphere turbulent? While the true reasoning is well beyond the scope of this book, we feel it important to give some perspective on the concept of turbulence. The atmosphere can be modeled as a fluid. Fluid is said to move in either a laminar or turbulent fashion. If we claim the atmosphere moves in a laminar way, we must be prepared to claim that all particles of the atmosphere are moving in the same direction without any deviation. This clearly cannot be the case, for example, the wind is not blowing in the same direction everywhere in the world at all times. By virtue of the atmosphere not moving as laminar flow, we may then reason that atmospheric motion is turbulent.\index{atmospheric turbulence}

We emphasize this point for a key reason: turbulence does not have to be ``aggressive'', it can be slow moving and calm such as the smoke that dances above a candle after being blown out. Furthermore, the effect the atmosphere has on an imaging system is not something that only happens \emph{sometimes}. Rather, it is happening all the time and is only easily observable in certain circumstances, such as ground-to-ground imaging over a long distance.

\subsection{Kolmogorov Power Spectral Density}
The statistics of a \keyword{random process}\index{random process} $X$ is often characterized by studying the spatial correlation between $X(\vr_1)$ and $X(\vr_2)$ where $\vr_1 \in \R^3$ and $\vr_2 \in \R^3$ are two spatial coordinates. We will denote correlation function of $X$ as $\Gamma_X(\vr_1,\vr_2) = \E[X(\vr_1)X(\vr_2)]$. A frequently used assumption for simplifying a random process is to assume it is \keyword{homogeneous}\index{homogeneity}. (In statistical signal processing, we call it \keyword{wide sense stationary\index{wide sense stationarity}}.) For a spatially varying random process, the correlation function of a homogeneous process takes the form of
\begin{equation}
\Gamma_X(\vr) = \E[X(\vr_1)X(\vr_1-\vr)],
\end{equation}
where $\vr = \vr_1-\vr_2$ is the spatial increment. In other words, homogeneity implies the autocorrelation function can be written as a function of the difference $\vr = \vr_1-\vr_2$ instead of the pair of absolute locations $(\vr_1,\vr_2)$.

An alternative to the correlation function is the \keyword{power spectral density} (PSD)\index{power spectral density}. For a homogeneous random process, the PSD is the Fourier transform of the autocorrelation function by the \keyword{Wiener–Khinchin theorem}\index{Wiener-Khinchin theorem}:
\begin{align}
\Phi_X(\vk) = \frac{1}{(2\pi)^2} \int_{-\infty}^{\infty} \Gamma_X(\vr) e^{j \vk^T\vr} d\vr,
\end{align}
where $\vk = [k_x,k_y,k_z]$ is the spatial \keyword{wavenumber} vector. The scalar wavenumber is $k = |\vk| = \sqrt{k_x^2 + k_y^2 + k_z^2}$.

With these concepts defined, we now turn to the fluctuations in the atmosphere. Our discussion of the index of refraction follows the one by Goodman \cite{Goodman_2015_a}. We will model the atmosphere's index of refraction as a function of spatial location $\vr = [x,y,z]$, time $t$ and wavelength $\lambda$:\index{index of refraction! atmospheric model}
\begin{equation}
n(\vr,t,\lambda) = \underset{\text{mean index}}{\underbrace{n_0(\vr,\lambda)}} + \underset{\text{fluctuation}}{\underbrace{n_1(\vr,t)}}.
\end{equation}
The mean refractive index $n_0(\vr,\lambda)$ can be assumed time \emph{independent} and may change with elevation or climate. The dependency of the wavelength is ignored in $n_1(\vr,t)$ because it is generally weakly varying in the visible spectrum. We will assume that the index of refraction $n_1(\vr)$ is a zero-mean Gaussian process so that $\E[n_1(\vr)] = 0$ for any $\vr$ \cite{Goodman_2015_a, Tatarski_1967_a} and denote its autocorrelation function as $\Gamma_n(\vr_1,\vr_2) = \E[n_1(\vr_1) n_1(\vr_2)]$.

\begin{figure}[h]
\centering
\includegraphics[width=0.8\linewidth]{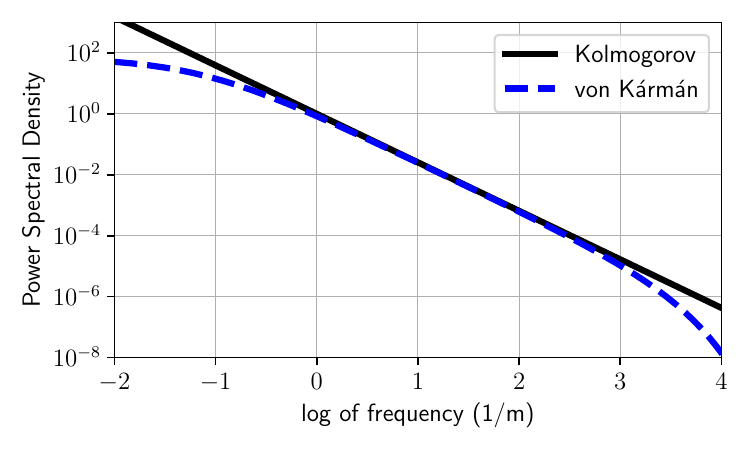}
\caption{Kolmogorov power spectral density and the von Kármán power spectral density, plotted as a function of the scalar wavenumber $k = |\vk|$.
}
\label{fig: ch2 Kolmogorov}
\end{figure}

This leads us to introduce the Kolmogorov model. The Kolmogorov model divides the fluctuations in $n_1$ into three regimes which are specified by the \keyword{outer scale} $L_0$ and the \keyword{inner scale}\index{turbulence! scale size} $\ell_0$. Scales can be thought of as structures that are \emph{on the order of} the specified size. The Kolmogorov model specifically models the energy behavior of the turbulent fluctuations of sizes that are between the outer and inner scales. Thus, for wavenumbers $2\pi/L_0 \le |\vk| \le 2\pi/l_0$, $\Phi_n(\vk)$ follows the Kolmogorov PSD\index{turbulence! Kolmogorov PSD} \cite{Tatarski_1967_a, Kolmogorov_1941_a, Kolmogorov_1941_b}:
\begin{equation}
\Phi_n(\vk) = 0.033 C_n^2 |\vk| ^{-11/3},
\label{eq: Ch2 Kolmogorov PSD}
\end{equation}
where $C_n^2$ is called \keyword{structure constant}\index{structure constant} of the index of refraction with units of $\left[\text{m}^{-2/3}\right]$. The Kolmogorov PSD is shown in \fref{fig: ch2 Kolmogorov}, along with a modified version known as the von Kármán PSD\index{turbulence! von Kármán PSD} spectrum \cite{Schmidt_2010_a, Roggemann_1996_a}:
\begin{equation}
\Phi_n(\vk) = \frac{0.033 C_n^2}{(|\vk|^2+k_0^2)^{11/6}} \exp\left\{-\frac{|\vk|^2}{k_m^2}\right\},
\end{equation}
where $k_0 = 2\pi/L_0$ and $k_m = 5.92/l_0$. The fluctuations are better modeled by the von Kármán PSD in the case of small wavenumbers $\vk$. For a thorough discussion of the varying PSDs (both Kolmogorov and non-Kolmogorov turbulence), we suggest the reader to Korotkova \cite{Korotkova_2017_a}.

\subsection{Structure Constant $C_n^2$}
The Kolmogorov spectrum presented in \eref{eq: Ch2 Kolmogorov PSD} introduces the \keyword{structure constant}\index{structure constant} $C_n^2$. The value of $C_n^2$ is an important parameter for modeling the turbulence strength. We can think of the structure constant as being analogous to the variance of a random variable. A higher value of $C_n^2$ suggests a greater strength of turbulence. $C_n^2$ is related to the altitude, location, time of day, etc. We will typically allow $C_n^2$ to vary over the path of propagation, thus typically referring to it as $C_n^2(z)$, which will call a \keyword{turbulence profile}\index{turbulence profile}.

In the literature, there exist many different models of $C_n^2$. For example, the Hufnagel-Valley profile\index{Hufnagel-Valley profile} \cite{Hufnagel_1978_a, Valley_1980_a, Parenti_1994_a, Roggemann_1996_a} models $C_n^2$ as a function of the altitude $h$ using
\begin{align*}
C_n^2(h) &= 5.94 \times 10^{-53}(v/27)^2 h^{10} e^{-h/1000} \\
         &\qquad + 2.7\times 10^{-16}e^{-h/1500} + Ae^{-h/100},
\end{align*}
where the typical value of $A$ is $A = 1.7 \times 10^{-14} \left[\text{m}^{-2/3}\right]$ and the value of $v$ is $v = 21$ [m/s]. The Submarine Laser Communication Day (SLCD) profile \cite{Parenti_1994_a, Roggemann_1996_a} states that
\begin{align*}
C_n^2(h) &=
\begin{cases}
0,                                  &\qquad 0 \text{ [m]} < h < 19 \text{[m]},\\
4.008 \times 10^{-13} h^{-1.054},   &\qquad 19 \text{ [m]} < h < 230 \text{ [m]},\\
1.3   \times 10^{-15} ,             &\qquad 230 \text{ [m]} < h < 850 \text{ [m]},\\
6.352 \times 10^{-7} h^{-2.966},    &\qquad 850 \text{ [m]} < h < 7000 \text{ [m]},\\
6.209 \times 10^{-18} h^{-0.6229},  &\qquad 7000 \text{ [m]} < h < 20000 \text{ [m]}.
\end{cases}
\end{align*}
For ground-to-ground imaging, it may be reasonable to assume a constant $C_n^2$ profile for simplicity, though it may certainly vary depending on many factors (e.g. a heat source along the path).

\begin{figure}[h]
\centering
\includegraphics[width=0.75\linewidth]{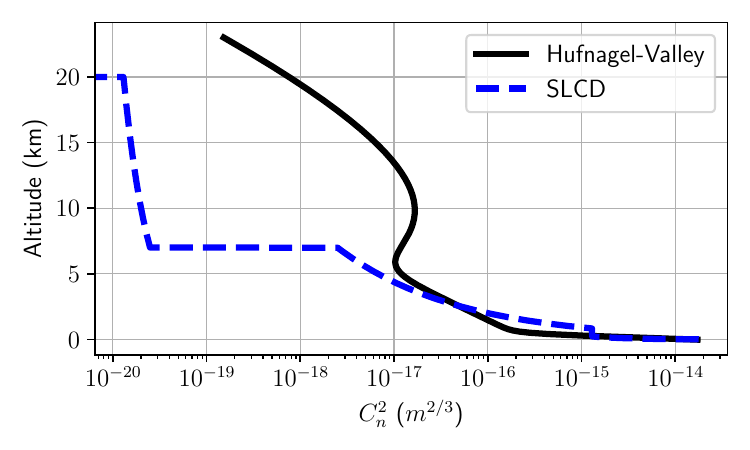}
\caption{Profile of the $C_n^2$ as a function of the altitude, where we are comparing the SLCD profile vs. the Hufnagel-Valley profile.
}
\label{fig: ch2 cn2}
\vspace{-2ex}
\end{figure}

\fref{fig: ch2 cn2} shows the $C_n^2$ profile according to the Hufnagel-Valley model. $C_n^2$ is generally large when the altitude is below 20 [m] and it starts to decrease sharply as the altitude increases. One explanation is that the ground has a higher temperature than air at a higher altitude. For ground-to-ground imaging, the range of $C_n^2$ is typically around $10^{-13}\left[\text{m}^{-2/3}\right]$ (for strong turbulence) to $10^{-17}\left[\text{m}^{-2/3}\right]$ (for weak turbulence).

\subsection{The Structure Function $\calD_n(\vr)$}
Examining \eref{eq: Ch2 Kolmogorov PSD}, we notice a particular difficulty, namely the singularity when $\vk = \mathbf{0}$. This will cause an issue for the evaluation of the correlation function by the Wiener–Khinchin theorem,
\begin{equation}
\Gamma_n(\vr) = \frac{1}{(2\pi)^2} \int_{-\infty}^{\infty} 0.033 C_n^2 |\vk| ^{-11/3} e^{-j \vk^T\vr} d\vk.
\end{equation}
This leads us to introduce a unique characteristic of the statistical turbulence literature, the \keyword{structure function}\index{structure function}. The structure function is an alternative to the correlation function as defined in \cite{Kolmogorov_1941_a}:
\begin{equation}
\calD_n(\vr) = \E[(n_1(\vr_1)-n_1(\vr_1-\vr))^2].
\end{equation}
With the assumption of stationarity, we can show that $\calD_n(\vr) = 2[\Gamma_n(0) - \Gamma_n(\vr)]$ and hence \cite{Tatarski_1967_a, Roggemann_1996_a}
\begin{align}
\calD_n(\vr)
&= 2\int \left[ \Phi_n(\vk) - \Phi_n(\vk)e^{-j \vk^T\vr}\right] d\vk \notag\nonumber \\
&= 2\int \left[ 1 - \cos(\vk^T\vr) \right] \Phi_n(\vk) d\vk. \label{eq: ch2 struct fun psd def}
\end{align}
The form of the structure function and PSD relationship aids in controlling the divergence near $\vk=\mathbf{0}$. It has been shown that this integral converges for a set of cases to which the Kolmogorov PSD belongs \cite{Tatarski_1967_a}.

Assuming that $n_1(\vr)$ is \keyword{isotropic}\index{isotropy} (spherically symmetric), and using the Kolmogorov PSD $\Phi_n(\vk)$, we can derive the following:
\boxedthm{
\begin{definition}[\keyword{Kolmogorov Structure Function of the Index of Refraction}]
Assuming the index of refraction is a homogeneous (spatially stationary) and isotropic (spherically symmetric) random process, the Kolmogorov structure function is defined as\index{structure function! Kolmogorov}\index{turbulence! structure function} \cite{Tatarski_1967_a, Goodman_2007_a, Roggemann_1996_a}
\begin{equation}
\calD_n(\vr) = C_n^2 |\vr|^{2/3},
\end{equation}
where $C_n^2$ is the structure constant of the index of refraction, and $\vr = [x,y,z]$ is the coordinate in the 3D space.
\end{definition}
}

The structure function is even well-defined even in the case of inhomogeneities. This is allowed by its invariance to the small wavenumber effects, which correspond to the large scale behavior. As a result of the definition of the structure function in \eref{eq: ch2 struct fun psd def}, the PSD will be attenuated near $\vert \vk \vert = 0$. Thus, even if the variance of $n_1$ were to fluctuate over space, the structure function will appropriately mitigate the resulting inhomogeneities. The structure function, as a result, will be homogeneous and isotropic even in the case of inhomogeneous atmospheric turbulence (for more details on this topic, we suggest the reader refer to Tatarskii \cite{Tatarski_1967_a}).

\section{Structure Function of the Phase}
\label{sec: sec2_2}
\begin{figure}[h]
\centering
\includegraphics[width=\linewidth]{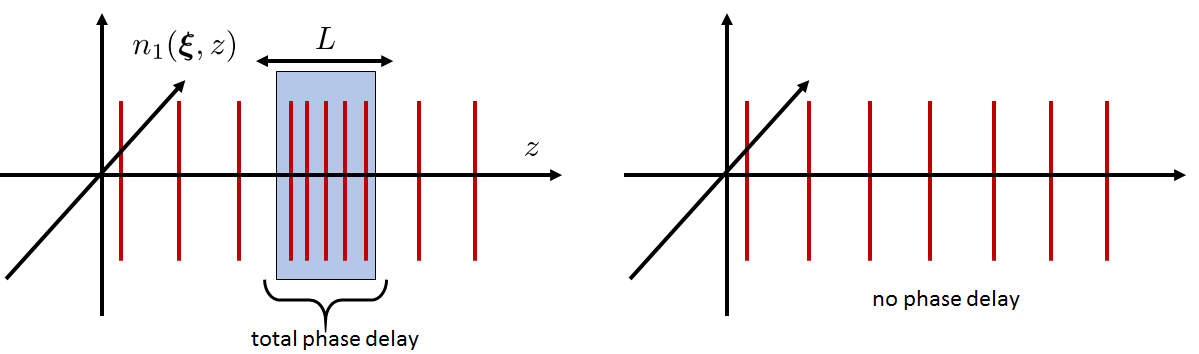}
\caption{Phase lead and lag caused by a medium with a high refractive index. As light passes through the medium, the speed of the light is slowed down because the refractive index is higher than the vacuum. This causes a phase delay. The longer the propagation path, the stronger the phase distortion will be.}
\label{fig: ch2 delay}
\vspace{-2ex}
\end{figure}

In \cref{sec: sec2_1}, we discussed the statistics of the index of refraction. For the purposes of imaging, we are interested in the effects the index of refraction has on a wave. We will focus specifically on how the atmosphere affects the phase of a propagating wave. Our plan of development is to first derive the statistics of the phase for a single layer of homogeneous turbulence, as shown in \fref{fig: ch2 delay}. We will then use this result for multiple layers, which will generalize our development to paths with varying $C_n^2$ profiles.

We wish to emphasize that the following analysis will be for a single point source. The extension to objects would only stand to complicate our approach at this time. We will introduce a way towards modeling the spatially varying nature towards the end of this Chapter regarding simulation. Chapter 3 will be dedicated to the image formation model more completely.

\subsection{A Single Turbulence Layer}
We will first begin with the propagation of a plane wave through a single layer of turbulence. It is first important to note our convention of measuring propagation distance: we choose to measure the distance \emph{from} the aperture plane, thus $z=L$ is in the object plane while $z=0$ is in the aperture plane. For a propagation distance of $L$, we segment the atmosphere into $M$ homogeneous segments partitioned by distances $L_i$. For simplicity, let us focus on the case of $z=0$ to $z=L_1$. At this point, we will drop the subscript $i$ as we are only focusing on one segment. We assume the turbulent medium of thickness $L$ is homogeneous, isotropic, and distributed according to the Kolmogorov spectrum.

Decomposing $\vr = [\vxi,z]$ where $\vxi = [x,y]$ is the coordinate in the 2D plane and $z$ is the coordinate along the propagation path, the phase distortion caused by the medium with a propagation distance $L$ follows the equation
\begin{equation}
\phi(\vxi, L) = k \int_{0}^{L} n_1(\vxi,z) \; dz,
\label{eq: Ch2 phi}
\end{equation}
where $k = |\vk| = 2\pi/\lambda$ is the scalar wavenumber. Note that this medium of refractive index $n_1(\vr)$ will cause phase effects that vary across the wavefront.

Following a similar development to the thin lens, let us denote the incident wave before the atmospheric layer to be $U(\vxi)$ and the output wave $U'(\vxi)$. Our focus is then on the distribution of the emergent wave, $U'(\vxi)$. For simplicity, let us assume that $U(\vxi)$ is a plane wave with $U(\vxi) = 1$. Accordingly, the input-output relationship is given by
\begin{equation}
U'(\vxi) = t_{\phi}(\vxi),
\label{eq: ch2 phase screen imparting}
\end{equation}
with $t_{\phi}(\vxi) = e^{j {\phi}(\vxi)}$.
In general, the distortions will consist of not only the phase term but also an amplitude term. However, for simplicity in this book, we shall skip the discussion of the amplitude term.

As we are interested in the wave's phase structure function, we begin with consideration of the autocorrelation function:
\begin{align}
\Gamma_{U'}( \vxi, \vxi')
&= \E[t_{\phi}(\vxi) t_{\phi}^*(\vxi')] \notag \\
&= \E[e^{j \phi(\vxi)} e^{-j \phi(\vxi') }]\notag \\
&\overset{(a)}{=} \E\left[ \exp\left\{jk \left(\int_{0}^{L} [n_1(\vxi,z) - n_1(\vxi', z)] \, dz \right) \right\}\right],
\end{align}
where $(\cdot)^*$ denotes the complex conjugate, where we've utilized substitution of \eref{eq: Ch2 phi} in (a). At this point, we need to make an assumption about $n_1(\vxi,z)$ or otherwise we cannot proceed further:
\boxedthm{
\begin{assumption}[\keyword{Assumption of $n_1(\vxi,z)$}]
We assume that the index of refraction $n_1(\vxi,z)$ is a zero-mean Gaussian random process. Moreover, we assume that $n_1(\vr)$ is homogeneous and isotropic for the region contained within $L$.
\end{assumption}
}
The normality (Gaussian statistics) is partially due to convenience because it is theoretically more tractable. We assume that the random process is zero mean because the index of refraction $n(\vxi,z)$ has a constant term $n_0(\vxi,z)$ and a random term $n_1(\vxi,z)$. The constant term can take care of the offset and so the random term can be zero mean. Our assumption of homogeneity and isotropy will again simplify our analysis. In the next subsection, we will move towards multiple layers, which will allow for large scale inhomogeneities.

With these assumptions in mind, we recall the following moment generating function property of a zero-mean Gaussian random variable.
\boxedthm{
\begin{lemma}[\keyword{Moment Generating Function}]\index{moment generating function}
\label{lemma: MGF Gaussian}
Let $X$ be a zero-mean Gaussian random variable, and let $s$ be any constant. Then,
\begin{equation}
\E[e^{sX}] = \exp\left\{\frac{s^2\sigma^2}{2}\right\},
\label{eq: moment gen func}
\end{equation}
where $\sigma^2 = \Var[X]$ is the variance of $X$.
\end{lemma}
}
Using the above lemma, it follows that $\E[e^{-jX}] = \exp\{-\sigma^2/2\}$. Therefore, we can show that
\begin{align*}
\Gamma_{t_\phi}( \vxi, \vxi')
= \exp\Bigg\{-\frac{1}{2}
\underset{\calD_{\phi}(\vxi,\vxi') \bydef \E[(\phi(\vxi)-\phi(\vxi'))^2]}{\underbrace{k^2 \E\left[  \left(\int_{0}^{L} [n_1(\vxi,z) - n_1(\vxi', z)] \, dz \right)^2 \right]}}
\Bigg\}.
\end{align*}
Notice that the usage of the moment generating function results in the structure function of the indices of refraction appearing in the exponential function. This term within the exponential function represents the structure function of the accumulated phase distortions and is accordingly referred to as the \keyword{phase structure function}.
\boxedthm{
\begin{definition}[\keyword{Phase Structure Function}]\index{structure function! turbulent phase}
Suppose light passes through a random medium of distance $L$ with a fluctuating index of refraction $n_1(\vxi,z)$. The structure function of the phase is defined as
\begin{equation}
\calD_{\phi}(\vxi, \vxi') \bydef k^2 \E\left[  \left(\int_{0}^{L} [n_1(\vxi,z) - n_1(\vxi', z)] \, dz \right)^2 \right],
\end{equation}
where the expectation is taken with respect to $n_1(\vxi,z)$.
\end{definition}
}

This serves as a useful result, however, there is a fair amount of simplification possible. Rewriting the square as a product of two integrals, we show that
\begin{align*}
\calD_{\phi}(\vxi, \vxi')
&= k^2 \E\Bigg[ \int_{0}^{L} \int_{0}^{L} [n_1(\vxi,z) - n_1(\vxi', z)] \\
&\qquad\qquad \times [n_1(\vxi,z') - n_1(\vxi', z')] \, dz \, dz' \Bigg].
\end{align*}
Expanding the inner terms algebraically and the fact that $D_n(\vxi,z) = 2[\Gamma_n(0,z)-\Gamma_n(\vxi,z)]$, recognizing that $D_n(\vxi,z) = D_n(-\vxi,z)$ we can rewrite the above equation as
\begin{align*}
\calD_{\phi}(\vxi, \vxi')
&= k^2 \int_{0}^{L} \int_{0}^{L} \Big[D_n(0,z-z') - \frac{1}{2}D(\vxi-\vxi',z-z') \\
&\qquad\qquad - \frac{1}{2}D_n(\vxi'-\vxi,z-z') \Big] dz \, dz'\\
&= -k^2 L \int_{-L}^{L} \left[D_n(0,z) - D_n(\vxi-\vxi', z)\right]\left(1-\frac{|z|}{L}\right) dz.
\end{align*}
Substituting the Kolmogorov structure function of the refractive index $D_n(\vxi,z) = D_n(\vr) = C_n^2|\vr|^{2/3} = C_n^2 (|\vxi|^2 + z^2)^{1/3} $ and using a key numerical observation by Fried \cite{Fried_1966_a}, it follows that
\begin{align}
\calD_{\phi}(\vxi, \vxi')
&= -k^2 L C_n^2 \int_{-L}^{L} \left[z^{2/3} - (|\vxi-\vxi'|^2 + z^2)^{1/3}\right]\left(1-\frac{|z|}{L}\right) dz \notag \\
&= 2.91 k^2 L C_n^2 |\vxi-\vxi'|^{5/3}.
\end{align}\index{phase structure function! single layer}
This expression constitutes a key result: we can now statistically describe the phase of a wave after propagation through a turbulent medium defined by the Kolmogorov spectrum. Furthermore, we can see it is homogeneous and isotropic. In what shall follow, we will see that we can use this result, combined with an assumption, to generalize to multiple slices of turbulence.

\subsection{Multiple Phase Screens}
Suppose $C_n^2(z)$ changes appreciably along the path of propagation such as in the case of astronomical viewing. We will then require our previous result to be extended to cases beyond a homogeneous layer of turbulence. For propagation of this variety, there may exist inhomogeneities in the distribution. A more rigorous wave analysis is beyond the discussion of this book, though we would point the interested reader to \cite{Goodman_2015_a, Tatarski_1967_a, Obukhov_1953_a, Korotkova_2017_a} or \cite[Ch. 13]{Born_1999_a}. We instead take a simpler approach described by \cite{Roggemann_1996_a} that will result in the same solutions as more complex methods, leading us to introduce the concept of the \keyword{layered model for turbulence}\index{turbulence! layered model} to model the atmosphere, which requires us to introduce an assumption:

\boxedmsg{
\vspace{2ex}
\textbf{Assumption}. We assume that the turbulent layers are statistically independent of one another.
\vspace{1ex}
}

We first define our resultant phase as
\begin{align}
\phi(\vxi) &= k \int_{0}^{L} n_1(\vxi,z) \; dz, \notag\nonumber\\
&= k \sum_{i=1}^{M-1} \int_{L_{i-1}}^{L_i}  n_1(\vxi,z) \; dz.
\end{align}
As previously, we are interested in the structure function of this phase realization.
As a result of our assumption of phase screen independence, we may write
\begin{align}
\calD_{\phi}(\vxi, \vxi') &= k^2 \E\left[ \left(\sum_{i=1}^{M-1}  \int_{L_{i-1}}^{L_i} [n_1(\vxi,z) - n_1(\vxi', z)] \, dz \right)^2 \right], \notag\nonumber\\
&\overset{(a)}{=} k^2 \E\left[ \sum_{i=1}^{M-1}\left(  \int_{L_{i-1}}^{L_i} [n_1(\vxi,z) - n_1(\vxi', z)] \, dz \right)^2 \right], \notag\nonumber\\
&= k^2 2.91 k^2 |\vxi-\vxi'|^{5/3} \sum_{i=1}^{M-1} C_n^2[i] \Delta L_i,
\label{eq: Ch3 struc fun multi}
\index{phase structure function! multiple layers}
\end{align}
where $\Delta L_i = L_{i+1} - L_i$ and (a) utilizes the independence of the atmospheric slices, thus dropping the cross terms in the squared summation. This result shows that the phase structure function is related to a weighted sum of the turbulence strength along the path of propagation.

Two additional extensions of this result will be of importance, the first of which is the extension of \eref{eq: Ch3 struc fun multi} to a continuous integration which we give without proof,
\begin{equation}
\calD_{\phi}(\vxi, \vxi') = 2.91 k^2 |\vxi - \vxi'|^{5/3} \int_0^L C_n^2(z) dz.
\index{phase structure function! planar wave}
\end{equation}
A second extension of our result will be to spherical waves. Thus far, we have only derived phase statistics as a result of plane wave propagation, which significantly simplified our analysis. The extension to spherical waves is done by the introduction of a term into the integrand,
\begin{equation}
\calD_{\phi}(\vxi, \vxi') = 2.91 k^2 |\vxi - \vxi'|^{5/3} \int_0^L C_n^2(z) \left(\frac{L - z}{L}\right)^{5/3} dz.
\label{eq: ch2 struct fun spherical}
\index{phase structure function! spherical wave}
\end{equation}
We again emphasize our convention of defining $z=L$ to be in the object plane, thus, the turbulence near the object contributes less than the turbulence near the imaging system due to the term $(L-z)/L$.

\subsection{Wave Structure Function}
Thus far we have commented on the phase structure function. For our purposes, this will be a sufficient description and indeed our chosen model for the rest of the book. However, it is worthwhile to mention that the phase structure function is a \emph{component} of the overall \keyword{wave structure function}.
\boxedthm{
\begin{definition}[\keyword{Wave Structure Function}]\index{Wave Structure Function}
The wave structure function is defined to be \cite{Fried_1967_a}
\begin{equation}
\calD(\vxi, \vxi') = \calD_l(\vxi, \vxi') + \calD_\phi(\vxi, \vxi'),
\end{equation}
where $\calD_l(\vxi, \vxi')$ and $\calD_\phi(\vxi, \vxi')$ are the structure functions of the \keyword{log-amplitude}\index{log-amplitude} and phase, respectively.
\end{definition}
}

The wave structure function leads us to briefly introduce the log-amplitude. The log amplitude quantifies the variation of the \emph{amplitude} of the wave (of course, the wave is comprised of phase and amplitude components.) In a paper by Fried \cite{Fried_1967_a}, the log-amplitude is defined to be
\begin{equation}
    l(\vxi) = \ln ( A(\vxi) / \overline{A} )
\end{equation}
where $A(\vxi)$ is the amplitude at a point $\vx$ and $\overline{A}$ is the root mean square (RMS) value of $A(\vxi)$. It is important to note that the log amplitude was studied in depth in Tatarskii's manuscript \cite{Tatarski_1967_a}, though is more easily referenced in Fried's paper with Fried providing a bit of additional exposition on his choice in usage of $\overline{A}$ in a footnote.

We will typically opt to describe only the phase structure function and not the wave structure function. The reason for this can be understood again by Fried \cite{Fried_1966_a} in which the following approximations are provided
\begin{align}
    \calD_\phi(\vxi, \vxi') &\approx \calD(\vxi, \vxi') &D \gg (L\lambda)^{1/2}, \label{eq: ch2 fried near} \\
    \calD_\phi(\vxi, \vxi') &\approx \frac{1}{2}\calD(\vxi, \vxi') &D \ll (L\lambda)^{1/2}. \label{eq: ch2 fried far}
\end{align}
We refer to the case of \eqref{eq: ch2 fried near} as the near field and \eqref{eq: ch2 fried far} as the far field.
This gives us \emph{some} sense of how to interpret the fact that the phase structure is typically sufficient. In the near field, the wave structure function is approximately the wave structure function whereas in the far field, it is approximately half of the wave structure function. Thus, in the near field, the phase structure function is sufficient in describing the perturbations.

At this point, we wish to remind the reader that the developed theory thus far, and the one which we will apply throughout the book, is valid for only weak to moderate turbulence. This limitation is most precisely shown in the context of Maxwell's equations for inhomogeneous media, though it is well beyond the scope or intent of this book. In Tatarskii's manuscript \cite{Tatarski_1967_a}, the inhomogeneous Maxwell's equations were simplified via the \keyword{Rytov approximation}\index{Rytov approximation}. The results which carry out from here are said to describe \keyword{weak fluctuations} and give rise to what has been presented thus far (weak fluctuations here refers to the amplitude fluctuations). Within this framework, one may analyze the amplitude behavior, also known as \keyword{scintillation}\index{scintillation}. For further details, we would refer the reader to Goodman \cite{Goodman_2015_a} as a starting point, though Tatarskii discusses these directly in his manuscript \cite{Tatarski_1967_a} along with Ishimaru \cite{Ishimaru_1978_a}.

Though one may analyze scintillation in this context (for example, as done by Goodman \cite{Goodman_2015_a}), there exist alternative approaches which utilize more complex mathematical concepts and provide more generality. One such approach is known as the \keyword{path integral formulation}\index{path integral} (alternatively known as a Feynman path integral) \cite{Feynman_1965_b}. The motivation for the path integral arose most famously from quantum mechanics, though the same methods have been applied to wave propagation through random media. These methods were applied to the case of turbulence in works from Tatarskii and collaborators \cite{Zavorotnyi_1977_a, Tatarskii_1980_a, Tatarskii_1993_a, Charnotskii_1993_a} along with others such as Dashen \cite{Dashen_1979_a}. To our knowledge, this framework represents the most general approach to describing the problem to date. These methods seek to describe the regime in which \keyword{strong fluctuations} exist (i.e. strong amplitude fluctuations).

\section{Important Applications of the Model}
\label{sec: sec2_3}
The results of \cref{sec: sec2_2} comprise some of the most significant results of turbulent imaging for our purposes. We now present a few general results such as the Fried parameter. After this, we move towards defining the isoplanatic angle, along with two optical transfer functions that describe various temporal aspects of imaging through the atmosphere.

\subsection{Fried Parameter and Isoplanatic Angle}
\begin{figure}[ht]
\centering
\includegraphics[width=\linewidth]{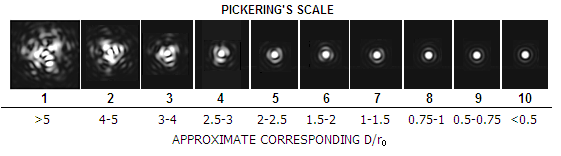}\index{Pickering's scale}
\caption{Impact of the $D/r_0$ to the observed point spread function. Here, $D$ is the aperture diameter and $r_0$ is the Fried parameter. Source: \url{https://eaae-astronomy.org/images/projects/catch-a-star/2015/18_How_to_measure_seeing.pdf}}
\label{fig: Ch2 D r0}
\end{figure}

The atmospheric coherence diameter $r_0$ (or as we shall call it, the \keyword{Fried parameter}) is defined to be a measure related to the resolution of an imaging system. We begin with defining this important parameter introduced by Fried \cite{Fried_1965_a}:
\boxedthm{
\begin{definition}[\keyword{Fried Parameter}]\index{Fried parameter}
Consider a turbulent medium with a structure constant of the index of refraction $C_n^2$, and propagation distance $L$. For a plane wave incident upon turbulence, the Fried parameter is defined as
\begin{equation}
r_0 = 0.185\left[\frac{4\pi^2}{k^2 \int_0^L C_n^2(z) dz}\right]^{3/5},
\end{equation}
and for a spherical wave:
\begin{equation}
r_0 = 0.185\left[\frac{4\pi^2}{k^2 \int_0^L \left(\frac{L - z}{L}\right) C_n^2(z) dz}\right]^{3/5}
\end{equation}
\end{definition}
}
The Fried parameter can be interpreted as an aperture size that is imposed by the limits of atmospheric viewing (analogous to the numerical aperture in image resolution). It is inversely proportional to the turbulence strength $C_n^2$ and the path length $L$. A longer path length and stronger turbulence will give a smaller $r_0$. This $r_0$ will then limit the resolution. If we define $D$ as the aperture diameter, then $D/r_0$ will then tell us the optical resolution of the image observed through turbulence. \fref{fig: Ch2 D r0} shows a few examples.

The Fried parameter is defined in careful accordance with the results of \eref{eq: Ch3 struc fun multi}. As a result, we may rewrite the structure function of the phase as follows:
\boxedthm{
\begin{definition}[\keyword{Kolmogorov Structure Function of the Phase}]
Assuming that index of refraction follows the Kolmogorov structure function, then the structure function of the phase is
\begin{equation}
\calD_{\phi}(\vxi,\vxi') = 6.88\left(\frac{|\vxi-\vxi'|}{r_0}\right)^{5/3},
\label{eq: structure function phase}
\end{equation}\index{phase structure function}
where $r_0$ is the Fried parameter.
\end{definition}
}
Due to isotropy and homogeneity, the structure function depends only on the magnitude of the difference $\vert \vx-\vx' \vert$ instead of the absolute positions. Thus, we can also write the structure function as $\calD_{\phi}(\vert \vxi-\vxi' \vert)$. We further note that \eref{eq: structure function phase} is valid for both planar and spherical waves (though, the definition of the Fried parameter should change accordingly!).

It should be noted that the Fried parameter varies in its definition from planar to spherical. Therefore, one must be aware of which definition is being considered. To mitigate this, many sources denote the spherical Fried parameter as $r_{0,sw}$ (though, we typically will not follow such notation). Additionally, for a constant $C_n^2$ profile, the spherical and planar Fried parameters are related by \cite{Roggemann_1996_a}
\begin{equation*}
r_{0,sw} = \left(\frac{8}{3}\right)^{3/5} r_0.
\end{equation*}

\boxedeg{
\vspace{2ex}
\textbf{Example}. In MATLAB, the Fried parameter can be determined using the command \texttt{integral}. If we let $C_n^2(z) = 1 \times 10^{-15}\mbox{m}^{-2/3}$ for all $z$, $\lambda = 525$nm, and $L = 7000$m, then the Fried parameter is $r_0 = 0.0478$m.  For an aperture of $D = 0.2034$m, the ratio $D/r_0$ is approximately $D/r_0 = 4.26$. This means the smallest spot (i.e., the diffraction PSF) increases 4.26 times due to turbulence.
}

A closely related quantity to the Fried parameter is the \keyword{isoplanatic angle}\index{isoplanatic angle}. The isoplanatic angle defines the field of view within which the turbulence effects are said to be similar. The definition of the isoplanatic angle arises from the adaptive optics community, wherein a guide star is used to ``calibrate'' the imaging system for another, nearby point. The isoplanatic angle quantifies this angle by the following definition \cite{Roggemann_1996_a}:
\boxedthm{
\begin{definition}[\keyword{Isoplanatic Angle}]
The isoplanatic angle is
\begin{equation}
\theta_0  = 58.1 \times 10^{-3} \lambda^{6/5} \left[\int_0^L z^{5/3} C_n^2(z) dz\right]^{-3/5},
\end{equation}
which quantifies the maximum angular separation by which adaptive optics methods will perform suitably well.
\end{definition}
}
For the simulation techniques discussed towards the end of this Chapter, both the Fried parameter and isoplanatic angle are the metrics that can be used to quantify how close our simulation matches with the specified turbulence profile and geometry.

\subsection{Instantaneous OTF}
Recalling the results of \cref{sec: sec1_6}, Theorem \ref{thm: ch1 incoherent} says that if the geometric image is $I_g(\vx)$, the observed image $I_i(\vx)$ will be
\begin{align}
I_i(\vx) &= |h(\vx)|^2 \circledast I_g(\vx).
\end{align}
In the Fourier domain where we define $\widetilde{I}_i(\vf) = \mathfrak{Fourier}[I_i(\vx)]$, it follows that
\begin{align}
\widetilde{I}_i(\vf) = \calH(\vf) \widetilde{I}_g(\vf).
\end{align}
Therefore, the resolution and quality of the observed image are determined by $\calH(\vf)$.

Consider now that there is turbulence present in the system. The turbulence effect is modeled as the product of amplitude distortion, phase distortion, and the pupil function. Substituting our understanding of the complex pupil function into the definition of the OTF, we can show that
\begin{align*}
\calH(\vf)
&= H(\vf) \circledast H^*(\vf)\\
&= \left(P(\lambda z \vf)e^{-j\phi(\lambda z\vf)}\right) \circledast \left(P(\lambda z \vf) e^{j\phi(\lambda z\vf)}\right),
\end{align*}
where $(\cdot)^*$ denotes complex conjugate, and $\circledast$ denotes convolution. We refer to this OTF as the \keyword{instantaneous OTF}. The instantaneous OTF is what the optical system sees at a particular instant. It is a \emph{random} OTF because the phase function $\phi(\lambda z\vf)$ is random.

\boxedthm{
\begin{definition}[\keyword{Instantaneous OTF}]
Let $P(\vu)$ be the aperture located at a distance $z$ from the object, and let $\phi(\lambda z \vf)$ be the random phase distortion. The instantaneous OTF is\index{optical transfer function! instantaneous}
\begin{equation}
\calH(\vf)
= \left(P(\lambda z \vf)e^{-j\phi(\lambda z\vf)}\right) \circledast \left(P(\lambda z \vf) e^{j\phi(\lambda z\vf)}\right),
\end{equation}
where $\circledast$ denotes the convolution.
\end{definition}
}

\boxedeg{

\vspace{2ex}
\textbf{Example}. (\keyword{Linear phase offset}.)
As an example of how the phase distortion can affect the observed image, we consider the case where
\begin{equation*}
\phi(\vxi) = -\frac{2\pi}{\lambda z}\valpha^T \vxi
\end{equation*}
for some \emph{random} vector $\valpha$. Assuming a circular aperture $P(\vxi)$, the amplitude transfer function is
\begin{equation*}
H(\vf) = P(\lambda z \vf)e^{j \phi(\lambda z \vf)} = P(\lambda z \vf) e^{-j 2\pi \valpha^T\vf},
\end{equation*}
where the inverse Fourier transform is
\begin{equation}
\calF^{-1}\left[P(\lambda z \vf)e^{-j 2\pi \valpha^T\vf} \right] = p\left(\frac{\vx-\valpha}{\lambda z}\right)
\end{equation}
where $p(\vx) = J_1(2\pi|\vx|)/|\vx|$. Therefore, the instantaneous PSF is
\begin{align}
h(\vx)
&= \left|p\left(\frac{\vx-\valpha}{\lambda z}\right)\right|^2 \notag \\
&= J_1\left(2\pi \frac{|\vx-\valpha|}{\lambda z}\right)^2 \Big/ \left(\frac{|\vx-\valpha|}{\lambda z}\right)^2,
\label{eq: example PSF}
\end{align}
which is simply a shifted version of the Airy disc.
Therefore, $\phi(\vxi) = -\frac{2\pi}{\lambda z}\valpha^T \vxi$ introduces a random tilt to the PSF. The amount of the random tilt is specified by $\valpha$.
}

\begin{figure}[ht]
\centering
\includegraphics[width=0.65\linewidth]{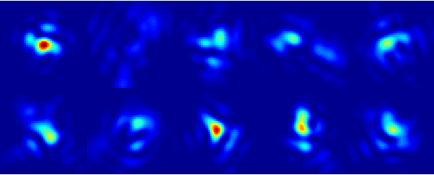}
\caption{Instantaneous point spread functions $|h(\vx)|^2$ generated from propagation through Kolmogorov phase screens. The PSFs shown in this figure are statistically independent. In practice, the spatial correlation needs to be taken into consideration to generate a realistic turbulence effect.}
\label{fig: Ch2 PSF}
\end{figure}

What does the instantaneous OTF look like? One way to visualize the effect is to consider the PSF by taking the inverse Fourier transform of the OTF:
\begin{equation}
\underset{\text{PSF}}{\underbrace{|h(\vx)|^2}} = \mathfrak{Fourier}^{-1}\{\calH(\vf)\}.
\end{equation}
Since $\phi(\vxi)$ is random, $|h(\vx)|^2$ is also random. \fref{fig: Ch2 PSF} shows a few snapshots of the PSFs generated from Kolmogorov theory. The shape of the PSF changes according to the strength of the turbulence.

The randomness of the OTF is governed by two effects:
\begin{itemize}
\item \keyword{Tilt}:\index{tilt} The random pixel displacement caused by the first-order effect of the phase $\phi(\vxi)$. The tilt will cause the pixels to depart from their ideal locations to another location.
\item \keyword{Aberration}:\index{phase aberration} The high-order effects such as blur, spherical aberration, elliptical aberration, etc. caused by the phase delays (and leads) in $\phi(\vxi)$. The aberration will cause a mixing of pixel values, thus making sharp edges become blurred.
\end{itemize}
The decomposition of the phase into tilt and aberration can be summarized by the equation:
\begin{equation}
\underset{\text{overall distortion}}{\underbrace{\phi(\vxi)}} = \underset{\text{aberration}}{\underbrace{\varphi(\vxi)}} + \underset{\text{tilt}}{\underbrace{\valpha^T\vxi}},
\end{equation}
where $\valpha$ is a vector defining the best-fitted first-order plane to the phase function. In the following discussions, we will introduce two OTFS: a \keyword{long exposure OTF} defined through $\phi(\vxi)$ and a \keyword{short exposure} OTF constructed using $\varphi(\vxi)$. The short exposure OTF is also known as the \keyword{tilt-free} OTF, which plays an important role in adaptive optics and the short-exposure analysis.

\subsection{Long Exposure OTF}
The long exposure function\index{long exposure OTF} can be seen as the OTF that will arise from leaving the camera exposure open infinitely long. To this end, we take the statistical expectation on both sides of the equation
\begin{align}
\E_\phi[\calH(\vf)]
&= \E_\phi\left[ \left(W(\lambda z \vf)e^{-j\phi(\lambda z\vf)}\right) \circledast \left(W(\lambda z \vf) e^{j\phi(\lambda z\vf)}\right) \right] \notag \\
&= \E_\phi\left[ \int W(\vxi) e^{-j\phi(\vxi)} \; \cdot \; W(\vxi- \lambda z \vf)  e^{j\phi(\vxi-\lambda z\vf)} d\vxi \right] \notag \\
&= \int W(\vxi) W(\vxi - \lambda z \vf) \E_\phi[e^{-j\phi(\vxi)} e^{j\phi(\vxi-\lambda z\vf)}] d\vxi,
\label{eq: Ch2 long exp 01}
\end{align}
where the subscript $\E_\phi[\cdot]$ emphasizes that the expectation is taken with respect to the random phase $\phi$.

Recognizing the expectation within the integrand, we may utilize the moment generating function \eqref{eq: moment gen func} as we did in \cref{sec: sec2_2} to directly write
\begin{align}
\E_\phi[e^{-j\phi(\vxi)} e^{j\phi(\vxi-\lambda z\vf)}] &=
\Gamma_{t_\phi}(\vxi,\vxi - \lambda z \vf), \\
&= \exp \left\{ -\frac{1}{2} D_\phi(\lambda z \vert \vf \vert) \right\}, \\
&= \exp \left\{ -\frac{1}{2} 6.88 \left(\frac{\lambda z |\vf|}{r_0}\right) \right\}.
\end{align}
where we have used the homogeneous property of the phase structure function. This results in the expected value of the OTF to be decomposed as
\begin{align*}
\E_\phi[\calH(\vf)] =
&\underset{\calH_{\text{diff}}(\vf)}{\underbrace{\int W(\vxi) W(\vxi- \lambda z \vf) d\vxi }}  \; \times \;
\underset{\calH_{\text{LE}}(\vf)}{\underbrace{\exp\left\{-\frac{1}{2} 6.88 \left( \frac{\lambda z |\vf| }{ r_0 } \right)^{5/3} \right\}}},
\end{align*}
which is a product of $\calH_{\text{diff}}(\vf)$ and $\calH_{\text{LE}}(\vf)$:
\boxedthm{
\begin{theorem}[\keyword{Decomposition of Average OTF}]
The average OTF consists of two terms:
\begin{equation}
\E_\phi[\calH(\vf)] = \calH_{\text{diff}}(\vf) \times \calH_{\text{LE}}(\vf),
\end{equation}
where $\calH_{\text{diff}}(\vf)$ denotes the diffraction-limited OTF and $\calH_{\text{LE}}(\vf)$ denotes the long exposure OTF of the atmosphere.
\end{theorem}
}
For a circular aperture, $\calH_{\text{diff}}(\vf)$ follows from \cref{sec: sec1_5}:
\boxedthm{
\begin{theorem}[\keyword{Diffraction limited OTF}]
For a circular aperture, the diffraction limited OTF is
\begin{align}
\calH_{\text{diff}}(\vf)
&= \frac{2}{\pi}\left[ \arccos\left(\frac{|\vf|}{2f_0}\right) - \frac{|\vf|}{2f_0}\sqrt{1- \left(\frac{|\vf|}{2f_0}\right)^2} \right],
\end{align}
for $f \le 2f_0$ where $f_0$ is the cutoff frequency of the coherent version of the system.
\end{theorem}
}

The atmospheric term $\calH_{\text{LE}}(\vf)$ is known as the long exposure OTF, which is what we would observe if we turn on the shutter of the camera for a prolonged period of time.
\boxedthm{
\begin{definition}[\keyword{Long Exposure OTF}]
The long exposure optical transfer function $\calH_{\text{LE}}(\vf)$ is defined as
\begin{equation}
\calH_{\text{LE}}(\vf) = \exp\left\{- 3.44 \left( \frac{\lambda z |\vf| }{ r_0 } \right)^{5/3} \right\},
\end{equation}
where $z$ is the path length, and $r_0$ is the Fried parameter.
\end{definition}
}
The difference between the long exposure OTF and the instantaneous OTF is that the instantaneous OTF is random whereas the long exposure OTF is deterministic. The randomness is absorbed by the structure function of the phase when evaluating the joint expectation.

\begin{figure}[h]
\centering
\begin{tabular}{cc}
\includegraphics[width=0.49\linewidth]{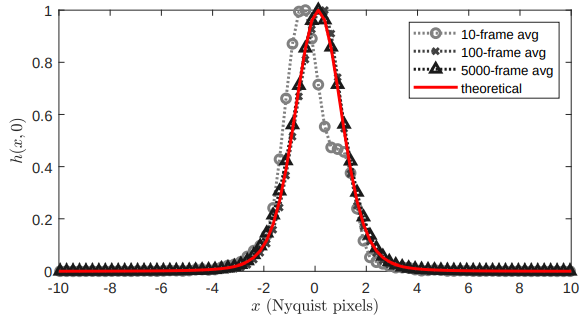}&
\hspace{-2ex}
\includegraphics[width=0.49\linewidth]{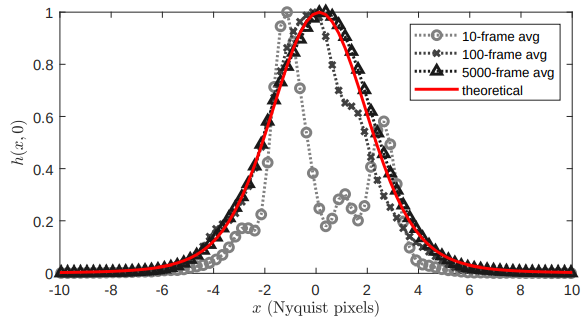}\\
$C_n^2 = 2.5\times10^{-16}$m$^{-2/3}$ & $C_n^2 = 1\times10^{-15}$m$^{-2/3}$\\
\end{tabular}
\caption{The theoretical long-exposure PSF compared with the empirical averaged instantaneous PSFs for $C_n^2 = 2.5\times10^{-16}$m$^{-2/3}$ and $C_n^2 = 1\times10^{-15}$m$^{-2/3}$. Shown in each sub-figure are the 10-frame average, 100-frame average, and 5000-frame average PSFs, all in black. The red curve is the theoretically predicted result. The $x$-axis denotes the Nyquist pixel, where one pixel corresponds to $\lambda z/D$ with $D$ being the aperture diameter. }
\label{fig: Long Exposure 2}
\end{figure}

As a visualization of the instantaneous OTF and the long-exposure OTF, in \fref{fig: Long Exposure 2} we show a comparison. Here, we plot the cross-section of the PSF (instead of the OTF) using a finite-sample ensemble average of 10, 100, and 5000 instantaneous PSFs. For the weaker turbulence case where $C_n^2 = 2.5\times10^{-16}$m$^{-2/3}$, we see that the random fluctuation of the instantaneous PSF becomes weak as soon as we use 100 instantaneous PSFs. If we consider a stronger turbulence case where $C_n^2 = 1\times10^{-15}$m$^{-2/3}$, the random fluctuation is stronger. Moreover, the spread of the long-exposure PSF is also greater when we use a higher $C_n^2$.

\begin{figure}[h]
\centering
\footnotesize{
\begin{tabular}{cccc}
\includegraphics[width=0.23\linewidth]{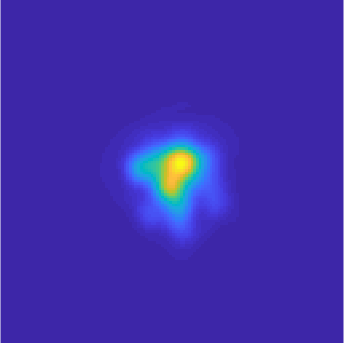}&
\hspace{-2ex}\includegraphics[width=0.23\linewidth]{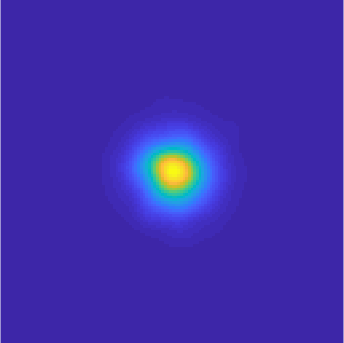}&
\hspace{-2ex}\includegraphics[width=0.23\linewidth]{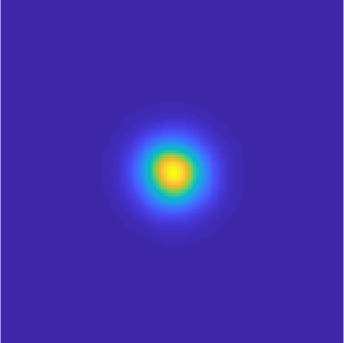}&
\hspace{-2ex}\includegraphics[width=0.23\linewidth]{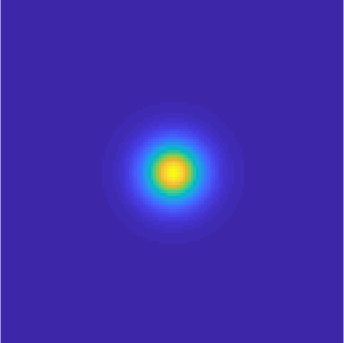}\\
(a) 50-avg &(b) 500-avg &(c) 5000-avg &(d) Theoretical\\
\end{tabular}}
\caption{Instantaneous PSFs for $C_n^2 = 1\times 10^{-15}$m$^{-2/3}$. (a) 50-frame average. (b) 500-frame average. (c) 5000-frame average. (d) Theoretical long-exposure PSF.}
\vspace{-2ex}
\label{fig: Long Exposure}
\end{figure}

\fref{fig: Long Exposure} shows another visual comparison between the finite-sample ensemble average and the statistical long-exposure. As we can see from the plots, the more instantaneous PSFs we use to compute the ensemble average, the closer the average will converge to the true long-exposure.

\subsection{Short Exposure OTF}
The long-exposure OTF is the statistical average of an infinite number of instantaneous OTFs observed over a very long period of time. In many advanced imaging systems, fast steering mirrors are installed to adaptively compensate for the \keyword{tilt} observed in the wavefront. The mathematical model removing the tilt from the phase is to subtract a linear term from the observed phase:
\begin{equation}
\varphi(\vxi) \bydef \phi(\vxi) - \valpha^T\vxi,
\end{equation}
where $\valpha^T\vxi$ is the best linear fit to $\phi(\vxi)$.

An image taken without the tilt is known as the \keyword{tilt-compensated} frame. If we can obtain many of these tilt-compensated frames, then we can take another statistical average known as the \keyword{short exposure}\index{short exposure} image. The short exposure image can be thought of \emph{freezing} the turbulence and re-aligning the pixels such that they are centered. The tilt contributes a majority of the energy of the distortions (such as in Table 4 of Noll \cite{Noll_1976_a} for example), and as such, being able to use hardware to compensate for the tilt can lead to a much better image.

Now consider the tilt-compensated phase realization. In this case, the phase is
\begin{align}
t_{\varphi}(\vxi)
&= \exp\left\{j\varphi(\vxi)\right\} = \exp\left\{j(\phi(\vxi) - \valpha^T\vxi)\right\}.
\end{align}
The autocorrelation of the tilt-correct wavefront is
\begin{align}
\Gamma_{t_{\varphi}}(\vxi,\vxi')
&= \E\left[ t_{\varphi}(\vxi)  t_{\varphi}^*(\vxi')  \right] \notag \\
&= \E\left[ \exp\left\{ j(\phi(\vxi) - \valpha^T\vxi) -  j(\phi(\vxi') - \valpha^T\vxi') \right\}\right] \notag \\
&= \E\left[ \exp\left\{ j(\phi(\vxi) -  \phi(\vxi') - \valpha^T(\vxi - \vxi')) \right\}\right].
\end{align}
Using the same trick of the moment generating function of a zero-mean Gaussian, we can show that
\begin{align}
\Gamma_{t_{\varphi}}(\vxi,\vxi') &= \exp\Bigg\{ -\frac{1}{2} \E\left[ (\phi(\vxi) - \phi(\vxi'))^2 \right] \\
&\;\; + \E\left[(\phi(\vxi) - \phi(\vxi'))\valpha^T(\vxi-\vxi')\right] - \frac{1}{2}\E\left[(\valpha^T(\vxi-\vxi'))^2 \right]  \Bigg\} \notag .
\end{align}
At this point, we aim to simplify the preceding expression. This leads us to separately analyze the three terms in the exponential:
\begin{itemize}
\item The first term is $\E\left[ (\phi(\vxi) - \phi(\vxi'))^2 \right]$, which is nothing but the structure function of the phase. Therefore, we have
\begin{align}
\E\left[ (\phi(\vxi) - \phi(\vxi'))^2 \right] = \calD_{\phi}(\vxi-\vxi'),
\end{align}
where we assumed that the phase is homogeneous so that the structure function can be written in terms of $\vxi-\vxi'$.
\item The second term is the cross correlation between the phase difference $\phi(\vxi) - \phi(\vxi')$ and the linear term $\valpha^T(\vxi-\vxi')$. Following the argument of Fried, we may assume that these two terms are statistically independent (though, this approximation has been considered more carefully by Heidbreder \cite{Heidbreder_1967_a}, showing the term may not always be so easily neglected). Therefore, by using the zero-mean property of the phase, we can show that
    \begin{align*}
     \E[(\phi(\vxi) - \phi(\vxi'))\valpha^T(\vxi-\vxi')] = \underset{=0}{\underbrace{\E[\phi(\vxi) - \phi(\vxi')]}} \E[\valpha^T(\vxi-\vxi')].
    \end{align*}
    Hence, the cross term can be dropped.
\item The third term is the second moment of $\valpha^T\vxi$, which is
\begin{align}
\E[(\valpha^T(\vxi-\vxi'))^2] = \E[|\valpha|^2] \cdot |\vxi-\vxi'|^2.
\label{eq: Ch2 short exp second moment}
\end{align}
According to Fried \cite{Fried_1966_a}, the expectation $\E[|\valpha|^2]$ is
\begin{equation}
\E[|\valpha|^2] = 6.88 r_0^{-5/3}D^{-1/3},
\end{equation}
where $D$ is the aperture diameter. Substituting this into \eref{eq: Ch2 short exp second moment} yields
\begin{equation*}
\E[(\valpha^T(\vxi-\vxi'))^2] = 6.88 r_0^{-5/3}D^{-1/3} |\vxi-\vxi'|^2.
\end{equation*}
\end{itemize}
Combining all three terms, we can show that
\begin{align}
\Gamma_{t_{\varphi}}(\vxi,\vxi')
&= \exp\Bigg\{ -\frac{1}{2} \calD_{\phi}(\vxi-\vxi') - \frac{1}{2}\E[(\valpha^T(\vxi-\vxi'))^2]\Bigg\} \notag \\
&= \exp\Bigg\{ -\frac{1}{2} 6.88\left(\frac{|\vxi-\vxi'|}{r_0}\right)^{5/3} - \frac{1}{2}6.88 \frac{|\vxi-\vxi'|^2}{r_0^{5/3}D^{1/3}}\Bigg\} \notag \\
&= \exp\Bigg\{ -3.44\left(\frac{|\vxi-\vxi'|}{r_0}\right)^{5/3} \left(1-\left(\frac{|\vxi-\vxi'|}{D}\right)^{1/3}\right)\Bigg\}.
\end{align}

Following the same argument as we did in \eref{eq: Ch2 long exp 01}, we can show that the short exposure OTF is
\begin{align*}
\E_\varphi[\calH(\vf)] =
&\underset{\calH_{\text{diff}}(\vf)}{\underbrace{\int W(\vxi) W(\vxi- \lambda z \vf) d\vxi }}  \\
&\; \times \; \underset{\calH_{\text{SE}}(\vf)}{\underbrace{
\exp\Bigg\{ -3.44\left(\frac{\lambda z |\vf|}{r_0}\right)^{5/3} \left(1-\left(\frac{\lambda z |\vf|}{D}\right)^{1/3}\right)\Bigg\}
}}.
\end{align*}
Thus, we arrived at another important result.
\boxedthm{
\begin{definition}[\keyword{Short Exposure OTF}]
The short exposure optical transfer function $\calH_{\text{SE}}(\vf)$ is defined as
\begin{equation}
\calH_{\text{SE}}(\vf) = \exp\Bigg\{ -3.44\left(\frac{\lambda z |\vf|}{r_0}\right)^{5/3} \left(1-\left(\frac{\lambda z |\vf|}{D}\right)^{1/3}\right)\Bigg\},
\end{equation}
where $D$ is the aperture diameter, $z$ is the path length, and $r_0$ is the Fried parameter.
\end{definition}
}

\begin{figure}[h]
\centering
\begin{tabular}{cc}
\hspace{-1ex}\includegraphics[width=0.49\linewidth]{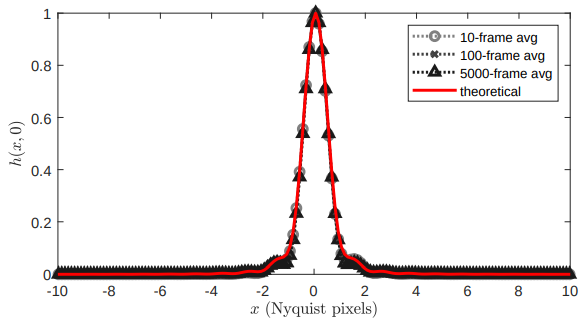}&
\hspace{-2ex}\includegraphics[width=0.49\linewidth]{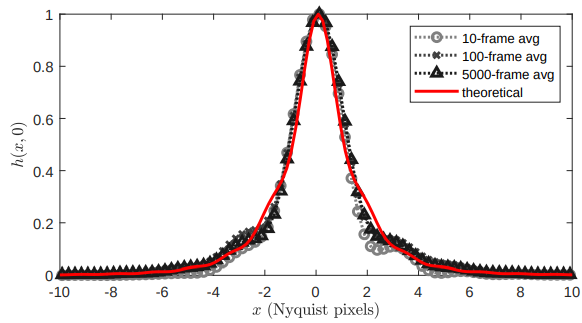}\\
$C_n^2 = 2.5\times10^{-16}$m$^{-2/3}$ & $C_n^2 = 1\times10^{-15}$m$^{-2/3}$\\
\end{tabular}
\caption{The theoretical short-exposure PSF compared with the empirical averaged instantaneous PSFs for $C_n^2 = 2.5\times10^{-16}$m$^{-2/3}$ and $C_n^2 = 1\times10^{-15}$m$^{-2/3}$. Shown in each sub-figure are the 10-frame average, 100-frame average and 5000-frame average PSFs, all in black. The red curve is the theoretically predicted result. The $x$-axis denotes the Nyquist pixel, where one pixel corresponds to $\lambda z/D$ with $D$ being the aperture diameter. }
\label{fig: Short Exposure 2}
\end{figure}

\fref{fig: Short Exposure 2} shows the short-exposure point spread functions for two levels of $C_n^2$. Compared to \fref{fig: Long Exposure 2}, the shape of the short exposure PSF is narrower. The reason is that when the tilts are compensated, the PSFs will be centered. The random fluctuation is thus reduced compared to the long-exposure case.

\subsection{Probability of Getting a Lucky Observation}
The atmosphere fluctuates in time, and so do the resultant PSFs that distort an image. This leads us to an interesting component of atmospheric imaging. When observing a point source, sometimes the point will look bad and sometimes it will look good! This is simply a result of the random nature of the atmosphere; sometimes we get \emph{lucky} and sometimes we don't.

The short exposure function introduced the tilt-compensated phase term known as the \keyword{effective wavefront}. The effective wavefront contributes blur from the higher order aberrations. If we are interested in taking a clean image of a star, whether or not it has shifted by a few pixels is hardly a consideration if it means having a non-blurry image. For this reason, Fried refers to $\varphi$ as the effective wavefront, as it is the component of the wavefront that will effectively limit us in resolution.

In 1978 Fried analyzed the probability of having a lucky observation\index{lucky probability} \cite{Fried_1978_a}. We may anticipate that a lucky observation is defined to be a PSF that is near a delta function. However, there is no closed form expression for a turbulent PSF, therefore we cannot analyze the PSF directly. Fried instead focuses on the mean square of the effective wavefront. Fried defines the mean square distortion to be
\begin{equation}
    \Delta^2(\varphi) = \left( \frac{\pi}{4} D^2 \right)^{-1} \int_{\text{Aperture}}  P\left(\frac{\vxi}{D}\right) |\varphi(\vxi)|^2 d \vxi,
\end{equation}
where we are normalizing by the area of the aperture. Fried's analysis ultimately results in the following approximate result:
\boxedthm{
\begin{definition}[\keyword{Probability of a Lucky Event}]
The probability of having a lucky observation is approximated as
\begin{equation}
    \Pb \left( \Delta^2(\varphi) \leq 1 \text{\emph{ rad}}^2 \right) \approx 5.6 \exp(-0.1557 (D/r_0)^2)
\end{equation}
for $D/r_0 \geq 3.5$ with $D$ as the imaging system's aperture diameter and $r_0$ as the Fried parameter.
\end{definition}
}

\begin{figure}
    \centering
    \includegraphics[width=0.85\linewidth]{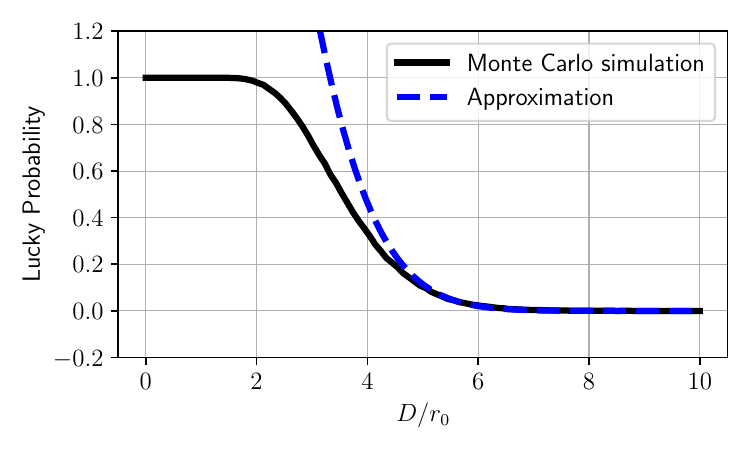}
    \caption{Lucky probability curve. As $D/r0$ increases the probability of a lucky observation drops with a sharp transition.}
    \label{fig: Ch2 lucky curve}
\end{figure}

This result showcases the fact that this $D/r_0$ ratio continually arises in the atmospheric turbulence literature. Fried additionally performs a Monte Carlo simulation, which we reproduce in \fref{fig: Ch2 lucky curve}. We can observe that at the critical threshold near $D/r_0 = 2$, the lucky probability begins its sharp transition. This can help to inform us as to the definition of the Fried parameter. When the Fried parameter is larger than the aperture diameter, the imaging system is near maximum efficiency. However, the moment $r_0$ drops below the size of the aperture, the system drops in its efficiency, dramatically attenuating as the ratio of $D/r0$ increases above 2.

We will revisit the discussions of the lucky effect in Chapter 5. There, we will take the perspective of image reconstruction algorithms.

\section{Split-Step Simulation}
\label{sec: sec2_4}
\index{split-step}With the theoretical descriptions of the atmospheric model presented, we now describe a way towards simulating these effects. The literature regarding the simulation of imaging through atmospheric turbulence presents a few varieties of simulations, with split-step being likely the most intuitive one. Implementing split-step such that the statistics are correct is a non-trivial task, thus we recommend the book by Schmidt \cite{Schmidt_2010_a} which provides an excellent, in-depth description of numerical optical simulations and split-step for atmospheric turbulence along with detailed analysis. This book pairs well with the book from Voelz \cite{Voelz_2011_a} which deals with general optical simulation principles.

\subsection{A History of Split-Step Simulation}
\index{split-step! history}The modality of split-step simulation is a general approach that has been used both computationally \cite{Hardie_2017_a, Roggemann_2012_a, Roggemann_1995_a} and in a laboratory sense with physical devices such as tunable phase modulators \cite{Wang_2016_a, Rodenburg_2014_a, Phillips_2005_a}, heat sources \cite{Vorontsov_2001_a, Yuandong_2012_a}, or plastic screens \cite{Pellizzari_2019_a}, where a series of physical devices are arranged in such a way that a propagating wave is perturbed by the device. A clear visualization of such an experimental setup with details is provided in a work by Pellizzari et al. \cite{Pellizzari_2018_a}. Computationally, the work generating the first phase screen simulator by use of Fourier-based \keyword{phase screens} was McGlamery \cite{McGlamery_1967_a}.

The scope of split-step is not limited to the propagation of atmospheric turbulence, with an example being an early application of this principle to sound waves through media \cite{Macaskill_1987_a} and its subsequent consideration as a general simulation tool \cite{Spivack_1989_a}. Applications towards wave propagation through turbulence were later presented \cite{Martin_1988_a, Coles_1995_a} which detailed both the approach and constraints in sampling upon properly performing a simulation of this variety. Proper sampling is critical in a split-step simulation. We ignore a majority of these details, opting for a simple rule, though a thorough explanation of the sampling criteria is given by Schmidt \cite{Schmidt_2010_a}.

The introduction of randomness and the Kolmogorov turbulent model into the simulation is by a set of turbulent layers which are numerically represented by phase screens. As a result, a great deal of effort has been dedicated to an accurate representation of these effects \cite{Bos_2015_a, Frehlich_2000_a, Burckel_2013_a, Herman_1990_a, Welsh_1997_a, Fried_2008_a, Lane_1992_a, Anzuola_2017_a}. We will briefly cover the prevailing approach towards accurately generating phase screens, as not including these more nuanced aspects leads to poor results in terms of statistical accuracy. Our coverage will be somewhat brief, thus our presentation should be considered as an introduction.

\subsection{Split-Step Simulation: Building Blocks}
Conceptually, split-step is a natural approach to the simulation of atmospheric turbulence effects. Our derivation of the distortions in the wave used phase screens to represent the atmosphere distributed across the path of propagation. Split-step is the numerical equivalence of this process: we will segment the propagation path into a set of screens which we will numerically propagate a wave through.

\begin{figure}[ht]
\centering
\includegraphics[width=0.85\linewidth]{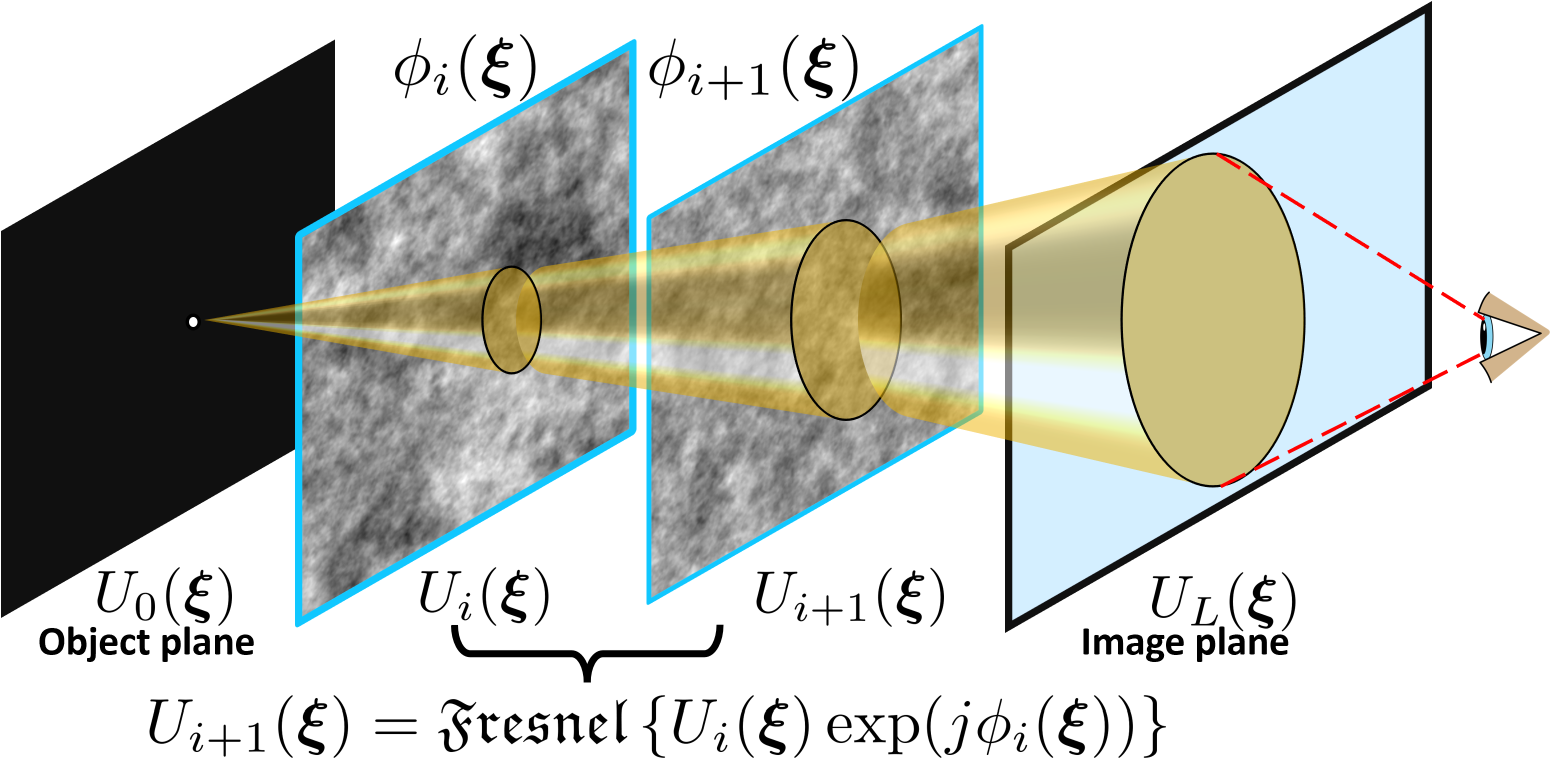}
\caption{Visualization of split-step. A point source (more generally a grid of point sources) is propagated through numerically generated phase screens, traveling between the phase screens by Fresnel propagation.}
\label{fig: Ch2 split step overview}
\end{figure}

We present a visualization of the split-step simulation in \fref{fig: Ch2 split step overview}. The split-step method is able to match atmospheric statistics and produce realistic generations of images for a wide variety of simulations. We present a step-by-step process of the simulation method, elaborating on the details throughout \cref{sec: sec2_4}.
\begin{enumerate}
\item \keyword{Phase screen generation}\index{phase screen}. The first and arguably most important step is the generation of the numerical phase screens. These phase screens set aspects of turbulence strength and imaging geometry. We will consider a set of phase screens $\{ \phi_i\}$.
\item \keyword{Numerical wave propagation}\index{numerical wave propagation}. Once the phase screens are generated, a point source is modeled and propagated through the series of phase screens. This is done by numerical wave propagation (via Fresnel or alternative methods) followed by a phase imparting step, directly following \eref{eq: ch2 phase screen imparting}. We emphasize a key point here: this point source generation and propagation is performed \emph{per-pixel}.
\item \keyword{PSF formation}. With the wave propagated to the aperture plane, we apply the PSF formation equation \eref{eq: Ch1 theorem lens convolution 2} and accordingly apply it in an incoherent or coherent fashion.
\end{enumerate}
As noted, the operation of numerical wave propagation is done per pixel. This utilizes the superposition property of wave propagation. This represents the main computational bottleneck of split-step.

\subsection{Generating Phase Screens}
\index{phase screen! generation of}The first step of the split-step propagation is the generation of the phase screens $\{\phi_i\}$. As the name suggests, a phase screen is a 2D function of the phase that should be imparted to the incident field to generate the field at the next step. Since the phase screen is a random field, it needs to be sampled from a probability distribution.

\begin{figure}[h]
\centering
\begin{tabular}{cc}
\hspace{-2ex}\includegraphics[width=0.4\linewidth]{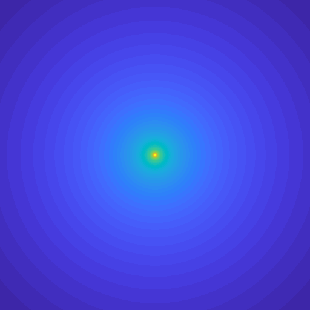}&
\hspace{-2ex}\includegraphics[width=0.4\linewidth]{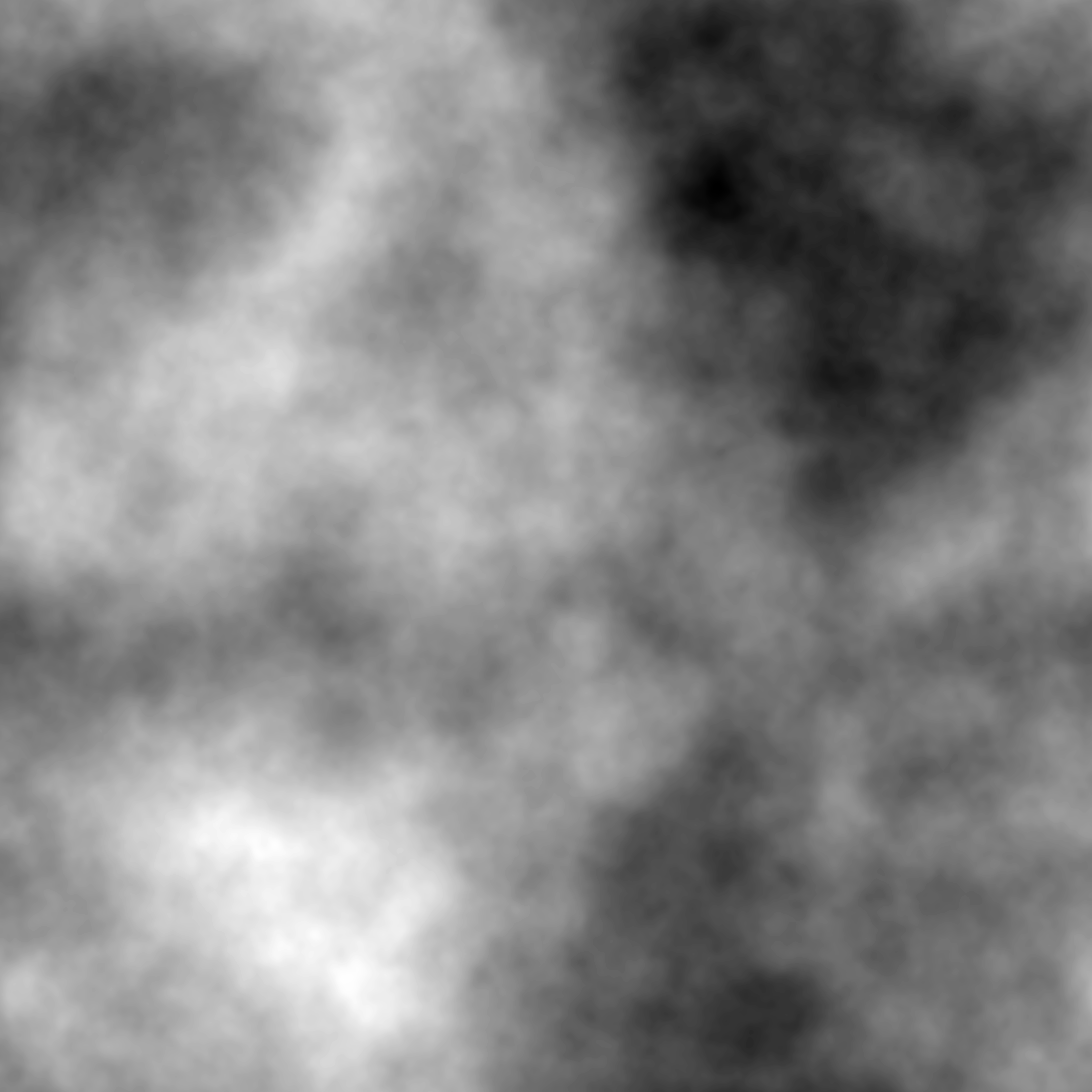}\\
(a) Power spectral density & (b) Random field \\
$\Phi_n(\vk) = 0.033C_n^2|\vk|^{-11/3}$ & Sample drawn from $\Phi_n(\vk)$
\end{tabular}
\caption{Kolmogorov power spectral density and a random realization of the random field generating from the Kolmogorov PSD.}
\label{fig: ch2 Kolmogorov PSD}
\end{figure}

The distribution for the phase screen, in most of the literature, is assumed to be a multivariate Gaussian \cite{Tatarski_1967_a}. If we discretize the continuous phase screen as a finite size array, then we are essentially drawing a random vector from a high-dimensional Gaussian distribution. Since a Gaussian is fully characterized by the correlation matrix (or the power spectral density in the Fourier domain), we can use the results such as the Kolmogorov power spectral density. An example of generating a phase screen is presented below. The result of a phase screen from this process is presented in \fref{fig: ch2 Kolmogorov PSD}.

\boxedeg{
\vspace{2ex}
\textbf{Example}. (\keyword{Drawing a Random Field from the Kolmogorov PSD}.) Like a one-dimensional random process, drawing samples from a given power spectral density can be done in the Fourier domain. Denote $\mGamma_n$ as the correlation matrix with the $(i,j)$th entry being $[\mGamma_n]_{ij} = \Gamma_n(\vr_i-\vr_j)$ for coordinates $\vr_i$ and $\vr_j$. The Fourier relationship $\Gamma_n(\vr) \overset{\mathfrak{Fourier}}{\longleftrightarrow} \Phi_n(\vk)$ implies that $\mGamma_n$ is diagonalizable using the discrete Fourier transform $\mF$:
\begin{equation*}
\mGamma_n = \mF \mPhi_n \mF^H,
\end{equation*}
where $\mPhi_n$ is a diagonal matrix representing the power spectral density. The $(i,j)$th entry is $[\mPhi_n]_{ij} = \Phi_n(\vk_i,\vk_j)$.

To draw a zero-mean Gaussian random field $\vy$ according $\mGamma_n$, we can start with a white noise vector $\vw \sim \text{Gaussian}(0,1)$, and multiply it with
\begin{equation*}
\vy = \mF \mPhi_n^{\frac{1}{2}} \vw.
\end{equation*}
Then, one can show that $\E[\vy\vy^T] = \mPhi_n$.

A piece of short MATLAB code to simulate the two-dimensional random field is given below. The generated random field is shown in \fref{fig: ch2 Kolmogorov PSD}. In this example, we assumed $C_n^2 = 10^{-16} \text{m}^{-2/3}$ and we use a normalized coordinate $|k| \in [0,1]$.
}

\begin{Verbatim}[frame=single,framerule=0.5mm, rulecolor=\color{NavyBlue}]
% MATLAB code
[kx, ky] = meshgrid(linspace(-1,1,512));
PSD_kolm = 0.033*1e-16*(sqrt(kx.^2+ky.^2)).^(-11/3);
w        = randn(size(kx));
phi      = ifft2(sqrt(PSD_kolm).*w);
imshow(abs(phi),[]);
\end{Verbatim}

Additionally, we must carefully consider the sampling of the phase screens. The sampling must be done so that (i) the sampling requirements for the phase screens are met and (ii) the sampling criterion for the aperture plane is also satisfied. A full discussion of this balance is provided in \cite{Schmidt_2010_a}. A standard approach towards meeting this criterion is the Voelz criterion for the phase screens \cite{Voelz_2011_a},\index{Voelz sampling criterion}
\begin{equation}
\Delta x = \frac{\lambda L}{sD},
\end{equation}
with $s \geq 4$.

These considerations are still not enough to properly sample a phase screen. The Kolmogorov PSD is often underrepresented in the low frequency terms due to sampling due to the decay near the small frequencies. The phase screens empirical structure function will not match the desired theoretical description (see Schmidt \cite{Schmidt_2010_a} Figure 9.3). To fix this, one may perform \keyword{subharmonic generation}\index{phase screen! subharmonic generation} which adds extra terms to the phase screen which models the low frequency components. The sampling of these phase screens, and which frequencies should be chosen, again must be done in coordination with other sample spacings in the system.

\subsection{Numerical Wave Propagation}
\index{numerical wave propagation}Wave propagation is the process by which the wave traverses the atmosphere, thus it must be a component of our simulation as well. Our modeling of numerical wave propagation will take place between each phase screen. This is the reasoning behind the name ``split-step'', the steps of wave propagation and phase imparting are split apart.

Consider a point source located in our object plane defined at location $\vu = \mathbf{0}$. We will model this by the assignment of this location to be $U_0(\vu) = \delta(\vu)$. With this, the Fresnel integral of \eref{eq: ch1 Fresnel diffraction main} for a distance of $z_i$ away is given by
\begin{equation}
U_1(\vxi) = \frac{e^{jk z_i}}{j\lambda z_i} e^{j \frac{k}{2 z_i} |\vxi|^2}.
\label{eq: ch2 point source fresnel u}
\end{equation}
This result is sufficient for a point source, however, what should we do with the field after we propagate it to our first phase screen when it is no longer a point source? This leads us to write the Fresnel integral in a slightly different way than before
\begin{equation}
U_{i+1}(\vxi) = \int_\Sigma U_i(\vu) \frac{e^{jk z_i}}{j\lambda z_i} e^{j \frac{k}{2 z_i} |\vxi - \vu|^2} d\vu,
\end{equation}
where we note that this \emph{also} applies to the original propagation of a point source.
Examining this equation, we notice this is simply a convolution. This gives us a new perspective of the Fresnel equation: Fresnel propagation is given by convolution with a spatially invariant kernel. This motivates us to define the Fresnel kernel,
\begin{equation}
h_\text{Fres}(\vxi; z) = \frac{e^{jk z}}{j\lambda z} e^{j \frac{k}{2 z} |\vxi|^2}.
\label{eq: ch2 point source fresnel}
\end{equation}

Viewing Fresnel propagation as a convolutional kernel simplifies our analysis. Specifically, for \emph{any} wave distribution at a particular propagation distance, we may compute the field distribution at a further distance by Fresnel kernel in \eref{eq: ch2 point source fresnel}. The wave propagation including phase screens will then be achieved by
\begin{equation}
U_{i+1} (\vxi) = [h_\text{Fres}(\vxi; z) \circledast U_i(\vxi)] e^{j\phi_i(\vxi)}.
\label{eq: ch2 split step main eq}
\end{equation}
It is worth noting that the propagation \eqref{eq: ch2 split step main eq} is notationally simplistic to write in this form, however, computationally we typically opt for \keyword{angular spectrum propagation}\index{angular spectrum propagation} \cite{Goodman_2005_a, Schmidt_2010_a}. Without going into too many details, the idea of the angular spectrum is to decompose a wave by a sum of plane waves. The propagation of a planar wave is simple to write, therefore, if we know the decomposition the propagation rules become easier. Mathematically this can be understood to be a Fourier decomposition.

\subsection{Image Formation}
After propagating our wave through a series of phase screens it will be incident upon our aperture. We will refer to this as $U_{M+1}(\vu)$. We must then model the wave passing through the lens-aperture system, then propagating to the sensor plane.

We can say our incident wave can be written as
\begin{equation}
    U_{M+1}(\vxi) = A(\vxi) \exp{j \left(\frac{k\vert \vxi \vert^2}{2L} + \phi(\vxi)\right)}.
\end{equation}
Here we have simplified the terminology somewhat through $A(\vxi)$ and $\phi(\vxi)$ which are the amplitude and phase of the \emph{resultant} split-step propagation steps. We wish to emphasize this point a bit further as it will be important for the next Chapter. We will refer to $\phi(\vxi)$ as a \keyword{phase realization}\index{phase realization} if it is representative of the phase of a wave propagated through turbulence. Notice this is slightly different than a phase \emph{screen}, which models a Chapter of the atmosphere's phase contribution.

At this point, we need to remove the Fresnel pattern in order to access the turbulence phase distortion $\phi(\vxi)$ directly. This leads us to recall the thin lens equation
\begin{equation}
    U_M'(\vxi) = t_{\text{lens}}(\vxi) U_M(\vxi),
\end{equation}
with our goal being that the response of the lens makes the following to be true,
\begin{equation}
    U_M'(\vxi) = A(\vxi) \exp{j\phi(\vxi)}.
\end{equation}
If the lens is chosen such that $t_{\text{lens}}$ is the complex conjugate of the Fresnel pattern, then it will cancel out this portion of the wavefront. Thus, we choose the lens response to be
\begin{equation}
    t_{\text{lens}}(\vxi) = \exp{-j \frac{k\vert \vxi \vert^2}{2L}}.
\end{equation}
Here we have assumed that the lens cancels the Fresnel pattern \emph{and} will focus the incident wave to the focal plane.
Typically, through careful construction of the point source, we can assume that $A(\vu) = 1$ \cite{Hardie_2017_a}. This leaves us with the standard PSF equation we have presented before,
\begin{equation}
    \vert h(\vx) \vert^2 = \mathfrak{Fourier} \{P(\vxi) e^{j\phi(\vxi)} \}.
\end{equation}
Throughout our derivation we have fixed $\vx = \mathbf{0}$, therefore, the generalization of this result is the familiar equation
\begin{equation}
    \vert h_{\vu}(\vx) \vert^2 = \mathfrak{Fourier} \{P(\vxi) e^{j\phi_{\vu}(\vxi)} \}.
\end{equation}

\begin{figure}[ht]
\centering
\includegraphics[width=0.75\linewidth]{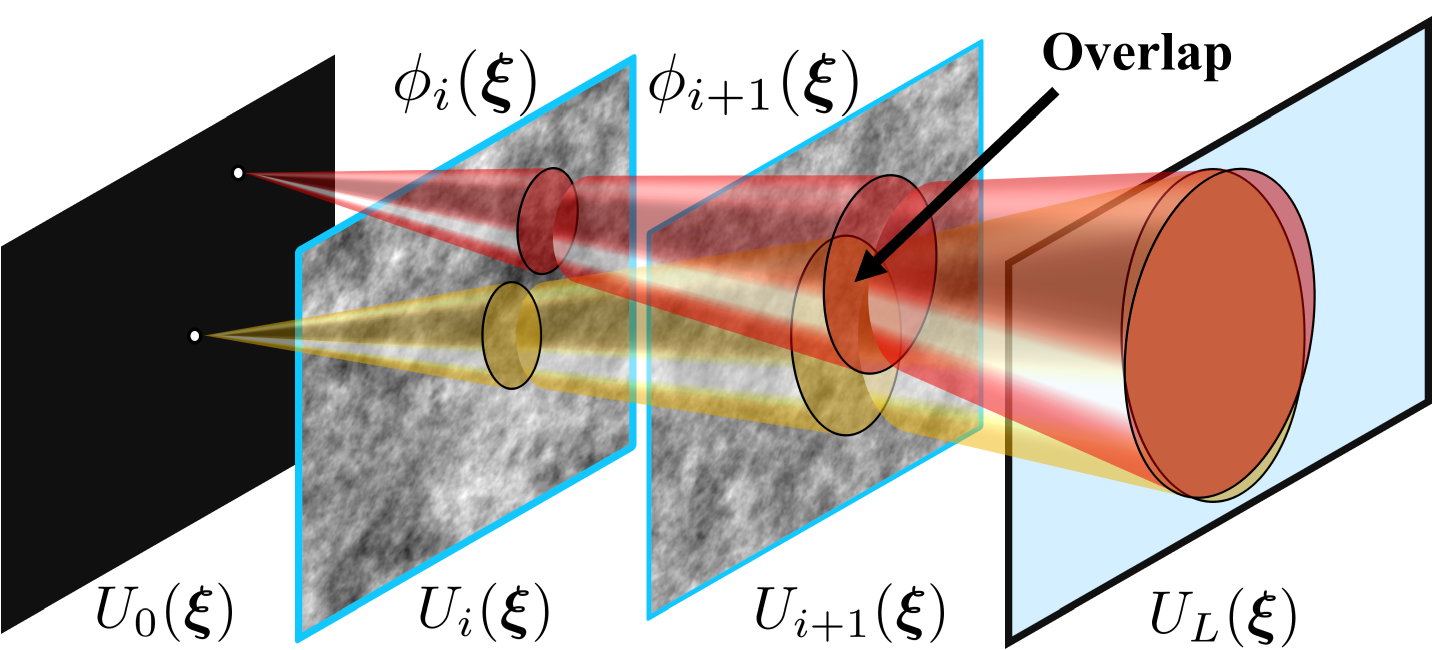}
\caption{Visualization of how correlations arise in split step. The more overlap between the propagation paths of two points, the more correlated their output phase realizations, and thus their PSFs, will be.}
\label{fig: Ch2 split step overlap}
\end{figure}

How should we model the phase's dependency on $\vu$? We provide a visualization in \fref{fig: Ch2 split step overlap}. This figure implies that nearby point sources will share components of the same phase screens. This is exactly the mechanism that will model the correlation. For practical implementation, the approach is often to generate large phase screens and propagate through paths over the same grid size whereas in our visualization the propagation window, represented by the cone, ``grows". We suggest the reader to \cite{Hardie_2017_a} for a sense of how this process is done in practice. This figure also showcases \emph{why} each point must be propagated individually. With the PSFs generated from this process, we can then take the spatially varying convolution
\begin{equation}
    I_i(\vx) = \vert h_{\vu}(\vu) \vert^2 \overset{\vu}{\circledast} I_g(\vx),
\end{equation}
to receive our output image.

\subsection{Pseudo-Code for Split-Step}
For the sake of clarity, we present a pseudo-code description of the split-step algorithm for a grid of point sources. For a full elaboration of the details and additional considerations, we point the reader towards \cite{Roggemann_2012_a, Hardie_2017_a, Schmidt_2010_a, Hardie_2022_a}.

\begin{algorithm}\index{split-step! pseudo-code}
	\caption{Split-Step Simulation}
	\begin{algorithmic}[1]
        \State Set sampling parameters for object plane space $\Omega_U$
		\State Determine the size in meters for each phase screen
		\State Generate $M$ phase screens at specified sizes according to PSD
		\For {$\vu \in \Omega_U$}
				\For {$i=1,2,\ldots,M$}
					\State $U_{i}(\vxi; \vu) = [h_\text{Fres}(\vxi; z) \circledast U_{i-1}(\vxi; \vu)]  e^{j\phi_i(\vxi; \vu)}$
				\EndFor
				\State Form PSF, $\vert h_{\vu}(\vx) \vert^2 = \mathfrak{Fourier}\{P(\vxi) \cdot U_M(\vxi)\}$
		\EndFor
		\State Perform spatially varying convolution $I_i(\vy) = \vert h_{\vu}(\vx) \vert^2 \overset{\vu}{\circledast} I_g(\vx)$
	\end{algorithmic}
\end{algorithm}

Summarizing this pseudo-code, we first generate phase screens of varying size, in accordance with \fref{fig: Ch2 split step overview}. As we move from point source to point source, we will propagate through different slices of the atmosphere. This will result in each PSF being different, though the overlap will contribute to the desired correlation. The simulation is completed once every PSF has been formed and the spatially varying convolution is computed.

\section{Summary}
\label{sec: sec2_5}
With the details of turbulence discussed, we presented a simplified flow of how images are formed through turbulence. The  process can be summarized into four parts:

\keyword{Part 1 Atmospheric model}: We began by discussing the turbulent nature of the atmosphere and characterized its fluctuations as a Gaussian process. The Kolmogorov PSD was presented, along with its according structure function $D_n(\vr) = C_n^2 |\vr|^{2/3}$. This additionally led us to describe $C_n^2$, a parameter that characterizes the strength of turbulent fluctuations.

\keyword{Part 2 Propagation through turbulence}: The layered atmospheric model was introduced and utilized to arrive at our main result of the phase structure function $\calD_{\phi}(\vert \vxi - \vxi' \vert)$ for propagation through a single layer of turbulence. This was then generalized to multiple screens, each of which may have different turbulence strength values. We also discussed the wave structure function, the more general version of the model for a wave propagated through turbulence.

\keyword{Part 3 OTF and other properties}: The result of our analysis produced the structure function of the phase, $\calD_{\phi}(\vxi,\vxi') = 6.88(|\vxi-\vxi'|/r_0)^{5/3}$. This led us to define the Fried parameter $r_0$ as well as the isoplanatic angle $\theta_0$, both of which are insightful statistics for turbulent imaging. Additionally, the LE and SE OTFs were presented, characterizing the effects within the context of Fourier optics.

\keyword{Part 4 Split-step simulation}: The method of simulation by split-step propagation was presented. Split-step works by assuming thin layers of turbulence, known as phase screens, to model the atmosphere in discrete slices. Waves are numerically propagated through the slices to form the resultant phase realizations and images.

\chapter{Propagation-Free Modeling and Simulation}
\vspace{-6ex}
\noindent\textcolor{myblue}{\rule{\textwidth}{4pt}}
\vspace{1ex}

\label{sec: sec3}

Thus far, this book has largely derived from more traditional optics considerations, which we've done intentionally. With the fundamentals established, the time to begin the ``translation'' of optics into computer vision has come. This will lead us to formulate the previous Chapter in terms of standard engineering concepts such as \keyword{random vectors}\index{random vector} and \keyword{basis representation}\index{basis representation}.

\section{Motivation}
In the previous Chapter, the model for atmospheric turbulence was presented along with the simulation approach of split-step. Split-step clearly arises as a natural extension of the model where we are discretizing the forward process of nature. The present Chapter, however, is dedicated to alternative models. Before presenting these alternatives, we would like to present why such a need exists depending on who you ask.

\subsection{What's Wrong with Split-Step?}
First, we must admit, there is really nothing significantly wrong with split-step. In fact, it is arguably the most general model; the alternatives we present will leave out effects which split-step can represent. One must remember that the authors are \emph{imaging} people -- we want to use split-step for something more than forward simulation! To us, split-step poses the following challenges:\index{split-step! computational limitations}
\begin{enumerate}
    \item \textbf{Generation of a dataset.} The cost of generality and accuracy is the speed at which split-step can run. One of our papers led us to estimate the time to generate a dataset suitable for training one of our networks, one only needs to wait approximately 150 years \cite{Chimitt_2022_a}! This is clearly unsuitable for generating training data with moderate computational resources.
    \item \textbf{How to plug into an end-to-end framework?} If we wanted to insert split-step into a network, how would we propagate the gradient for the loss term? Following a gradient through the method would offer many challenges and, if even possible, slow the training tremendously. As a result, split-step cannot be used as an innate prior at present time.
\end{enumerate}

From our perspective, a perfect simulator would be one which has generality, the ability to generate a dataset, easily allows gradient computation, and maintains accuracy. It is likely that one must compromise on at least one of these properties.

\subsection{Diagnosing the Problem and a Solution}
Is it possible to point at one aspect of split-step and blame it for the pain we are suffering on the applications side? The authors have considered this point in detail over the last 5 years and have come to the following conclusion: the numerical wave propagation is to blame!\footnote{Our confidence in blaming numerical propagation now is much more certain than it was in 2018. We also say this in good humor, the reasoning for doing wave propagation is of course well justified!}

Split-step requires each point source to be numerically propagated through a series of phase screens. This requires $M+1$ 2D FFTs for each point source on an $N \times N$ grid. Then, we must model the PSF formation, requiring an additional FFT. This results in $N^2(M+2)$ 2D FFTs \emph{and} a spatially varying convolution. If we can avoid taking so many FFTs, this will save us a great deal of computation.

This puts us in a difficult predicament. We have identified the problem as numerical wave propagation. However, the problem we want to study is \emph{wave propagation} through turbulence. How can we resolve these two seemingly conflicting desires?

This leads us to make the following problem statement: \emph{if} we can generate a set of phase functions without numerical wave propagation \emph{then} our problems would be overwhelmingly solved. As one may anticipate, this introduces a whole new set of issues, however, it will avoid our main issues with numerical wave propagation.

This ``skipping'' of numerical wave propagation is exactly what the simulator proposed by the authors and collaborators achieves. What is more is that the accuracy is still guaranteed to be high, though there are a few carefully considered approximations that are unavoidable at this time. Our model also does not incorporate amplitude effects such as scintillation, however, this is in line with the previous Chapter's discussions. We are not alone in such an endeavor, as there are many other alternatives to split-step which in many ways go about solving the same or related problems. Some rely on the split-step methodology more than others, though overwhelmingly the idea is the same: the goal is to skip straight to the phase or image effects and still maintain accuracy. This leads us to define these methods as \keyword{propagation-free methods}, methods that do not require numerical wave propagation.

\section{A Survey of Propagation-Free Methods}
\label{sec: sec3_1}
\index{propagation-free methods}The source of propagation-free methods is a combination of the optics and computer vision communities. This is in stark contrast to split-step, which was primarily developed within the physics/optics community. A purely optics-inspired method may present a disadvantage to someone interested in computer vision due to speed. If one desires a large amount of training data for the purposes of machine learning, split-step will likely not be suitable for this task \cite{Chimitt_2022_a, Mao_2021_a}. Propagation-free methods are typically motivated by speed, though the end applications may vary. These different end goals inspire different varieties of propagation-free simulations.

\subsection{Simplistic Models}
\begin{figure}
    \centering
    \includegraphics[width=0.95\linewidth]{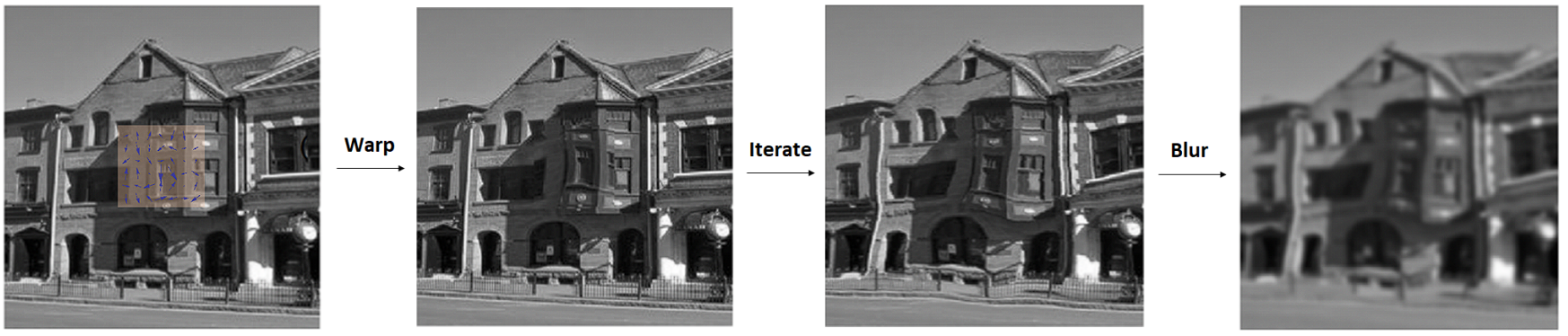}
    \caption{In these methods, a spatially variant pixel shifting and spatially invariant blur are applied to produce an image that shares some similarity with turbulent imagery. Source: \cite{Chak_2021_a}}
    \label{fig: chak}
\end{figure}

A natural starting point for alternative simulations is what we shall refer to as \keyword{zero-order approximations}\index{propagation-free methods! zero-order}. These methods are often used as a proof-of-concept type simulation in place of a more sophisticated simulator. These methods are inspired by the fact that turbulent images have two characteristics: they are distorted geometrically and blurry. The standard way to do this is that given an image that represents $I_g(\vu)$, the pixels are shifted randomly (per pixel) by a 2D Gaussian vector. After this, a Gaussian blur is applied to the image. The methods by Chak et al. \cite{Chak_2021_a} and Lau et al. \cite{Lau_2019_a} both utilize such a model, though the contributions of these papers are more on the restoration side, with their simulation methodology being used more for proof-of-concept. We present a figure from Chak et al. \cite{Chak_2021_a} in \fref{fig: chak}.

These methods may indeed produce images that are geometrically warped and blurred. However, the same statistics that arise in the Kolmogorov model are clearly not represented in this framework. For these reasons, we typically suggest other methods to be utilized for simulation. That being said, the concept of pixel shifting and blur is useful in turbulence restoration as in the case of \cite{Lau_2019_a, Chak_2021_a} and Zhu and Milanfar \cite{Milanfar_2013_a}. We should note that this concept of splitting the operators of pixel shifting and blurring in itself and keeping the operators to be general is valid, such as done by Zhu and Milanfar \cite{Milanfar_2013_a} and further elaborated on from the side of simulation in \cite{Chan_2022_a}. We'll return to the topic of restoration and the use of the geometric warping and blurring model in Chapter 5.

\subsection{Direct-to-Image Methods}
The simplistic model uses the concept of pixel shifting and blur, though the ways in which these two elements are modeled are minimally dependent on the underlying physics. There are, however, ways to introduce the physics into the image formation process \emph{without} going to the phase domain. By this, we mean that our goal is not to model $\phi_{\vx}$, but to model its resultant PSF $\vert h_{\vx} \vert^2$ directly. This may feel as if it contradicts some of the previous discussions in Chapter 3 -- we said there was no straightforward model of the PSF! However, we do know certain things about it, such as the long and short exposure functions which are closely related to the average PSFs. For these reasons, we refer to these methods as \keyword{direct-to-image methods}.\index{propagation-free methods! direct-to-image}

\begin{figure}
    \centering
    \begin{tabular}{c}
        \includegraphics[width=0.6\linewidth]{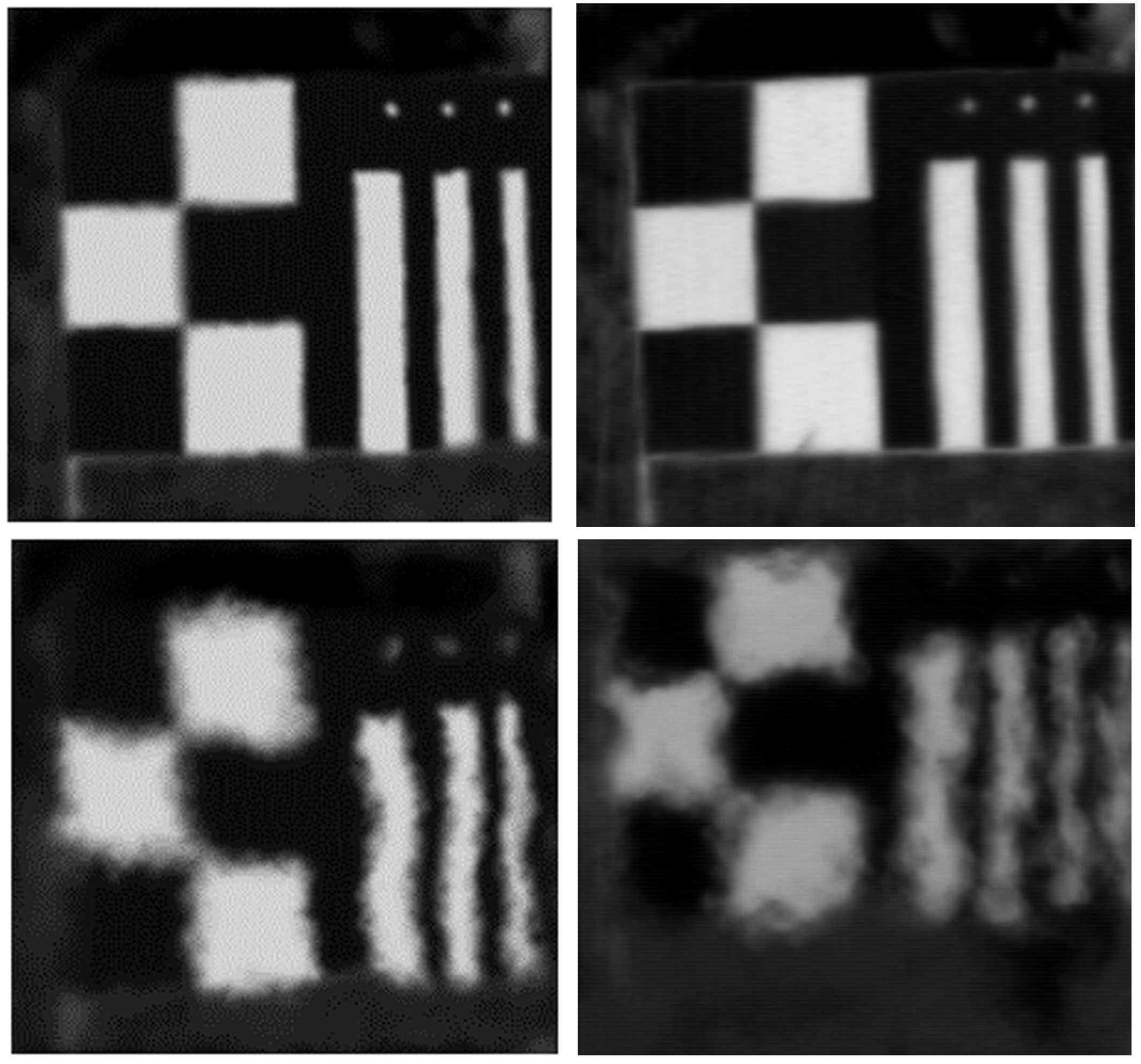} \\
    \end{tabular}
    \caption{[Left] Two simulated images and [Right] real images with corresponding turbulence strength from the NATO dataset. Source: \cite{Repasi_2011_a}}
    \label{fig: repasi}
\end{figure}

There exist some simulations within this space such as that by Repasi and Weiss \cite{Repasi_2008_a, Repasi_2011_a}, Leonard et al. \cite{Leonard_2012_a}, or Potvin et al. \cite{Potvin_2011_a} which use a blend of analytic and empirical properties (the empirical aspects partially based on the NATO RTG-40 dataset\index{NATO RTG-40} \cite{Tofsted_2006_a, Tofsted_2007_a}) to simulate PSFs and the subsequent images directly. The NATO dataset used various measurement devices, with one example being \keyword{scintillometers}\index{scintillometer} to measure the effective turbulence level by providing an estimate for $C_n^2$. This makes it an incredibly useful collection of data for the sake of simulator verification.

\begin{figure}
    \centering
    \includegraphics[width=0.6\linewidth]{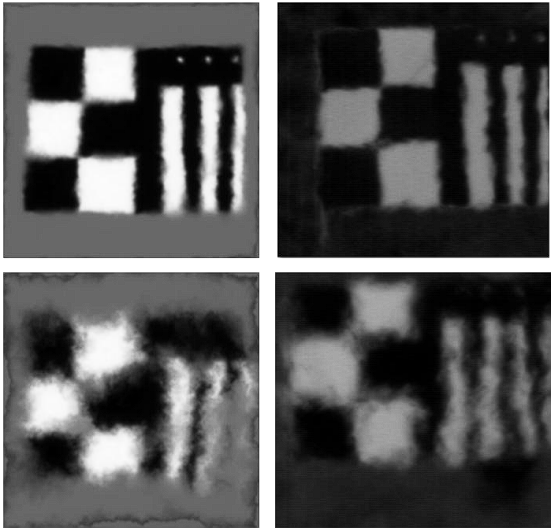}
    \caption{[Left] Simulated results for different turbulence strengths which correspond to [Right] field data. These results also utilize the NATO dataset. See \cite{Leonard_2012_a} for more one-to-one comparisons and $C_n^2$ information. Source: \cite{Leonard_2012_a}.}
    \label{fig: leonard}
\end{figure}

We present some images from Repasi and Weiss's paper that showcase some results from the NATO project (\fref{fig: repasi}) in which real and simulated images are compared (a digital ground truth is used for producing the simulated images). Leonard et al. \cite{Leonard_2012_a} uses a similar approach that utilizes empirical results from the NATO dataset which we present in \fref{fig: leonard}.

These methods are similar in their approach, so we will speak a bit generally to include them under the same description. These methods use pixel displacements, similar to the simplistic models, however, the variance of the pixel displacement is given as a function of $C_n^2$, wavelength, distance, etc. The result is displacements that more closely match true turbulent cases. The pixel shifts are also drawn in a correlated fashion. With respect to the blur, both methods use the short exposure function as a mean width value for the blur size (i.e. the SE transfer function can be mapped to a SE PSF which is the average blur of centered PSFs) which is approximated as a Gaussian. The width is then allowed to vary in a patch or pixel-wise fashion. The result is a spatially varying blur that fluctuates similarly to turbulent cases.

These simulation methodologies have been revisited more recently by Miller et al. \cite{Miller_2019_a, Miller_2021_a}. We would point the interested reader to Figure 1 of Miller et al. \cite{Miller_2019_a} for a clear and concise visual description of their processing pipeline. The concept is similar, but we wish to emphasize a few key elements of these works. Firstly, a camera model is introduced in \cite{Miller_2019_a} which allows for the incorporation of temporal information in the imaging process (such as longer / shorter exposures) as well as sensor and post-processing effects, effectively modeling the entire image signal processing (ISP) pipeline. These papers also demonstrate the dependence of the image realization on a random seed, which produces three \keyword{random fields}, two for pixel shifts and one for the PSF blur width. This work also models temporal effects, which allows turbulence to vary over the acquisition time of an image which introduces additional blurring that is present in real imagery.

\subsection{Ray Tracing Methods}
\index{propagation-free methods! brightness function}There are other propagation-free methods which we refer to as ray-tracing, the first of which is known as the brightness function simulation\index{brightness function} \cite{Vorontsov_2005_a, Lachinova_2007_a, Lachinova_2017_a}. The brightness function model is faster than split-step as a result of its lack of wave propagation evaluation, instead propagating ``bundles'' of rays through a perturbing medium. These bundles of rays are then distributed across the imaging plane for each pixel as a function of the medium, resulting in spatially varying effects as a function of the phase screens. Given certain choices in simulation parameters, the approach may be considerably faster than split-step.

\begin{figure}
    \centering
    \includegraphics[width=0.9\linewidth]{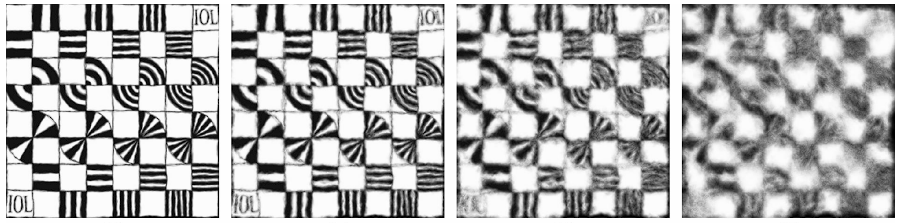}
    \caption{Results from the brightness function model with increasing $D/r_0$ (all simulated). Source \cite{Lachinova_2007_a}.}
    \label{fig:my_label}
\end{figure}

We will not provide such a detailed account of this model as in the case of the previous discussion. The reason for this is that it is mostly similar to split-step, with the propagations being replaced with ray tracing. However, we would like to emphasize that this is more easily said than done with the main theoretical basis of this simulation with Vorontsov et al. \cite{Vorontsov_2005_a} being a testament to this fact.

More ``standard'' ray tracing (i.e., those other than the brightness function) have been applied for the simulation and modeling of turbulent effects on an imaging system. Voelz et al. \cite{Voelz_2018_a} provides an analysis of standard ray tracing approaches, with carefully performed ray tracing matching wave optics simulations to a suitable degree of accuracy for most applications. Additionally, a comprehensive work on a similar simulation modality is described by \cite{Peterson_2015_a} and made publicly available (see paper for details).

\subsection{Other Methods}
There exist methods outside of these clear boundaries. For example, Hunt et al. \cite{Hunt_2018_a} suggested an empirical PSF generation method by a dictionary of PSFs obtained by SVD. This particular approach is unique in that the PSFs decomposed by SVD were obtained in a real imaging scenario, thus incorporating real data into the generation step.

In recent years, deep learning methods are becoming popular. Generative models such as generative adversarial networks (GAN) have shown impressive results in synthesizing various image distortion effects, even within the field of atmospheric turbulence such as Feng et al. \cite{Feng_2022_a}. Our argument against GAN is that the GANs reported in the literature so far are still black box algorithms. Besides being not easily interpretable, these methods also lack robustness and generalization capability when handling unseen turbulence conditions. Furthermore, verifying the turbulence statistics remains a challenge because we cannot assess the validity of the steps inside the algorithm. The statistical model is in the wave space, while GAN models typically are designed to produce effects in the image space. Therefore, theoretical comparison is not feasible for a GAN of this variety as the turbulent model is given for a wave and not an image, with the exception of a few cases as a result of temporal averaging.\index{propagation-free methods! generative adversarial network}

\subsection{Phase-Based Simulation}
In direct-to-image simulation, generated random fields described properties of the distortions in the image domain. \keyword{Phase-based simulation}\index{propagation-free methods! phase-based simulation} is similar in this sense, however, instead the distortions model the \keyword{phase domain}\index{phase domain}. One may wonder why would this be preferred over the direct-to-image approach. The phase domain requires a conversion through the image formation equation, whereas direct-to-image requires no extra computation.

The answer is that the statistics are described not in the image domain, but in the phase domain! Therefore, if we wish to derive various properties and verify directly with the developments stemming from Tatarskii, it will be most easily done in the phase domain. These methods have been used for empirical study \cite{Roddier_1990_a, Rucci_2021_a}, however, often only for a \emph{single} point. This won't do us much good for simulating images under \keyword{anisoplanatic}\index{anisoplanticism} conditions, i.e., conditions that are not isoplanatic.

One may suspect that if we know the Kolmogorov power spectral density, the phase realizations can be drawn directly from the correlation matrix (or the structure function $\calD_{\phi}(\vxi,\vxi')$ as we presented in \cref{sec: sec2}). However, the structure function of the phase only describes the phase realization which emerges from a single point $\vu$ in the object plane. It does not tell us how the phase functions $\phi_{\vu}(\vf)$ and $\phi_{\vu'}(\vf)$, located at $\vu$ and $\vu'$, are related to one another. Thus, moving towards simulating images will require additional considerations.

This is the exact problem that was considered by the authors and collaborators in a series of publications \cite{Chimitt_2020_a, Mao_2021_a, Chimitt_2022_a, Chimitt_2023_b}. As a result, we are intimately familiar with the ideas that generated this simulator, which we refer to as \keyword{Zernike-based simulation}\index{propagation-free methods! Zernike-based}. The rest of this Chapter will be dedicated largely to understanding this variety of simulation, partly due to our familiarity with it but more for another reason. This method is fundamentally motivated for use in image processing, therefore, the tricks and language used to describe it comes from this lens. Therefore, describing this simulation will serve as a tool for interpreting Chapter 3 through a more image processing/computer vision lens. Our coverage will be reasonably detailed, but not complete; if one requires more details, we would refer them to our previously mentioned publications.

Moving to the high-level details, Zernike-based simulation wishes to simulate the familiar image formation process
\begin{equation}
I_i(\vx) = \left( \vert \mathfrak{Fourier} \{ P(\vxi) e^{j \phi_\vu(\vxi)} \} \vert^2 \overset{\vu}{\circledast} I_g \right)(\vx),
\label{eq: ch3 big idea 1}
\end{equation}
where $\phi_\vu(\vxi)$ is a spatially varying phase realization that arises from the propagation. It is important to note that a phase realization (or phase function) is different from a phase screen, therefore we will reserve the term ``phase screen'' to describe the phase screens used in split-step. The phase function is defined across the aperture, whereas a phase screen represents a section of the atmosphere along the propagation path.

These phase realizations have two important properties in the case of atmospheric turbulence: (1) Their structure function is described through the Kolmogorov model (or its related models); (2) They are correlated spatially. Our goal, as previously stated, is to maintain these two properties \emph{without} numerical wave propagation. For each point $\vu$ in an image, there is an associated point spread function
\begin{equation}
\vert h_{\vu}(\vx) \vert^2 = \left|\mathfrak{Fourier}\left\{e^{j \phi_{\vu}(\vxi)}\right\}\right|^2,
\label{eq: Ch3 PSF from phase}
\end{equation}
where we have dropped some constants for brevity. This means for each point in the image, there is a corresponding phase function $\phi_{\vu}(\vxi)$ and point spread function $\vert h_{\vu}(\vx) \vert^2$. Using \eref{eq: Ch3 PSF from phase} to interpret our desired statistical properties, if we can generate the per-pixel phase functions $\phi_{\vu}(\vxi)$ and $\phi_{\vu'}(\vxi)$ from the specified distribution then we will have achieved our goal.

\begin{figure}[th]
\centering
\includegraphics[width=\linewidth]{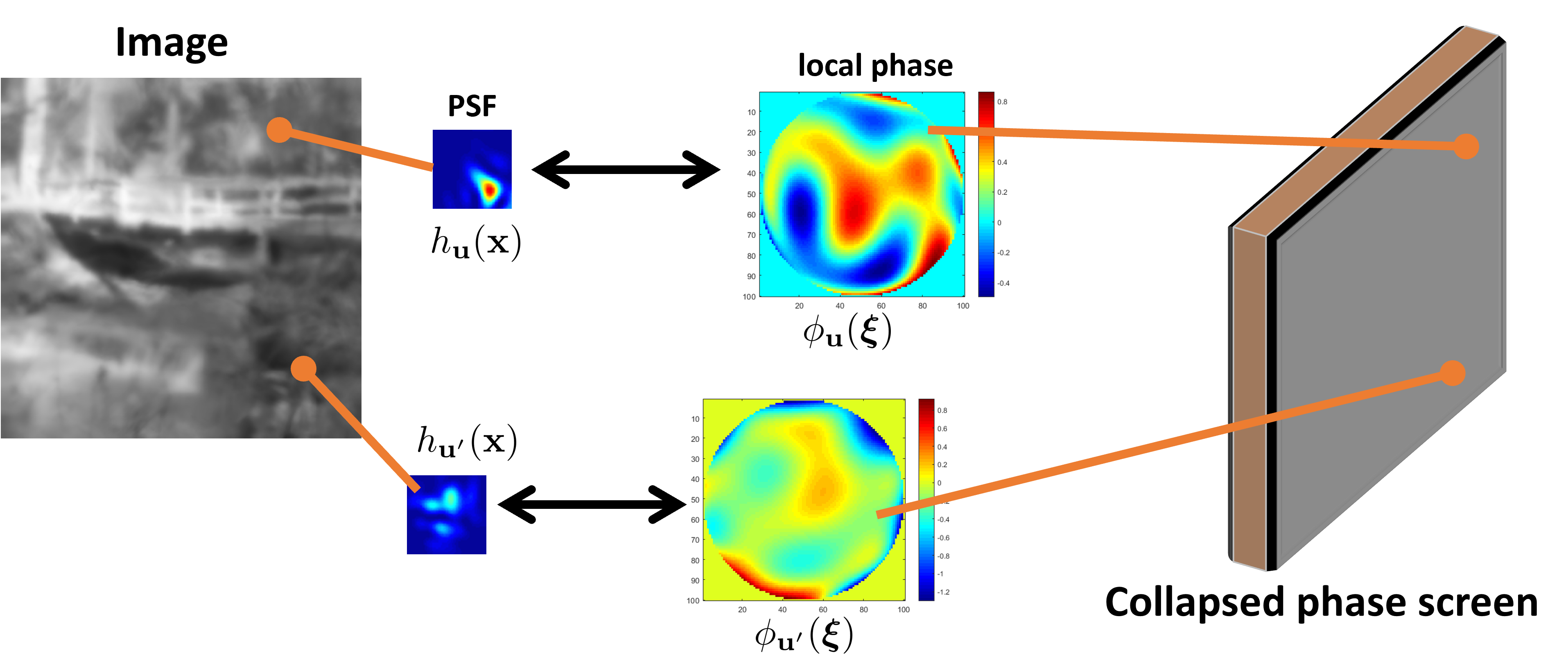}
\caption{The collapsed phase-over-aperture model by Chimitt and Chan \cite{Chimitt_2020_a}. The idea is to compress the propagation path into a single abstract volume where each location has its own phase function and corresponding PSF. We note this differs conceptually from a phase screen.}
\label{fig: Ch3 collapsed phase over aperture}
\end{figure}

The above argument leads to a conceptual diagram shown in \fref{fig: Ch3 collapsed phase over aperture}. In this figure, the information contained in the set of phase realizations is embedded into the \keyword{Zernike space}\index{Zernike! space} (referred to as a collapsed phase screen in \cite{Chimitt_2020_a}). For each location $\vu$ in the Zernike space there is a phase realization $\phi_{\vu}(\vf)$, related to the PSFs $\vert h_\vu(\vx) \vert^2$ by \eref{eq: Ch3 PSF from phase}. The resulting image can be formed by performing a spatially varying convolution:
\begin{equation}
I(\vx) = \vert h_{\vu}(\vx) \vert^2 \overset{\vu}{\circledast} I_g(\vx),
\label{eq: Ch3 spatially varying convolution}
\end{equation}
where $I_g(\vx)$ is the geometric image and $I(\vx)$ is the turbulence distorted image.

This Chapter serves to address the following questions which must be answered to enable this variant of simulation:
\begin{enumerate}
\item How do we generate the phase functions $\phi_{\vu}(\vxi)$ at each $\vu$? This is the premise of the Zernike space model. If we cannot generate the phase functions $\phi_{\vu}(\vxi)$, then there is no way that we can construct the PSFs $h_{\vu}(\vx)$. We shall address this problem using a representation technique known as the \keyword{Zernike polynomials}\index{Zernike! polynomials}\index{Zernike}.
\item How do we enforce the spatial correlation of two phase functions $\phi_{\vu}(\vxi)$ and $\phi_{\vu'}(\vxi)$? Again, this is a critical part of the method. If we cannot draw spatially correlated phase functions, we will defeat the purpose of skipping the wave propagation steps. To articulate this question, we will introduce few powerful results in the literature about the angle of arrival statistics, together with a mirroring technique, to develop a so-called \keyword{multi-aperture approach}\index{multi-aperture model}.
\item Given that we may develop the model to represent turbulent distortions, how should we implement the method to be \emph{fast}? Our main issue with split-step is its speed, therefore, we'd want Zernike-based simulation to give us a significant increase in speed.
\end{enumerate}
Thus, \cref{sec: Zernike basics} and \cref{sec: Zernike space} will be dedicated to these more theoretical considerations while \cref{sec: Zernike simulation} addresses the final question.

\section{Representing Turbulent Phase Realizations}
\label{sec: Zernike basics}

\subsection{Phase as a Basis}
The last Chapter presented two ways to represent the phase:
\begin{align*}
    \phi(\vxi) &= k \int_{0}^{L} n_1(\vxi,z) \; dz &\text{(the definition)}, \\
    \phi(\vxi) &= \mathfrak{Fresnel}[L/N]\{ e^{\phi_N} \mathfrak{Fresnel}[L/N] \{ \cdots \} \} &\text{(via split-step)},
\end{align*}
with $L$ as the propagation distance and $N$ as the number of phase screens in split-step. Note that the split-step propagation expression is a sequential application of free-space propagation followed by phase imparting. Both of these representations were utilized in the last Chapter, though they present a few challenges: (1) lack of clarity for informing a prior model; (2) the statistical properties of each are a bit unfamiliar to the standard computer vision researcher, such as the structure function; (3) the extension to multiple phase realizations is unclear.

This leads us to instead formulate the phase in a different way through the concept of \keyword{basis representation} of a function. A basis representation allows us to write $\phi_{\vu}(\vxi)$ as\index{basis representation! definition}\index{basis representation! of phase}
\begin{equation}
\phi_{\vu}(\vxi) = \sum_{j=1}^N a_{\vu,j} Z_j(\vxi),
\label{eq: Ch3 phase linear combination}
\end{equation}
for some basis functions $\{Z_j(\vxi) \,|\, j = 1,\ldots,N\}$ and some basis coefficients $\{a_{\vu,j} \,|\, j=1,\ldots,N\}$. Equations of this form are ubiquitous in computer vision. If we recall the classical Fourier expansion of a function, a continuously differentiable function $f(t)$ with bounded energy can be written as a linear combination of complex exponentials: $f(t) = \sum_{n=1}^N a_n Z_n(t)$, where $Z_n(t) = \exp\{j2\pi nt /N \}$ is the Fourier basis, and $a_j = \langle f(t), Z_n(t) \rangle$ is the Fourier coefficient. The benefit of expressing the function $f(t)$ in terms of the Fourier basis is that it allows us to describe the function using a low-dimensional representation. Thus, if we can generate the coefficients $\{a_n\,|\,n=1,\ldots,N\}$, we can represent the function $f(t)$.

Going back to the phase function $\phi_{\vu}(\vxi)$, the function $Z_j(\vxi)$ is the basis representation of the phase. Unlike the phase function $\phi_{\vu}(\vxi)$ which is location dependent (hence spatially varying), the basis function $Z_j(\vxi)$ is spatially \emph{invariant}. Two phase functions $\phi_{\vu}(\vxi)$ and $\phi_{\vu'}(\vxi)$ differ in their basis representation only by the coefficients $a_{\vu,j}$ and $a_{\vu',j}$. Addressing the notation $a_{\vu,j}$, the first index $\vu$ denotes the location whereas the second index $j$ tells us which basis function are we using. We will refer to the vector $\mathbf{a}_{\vu} = [a_{\vu,1},\ldots,a_{\vu,N}]^T$ to describe the vector of basis coefficients at location $\vu$.

\subsection{Zernike Basis}
How do we pick an appropriate basis function $Z_j(\vxi)$? The phase function $\phi_{\vu}(\vxi)$ is defined over the aperture, and the aperture is usually circular in shape. As such, the basis functions $Z_j(\vxi)$ must also be a function defined over a circle.

Our choice of the basis function $Z_j(\vxi)$ is the \keyword{Zernike polynomials}\index{Zernike! polynomials}. The Zernike polynomials are two-dimensional orthogonal functions defined over the unit disk. The Zernike polynomials are mathematically well defined and there are tools to handle these mathematical objects. Furthermore, the work of Noll \cite{Noll_1976_a} utilized the Zernike polynomials to represent the phase with some important additional comments and small corrections by Roddier \cite{Roddier_1990_a} and Wang et al. \cite{Wang_1978_a}. Noll's paper represents the starting point of the simulator developed by the authors and collaborators.

To work with Zernike polynomials, it is more convenient to switch from Cartesian coordinate $\vxi = [\xi_x, \xi_y]^T$ to the polar coordinate $\vvarrho = [r,\theta]^T$, where $\xi_x = r\cos\theta$ and $\xi_y = r\sin\theta$. Moreover, with Zernike polynomials, we prefer to work with unit norm vectors because the domain of the Zernike polynomials is the unit disk. To this end, we define $R \bydef D/2$ as the aperture radius and scale $r$ to $r = R \rho$ so that $0 \le \rho \le 1$. With these notations, the phase function $\phi(\vxi)$ can be written as $\phi(R\vrho)$ where $\vrho = [\rho,\theta]^T$ and $R\vrho = [R\rho,\theta]^T$. In terms of mathematical expressions, we can write
\begin{equation}
    \underset{=\phi_{\vu}(\vxi)}{\underbrace{\phi_{\vu}(R\vrho)}} = \sum_{j=1}^{N} a_{\vu,j} \underset{=Z_j(\vxi)}{\underbrace{Z_j(\vrho)}}.
\end{equation}

\begin{figure}[h]
\centering
\begin{tabular}{cc}
\includegraphics[height=4.5cm]{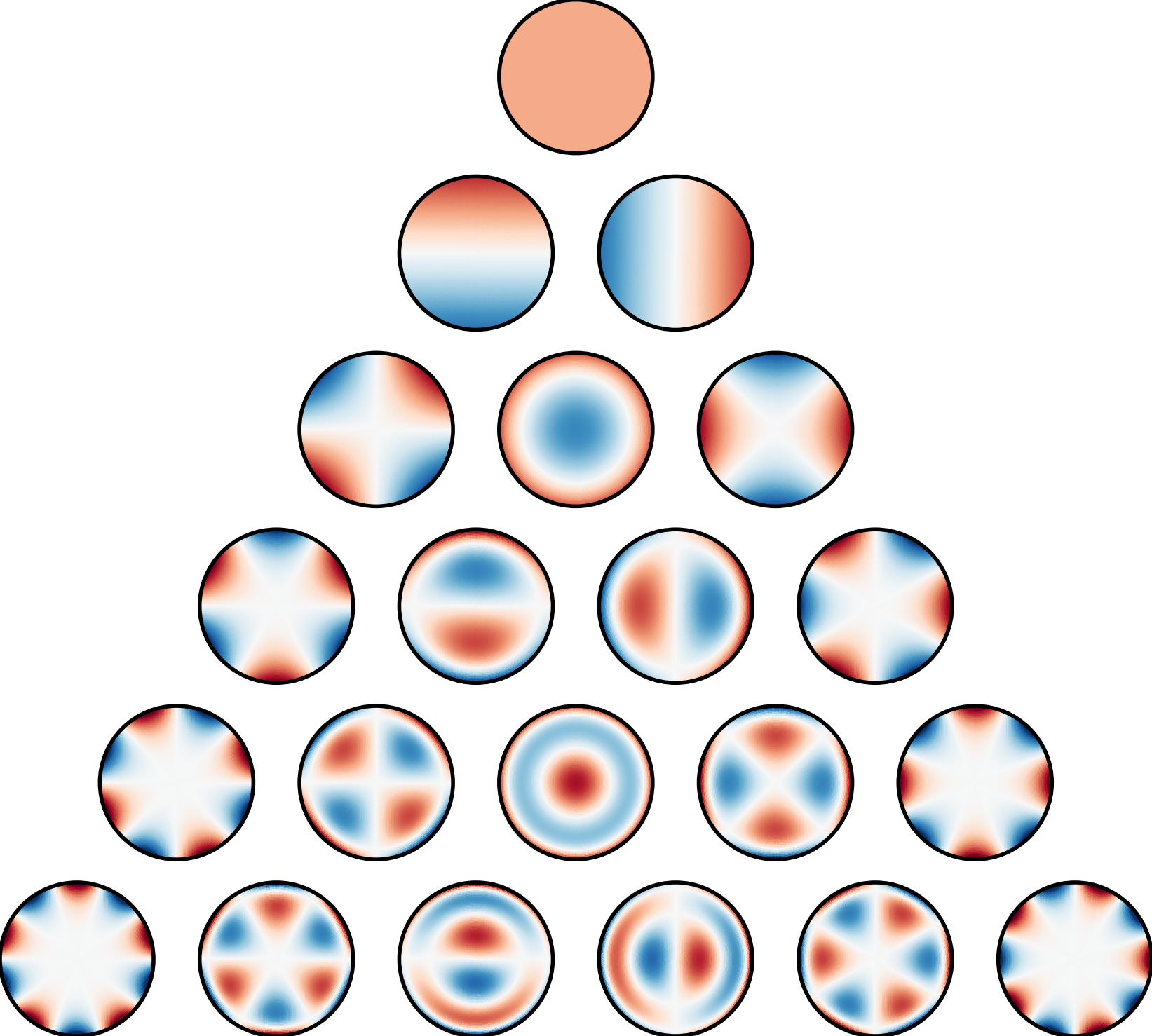}&
\includegraphics[height=4.5cm]{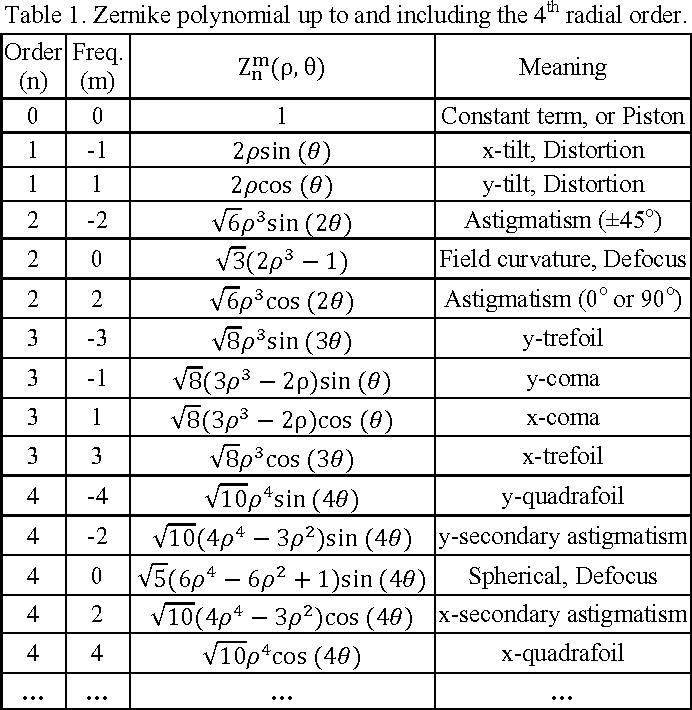}
\end{tabular}
\caption{The shape of the Zernike basis functions, and their corresponding equations. Source: \url{https://en.wikipedia.org/wiki/Zernike_polynomials}.}
\label{fig: Ch3 Zernike}
\end{figure}

Before we look at the equations of the Zernike polynomials, we first show the shape in \fref{fig: Ch3 Zernike}. Among the different orders, the base  $Z_0(\vrho)$ is the constant offset which can be neglected for incoherent imaging. The first and the second Zernike polynomials $Z_1(\vrho)$ and $Z_2(\vrho)$ are known as the \keyword{$x$-tilt} and the \keyword{$y$-tilt}\index{tilt} respectively. It is worth noting in some literature these are referred to as tip and tilt, respectively, though for clarity we prefer the former terminology. Other high order Zernike polynomials include astigmatism, coma, spherical aberrations, etc.

As indicated by the table in \fref{fig: Ch3 Zernike}, the Zernike polynomials are indexed by their \keyword{azimuthal} order $m$ and their \keyword{radial} order $n$. Thus, the pair $(m,n)$ specifies the particular Zernike polynomial we are describing; $m$ tells you how far out from the ``center line'' to go and $n$ tells you how many rows down from the top. This leads to the notation $Z_n^m(\vrho)$, however, it is often easier to work with a single index $j$ with notation $Z_j(\vrho)$ as we've written previously. In this book, we follow Noll's paper \cite{Noll_1976_a} that maps $(m,n) \rightarrow j$\index{Zernike! Noll's indexing} whose formula is skipped for brevity. With this mapping, the Zernike polynomial is defined as (recall that $\vrho = [\rho,\theta]$):\index{Zernike! polynomial definition}
\begin{align*}
Z_{\text{even }j}(\vrho) &= \sqrt{2(n+1)} R_n^m(\rho)\cos(m\theta), &\qquad m \neq 0,\\
Z_{\text{odd }j}(\vrho) &= \sqrt{2(n+1)} R_n^m(\rho)\sin(m\theta), &\qquad m \neq 0,\\
Z_j(\vrho) &= \sqrt{n+1} R_n^m(\rho)\sin(m\theta), &\qquad m = 0,\\
\end{align*}
where $R_n^m(\rho)$ is
\begin{align*}
R_{n}^m(\rho) = \sum_{s=0}^{(n-m)/2} \frac{(-1)^s(n-s)!}{s![(n+m)/2-s]![(n-m)/2-s]!}\rho^{n-2s}.
\end{align*}
There are many other such indexing, thus we do not feel it necessary to present exactly how the mapping is performed. Though, it is important to re-emphasize that in this book we indeed use Noll's indexing scheme as is common in the atmospheric turbulence literature.

\subsection{Zernike Coefficients}
The Zernike polynomials are defined over the unit circle, therefore, the inner product must also be defined on the unit circle. To facilitate our discussions, we denote the circular aperture as
\begin{equation*}
P(\vrho) =
\begin{cases}
1, &\quad |\vrho| \le 1,\\
0, &\quad \text{otherwise}.
\end{cases}
\end{equation*}
With this circular aperture defined, we can then consider the orthogonality principle over the unit circle. If we are given a phase function $\phi_{\vu}(\vrho)$, we can define the $j$th Zernike coefficient at location $\vu$ as
\begin{align}
a_{\vu,j}
&= \langle \phi_{\vu}(R\vrho), Z_j(\vrho) \rangle_{P} \notag \\
&\bydef \int_{0}^{2\pi}\int_0^1 P(\vrho) \phi_{\vu}(R\vrho) Z_{j}(\vrho) d\rho d\theta,
\label{eq: Ch3 Zernike coefficient}
\end{align}
where $\langle \cdot, \cdot \rangle_{P}$ denotes the inner product using $P(\vrho)$ as a weight. Statistically, Tatarskii \cite{Tatarski_1967_a} argued that $\phi_{\vu}(R\vrho)$ is zero-mean Gaussian. This is partially a physics observation, and partially for mathematical convenience. If we agree with Tatarskii's assumption, then the Zernike coefficients $\{a_{\vu,j}\,|\, j=1,\ldots,N\}$ are zero-mean Gaussians because they are obtained through linear projections.

Generating the Zernike polynomials and subsequently the phase function can be done on a computer using customized libraries. In the example below, we use the Zernike polynomial MATLAB code developed by Gray \cite{Gray_2022_a}.

\begin{figure}[h]
\centering
\begin{tabular}{cc}
\includegraphics[height=3.5cm]{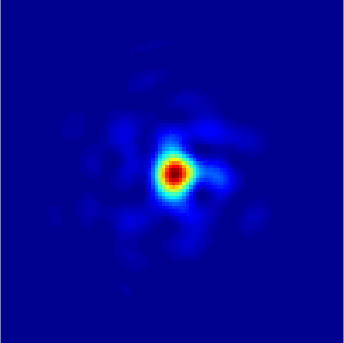}&
\includegraphics[height=3.5cm]{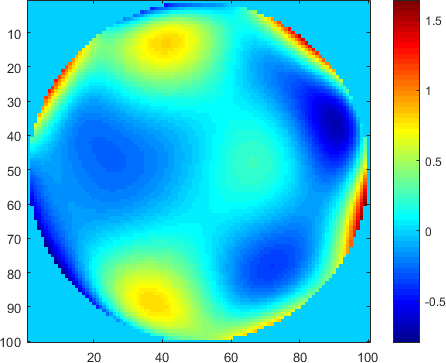}\\
(a) $h_{\vu}(\vx)$ & (b) $\phi_{\vu}(R\vrho)$
\end{tabular}
\caption{The random phase distortion (b) generated from the random Zernike polynomial in a grid of $100 \times 100$ (corresponding to a disk of diameter 2), and its associated PSF (a).}
\label{fig: Zernike and PSF}
\end{figure}

\boxedeg{
\vspace{2ex}
\noindent\textbf{Example}. To make this example simple, we fix a location $\vu$. We randomly draw 36 \emph{independent} Gaussian coefficients $\{a_{\vu,j}\}_{j=1}^{36}$ from $a_{\vu,j} \sim \calN(0,1/\sqrt{36})$. The example here is merely for demonstration of how the coefficients can affect the shape of the phase.

The phase $\phi_{\vu}(R\vrho)$ is generated using \texttt{ZernikeCalc} \cite{Gray_2022_a}. \fref{fig: Zernike and PSF} illustrates one random realization of the phase distortion function $\phi_{\vu}(R\vrho)$ and the corresponding PSF $h_{\vu}(\vx)$. The MATLAB code is shown below. Here, we over-sample the Fourier spectrum using \texttt{fftK = 8} for anti-aliasing. The division \texttt{ph/2} is due to the fact that the diameter is 2.
}

We emphasize that the phase function and the point spread function shown in \ref{fig: Zernike and PSF} do not correspond to a wave propagated through Kolmogorov turbulence. The reason is that the coefficients are \emph{independently} sampled -- we haven't put the Kolmogorov model anywhere! In reality, we shall sample the Zernike coefficients according to some known statistics which arise from the model of Chapter 3. Only then will the generated phase realizations and associated PSFs have the desired properties.

\begin{Verbatim}[frame=single,framerule=0.5mm, rulecolor=\color{NavyBlue}]
N = 100;
fftK = 8;
[P, ~]  = ZernikeCalc(1,1);
[ph, ~] = ZernikeCalc(1:36, (1/6)*randn(36,1), ...
            N, 'STANDARD');
U   = exp(1i*ph/2).*P;
u   = abs(fftshift(ifft2(U, N*fftK, N*fftK).^2));
\end{Verbatim}

\subsection{Intermodal Correlation}
\begin{figure}[h]
\centering
\includegraphics[width=\linewidth]{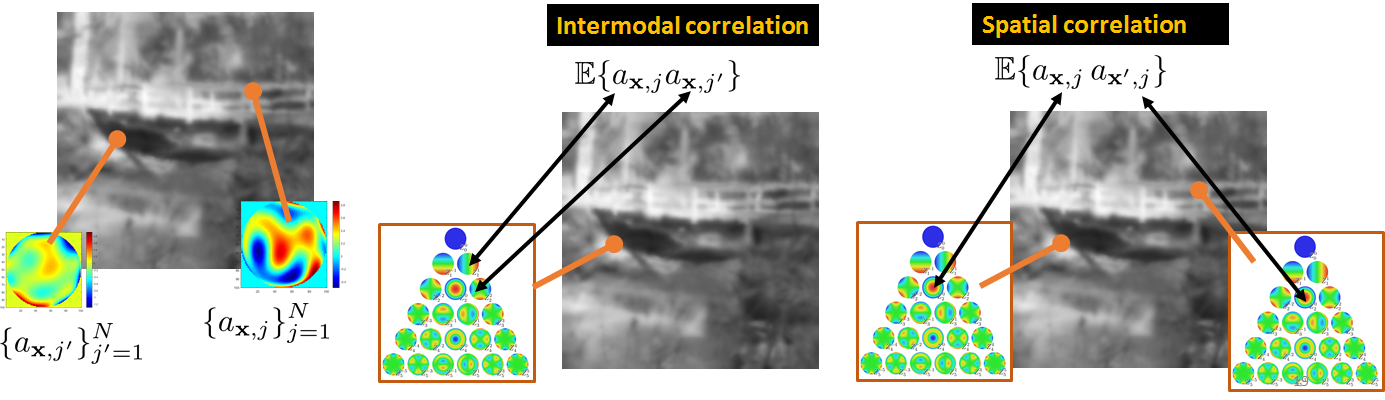}
\caption{Intermodal and spatial correlation of the Zernike coefficients.}
\label{fig: Ch3_intermodal_spatial}
\end{figure}

As mentioned previously, Noll studied the use of the Zernike polynomials in representing a turbulent phase distortion \cite{Noll_1976_a}. Specifically, Noll asked the question: For a phase realization $\phi_{\vu}(R\vrho)$, what is the correlation $\E[a_{\vu, i} a_{\vu, j}]$? Due to the terminology of ``modes", this is referred to as \keyword{intermodal correlation}\index{Zernike! intermodal correlation}. This allows one to write the covariance matrix of the basis coefficients for a single phase realization, and thus we can draw phase realizations that \emph{are} in accordance with Kolmogorov's theory. We additionally note that there exists another possible correlation, which is for two \emph{distinct} points $\vu$ and $\vu'$, we could ask the \keyword{spatial correlation} of $\E[a_{\vu, i} a_{\vu', i}]$. We illustrate the difference in \fref{fig: Ch3_intermodal_spatial}.

\begin{figure}[h]
\centering
\includegraphics[height=4.6cm]{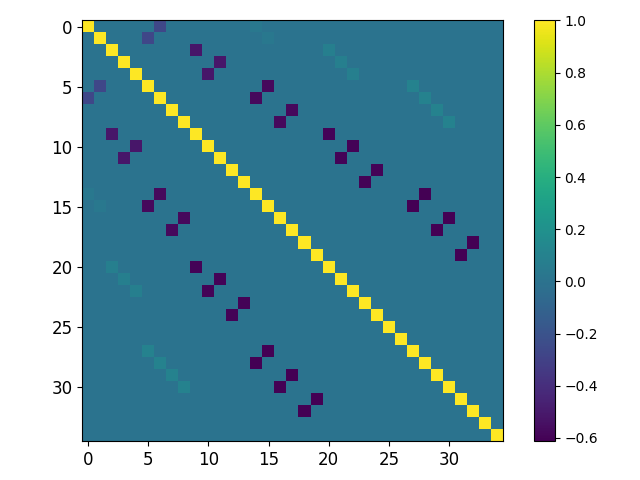}
\caption{Inter-modal correlation of the Zernike coefficients $\E[a_{\vu,i} a_{\vu,j}]/\sqrt{\E[a_{\vu,i}^2]\E[a_{\vu,j}^2]}$ for a fixed spatial location $\vu$.}
\label{fig: inter mode}
\end{figure}

Recalling \eref{eq: Ch3 Zernike coefficient}, we can show that the joint expectation $\E[a_{\vu,i}\,a_{\vu,j}]$
(at the same location $\vu$) is:
\begin{align}
\E[a_{\vu,i} \, a_{\vu,j}]
&= \frac{1}{\pi^2} \iint  P(\vrho) P(\vrho') Z_i(\vrho) Z_j(\vrho') \notag \\
&\qquad \qquad \times \E[\phi_{\vu}(R\vrho) \phi_{\vu}(R\vrho')] \, d\vrho d\vrho'.
\label{eq: Ch3 correlation of coefficient}
\end{align}
The main result of Noll's paper is to evaluate \eref{eq: Ch3 correlation of coefficient}. Doing so through the Fourier Transform and usage of the Kolmogorov spectrum, the joint expectation can be explicitly evaluated as \cite{Roddier_1990_a}:\index{Zernike! intermodal correlation}
\begin{align}
\E[a_{\vu,i} \; a_{\vu,j}]
&= 2.2698 (-1)^{(n_i + n_j - 2m)/2}\sqrt{(n_i + 1)(n_j + 1)}  \mathbb{I}_{m_im_j} \notag\\
&\times \frac{(D/r_0)^{5/3}}{\Gamma[(n_i+n_j+23/3)/2]} \notag \\
&\times \frac{\Gamma[(n_i+n_j-5/3)/2]}{\Gamma[(n_j-n_i+17/3)/2]\Gamma[(n_i-n_j+17/3)/2]}, \label{eq: Ch3 intermodal}
\end{align}
if $i-j = \text{even}$ and $\E[a_{\vu,i} \; a_{\vu,j}] = 0$ if $i-j = \text{odd}$. The indices $m_i$ and $n_i$ are the azimuthal and radial orders associated with the $i$th Zernike polynomial, and $\mathbb{I}_{m_im_j}$ is an indicator function that gives $\mathbb{I}_{m_im_j} = 1$ if $m_i = m_j$ and $\mathbb{I}_{m_im_j} = 0$ if $m_i \not= m_j$. The normalized correlation matrix, which we will refer to as the \keyword{Noll matrix}\index{Noll matrix}, is shown in \fref{fig: inter mode}, where we note most of the other entries are \emph{sparse}.

Through standard random sampling methods, we can properly sample individual turbulent phase realizations. Consider a multivariate Gaussian random variable with mean $\E[\vy] = \mathbf{0}$ and the correlation matrix $\mSigma = \E[\vy\vy^T]$. In our case, $\mSigma$ will represent the Noll matrix. To draw a random vector $\vy$ from this distribution, we decompose the correlation matrix $\mSigma = \mL\mL^T$ via \keyword{Cholesky decomposition}\index{Cholesky decomposition}. Then, starting with a white noise vector $\vn \sim \text{Gaussian}(\mathbf{0},\mI)$, the transformed vector $\vy = \mL \vn$ will satisfy the desired property that $\E[\vy] = \mathbf{0}$ with covariance:
\begin{align*}
	\E[\vy\vy^T] &= \E[\mL \vn \vn^T \mL], \\
			&= \mL \E[\vn \vn^T] \mL, \\
			&= \mSigma.
\end{align*}
The Cholesky decomposition is a standard method in various packages such as MATLAB or NumPy/PyTorch/etc., and thus is readily applied to the Noll matrix.

The Noll matrix contains off-diagonal entries. One may wonder if we may diagonalize the matrix by an eigen-decomposition. This is possible, however, the matrix is \emph{paired} with the fact that it describes the coefficients for our Zernike basis. If we wish to diagonalize the matrix, we will require different basis functions. This type of decomposition is known as the \keyword{Karhunen–Loève (KL)}\index{Karhunen–Loève decomposition} decomposition where the bases are the eigenfunctions of the covariance function. This has been considered in papers such as Roddier \cite{Roddier_1990_a}. However, this was also done earlier \emph{by hand} for analyzing the probability of getting a lucky observation by Fried \cite{Fried_1978_a}. The KL basis functions are in non-closed form and accordingly more difficult to work with, therefore, we choose to continue with the Zernike polynomials.

\subsection{What's Missing?}
If we wish to describe or simulate the effects on an object observed through turbulence, it may be tempting to jump to the conclusion that we are ready to do so with Noll's result. However, we should recall that in the case of turbulent PSFs which span an image we require a spatially variant representation,
\begin{equation}
\abs{h_{\vu}(\vx)}^2 = \abs{\mathfrak{Fourier} \{ P(\vrho) e^{j \phi_{\vu}(R\vrho)} \}}^2_{\vf = \vx / (\lambda z)}.
\label{eq: ch3_turb_psf}
\end{equation}
Here we emphasize the spatial dependency of the phase realizations. As a result, we must have \emph{multiple} realizations of $\phi_{\vu}$ due to the dependency on the source location. Noll's result does not allow such a possibility because the spatial correlation is missing!

The work of Noll is a great starting point, however, more work will be required to draw Zernike samples that are sampled properly in both intermodal and spatial axes. If we can find this representation, then we may successfully cut out the numerical wave propagation from the simulation procedure and allow us to generate samples much faster.

\section{The Zernike Space}
\label{sec: Zernike space}
This leads us to formally introduce the Zernike space\index{Zernike space}. The Zernike space generalizes the concept of a basis function for a single phase realization to a \keyword{random vector field}\index{random vector! field} which is a collection of basis coefficient vectors. While that is easy to say, proper sampling will prove to be computationally infeasible at present \cite{Chimitt_2022_a}. We will set this issue aside for now and focus on the concept itself, concluding this discussion with these computational issues and an approximation that avoids them.

\subsection{Overview}
The Zernike space arises from the consideration that for each point in the object plane, there exists a phase realization $\phi_{\vu}$, each with a Zernike representation $\mathbf{a}_{\vu}$. Furthermore, each phase realization describes the PSF $\vert h_{\vu}(\vx) \vert^2$. However, by our basis representation, the vector $\mathbf{a}_{\vu}$ is \emph{also} sufficient in describing the PSF. We refer to this set $\{ \mathbf{a}_{\vu} \}$ as the Zernike space.

\begin{figure}[h]
\centering
\vspace{-2ex}
\includegraphics[width=0.8\linewidth]{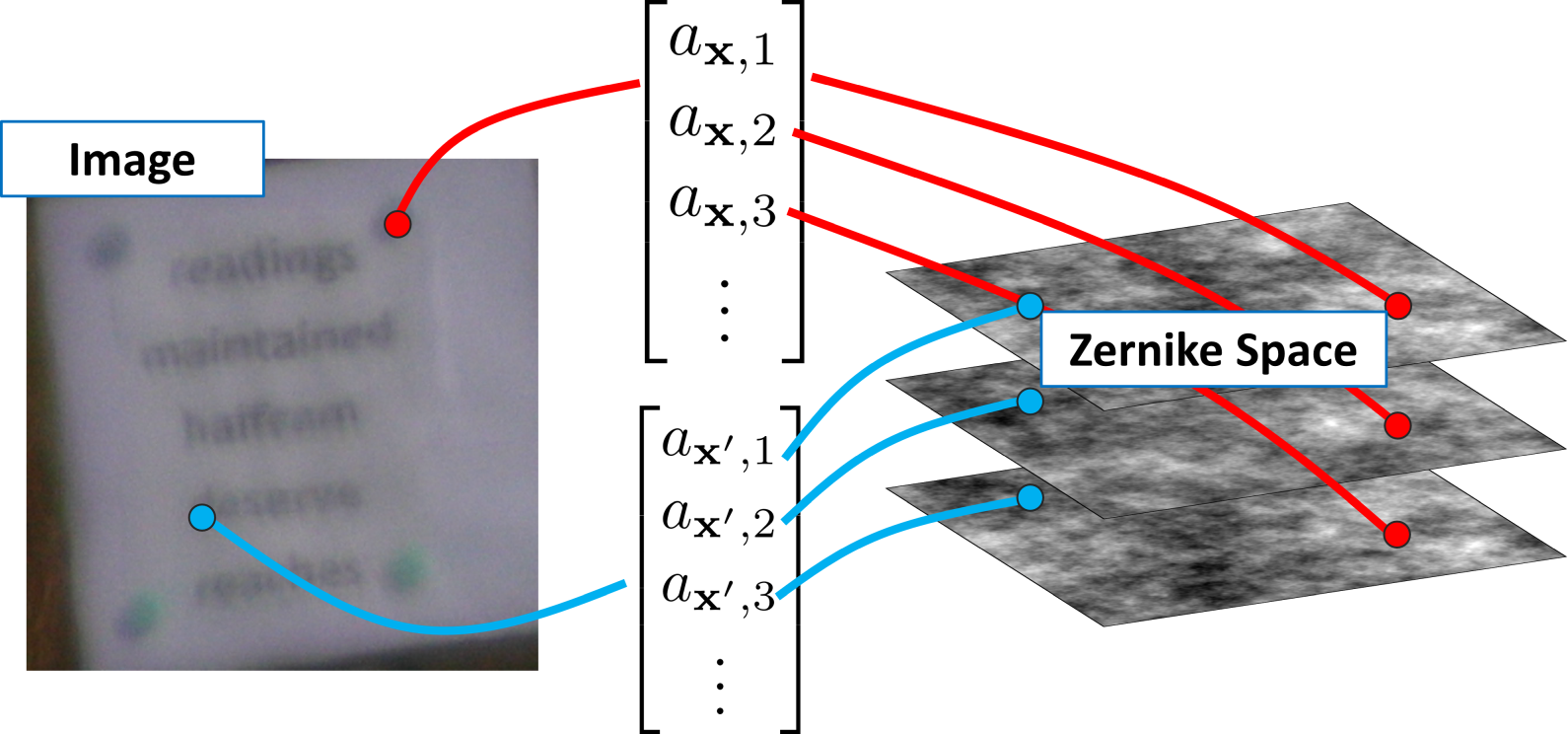}
\caption{Visualization of the Zernike space. For each point in the image, there exists a Zernike vector, which further defines a vector field.}
\label{fig: Ch3 zernike space}
\vspace{-2ex}
\end{figure}

We present a visualization of the Zernike space in \fref{fig: Ch3 zernike space}. For each point in the image, there exists a vector that is correlated intermodally and spatially to its neighbors. Thus, the Zernike space is a random vector field. The two axes of the random fields are spatial axes while the third axis is the Zernike coefficient axis. Thus, it is incorrect to say that the Zernike space is a volume, as the notion of locality is lost in the coefficient axis. It is this problem that will give rise to a majority of our issues when attempting to generate realizations of the Zernike space.

The Zernike space allows us to simply state the model that captures all of the turbulent statistics \emph{without} having to describe the propagation. Because the turbulence is distributed as a random process, the Zernike space will accordingly be random. The way we will need to describe the Zernike space will be through its correlations. This will require work to ``bake in'' the required properties, though the gain will be eliminating numerical wave propagation in our computation.

\subsection{Mirroring}
A critical piece of information that would be needed to enable the Zernike-based approach is the \keyword{spatial correlation}\index{Zernike space! spatial correlation} of the basis coefficients. That is, for a fixed Zernike polynomial order $i$, we want to find the joint expectation $\E[a_{\vu,i}\,a_{\vu',i}]$ for two different coordinates $\vu$ and $\vu'$. The derivation here is through an approximate analysis by the authors \cite{Chimitt_2020_a} which is conceptually a bit easier to understand. We later present an overview of a more general solution.

To begin our analysis, we first generalize the structure function. To begin, let us recall the structure function, though slightly re-arranged:
\begin{align}
\calD_{\phi}(R\vrho - R\vrho') &= 2.91 k^2 \int_0^L C_n^2(z) \left|R(\vrho - \vrho')\left(1-\frac{z}{L}\right)\right|^{5/3}dz.
\label{eq: Ch3 struct_fun_one_point}
\end{align}
Note \eref{eq: Ch3 struct_fun_one_point} is equivalent to our continuous spherical wave structure function definition from \eref{eq: ch2 struct fun spherical}. The structure function $\calD_{\phi}(R\vrho - R\vrho')$ is defined just by the two coordinates $R\vrho$ and $R\vrho'$ which are two points of the \emph{same} phase function $\phi_{\vu}(R\vrho)$. However, what we are interested in is the pair $\phi_{\vu}(R\vrho)$ and $\phi_{\vu'}(R\vrho)$. Therefore, a key step here is to redefine the structure function $\calD_{\phi}(R\vrho - R\vrho')$ so that it depends on $\vu$ and $\vu'$. This will allow us to model the case of two separate phase realizations.

\begin{figure}[h]
    \centering
    \begin{tabular}{cc}
    \includegraphics[width=0.45\linewidth]{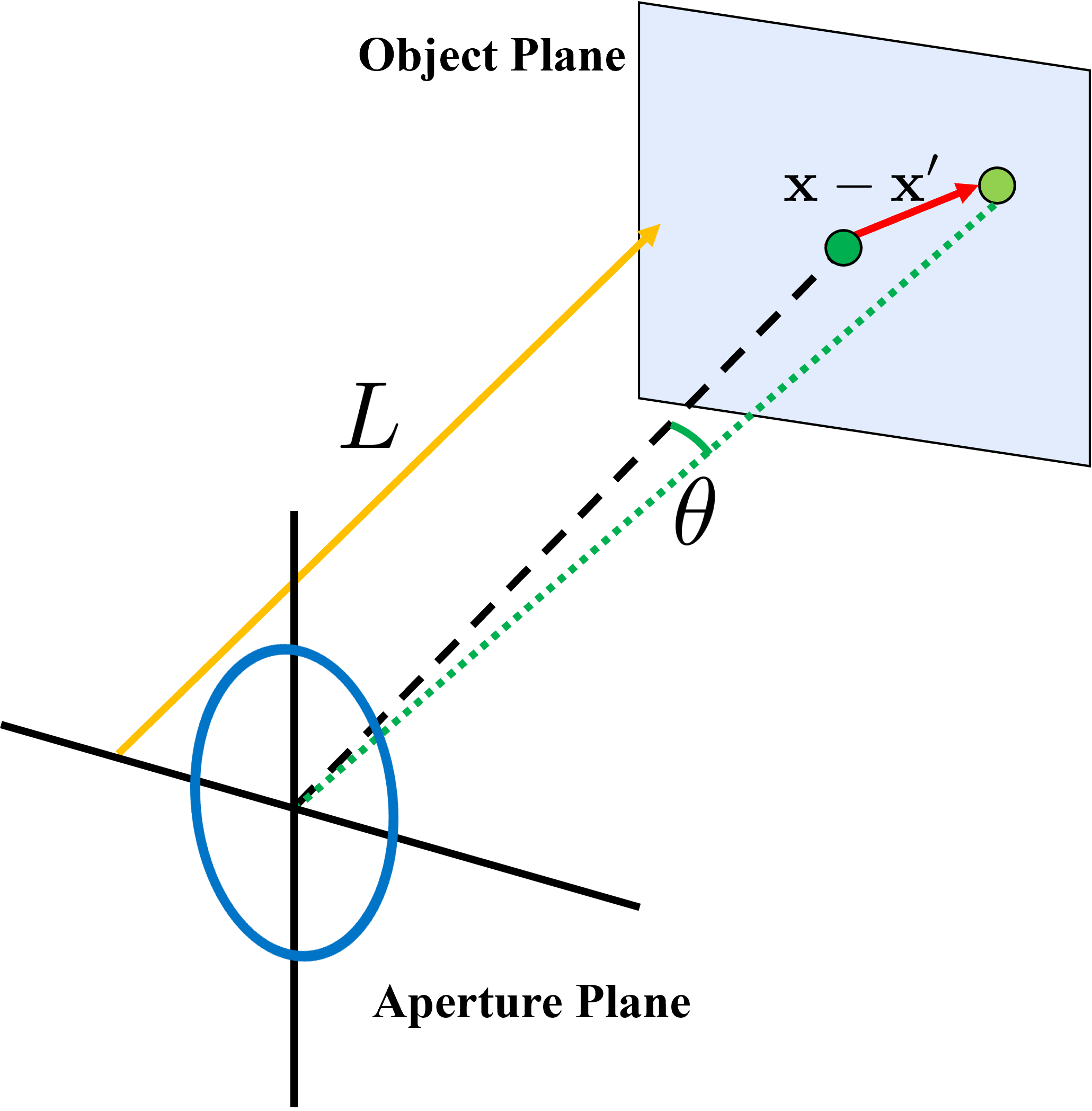}&
    \includegraphics[width=0.45\linewidth]{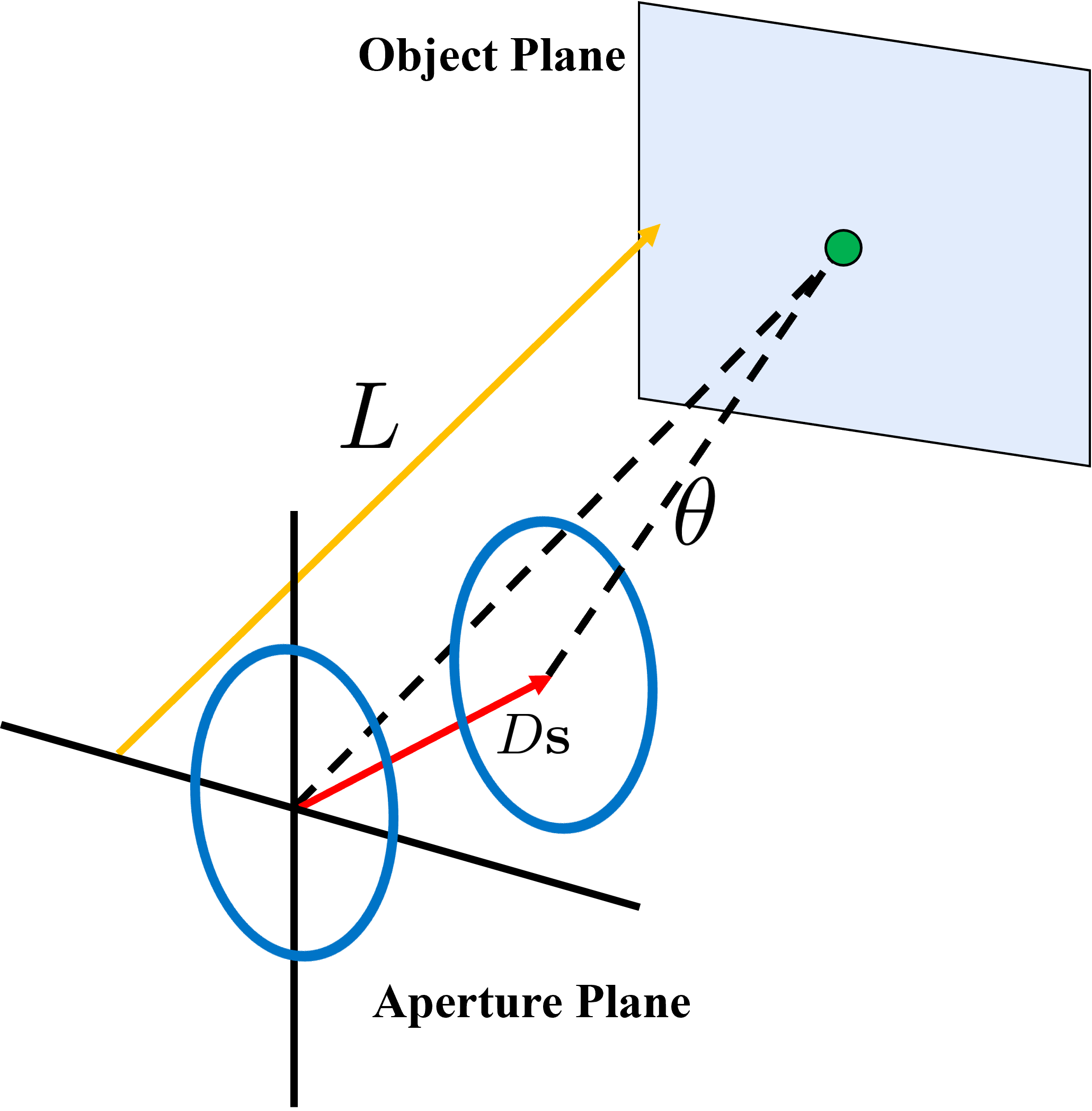}\\
    (a) Our problem:  & (b) Chanan \cite{Chanan_1992_a} and Takato \cite{Takato_1995_a} \\
    Two points in the scene & Two different apertures\\
    \end{tabular}
    \caption{Visualization of the geometries of the two types of problems. The problem we are primarily concerned with is shown in (a), however, we would like to leverage the results of analysis of the geometry considered in (b).}
    \label{fig: Ch3 fried_vs_takato}
\end{figure}

To tackle the problem, we start by considering the geometry shown in \fref{fig: Ch3 fried_vs_takato}. The problem of interest is the one shown in \fref{fig: Ch3 fried_vs_takato}(a) where we use one aperture to observe two different points in the scene. The separation of the two points in the scene is $\vu-\vu'$. If the distance from the aperture to the object plane is $L$, the normal vector pointing out of the image plane will form an angle with $\vu$ (and another angle with $\vu'$). These two angles are the \keyword{angle-of-arrival}\index{angle-of-arrival}. Though the angles are typically of interest, we prefer to consider the separation in the object plane $\vu-\vu'$. Let us then extend the structure function to include a separation in the object plane, as well as the aperture plane. This leads us to define a more general structure function \cite{Basu_2015_a, Roggemann_1996_a},
\begin{align}
    \calD(R\vrho - R\vrho', \vu - \vu') &= 2.91 k^2 \int_0^L dz C_n^2(z) \notag \\
    &\times \left|R(\vrho - \vrho')\left(\frac{L - z}{L}\right) + z/L(\vu - \vu')\right|^{5/3}.
    \label{eq: struct_fun_sphere_two_points}
\end{align}
The magnitude term within the integrand can be understood through simple geometry: it is an interpolation between the vector characterizing the separation in the object plane to the difference vector in the aperture plane.

Using geometric optics and symmetry, the roles of the object plane and the aperture can be flipped as shown in \fref{fig: Ch3 fried_vs_takato}(b). We refer to such a technique as \keyword{mirroring}. In this new configuration, the separation $\vu-\vu'$ is translated to the separation of two points in the aperture plane. Assuming that each aperture has a diameter $D$, we can define a vector $\vs$ such that
\begin{equation}
D\vs = \vu-\vu',
\end{equation}
according to the symmetry of the mirroring. The vector $\vs$ tells us how many units of $D$ is needed for the separation of two points.

Readers at this point may wonder: Why construct such a mirror and map the original problem to the mirrored space? The short answer is that the spatial correlation in \fref{fig: Ch3 fried_vs_takato}(b) has been thoroughly studied whereas the original one in \fref{fig: Ch3 fried_vs_takato}(a) is not. In the prior work of Chanan \cite{Chanan_1992_a} and Takato and Yamaguchi \cite{Takato_1995_a}, it was shown that if there are two \emph{different} imaging systems looking at one point in the space (i.e., \fref{fig: Ch3 fried_vs_takato}(b)), there exists closed-form expressions of the spatial correlation (of the Zernike coefficients). They were interested in this problem because for astronomy, large telescope cameras with \keyword{multiple apertures}\index{multiple apertures} are used to observe stars. Therefore, if we can find this mapping, we can simply leverage the existing results of Chanan \cite{Chanan_1992_a} and Takato and Yamaguchi \cite{Takato_1995_a}.

Going back to the spatial correlation, our intention is to replace $\E[\phi_{\vu}(R\vrho) \phi_{\vu}(R\vrho')]$ with $\E[\phi_{\vu}(R\vrho) \phi_{\vu'}(R\vrho')]$. We will then leverage the two aperture results to solve our problem. Beginning with the expectation, we may write
\begin{align}
\E[a_{\vu,i} \, a_{\vu',j}]
&= \frac{1}{\pi^2} \iint P(\vrho) P(\vrho') Z_i(\vrho) Z_j(\vrho') \notag \\
&\qquad \qquad \times \E[\phi_{\vu}(R\vrho) \phi_{\vu'}(R\vrho')] \, d\vrho d\vrho'.
\end{align}
Through the definition of the structure function, we may write this as
\begin{align}
    \E [a_{\vu,i} a_{\vu',j}] = \frac{-1}{2\pi^2} \iint &d\vrho d\vrho' P(\rho) P(\rho') Z_i(\vrho) Z_j(\vrho') \notag\\
    &\times\calD(R\vrho - R\vrho', \vu - \vu'). \label{eq: our_problem_with_struct}
\end{align}
Interestingly, it was shown in \cite{Chimitt_2020_a} that if a first-order Taylor approximation was made upon the magnitude expression at $z = L/2$, combined with an assumption of a constant $C_n^2$ profile, then the expression may be written through mirroring as
\begin{align}
    \E [a_{\vu,i}a_{\vu',j}]
    &= \frac{L}{2^{5/3}\pi^2} \iint P(\vrho) P(\vrho') \label{eq takatos_problem}\\
    &\qquad \times Z_i(\vrho) Z_j(\vrho') \E[\phi_{\vu}(R\vrho) \phi_{\vu}(R\vrho' + D\vs)] \, d\vrho d\vrho' \notag.
\end{align}
The impact of the mirroring is that the two phase functions $\phi_{\vu}(R\vrho)$ and $\phi_{\vu'}(R\vrho')$ are now changed to $\phi_{\vu}(R\vrho)$ and $\phi_{\vu}(R\vrho' + D\vs)$. The latter is due to the mirroring analysis.

The result of this is that the problem of angle-of-arrival correlation has been mapped to the multiple aperture problem. Therefore, the results of Chanan \cite{Chanan_1992_a} and Takato and Yamaguchi \cite{Takato_1995_a} can be applied directly for our case. Using the Kolmogorov power spectral density results in an expression that we present in the following theorem:
\boxedthm{
\begin{theorem}[\keyword{Zernike Spatial Correlation via Multi-Aperture}]\index{Zernike space! spatial correlation (approximation)}
\label{thm: Ch3 Zernike spatial 01}
Let $\vu$ and $\vu'$ be two points in the object plane. Consider a two-aperture model with $D\vs = \vu-\vu'$ where $D$ is the aperture diameter, and $\vs$ is a vector defining the separation. The spatial correlation of two Zernike coefficients $a_{\vu,i}$ and $a_{\vu',j}$ is given by
\begin{align}
    \E [a_{\vu,i} a_{\vu',j}]
    &\approx \frac{-2.91 k^2 C_n^2 L}{2\pi^2 2^{5/3}} \iint  P(\vrho) P(\vrho') Z_i(\vrho) Z_j(\vrho')  \nonumber\\
    &\quad \times  \left|R(\vrho - \vrho') + D\vs\right|^{5/3} d\vrho d\vrho',
\end{align}
\end{theorem}
}

\subsection{Spatial Correlation of the Tilts}
Theorem~\ref{thm: Ch3 Zernike spatial 01} is a general result for \emph{all} Zernike coefficients. To gain some insights into how this equation can be used to draw random turbulence samples, we focus on two Zernike coefficients: the horizontal tilt (characterized by $a_{\vu,2}$) and the vertical tilts (characterized by $a_{\vu,3}$). We are interested in these two coefficients because the tilts occupy the majority of the energy in the Zernike basis representation. In addition, the tilt statistics are slightly more manageable to track analytically.

When restricted to $a_{\vu,2}$ and $a_{\vu,3}$, the Zernike basis $Z_2(\vrho)$ and $Z_3(\vrho)$ will take a relatively form. Using Chanan's result  \cite[Eq. 11]{Chanan_1992_a}, we can show that for $j \in \{2,3\}$, the spatial correlation is given as follows.

\begin{figure}[h]
\centering
\vspace{-2ex}
\includegraphics[width=0.6\linewidth]{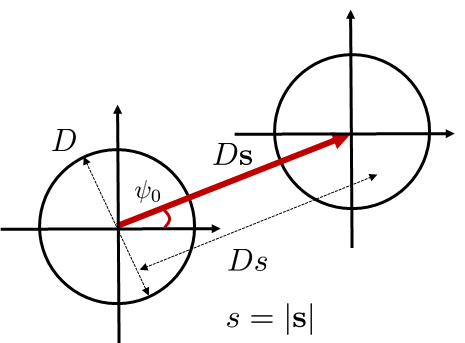}
\caption{Notation of the coordinates. The magnitude of $\vxi$ is measured in the unit of the diameter $D$.}
\label{fig: Ch3 coordinate}
\vspace{-2ex}
\end{figure}

\boxedthm{
\begin{theorem}[\keyword{Spatial correlation of tilts}]\index{Zernike space! tilt correlation (approximation)}
\label{thm: spatial correlation tilt}
The spatial correlation of the 2nd and the 3rd Zernike coefficients (the horizontal and the vertical tilts) are approximated by
\begin{align}
&\E[a_{\vu,j} a_{\vu',j}] \approx \frac{c_2}{2^{5/3}} \left(\frac{D}{r_0}\right)^{5/3} \left[I_{0}(s) \mp \cos 2\psi_0 I_2(s)\right], \;\; j = 2,3,
\label{eq: correlation 3}
\end{align}
where $c_2 = 7.7554$, and the minus sign is for $a_{\vu,2}$ and the plus sign is for $a_{\vu,3}$. The coordinate $[s, \psi_0]$ is illustrated in \fref{fig: Ch3 coordinate}. The length of $D\vs = \vu-\vu'$ is the scalar $Ds$, where $s = |\vs|$ and the angle is $\psi_0$.
\end{theorem}
}

The integrals $I_0(s)$ and $I_2(s)$ in \eref{eq: correlation 3} are defined through the Bessel functions of the first kind:
\begin{align}
\calJ_0(s) &= \int_0^{\infty} \zeta^{-14/3}J_0(2s\zeta)J_2^2(\zeta)d\zeta, \\
I_2(s) &= \int_0^{\infty} \zeta^{-14/3}J_2(2s\zeta)J_2^2(\zeta)d\zeta,
\end{align}
where $J_0$ and $J_2$ are the Bessel functions of the first kind. On computers, these can be defined using built-in libraries. In MATLAB, for example, we can use commands \texttt{besselj} and \texttt{integral} as shown below.

\begin{Verbatim}[frame=single,framerule=0.5mm, rulecolor=\color{NavyBlue}]
% MATLAB code to evaluate integrals I_0 and I_2
s   = linspace(0,smax,N);
f   = @(z) z^(-14/3)*besselj(0,2*s*z)*...
                    besselj(2,z)^2;
I0  = integral(f, 1e-8, 1e3, 'ArrayValued', true);
g   = @(z) z^(-14/3)*besselj(2,2*s*z)*...
                    besselj(2,z)^2;
I2  = integral(g, 1e-8, 1e3, 'ArrayValued', true);
\end{Verbatim}

How do we draw the random tilts then? Considering \eref{eq: correlation 3} again, we recognize that it defines the correlation matrix.
\begin{equation}
[\mC_j]_{\vu,\vu'} = \frac{\E[a_{\vu,j} a_{\vu',j}]}{\E[a_{\vu,j} a_{\vu,j}]} = \frac{I_{0}(s) \mp \cos (2\psi_0) I_2(s)}{I_0(0)}, \; j =2,3.
\end{equation}
In this equation, $[\mC_j]_{\vu,\vu'}$ denotes the correlation matrix $\mC_j$ for $j = 2, 3$. The subscript $(\vu,\vu')$ denotes that we are looking at the coordinate $(\vu,\vu')$. Since the right-hand side of the correlation matrix is described by the polar coordinate $[s,\psi_0]$ of the difference vector $\vs = (\vu-\vu')/D$, the matrix $\mC_j$ is Toeplitz\index{Toeplitz matrix}. This should not be a surprise because the correlation matrix is \keyword{homogeneous} (aka wide-sense stationary).

\begin{figure}[h]
\centering
\begin{tabular}{cc}
\includegraphics[height=4cm]{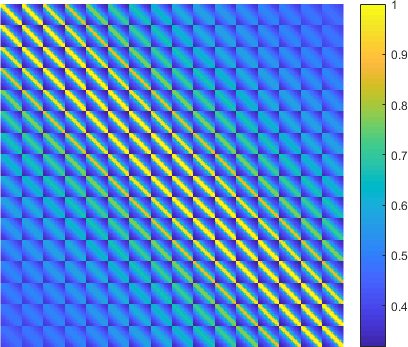}&
\includegraphics[height=4cm]{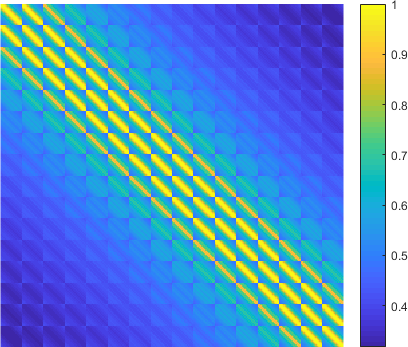}\\
(a) $[\mC_2]_{\vu,\vu'}$ &(b) $[\mC_3]_{\vu,\vu'}$
\end{tabular}
\caption{The covariance matrix $\mC_2$ and $\mC_3$ for an image of $16 \times 16$ pixels. The size of the matrices are $16^2\times 16^2$.}
\label{fig: Ch3 SigmaX and SigmaY}
\vspace{-2ex}
\end{figure}

A visualization of the correlation matrices $[\mC_2]_{\vu,\vu'}$ and $[\mC_3]_{\vu,\vu'}$ is shown in \fref{fig: Ch3 SigmaX and SigmaY}. The matrices demonstrate a Toeplitz block structure because the correlation is two-dimensional. Nevertheless, the Toeplitz structure is preserved due to its homogeneity.

Thus far, we have described how the Zernike coefficients $a_{\vu,2}$ and $a_{\vu,3}$ can be evaluated statistically. To finally draw the random tilt, we still need proper scaling so that the Zernike coefficient can correspond to the number of pixels. As mentioned in \cite[Eq. 2]{Chanan_1992_a}, the relationship between the tilt angle $\alpha_x$ and the Zernike coefficient is given by
\begin{equation}
\alpha_{\vu,2} = \frac{2\lambda}{\pi D}a_{\vu,2}, \quad\mbox{and}\quad \alpha_{\vu,3} = \frac{2 \lambda}{\pi D}a_{\vu,3}.
\end{equation}
Since the Nyquist sampling between two adjacent pixels in the object plane is $L\lambda/2D$, it holds that the tilt angles converted to the number of pixels in the object are
\begin{align*}
\alpha_{\vu,2} \mbox{ [pixels]} =  \frac{L  \cdot \frac{2\lambda}{\pi D} a_2}{\frac{L\lambda}{2D}} = \frac{4}{\pi}a_{\vu,2}, \quad\mbox{and}\quad \alpha_{\vu,3} \mbox{ [pixels]} = \frac{4}{\pi} a_{\vu,3}.
\end{align*}
Putting everything together, we can show that the un-normalized correlation coefficient of the tilt angles is
\begin{align}
&\E[\alpha_{\vu,2} \alpha_{\vu',2}] = \frac{16}{\pi^2} \E[a_{\vu,2} a_{\vu',2}] \notag \\
&\quad\quad = \underset{\kappa^2}{\underbrace{\frac{16}{\pi^2} \frac{c_2}{2^{5/3}} I_0(0)}} [\mC_2]_{\vu,\vu'},
\end{align}
where we defined $\kappa$ as the constant preceding the normalized correlation function $[\mC_2]_{\vu,\vu'}$. For vertical tilts, the relationship is $\E[\alpha_{\vu,3} \alpha_{\vu',3}] = \kappa^2 [\mC_3]_{\vu,\vu'}$.

We are now in the position to draw random tilts. Since the correlation matrix represents a homogeneous process and so that matrix is block Toeplitz, drawing random samples can be done in the Fourier domain. The program below is a MATLAB implementation of drawing random horizontal tilts from the statistics.

\begin{Verbatim}[frame=single,framerule=0.5mm, rulecolor=\color{NavyBlue}]
kappa2 = I0(1)*7.7554*(D/r0)^(5/3)/(2^(5/3))*...
        (2*lambda/(pi*D))^2*(2*D/lambda)^2;
Cx     = kappa2*C2;
Cxfft  = fft2(Cx);
S_half = sqrt(abs(Cxfft));
b   = randn(N,N);
MVx = real(ifft2(S_half.*b))*sqrt(2)*N;
\end{Verbatim}

\begin{figure}[h]
\centering
\begin{tabular}{cc}
\includegraphics[height=5cm]{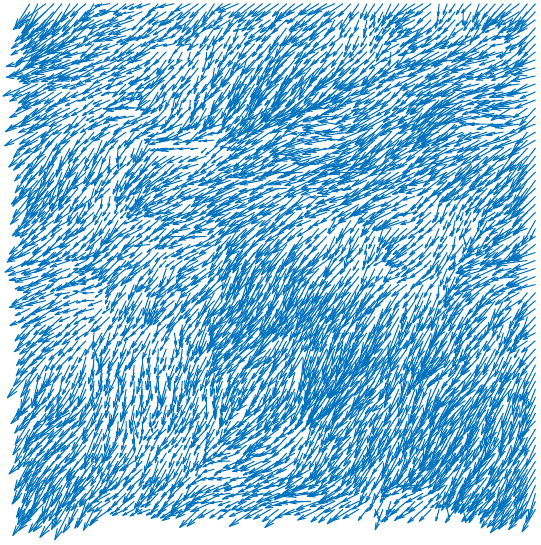}&
\includegraphics[height=5cm]{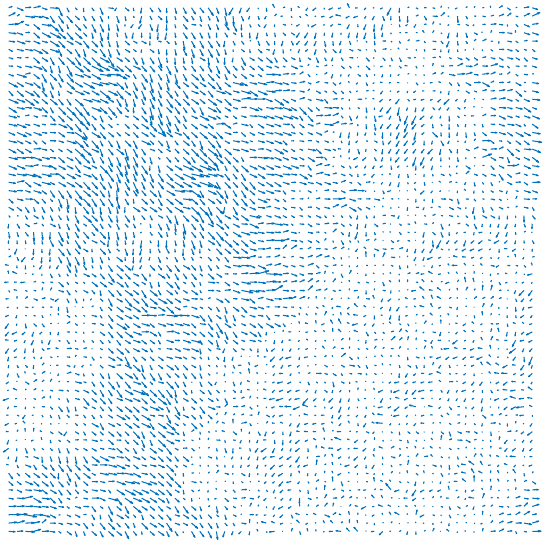}\\
$C_n^2 = 1\times 10^{-15}$m$^{-2/3}$ & $C_n^2 = 2.5\times 10^{-16}$m$^{-2/3}$
\end{tabular}
\caption{Simulated tilt maps for $C_n^2 = 1\times 10^{-15}$m$^{-2/3}$, and $C_n^2 = 2.5\times 10^{-16}$m$^{-2/3}$. The tilts are scaled by $2\times $ and skipped per every 8 pixels for display.}
\label{fig: Ch3 tilt}
\vspace{-2ex}
\end{figure}

Two examples of the randomly sampled tilt vectors are shown in \fref{fig: Ch3 tilt}. For the stronger turbulence condition $C_n^2 = 1\times 10^{-15}$m$^{-2/3}$, we can observe a strong correlation from one corner of the image to the other corner of the image. This explains the necessity of drawing the turbulence according to the theoretical model instead of a set of i.i.d. Gaussian random samples smoothed by heuristic operations.

A typical question for a numerical simulator is how do we know that the random samples we drew are correct? Since these random tilts are the particular realization of a two-dimensional random process, it is impossible to take a real image in the field and ask the simulator to match exactly. The best we can do is to argue from a statistical average point of view. If we take enough measurements, will the simulator produce something that would match the known theoretical statistics?

\begin{figure}[h]
\centering
\footnotesize{
\begin{tabular}{cc}
\includegraphics[width=0.475\linewidth]{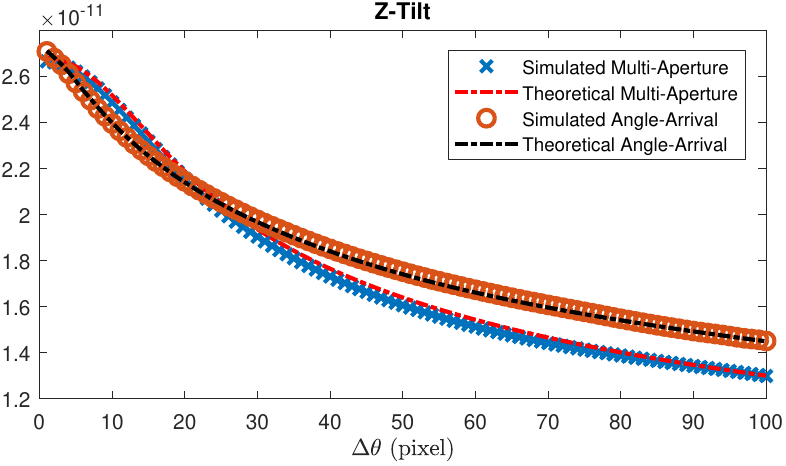}&
\includegraphics[width=0.475\linewidth]{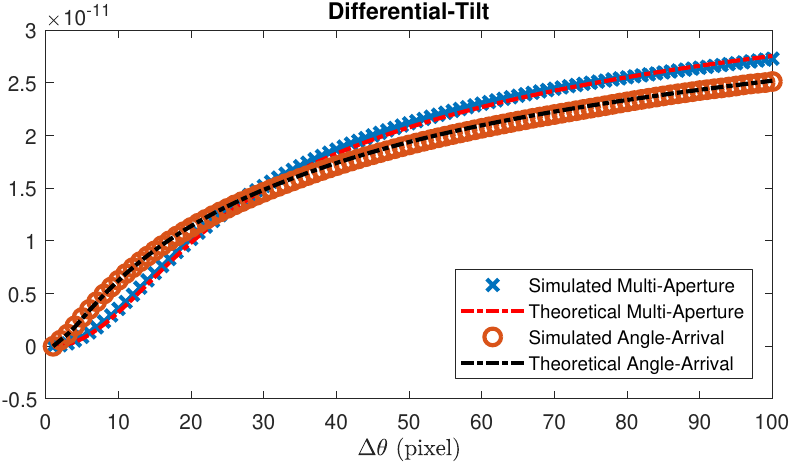}
\end{tabular}}
\caption{The theoretical and simulated tilt statistics at $C_n^2 = 1 \times 10^{-15}$m$^{-2/3}$. The curves are plotted versus the number of Nyquist pixels. The angle-of-arrival is due to Basu et al. \cite{Basu_2015_a} as well as Fried \cite{Fried_1976_a}, and the multi-aperture is our approximation based on Chanan \cite{Chanan_1992_a}. Additionally, we note $20,000$ random realizations were used in generating the simulated curves.}
\label{fig: Ch3 Ztilt and DTV}
\end{figure}

In the turbulence literature, the two commonly used statistics are the $Z$-tilt statistics and the differential-tilt statistics. The \keyword{$Z$-tilt}\index{z-tilt} is defined as
\begin{equation}
\texttt{Z-tilt}(\vu,\vu') = \E[\alpha_{\vu,j} \alpha_{\vu',j}], \qquad j = 2, 3,
\end{equation}
whereas the \keyword{differential tilt}\index{differential tilt} is defined as
\begin{equation}
\texttt{D-tilt}(\vu,\vu') = \E[(\alpha_{\vu,j}-\alpha_{\vu',j})^2], \qquad j = 2, 3.
\end{equation}
For the numerical simulation we just described, we can verify the validity of the simulator by comparing it against the theoretical statistics. In \fref{fig: Ch3 Ztilt and DTV} we show the comparison between the simulator (and its theoretical limit) and the one derived by Basu et al. \cite{Basu_2015_a}. Overall the match between what the multi-aperture simulator can offer and what we expect it from the angle of arrival analysis in \cite{Basu_2015_a, Fried_1975_a}. Therefore, we can conclude that drawing the tilt without wave propagation has been successfully implemented.

\subsection{High-Order Correlations and a General Solution}
The previous discussion focused on the tilt coefficients. This is for two reasons: (1) The tilt coefficients are easy to compare to existing results which are already due to the study of angle-of-arrival correlations; (2) the expression for tilt is a bit simpler to write down. To generalize the approximation, we want to include any pair $\E[a_{\vu, i} a_{\vu', j}]$, leading us to present the following theorem which gives the correlations in a slightly more compact form:
\boxedthm{
\begin{theorem}[\keyword{Spatial correlation of all Zernike coefficients}]\index{Zernike space! spatial correlation (approximation)}
\label{thm: spatial correlation, high order Zernike}
The spatial correlation of all Zernike coefficients $i,j \ge 1$ is approximated by
\begin{align}
\E [a_{\vu,i} a_{\vu',j}]
&\approx \calA \sqrt{(n_i+1)(n_j+1)}\left(\frac{1}{2}\right)^{5/3} C_n^2 \;  L  \; f_{ij}\left(\vs,k_0\right),
\label{eq: chim_chan_main_result_1}
\end{align}
where $\calA = 0.00969k^2 2^{14/3}\pi^{2/3}R^{5/3}$, $n_i$ and $n_j$ are the Noll indices, $\vs = (\vu-\vu')/D$, and $f_{ij}(\vs,k_0)$ is the integral defined in Takato and Yamaguchi \cite{Takato_1995_a}.
\end{theorem}
}
This theorem represents the types of correlations utilized in the Zernike-based simulations \cite{Chimitt_2020_a, Mao_2021_a, Chimitt_2022_a}, though we note that it is \emph{still} an approximation.

Though we do not list $f_{ij}$ directly, it is useful to observe a few properties of \eref{eq: chim_chan_main_result_1}. Firstly, the correlation depends on the difference $\vu - \vu'$, thus the field for a single coefficient is WSS. There is also an angular dependence, thus the correlation functions are anisotropic in nature. The aperture $D$ also affects the length of correlation -- a larger aperture will induce more correlations while a smaller aperture will create a more anisotropic effect. Finally, as expected, the propagation distance $L$ and $C_n^2$ value influence the variance of the result, with a higher constant $C_n^2$ or longer propagation distance making the turbulence stronger.

To close our theoretical description of the correlations, the above problem of Zernike coefficient correlations has been solved without such approximation in \cite{Chimitt_2023_b} and is written in the following way:\index{Zernike space! spatial correlation (exact)}
\begin{align}
    \E[a_{\vu,i} a_{\vu',j}] =  \calA &\sqrt{(n_i+1)(n_j+1)} \int_0^L \left(\frac{L - z}{L}\right)^{5/3} C_n^2(z) \nonumber\notag \\ &\times
     f_{ij} \left( \left(\frac{z}{D(L - z)}\right)(\vu - \vu'), k_0 \right) dz.
    \label{eq: main_result_continuous2}
\end{align}
This represents a continuous integration over the path, which is a significant improvement over the approximation provided in \cite{Chimitt_2020_a}. It can be shown that \eref{eq: chim_chan_main_result_1} is a special case of \eref{eq: main_result_continuous2} as described in Chimitt and Chan \cite{Chimitt_2023_b}. These results are similar to a paper by Whiteley et al. \cite{Whiteley_1998_a} thus we would suggest the interested reader to either paper for a discussion of the more general approach.

Fortunately, the correlation function one chooses to use (approximate or exact) doesn't affect \emph{how} they are used in simulation.

\subsection{Image Formation and the Zernike Space}
\index{Zernike space! image formation}The previous discussions highlight how we can describe correlations in the Zernike space, either through the approximation of \eqref{eq: chim_chan_main_result_1} or \eqref{eq: main_result_continuous2}. We now turn to answering a main question of this book: How can we understand image formation through the Zernike space?

To begin, we assume that phase distortions are sufficient in describing the effects. Recalling our familiar image formation equation,
\begin{equation}
I(\vx) = \vert h_{\vu}(\vx) \vert^2 \overset{\vu}{\circledast} I_g(\vx),
\end{equation}
and PSF formation equation (utilizing the Zernike representation)
\begin{equation}
\abs{h_{\vu}(\vx)}^2 = \abs{\mathfrak{Fourier}\left\{ P(\vrho) e^{j   \sum_{j=1}^{N} a_{\vu,j} Z_j(\vrho/R)  } \right\}}^2_{\vf = \vx / (\lambda z)},
\end{equation}
we can combine the two to write
\begin{equation}
I_i(\vx) = \abs{\mathfrak{Fourier}\left\{ P(\vrho) e^{j   \sum_{j=1}^{N} a_{\vu,j} Z_j(\vrho/R)  } \right\}}^2 \overset{\vu}{\circledast} I_g(\vx),
\label{eq: Z space image formation}
\end{equation}
dropping the resizing terms. Equation \eqref{eq: Z space image formation} represents a compact form of image formation using the Zernike space. Thus, the entire process can be described with basis decomposition, random vectors, and spatially varying convolution.

While \eqref{eq: Z space image formation} is fairly compact in some sense, we can go a step further. That is, what if we were able to decompose the PSF into a set of basis functions, similar to what we did with the phase? If we were to write the PSF as
\begin{equation}
\vert h_{\vu}(\vx) \vert^2 = \sum_{m=1}^M \beta_{\vu,m} \varphi_m(\vx),
\end{equation}
we could then approximate the spatially varying convolution as a sum of invariant convolutions. The reasons and methodology for doing so will be described shortly, but the result is the following:
\begin{equation}
I_i(\vu) = \sum_{m=1}^M (\varphi_m \circledast (\beta_m \odot I_g ) )(\vu),
\end{equation}
where $\varphi_m$ is our invariant basis representation of the PSF and $\beta_m$ are the coefficients for the PSF basis (which, of course, are related to the Zernike space). While stating $\varphi_m$ or the transform from the Zernike space to $\beta_m$ will prove to be analytically intractable, we will discuss how it can be utilized in the context of reconstruction in Chapter 5.

\section{Zernike-Based Simulation}
\label{sec: Zernike simulation}\index{Zernike-based simulation}\index{propagation-free methods! Zernike-based}
Much like split-step simulation, it is a numerical interpretation of the true image formation process. In a similar fashion, the Zernike space is a theoretical way of describing the turbulence distortions which we are presently interested in considering the numerical implementation of. Therefore, the following discussion is motivated by the computational difficulties one encounters when attempting to actualize this form of simulation.

\subsection{Numerical Challenges}
It may seem that we can use Cholesky-based generation to generate samples of the Zernike space as we did with a single coefficient vector. When the size of the Zernike spatial grid is small, its covariance is accordingly small and may therefore be generated using Cholesky decomposition. However, for a large grid of points, the size of the covariance matrix may result in the decomposition becoming infeasible. The size of the correlation matrix for an image of size $W \times H$ and $36$ Zernike coefficients requires the construction of a matrix that is $36HW \times 36HW$. For a standard $256\!\times\!256$ image this results in a matrix that is over 2 million by 2 million entries in size! We would have to decompose and then multiply by this matrix for \emph{one} realization, motivating us to look for an alternative.

One important case of numerical generation is that if the correlation structure is WSS, the matrix $\mSigma$ is \emph{circulant} and thus the eigendecomposition is equivalent to the Fourier transform. In this case, generating the random vector $\vy$ can be implemented via
\begin{equation*}
    \vy = \mU\mS^{\frac{1}{2}}\mU^{T} \ve = \calF^{-1}( \mS^{\frac{1}{2}} \calF(\ve) ),
\end{equation*}
where $\calF$ denotes the discrete-time Fourier transform, and the diagonal matrix $\mS$ is the Fourier spectrum of one row (or column) of the correlation matrix $\mSigma$.  The significance of the homogeneity is that it allows us to speed up the sampling process by performing all computations in the Fourier space. In addition, the memory bottleneck is resolved because we don't need to construct the full correlation matrix $\mSigma$ and run the Cholesky factorization. This is exactly the property we utilized for generating phase \emph{screens} in Chapter 3.

This brings us to present two issues. For a single field, the correlation is WSS. Although \eref{eq: chim_chan_main_result_1} is a function of the spatial separation $\vu - \vu'$ in the case of a single $f_{ii}$, the \emph{entire} Zernike space is not WSS due to the fact that $f_{ij}$ is not a function of index difference $i - j$. On top of this, we will still need to perform an FFT for converting the phase realizations into PSFs and spatially varying convolution. Therefore, there are many numerical issues that have a non-trivial path forward to getting this simulation to where it needs to be for large dataset generation. We will tackle these one at a time, starting with this issue of a lack of WSS.

\subsection{Fourier Sampling the Approximate Zernike Space}
Although the Zernike space as a whole is non-WSS, there is an approximation that largely solves this issue \cite{Chimitt_2022_a}. This method utilizes the insight that for a single coefficient field, the correlation \emph{is} WSS. This means that one may generate individual correlation fields correctly, however, they will be independent from one another. This alone will not solve our problem as this neglects the intermodal correlation.

To introduce the approximation, we will refer to the matrix which represents the correlational structure of the $i$th field with the $j$th field as $\mA_{i,j}$. In other words, $\mA_{i,j}$ corresponds to a matrix form of $f_{i,j}$. The entire Zernike space correlation, which is in the form of a \keyword{tensor}, can then be written as
\begin{equation*}
\mA =
\begin{bmatrix}
    \mA_{1,1} &\cdots &\mA_{1,N}\\
    \vdots &\ddots &\vdots\\
    \mA_{N,1} &\cdots &\mA_{N,N}
\end{bmatrix}
= \begin{bmatrix}
    \mA_{\vs_0} &\cdots &\mA_{\vs_p}\\
    \vdots &\ddots &\vdots\\
    \mA_{\vs_p} &\cdots &\mA_{\vs_0}
\end{bmatrix},
\end{equation*}
where here we are showing two ``projections'' of the same tensor.
To motivate our development, we pose the following question: Can we generate the fields independently, utilizing the single coefficient WSS property, then mix them according to \emph{some} matrix? This question eventually led to the following scheme for sampling the Zernike space in an approximate manner:\index{Zernike-based simulation! WSS approximation}
\begin{enumerate}
    \item Generate $i=\{1,2,\ldots, N\}$ unit-variance, spatially correlated random fields according to $\mA_{i,i}$. Note this generation uses only \emph{auto}covariance functions. At this stage, our $N$ fields are independent, thus utilizing the FFT-based method based on their homogeneity;
    \item Perform a point-wise mixing of the random fields according to the Noll matrix $\mSigma$. This mixing is done pixel-wise (across the coefficient index dimension) per pixel.
\end{enumerate}
This generation process can be written mathematically in the following way
\begin{equation}
    \widetilde{\mA} = \mL\begin{bmatrix}
    \mA_{1,1} & 0 & \dots & 0 \\
    0 & \mA_{2,2} & \dots & 0 \\
    \vdots & \vdots & \ddots & \vdots \\
    0 & 0 & \cdots & \mA_{N,N}
    \end{bmatrix}
    \mL^T,
\end{equation}
where $\mL \mL^T = \mSigma$ and $\widetilde{\mA}$ is regarded as the approximation to the Zernike space.

\begin{figure}
    \centering
    \includegraphics[width=0.8\linewidth]{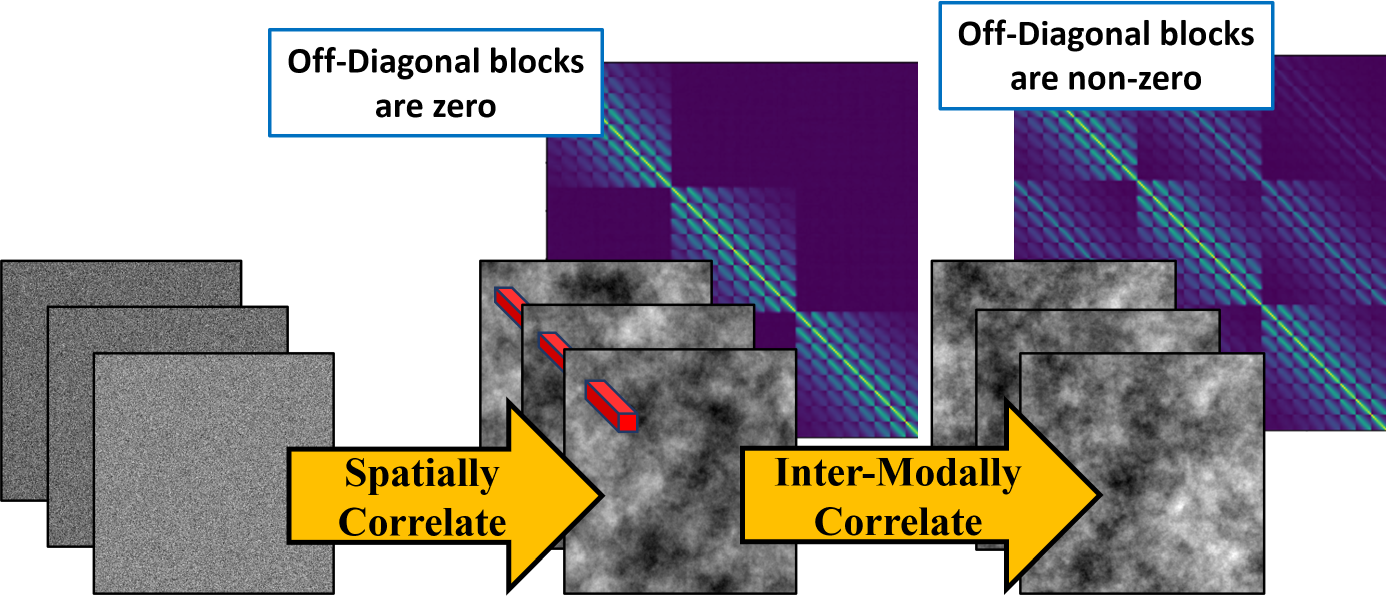}
    \caption{A visual representation of the realizations and covariance structure as it changes through the generation process. White noise is used to generate independent random fields which are then mixed according to the index axis covariance. Here, the covariance structure is of a smaller grid than the random field shown for ease of interpretation. Source: \cite{Chimitt_2022_a}.}
    \label{fig: rand_gen}
\end{figure}

We note the resultant matrix is no longer diagonal, however, the off-diagonal entries will be approximations of the true correlation structure. This approximation and its impact have been analyzed in \cite{Chimitt_2022_a}. A visualization of the proposed generation process is provided in \fref{fig: rand_gen} using a simplistic example in the case of an $8\times8$ image with 3 coefficient fields for visualizing the covariance structure. The same concept presented in this Figure can be extended to the case of an $W\times H$ image with $N$ coefficient fields.

\subsection{Approximating Spatially Varying Blur}
\index{spatially varying blur! approximation}Invariant convolution may be implemented through the FFT and is accordingly fast to compute. However, spatially variant convolution does \emph{not} have such a simple mathematical relationship, and thus is accordingly far slower than its invariant counterpart. Therefore, we are interested in representing spatially variant convolution and a summation of \emph{invariant} convolutions. To this end, we follow the approach outlined in Mao et al. \cite{Mao_2021_a}.

To begin, we recall that the PSF $\vert h_{\vu}(\vx) \vert^2$ is defined per pixel. This is analogous to the phase, which is also defined per pixel. For the phase, we decomposed it via the Zernike polynomials, for the PSF there is no such easy decomposition. However, if it \emph{were} true, we could write the following \keyword{PSF basis representation}\index{basis representation! of PSF} via basis functions $\varphi_1(\vx),\ldots,\varphi_M(\vx)$ as
\begin{equation}
\vert h_{\vu}(\vx) \vert^2 = \sum_{m=1}^M \beta_{\vu,m} \varphi_m(\vx),
\label{eq: Ch3 PSF basis}
\end{equation}
where $\beta_{\vu,m}$ is the $m$th basis coefficient at pixel location $\vu$. Here, $\varphi_m(\vx)$ is the basis function that will be shared across \emph{all} the pixel locations. The coordinate $\vu$ is embedded in the coefficient $\beta_{\vu,m}$ which is $\vu$-dependent.

Since we are interested in spatially varying convolution, we are inevitably led back to our discussion of scattering versus gathering. As described in Chimitt et al. \cite{Chimitt_2023_a}, gathering results in the following form:
\begin{align}
I_i(\vx)
&= \underset{\text{linear combination of invariant blurs}}{\underbrace{\sum_{m=1}^M \beta_{\vx,m} \;\; \underset{\text{invariant blur}}{\underbrace{(\varphi_m \circledast I_g)(\vu)}}}}, \label{eq: I approx gath}
\end{align}
while scattering results in
\begin{align}
I_i(\vx) &= \underset{\text{linear combination of invariant blurs}}{\underbrace{\sum_{m=1}^M \bigg( \varphi_m \quad \circledast \underset{\text{pixelwise weighted input}}{\underbrace{\Big(\beta_m \odot I_g \Big)}} \bigg)(\vx)}} \label{eq: I approx scat}.
\end{align}
The difference between these two is subtle, coming down to the order of operations that are not commutable. That is, do we multiply by the coefficients first, or do we convolve first? The answer to this question is given in \cite{Chimitt_2023_a} where scattering was found to be the proper choice for simulation.

\begin{figure}[h]
\centering
\includegraphics[width=0.9\linewidth]{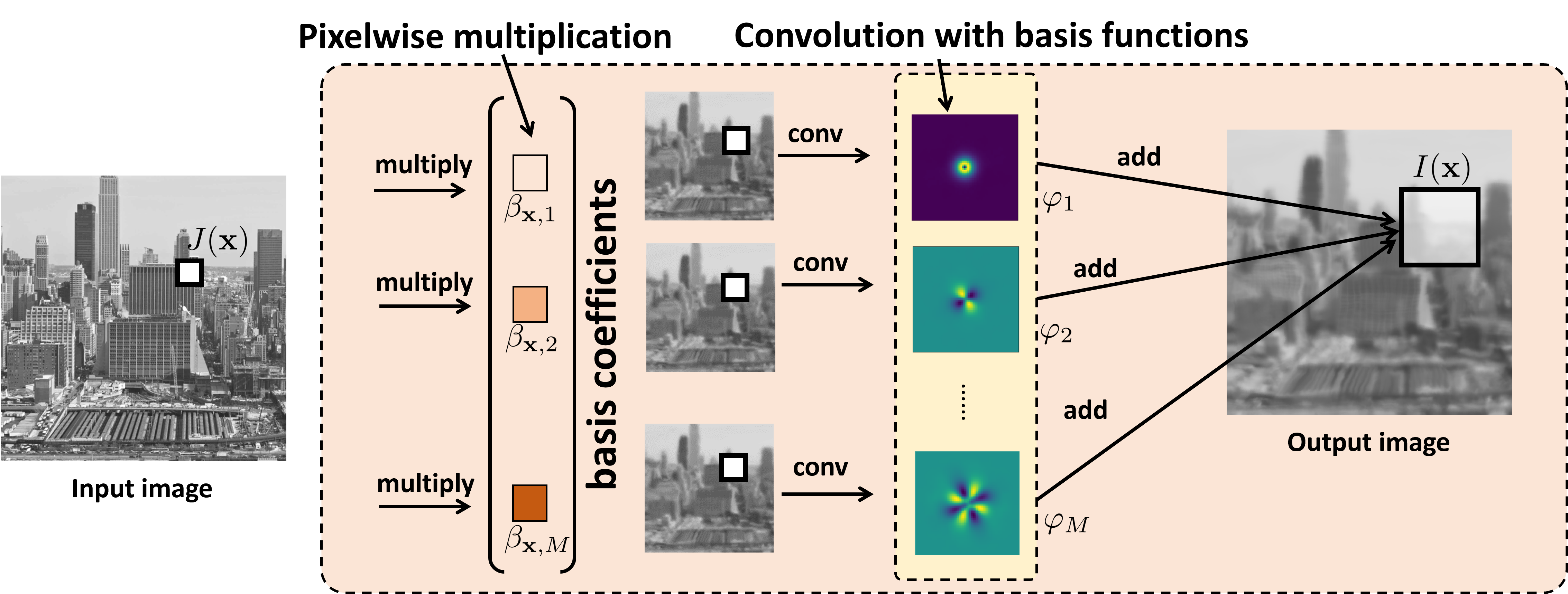}
\caption{A block diagram that depicts how a pixel is multiplied by its coefficient vector and convolved with its corresponding basis functions. Their contribution is then accordingly distributed across the output image. Note that the ``receptive field'' is larger in the output image due to the interpretation of scattering.}
\label{fig: Ch3 conv}
\vspace{-2ex}
\end{figure}

Through \eqref{eq: I approx scat}, the convolutions within the summation are spatially invariant. Thus, we can multiply the image $I_g(\vx)$ by each random field $\beta_m$ and convolve with the associated basis functions $\varphi_m$. After this, summation gives us a spatially varying image, which we depict in \fref{fig: Ch3 conv}. The result is a reduction of the computation, now we only require $M$ 2D FFTs where in Mao et al. \cite{Mao_2021_a} $M$ was chosen to be 100.

This argument rests upon the assumption that \eqref{eq: Ch3 PSF basis} is possible. This leaves us with two fundamental questions in order to facilitate this approach: (1) How do we construct the PSF basis functions? (2) How do we determine the basis coefficients from the Zernike space?

\begin{figure}[h]
\centering
\includegraphics[width=0.9\linewidth]{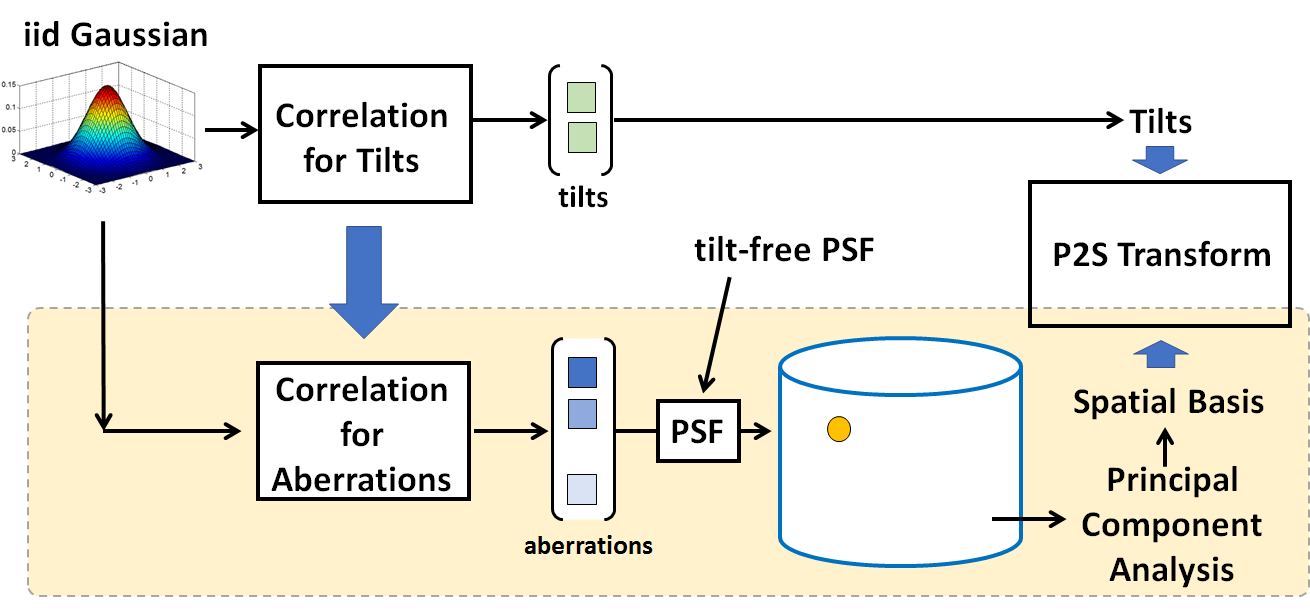}
\caption{To generate the basis functions $\{\varphi_{m}(\vx)\,|\,m=1,\ldots,M\}$, we first simulate a large set of PSFs $h_{\vu_1}(\vx),\ldots,h_{\vu_P}(\vx)$. By running the principal component analysis, we can determine the basis functions. Source: \cite{Mao_2021_a}}
\label{fig: Ch3 PCA}
\vspace{-2ex}
\end{figure}

For the PSF basis functions, we can construct them \emph{offline} via a \keyword{supervised learning} approach as shown in \fref{fig: Ch3 PCA}. Suppose that we have run the simulation offline to generate a large set of PSFs $\vert h_{\vu_1}(\vx) \vert^2,\ldots, \vert h_{\vu_P}(\vx) \vert^2$ where $P$ is a very large number. These PSFs can be drawn independently because the goal here is to learn the common structure of all of them. To ensure sufficient variability, when generating these PSFs, we can use a wide range of $C_n^2$ values so that the PSFs cover the turbulence conditions from weak to strong.

Given the simulated PSFs $\vert h_{\vu_1}(\vx) \vert^2,\ldots, \vert h_{\vu_P}(\vx) \vert^2$, we can perform a standard \keyword{principal component analysis} (PCA) to determine the basis functions $\{\varphi_{m}(\vx)\,|\,m=1,\ldots,M\}$. Implementations of the PCA are available in most of the modern computing libraries and so we skip it for brevity. However, one thing to note is that the PSFs are the results of both the \emph{tilts} and the \emph{high-order aberrations}. For tilts, using the PCA to encode the tilt is not cost-effective because the tilts are simple delta functions in the two-dimensional space. All they do is to point a pixel to the new location. For such a simple operation, we can just record them directly without wasting the principal components. Decoupling the tilts from the high-order aberrations also decouples the training data from the tilts. This will then reduce the number of training samples for the PCA. The idea is that for the tilts, we save them and record their statistics. For the high-order aberrations, we send the training samples to the PCA to learn the PSF structure.

\subsection{The Phase-to-Space Transform}
\index{phase-to-space transform}With the basis functions $\varphi_{m}(\vx)$ able to be found computationally, we now discuss the basis coefficients $\{\beta_{\vu,m} \,|\, m = 1,\ldots,M\}$. The definition of the basis coefficients is that
\begin{equation}
\beta_{\vu,m} = \langle \vert h_{\vu}(\vx) \vert^2,  \varphi_{m}(\vx) \rangle  = \int \vert h_{\vu}(\vx) \vert^2  \varphi_{m}(\vx) \, d\vx.
\label{eq: Ch3 beta}
\end{equation}
However, in the absence of the PSF $\vert h_{\vu}(\vx) \vert^2$, we need a mechanism to generate the coefficients $\beta_{\vu,m}$. If we knew the $\vert h_{\vu}(\vx) \vert^2$ then there is no problem to solve because the goal here is to generate the PSF!

The correspondence between the Zernike coefficients and a set of PSF basis coefficients has been considered in the form of the extended Nijboer-Zernike diffraction theory\index{Nijboer-Zernike diffraction theory} \cite{Janssen_2002_a, vanHaver_2010_a}. This theory is rather elegant, however, our PCA basis functions $\varphi_m$ will be of no use if we adopt it. Furthermore, it is unclear whether or not the basis is optimal for turbulence. Therefore, in \cite{Mao_2021_a}, Mao et al. produced a learning-based approach to provide the mapping from $\{a_{\vu,n} \,|\, n = 1,\ldots,N\}$ to $\{\beta_{\vu,m} \,|\, m = 1,\ldots,M\}$. This network is called the \keyword{Phase-to-Space (P2S) Transform}. The name Phase-to-Space means that we are converting the phase coefficients to spatial PSF coefficients.

\begin{figure}[h]
\centering
\includegraphics[width=0.9\linewidth]{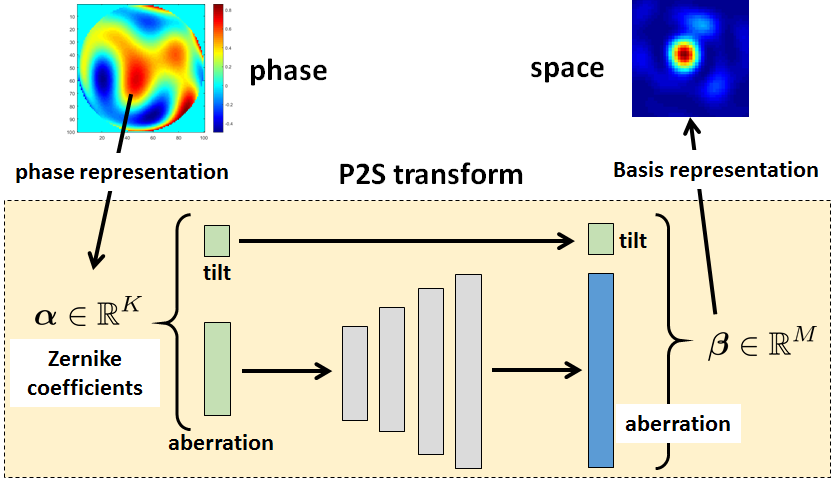}
\caption{A visualization of the Phase-to-Space (P2S) Transform network. A shallow neural network maps the higher-order Zernike coefficients to a set of PSF coefficients. Source: \cite{Mao_2021_a}.}
\label{fig: Ch3_p2s}
\vspace{-2ex}
\end{figure}

The P2S framework is shown in \fref{fig: Ch3_p2s}. The idea is train a \emph{shallow} neural network that maps the vector $\mathbf{a} = [a_{\vu,1},\ldots,a_{\vu,N}]^T$ to another vector $\vbeta = [\beta_{\vu,1},\ldots,\beta_{\vu,M}]^T$. The dimension of the vector $\mathbf{a}$ is typically in the order $N = 36$ whereas the dimension of the vector $\vbeta$ is typically around $M = 100$. Since the dimension is so small, a simple fully-connected neural network would suffice for learning the mapping.

How can we train such a transform?\index{phase-to-space transform! training of} If we had a large set of training data consisting of pairs of Zernike coefficient vectors and PSF basis coefficient vectors, we could simply train a network to find the mapping. Here we drop the subscript of position as we are interested in realizations that are completely independent. Each training sample will consist of the pair of vectors $(\mathbf{a},\vbeta)$, with the Zernike basis coefficient vector $\mathbf{a}$ as the input and the PSF basis coefficient vector $\vbeta$ as the output. To prepare the vectors $\vbeta$, we first construct the PSF $\vert h_{\vu}(\vx) \vert^2$ associated with the vector $\mathbf{a}$. Then, we use \eref{eq: Ch3 beta} to compute $\vbeta$. The entire process can be done offline. When the training samples are prepared, we train the shallow neural network until the convergence criteria are met. The resulting neural network will then be able to perform the mapping $\mathbf{a}\to\vbeta$. Concerning the exact network architecture, Table~\ref{tab: network} shows the network reported in \cite{Mao_2021_a}.

\begin{table}[h]
\centering
\begin{tabular}{cccc}
\hline\hline
input layer     & layer 1 & layer 2 & output layer \\
\hline
     & full conn. & full conn. & full conn. \\
34   & 34  & 100  & 100\\
\hline
\end{tabular}
\vspace{1ex}
\caption{Network architecture of the P2S transform network.}
\label{tab: network}
\end{table}

The advantage of using a neural network for the Phase-to-Space transform is that it can be parallelized. For a high-resolution image with many pixels, the parallelism enables us to generate the $\vbeta$'s simultaneously across all pixels in the image in \emph{a single pass}. Such a speed up can be significant for large images.

\subsection{Summary}
\begin{figure}[h]
\centering
\vspace{-2ex}
\includegraphics[width=0.95\linewidth]{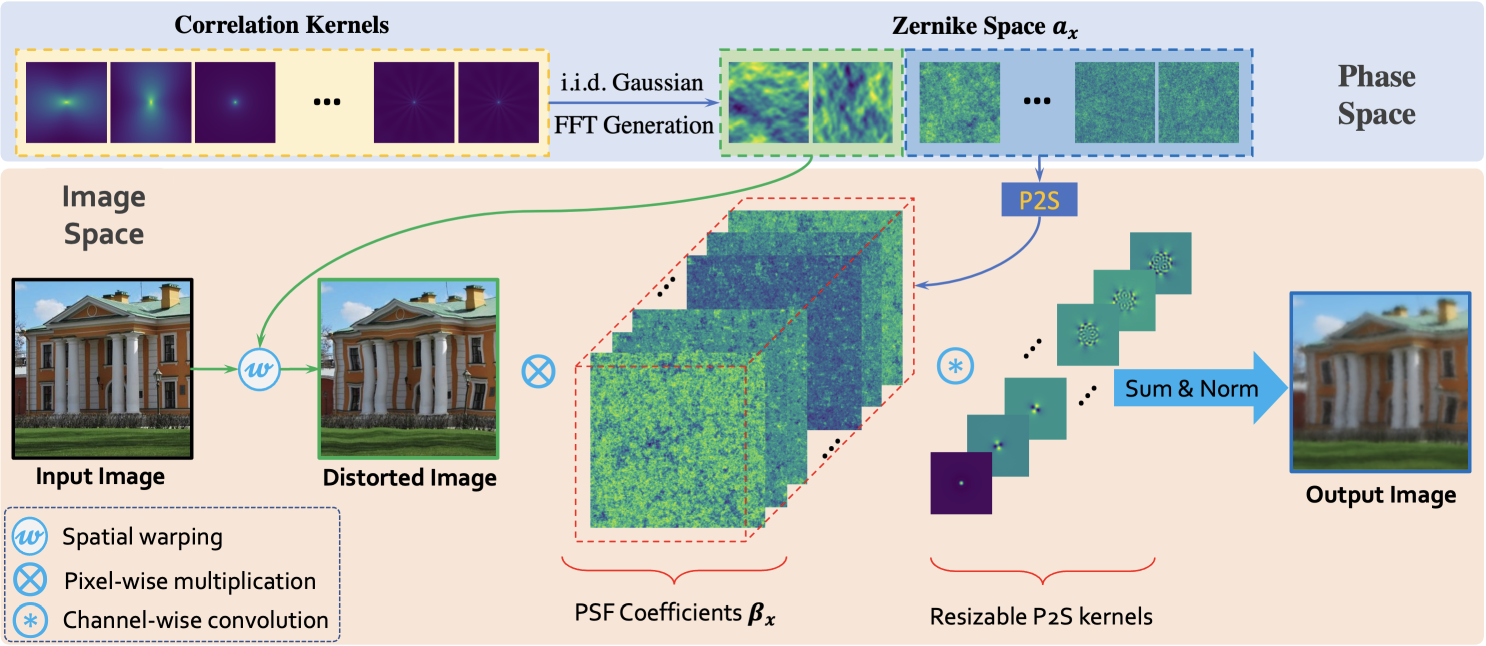}
\caption{An overview of Zernike-based simulation as described in the series of papers \cite{Chimitt_2020_a, Mao_2021_a, Chimitt_2022_a}.}
\label{fig: Ch3 zernike simulation}
\vspace{-2ex}
\end{figure}

\begin{figure}
    \centering
    \begin{tabular}{cc}
    \includegraphics[width=0.45\linewidth]{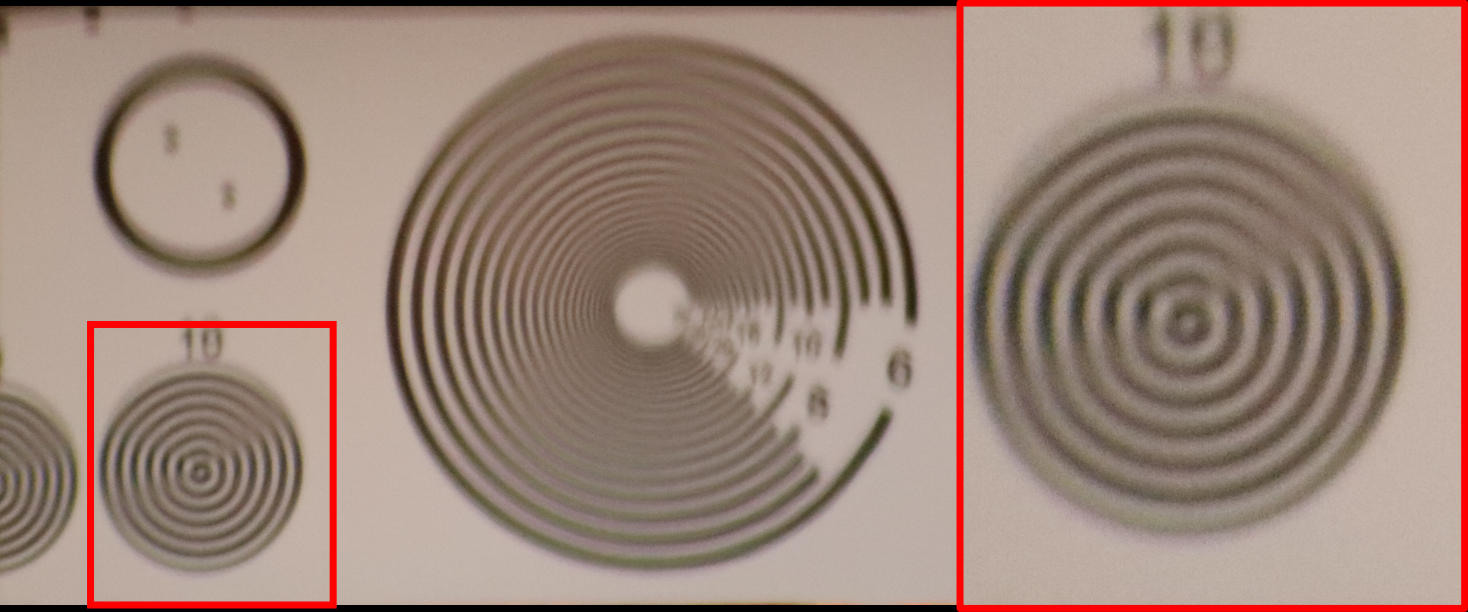} &
    \includegraphics[width=0.4\linewidth]{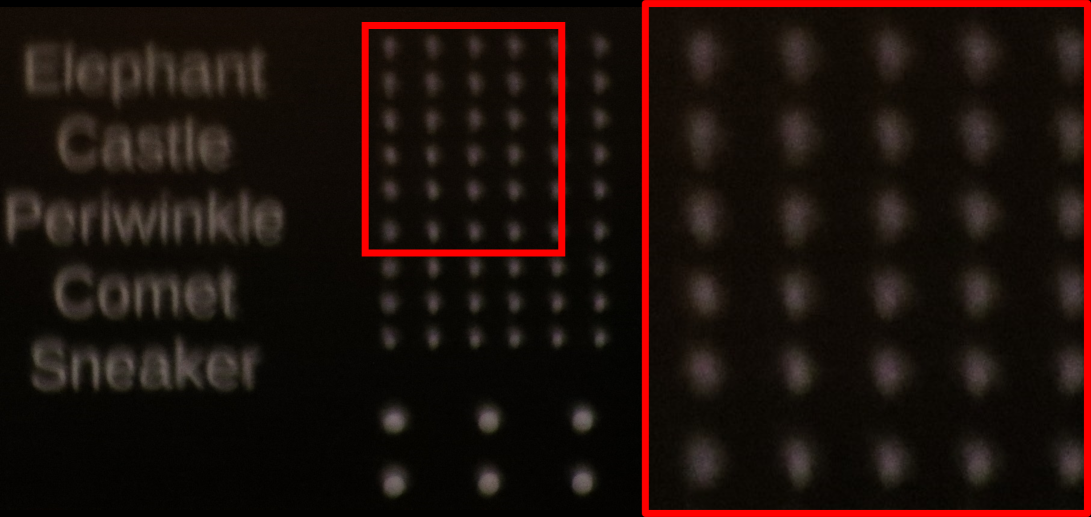} \\
    \includegraphics[width=0.45\linewidth]{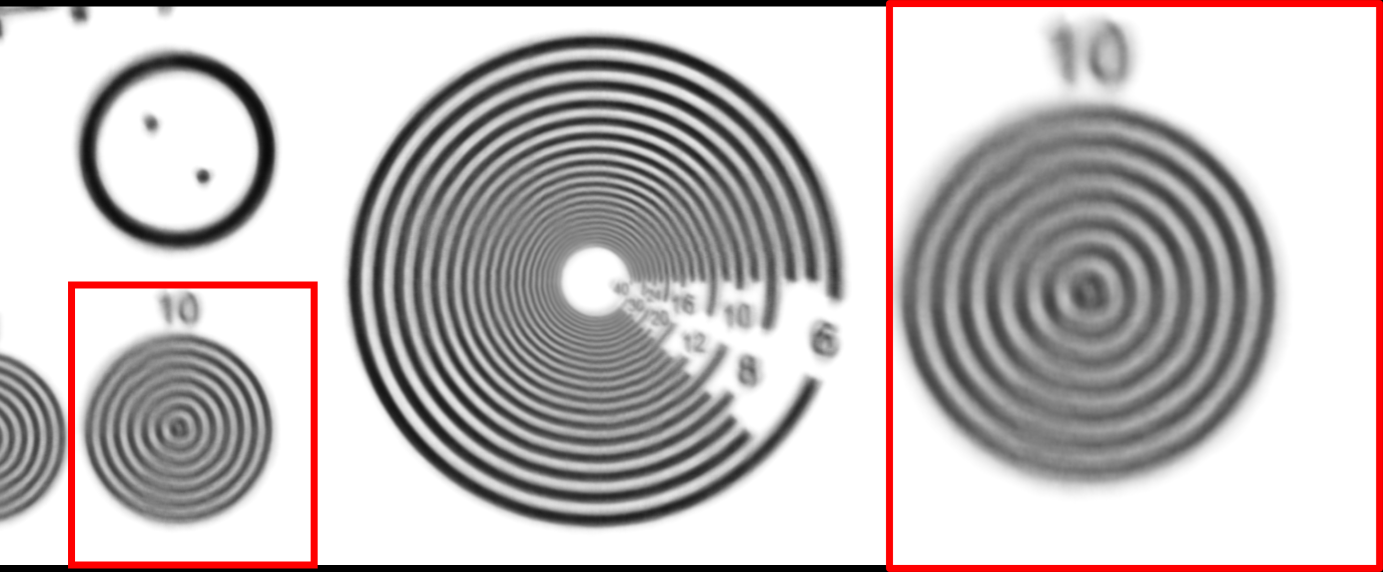} &
    \includegraphics[width=0.4\linewidth]{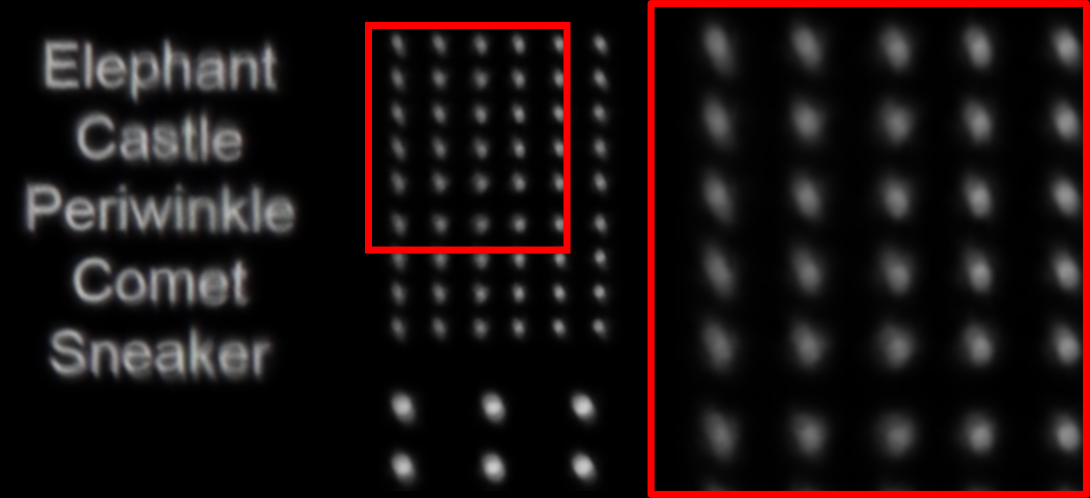} \\
    \includegraphics[width=0.45\linewidth]{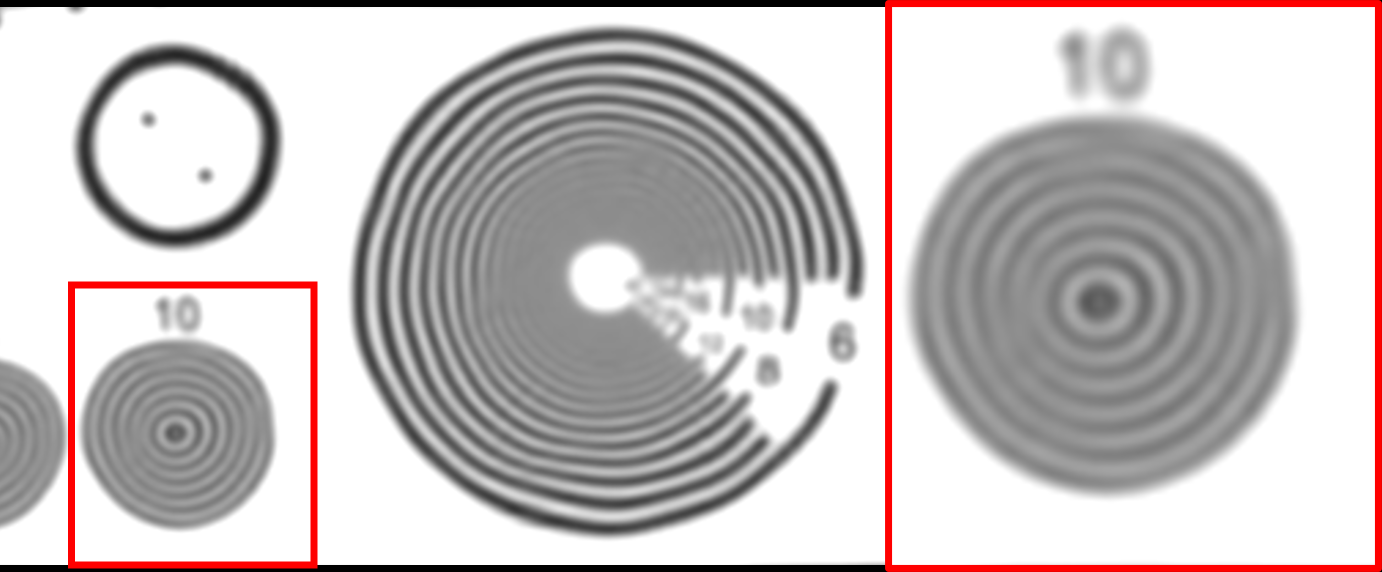} &
    \includegraphics[width=0.4\linewidth]{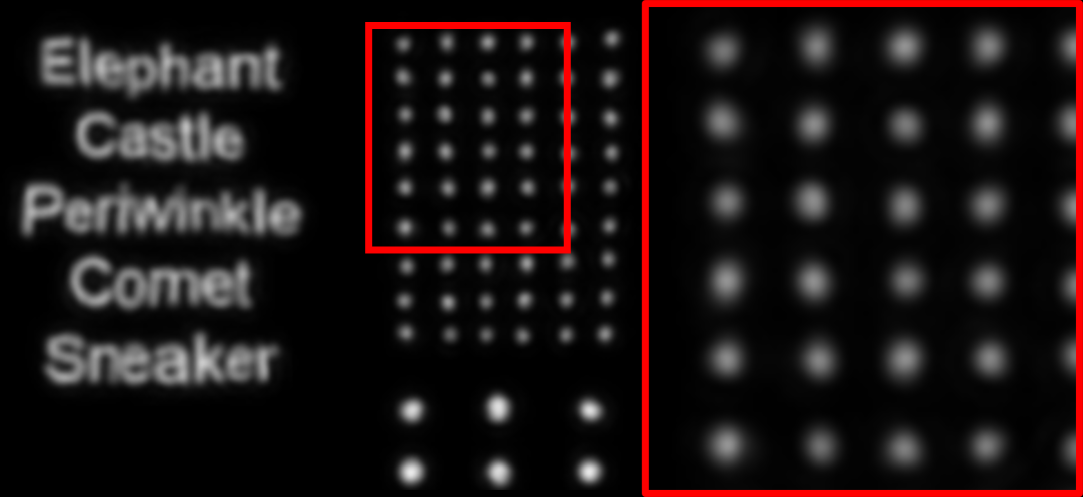}
    \end{tabular}
    \caption{A comparison of [Top] Ground truth data [Middle] Zernike-based simulation and [Bottom] Split-step simulation. Source: \cite{Chimitt_2022_a}.}
    \label{fig: visual_comparisons}
\end{figure}

At this point, we have described various aspects of Zernike-based simulation, now we wish to put it all together. At a high level, Zernike-based simulation follows the pipeline of generating Zernike space samples and applying them to an image via the P2S transform (also utilizing \eqref{eq: I approx scat}).  A depiction of Zernike-based simulation is presented in \fref{fig: Ch3 zernike simulation}. Let us spend some effort in writing the recipe down to accompany this depiction which should also help to unify the topics we've discussed:\index{Zernike-based simulation! short summary}
\begin{enumerate}
	\item \textbf{Zernike space sampling.} Using the approximation of Chimitt et al. \cite{Chimitt_2022_a}, we can generate realizations of the Zernike space quickly. To do so, we must have precomputed either of the correlation kernels \eqref{eq: chim_chan_main_result_1} or \eqref{eq: main_result_continuous2}. It is possible to compute these correlation kernels once and sample them as required -- thus the integration need only be performed computationally one time. Thus, the correlation kernels shown are sampled from the precomputed integrals and used to generate the realization (through FFT-based random signal generation) of the Zernike space $\mathbf{a}_\vu$.
	\item \textbf{Image warping.} With a random realization of the Zernike space, we can first distort the input image via warping. To do this, we take the tilt Zernike coefficients and send them to a warping module which distorts the image accordingly. It is important to note that one must convert from the Zernike coefficient values to pixels (which requires knowledge of the sample spacing).
	\item \textbf{P2S transform.} The P2S network then receives the higher order Zernike coefficients and converts them to $\vbeta_\vu$. This represents the $\beta_m$s in \eqref{eq: I approx scat} that we will multiply the warped image by. Note that this step saves us a great deal of time -- we don't have to take an FFT to form the PSFs \emph{and} we avoid a direct implementation of spatially variant convolution, instead opting for an approximation.
	\item \textbf{Spatially variant convolution.} Finally, we wish to form the output image by the usage of the approximation to spatially variant convolution. The warped image is multiplied by each ``plane'' of coefficients $\vbeta_\vu$ and is convolved by its corresponding basis function $\varphi_m$ which we refer to as the P2S kernels. From here, we can sum and normalize them, forming the final output image.
\end{enumerate}
We compare the results by Zernike-based simulation with split-step simulation along with simulation times in \fref{fig: visual_comparisons}. Although the results seem somewhat similar, the Zernike-based simulation finishes in under 1 second whereas a full dense-field split-step simulation (i.e. one complete split-step propagation per pixel) would take approximately 20 minutes.

\begin{figure}
    \centering
    \includegraphics[width=0.8\linewidth]{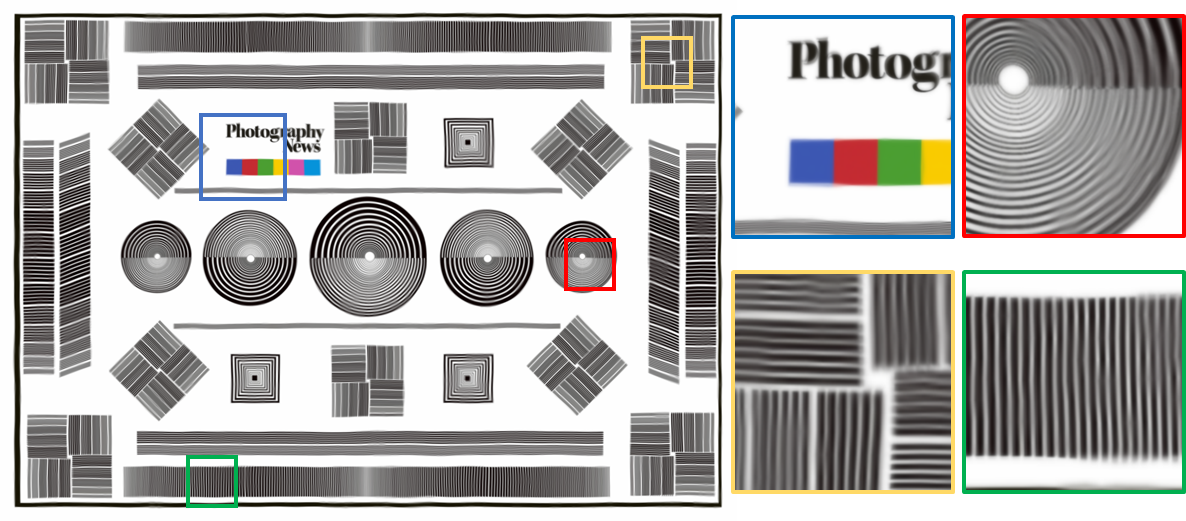}
    \caption{A 4k image simulated using the Zernike-based approach (here at a slightly lower resolution to save on PDF rendering time). Source: \cite{Chimitt_2022_a}.}
    \label{fig: 4k}
\end{figure}

To highlight the gain in speed offered by Zernike-based simulation, this process can be done for a 4k image in 1 minute \cite{Chimitt_2022_a} with a basis vector for \emph{every} pixel. While this may seem slow to some (one minute for a single image!) using split-step with 10 phase screens in the same fashion for a 4k image (that is, using a propagation for every pixel in the image) would take approximately 13 hours! We show a 4k image simulated in this fashion in \fref{fig: 4k} (at a lower resolution here to save on PDF rendering time).

To mirror the presentation of split-step, we also provide a pseudo-code version of Zernike-based simulation as follows:
\begin{algorithm}\index{Zernike-based simulation! pseudo-code}
	\caption{Zernike-based simulation}
	\begin{algorithmic}[1]
		\State Determine the size of each Zernike field (equivalent to the size of an image in pixels)
		\State Generate $M$ independent Zernike fields
        \State Mix along index-dimension according to the Noll matrix
        \State Apply warping transformation to input image $I_g$
        \State Perform P2S transform
		\State Multiple $I_g$ by the various basis coefficient fields, producing $\beta_m I_g$ for $m \in 1,\ldots, M$.
        \State Mix according to scattering convolution approximation
	\end{algorithmic}
\end{algorithm}

We note that the spatially varying convolution approximation step may be utilized in split-step, however, the phase realization $\phi_\vu$ would need to be decomposed by the Zernike polynomials, thus requiring an additional step of inner products. However, there would still likely be a gain in speed due to the vast advantage of Fourier-based convolution. With the main recipe of this variety of simulation presented, we turn to a few additional considerations we wish to highlight.

\section{Additional Topics}
To not distract us from the critical details, we have avoided a few additional considerations. To close the loop on these, we present them now to be applied in hindsight. These do not dramatically impact the understanding of the model or simulation, however, they are important considerations when doing simulation. We additionally present a real-time application of this variety of simulation.

\subsection{Wavelength}
In principle, the spectral response of the turbulent medium is wavelength dependent, and the distortion must be simulated for a dense set of wavelengths. However, as we discussed in the beginning of \cref{sec: sec2}, the index of refraction can be decoupled into the static part $n_0(\vr,\lambda)$ and the varying part $n_1(\vr,t)$. The static part $n_0(\vr,\lambda)$ has some dependencies on the wavelength but it does not contribute to the turbulence distortion. It is mostly the refractive index in the ambient medium. The varying part $n_1(\vr,t)$ changes with respect to the location $\vr$ and time $t$. For the typical visible spectrum (roughly 400nm to 700nm), the influence due to the wavelength is insignificant compared to the time and location. Therefore, it is sometimes justifiable to use one wavelength for all three color channels (RGB) when constructing the PSFs.

\begin{figure}[ht]
	\centering
	\begin{tabular}{c c c}
\multicolumn{3}{c}{\includegraphics[width=0.9\linewidth]{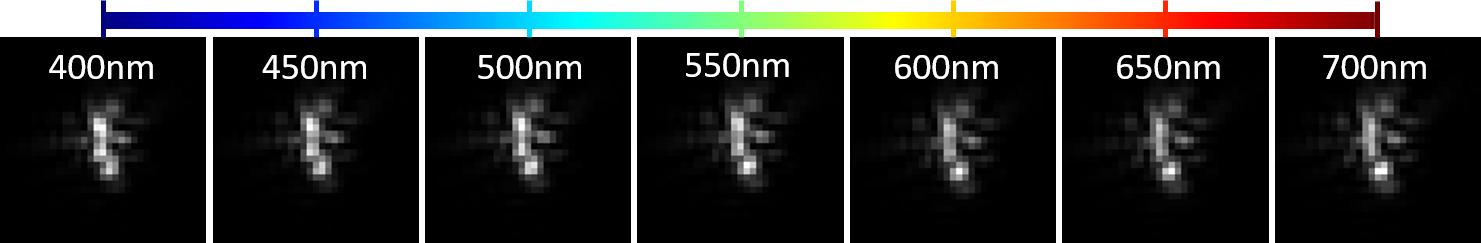}}\\
\multicolumn{3}{c}{(a)}\\
\includegraphics[width=0.3\linewidth]{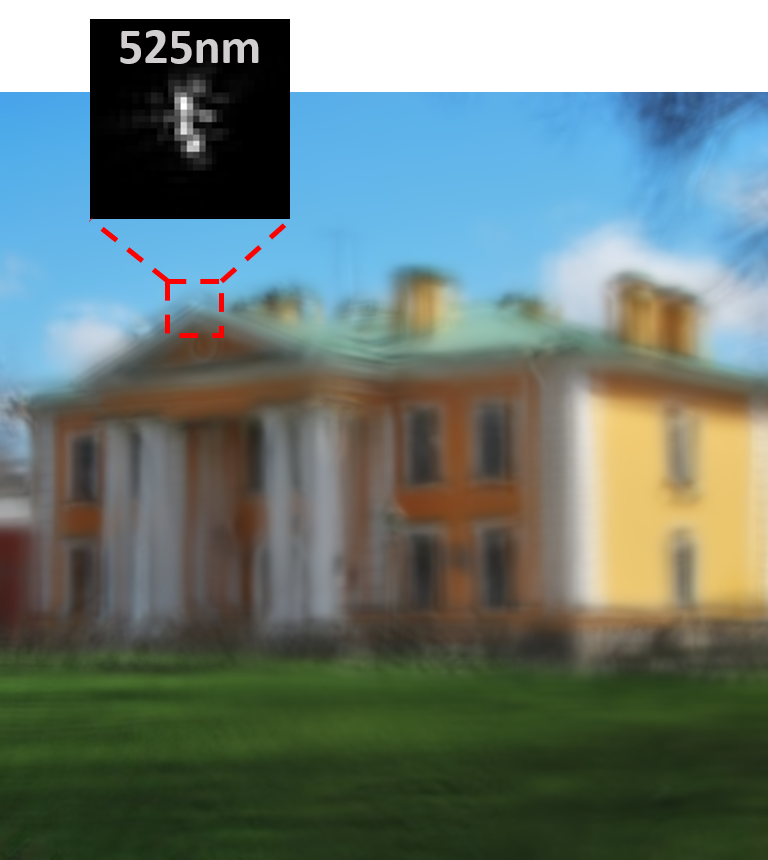}&
\hspace{-2ex}\includegraphics[width=0.3\linewidth]{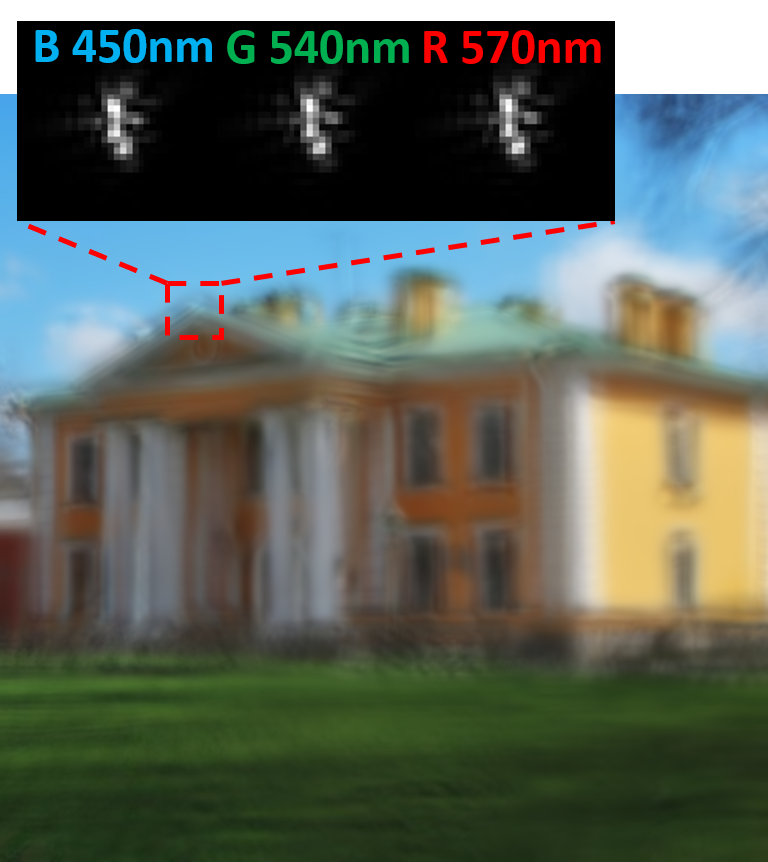}&
\hspace{-2ex}\includegraphics[width=0.3\linewidth]{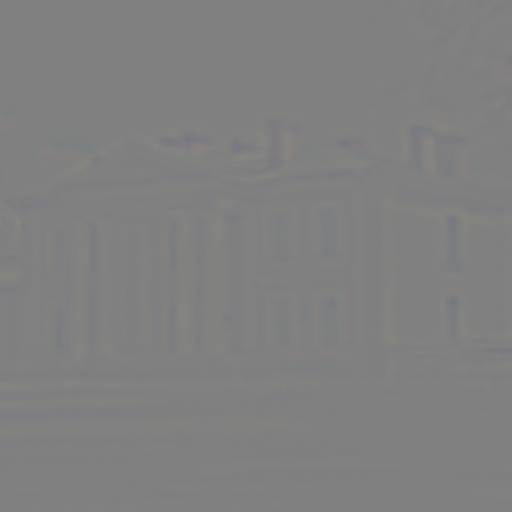}\\
(b)  & \hspace{-2ex}(c)  & \hspace{-2ex}(d)
	\end{tabular}
	\caption{(a) PSFs across the visible spectrum. (b) The same distortion applied to three channels using the center wavelength of the visible spectrum. (c) Wavelength-dependent distortions applied to three channels. (d) Difference map between (b) and (d). Source: \cite{Mao_2021_a}.}
	\label{fig: Ch3color}
\end{figure}

To illustrate the situation, Mao et al. \cite{Mao_2021_a} showed an example (re-shown in \fref{fig: Ch3color}) of the individual PSFs for several wavelength from 400nm (blue) to 700nm (red). By observing the figure, we notice that the shape of the PSFs barely changes from one wavelength to another. In the same figure, we simulate two color images. The first image is simulated by using a single PSF (525nm) for the color channels (and displayed as an RGB image). The second image is simulated by considering 3 PSFs with wavelengths 450nm, 540nm, and 670nm. We note that (c) is a more realistic simulation but requires $3\times$ computation. However, the similar PSFs across the color make the difference visually indistinguishable, as seen in (d).

The small gap demonstrated in \fref{fig: Ch3color} suggests that we can simulate the RGB channels identically. In Mao et al. \cite{Mao_2021_a}, the authors acknowledge that this approximation does not hold in the worst-case scenarios when the optical system demonstrates substantial chromatic aberrations. 
This in itself is not a \emph{proof} that regardless of the situation one is simulating, one is justified to simulate one wavelength and apply it to an image as if it were valid for all wavelengths. We do not intend the previous discussion as a suggestion that this applies in all cases \emph{or} that it is even a proof for the particular case considered. It is always important to consider whether or not this approximation is valid for the situation one is modeling. We refer the reader to Hardie et al. \cite{Hardie_2022_a} for a detailed discussion of multi-spectral simulation, which addresses this problem in the context of split-step simulation.

\subsection{Elements of Realistic Simulation}
With publicly available code for the simulator by Mao et al. \cite{Mao_2021_a}, it seems natural to download it, vary the parameters, and receive a large amount of perfectly accurate training data. We emphasize that, unfortunately, this is \emph{not} the case.

The careful application of an imaging simulation is not as simple as it may appear. Consider an image $I_g$ passed to the simulator. Firstly, it must be noted that \emph{no} image, in reality, is truly equal to $I_g$. A perfectly in-focus image, however, may serve as a suitable approximation. A more significant difficulty arises when we consider the sampling of $I_g$, which is potentially unknown. Our image formation equation depends on knowing the spatial parameters so that the PSF may be properly applied.

To elaborate further, let us consider an image that is chosen out of a dataset. Applying the simulator naively will almost certainly result in a mismatch between the generated PSFs and the image. In other words, our $\vert h_\vu \vert^2$ will be either too big or too small. How do we account for this? We must know or assume the sampling of the image to be simulated as it will impact the perceived strength of turbulence, correlation distance, pixel shifting, and so on. It cannot be ignored.

As a closing note, we find it to be permissible to perform this ``naive'' application of a simulator to a dataset if the parameters are chosen somewhat reasonably. This will allow one to train a network for some downstream task which, in our experience, can be used on real data reasonably well. However, to say that any image out of the dataset is representative of turbulence for the chosen parameters is simply wrong \emph{unless} the user took the care of ensuring the sampling measurements were correct, justifying any assumption of wavelength-invariance, and so on. As a rule of thumb, naive application to a dataset is acceptable if a downstream task stands to benefit from it; it is not acceptable in the case of any analysis of the nature of turbulence.

\subsection{Real-Time Turbulence Generation}
\begin{figure}[ht]
    \centering
    \includegraphics[width=\linewidth]{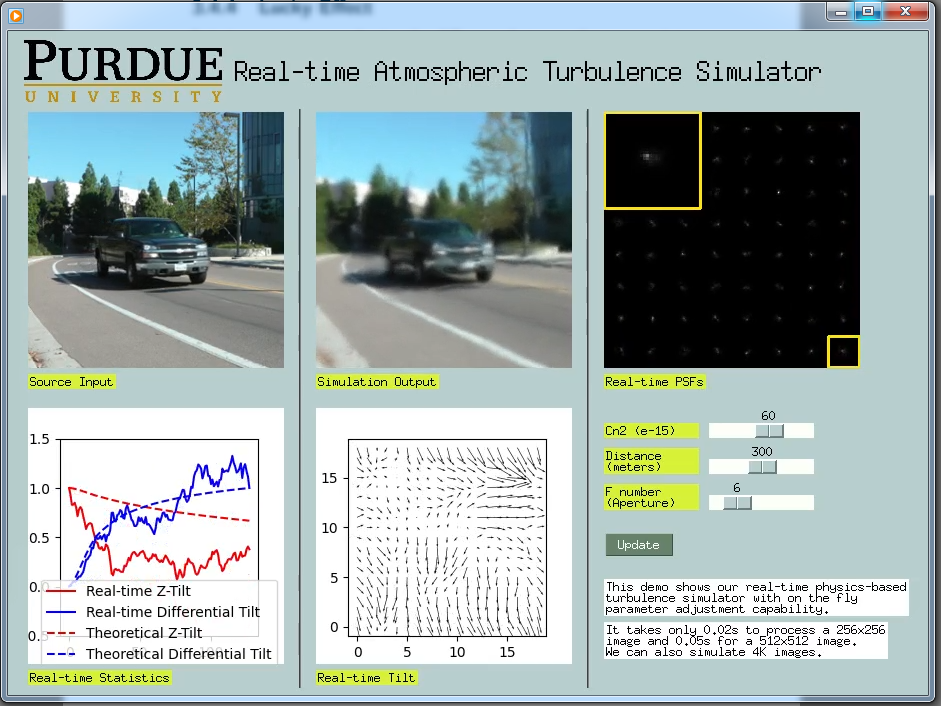}
    \caption{The GUI for turbulence simulation developed by Purdue University, with tunable knobs for turbulence parameters and real-time turbulent realization graphics.}
    \label{fig: Ch3 GUI}
\end{figure}

We conclude the Chapter with a graphical user interface (GUI) for real-time turbulence simulation. \fref{fig: Ch3 GUI} shows a GUI built by Purdue University. In this GUI, we can feed the simulator with a camera source such as a webcam. The resolution of the camera feed for this particular GUI is limited to $512 \times 512$. The simulation is run on a host machine with an nVidia RTX 3080Ti GPU. The throughput of the simulator is approximately 7 frames per second.

As we can see in the demonstration, the GUI will generate in real time the turbulence-distorted image. Simultaneously, it displays the real-time point spread functions as well as the real-time tilt map. The subfigures on the bottom left are the real-time tilt statistics, including the Z-tilt and the D-Tilt. Of course, they don't match as they are individual realizations, on expectation they do indeed match!

\section{Summary}
\label{sec: sec3_7}
This Chapter has described a more computer vision and computational imaging view of the simulation and modeling of the phase distortions caused by atmospheric turbulence. The ideas discussed frame the image formation process as a sampling problem, with the Zernike basis coefficients forming a field from which the image is formed.

\keyword{Part 1 Split-step alternatives}: We began by introducing the problems which arise from split-step when we wish to use it for generating training data. This led us to survey the existing alternatives to split-step, ranging from ray tracing to GANs. This brought us to introduce phase-based simulation to which the described Zernike-based simulation modality belongs.

\keyword{Part 2 The phase as a basis}: The phase-over-aperture model and Zernike representation introduce an alternative way of viewing the process of modeling the turbulent image formation process. Specifically, the distortions may be viewed as arising from a basis coefficients field, which we refer to as the Zernike space $\{\mathbf{a}_\vu\}$.

\keyword{Part 3 The Zernike space}: The concept of the Zernike space and its application to simulation was introduced by the development of the multi-aperture simulation. Utilizing classical results from astronomical telescope correlations, the multi-aperture simulation extends these concepts to simulating spatially correlated ground-to-ground images.

\keyword{Part 4 Zernike-based simulation}: The P2S approach \cite{Mao_2021_a} combines the theoretical insight of the multi-aperture simulation with a forward transformation which is a neural network-based approach. This allows for the multi-aperture simulation to be applied to a broader class of problems, such as generating data for training machine learning restoration methods. 
\chapter{Image Restoration}
\vspace{-6ex}
\noindent\textcolor{myblue}{\rule{\textwidth}{4pt}}
\vspace{1ex}

\newcommand\nwidth{0.14}
\newcommand\nspace{-1}
\newcommand\nspacetwo{0}
\label{sec: sec4}

In long-range imaging, one of the biggest challenges is to recover distorted images. There are hardware and software solutions, but they each have pros and cons. Hardware solutions are often faster, but they are constrained by size, weight, and power. Software solutions are not as customized although they have a larger degree of freedom. In this Chapter, we will focus on the software (algorithm) solutions.

Developing image restoration algorithms for atmospheric turbulence is not an easy task. For the most part, the challenges are associated with the uncertainty and complexity of the forward model which is random. This is different from classical inverse problems such as \keyword{deconvolution} where the blur is unknown but structured and deterministic.

\subsection*{Notations}
Discussing the inverse problems requires some changes in notations. Since all inverse problems are solved on a computer, any continuous function has to be discretized. On the object plane, the continuous function $J(\vx)$ is now represented by a $N$-dimensional vector
\begin{equation*}
(\text{object plane}) \qquad \mJ = [J(\vx_1),J(\vx_2),\ldots,J(\vx_N)]^T \in \R^N,
\end{equation*}
where $\vx_1,\ldots,\vx_N$ are the $N$ coordinates the digital image is defined upon. Similarly, on the image plane, the function $I(\vx)$ is now represented as
\begin{equation*}
(\text{image plane}) \qquad \mI = [I(\vx_1),I(\vx_2),\ldots,I(\vx_N)]^T \in \R^N.
\end{equation*}
The discretized PSFs $h_\vu(\vx)$ will be written as
\begin{equation*}
(\text{image plane}) \qquad \vh_\vu = \left[ h_\vu(\vx_1), h_\vu(\vx_2),\ldots, h_\vu(\vx_N) \right]^T \in \R^N,
\end{equation*}
where we note for simplicity in notation we will opt to use $h_\vu(\vx_N)$ and $\vh_\vu$ for the PSFs in this Chapter.

The relationship between the clean image and the distorted image is given by a nonlinear mapping $\calH_{\vtheta}: \R^N \rightarrow \R^N$:
\begin{equation}
\mI = \calH_{\vtheta}(\mJ),
\end{equation}
where $\vtheta$ denotes some underlying hyper-parameters. Putting this in the context of atmospheric turbulence, $\calH_{\vtheta}$ is the wave propagation equations and $\vtheta$ represents the turbulence parameters. Or, if we use the Zernike space model, then $\vtheta$ would be the Zernike coefficients defined by the atmospheric turbulence. In the simplest case where $\calH_{\vtheta}$ represents an isoplanatic turbulence so that the blur is spatially invariant, the equation can be simplified to $\mI = \vh \circledast \mJ$ using a blur kernel $\vh$ and the convolution operator $\circledast$. If this happens, it resembles the classical deconvolution problem.

In non-local image processing algorithms, we sometimes need to process \emph{patches} instead of the whole image. An image patch is a small 2D neighborhood around a certain pixel. Suppose we are looking at coordinate $\vx_i$, we define a $p$-dimensional patch as
\begin{equation*}
(\text{patch at pixel $\vx_i$}) \qquad\qquad \underline{\mI}(\vx_i) = [I(\vx_i+\Delta \vx_1), \ldots, I(\vx_i+\Delta \vx_p)]^T,
\end{equation*}
where $\Delta \vx_1,\ldots,\Delta \vx_p$ define a neighborhood of $p$ pixels surrounding $\vx_i$. The neighborhood is typically square.

If we need to extend the definitions to videos instead of a single image, we can associate the time axis to either the image or a patch. Using the image as an illustration, we say that we are looking at a frame located at time $t$ as
\begin{equation*}
(\text{image at time $t$}) \qquad \mI(t) = [I(\vx_1,t),I(\vx_2,t),\ldots,I(\vx_N,t)]^T.
\end{equation*}
To specify a patch located at pixel $\vx$ and time $t$, we write $\underline{\mI}(\vx,t)$.

\subsection*{Inverse Problems}\index{inverse problems}
As we shall elaborate in detail later in this Chapter, the inverse problem associated with atmospheric turbulence is often formulated as
\begin{equation}
\widehat{\mJ} = \argmin{\mJ} \; \|\mI - \calH_{\vtheta}(\mJ)\|^2 + \lambda\; g(\mJ),
\label{eq: ch4 inverse problem main}
\end{equation}
where $\mI$ is the observed distorted image, $\mJ$ is the latent clean image, $\calH_{\vtheta}$ is the turbulence forward degradation model, and $g(\mJ)$ is a \keyword{regularization function}\index{regularization function} constraining the search space of the solution. This least-squares fitting problem is \emph{not} necessarily the only way to formulate the problem. In fact, many deep learning algorithms would directly aim to recover $\mJ$ without even worrying about if we are using the least-square function or which regularization function $g$ to use. The purpose of introducing \eref{eq: ch4 inverse problem main} is to help readers who are familiar with the classical optimization-based inverse problems (such as total variation) to appreciate the connection. Even for readers coming from a deep learning perspective, we often find \eref{eq: ch4 inverse problem main} useful in terms of elaborating insights.

The difficulty of \eref{eq: ch4 inverse problem main} is when $\calH_{\vtheta}$ is complicated, e.g., $\calH_{\vtheta}$ is the turbulence model of which the instantaneous random realization is not known. For this problem, we need to simultaneously estimate $\calH_{\vtheta}$ while recovering $\mJ$. This is a strongly ill-posed optimization because the number of parameters needed to describe $\calH_{\vtheta}$ can be even larger than the number of pixels in $\mJ$. Thus, unless the structures of the image content and turbulence statistics are properly utilized, recovering $\mJ$ from the above optimization can be extremely challenging if not impossible.

\subsection*{Plan for this Chapter}
Our goal in this Chapter is to highlight the key principles behind atmospheric turbulence restoration algorithms. We acknowledge the exponential growth of papers in the field over the past three years thanks to the promising results of deep learning. However, given the sheer volume of the algorithms, instead of listing every single method, we believe it is more meaningful to focus on a few main concepts.

\section{Understanding the Forward Model}
\label{sec: sec4_1}
Since the core difficulty in solving the inverse problem is the complexity of the forward model, in this Chapter, we take a closer look at the operator that defines the turbulence forward model. For simplicity of the discussion, we shall focus on the spatial domain.

\subsection{Turbulence Operator $\calH_{\theta}$}\index{turbulence operator}
The distortion caused by turbulence is spatially varying. Therefore, any convolution that defines the blur needs to be spatially varying too. Let's start by considering the spatially varying convolution. If the ground truth clean image intensity (at an input pixel coordinate $\vu$) is $J(\vu)$ and the spatially varying point spread function (PSF) is defined by $h_{\vu}(\vx)$ (where $\vu$ specifies the source location), the observed image intensity (at an output coordinate $\vx$), $I(\vx)$, will be defined according to
\begin{align}
I(\vx)
&= \int h_{\vu}(\vx - \vu) J(\vu) \; d\vu
\label{eq: ch4 image formation}
\end{align}
where $\circledast$ denotes convolution.

Although the model in \eref{eq: ch4 image formation} is spatially varying, it is linear due to the linearity of convolution. This can be shown by noting that if there are two clean images $J_1(\vx)$ and $J_2(\vx)$ (correspondingly two distorted images $I_1(\vx)$ and $I_2(\vx)$), then for any constants $a$ and $b$, it holds that
\begin{align*}
I(\vx)
&= \int h_{\vu}(\vx - \vu) \Big(a J_1(\vu) + b J_2(\vu)  \Big) \; d\vu \\
&= a \underset{I_1(\vx)}{\underbrace{\int h_{\vu}(\vx - \vu) J_1(\vu) \; d\vu }} + b \underset{I_2(\vx)}{\underbrace{\int h_{\vu}(\vx - \vu) J_2(\vu) \; d\vu}}.
\end{align*}
Therefore, we have shown the linearity of the operator.

\begin{figure}[h]
\centering
\includegraphics[width=0.8\linewidth]{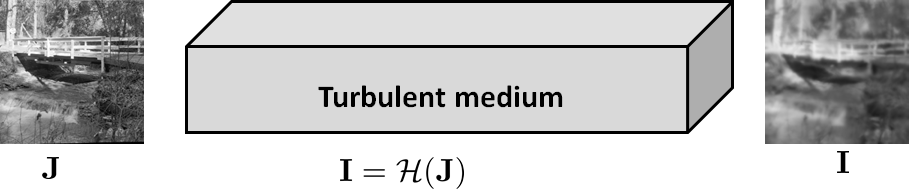}
\caption{The forward model of the atmospheric turbulence can be written as a general operator $\calH_{\vtheta}$ that takes a clean image $\mJ$ and maps it to a distorted image $\mI$.}
\label{fig: ch4_forward_model}
\end{figure}

In matrix and vector form, the convolution equation for each of the coordinates $\vx_1,\ldots,\vx_N$ can be expressed in terms of a linear combination of the input pixels
\begin{equation}
I(\vx_i) = \sum_{j=1}^N h_{\vu_j}(\vx_i - \vu_j) J(\vu_j), \quad i = 1,\ldots,N,
\label{eq: I = h J}
\end{equation}
where $h_{\vu_j}(\vx_i)$ denotes the $\vx_i$th pixel of the point spread function from pixel $\vu_j$, and $I(\vx_i)$ and $J(\vu_j)$ are the pixel values of the observed and the clean images, respectively. This leads us to write
\begin{equation}
\mI = \calH_{\vtheta}(\mJ),
\end{equation}
where the operator $\calH_{\vtheta}$ captures the convolutions imparted by the spatially varying point spread functions $h_{\vu_1},\ldots,h_{\vu_N}$. If we consider all the $N$ pixels in \eref{eq: I = h J}, $\calH_{\vtheta}$ takes on a familiar form:
\begin{equation}
\mI
=
\underset{\calH_{\vtheta}}{\underbrace{\begin{bmatrix}
h_{\vu_1}(\vx_1-\vu_1) & h_{\vu_2}(\vx_1-\vu_2) & \ldots & h_{\vu_N}(\vx_1-\vu_N)\\
h_{\vu_1}(\vx_2-\vu_1) & h_{\vu_2}(\vx_2-\vu_2)  & \ldots & h_{\vu_N}(\vx_2-\vu_N)\\
\vdots           & \vdots           & \ddots & \vdots \\
h_{\vu_1}(\vx_N-\vu_1) & h_{\vu_2}(\vx_N-\vu_2) & \ldots & h_{\vu_N}(\vx_N-\vu_N)
\end{bmatrix}}}
\mJ.
\end{equation}
A pictorial illustration is shown in \fref{fig: ch4_forward_model}.

As a special case, if the point spread function is spatially invariant, we have $h_{\vu_j}(\vx_i - \vu_j) = h(\vx_i - \vu_j)$. This will recover the familiar spatially invariant convolution we see in the image deconvolution literature, i.e.,
\begin{equation}
I(\vx_i) = \sum_{j=1}^N h(\vx_i - \vu_j) J(\vu_j), \quad i = 1,\ldots,N.
\end{equation}
The resulting convolution matrix is then a \emph{circulant} matrix that is diagonalizable by the discrete Fourier transform matrix.

Inspecting the operator $\calH_{\vtheta}$, we recognize that any \emph{inverse problem} has to involve some kind of inverse mapping $\calH_{\vtheta}^{-1}$. However, directly inverting $\calH_{\vtheta}$ can be difficult because $\calH_{\vtheta}$ consists of not only blurs but also tilts. The question we want to address here is how to decouple the blur and tilt as we model the turbulence.

\subsection{Decoupling Tilt and Blur}
There are two options we can consider when decoupling $\calH_{\vtheta}$. The first one is the \keyword{blur-then-tilt} model, where we first blur the image using spatially varying blurs and then add tilts to the blurred image:
\begin{align*}
\mI = \calH_{\vtheta}(\mJ) &= [\calT \circ \calB](\mJ) = \calT(\calB(\mJ)).
\end{align*}
Here, ``$\circ$'' denotes the function composition of two operators.

The second one is the \keyword{tilt-then-blur}\index{tilt-then-blur model} model, where we first add tilts to the image pixels and then blur the image using spatially varying blurs:
\begin{align*}
\mI = \calH_{\vtheta}(\mJ) &= [\calB \circ \calT](\mJ) = \calB(\calT(\mJ)).
\end{align*}
The difference between the \keyword{tilt operator} $\calT$ and the \keyword{blur operator} $\calB$ is the Zernike modes we use to construct the phase distortion:
\begin{itemize}
\item Tilt $\calT$: The phase is distorted by the first two Zernike coefficients via
\begin{equation*}
\phi_{\vu}(\vrho) = \sum_{m=2}^3 a_{\vu,m}Z_m(\vrho).
\end{equation*}
Recall that the first two Zernike modes (ignoring the DC term) represent the horizontal and the vertical displacements, i.e., $Z_2(\vrho) = 2\rho_x$ and $Z_3(\vrho) = 2\rho_y$ where $\vrho = [\rho_x,\rho_y]^T$ is the coordinate. Thus, if we define $\valpha_{\vu} = [\lambda a_{\vu,2} / R, \lambda a_{\vu,3} / R ]^T$, then $\phi_{\vu}(\vrho)$ can be written as $\phi_{\vu}(\vrho) = \valpha_{\vu}^T\vrho$.
\item Blur $\calB$: The phase is distorted by all Zernike coefficients except the first two:
\begin{equation*}
\phi_{\vu}(\vrho) = \sum_{m=4}^\infty a_{\vu,m}Z_m(\vrho).
\end{equation*}
If we subtract the overall phase from the tilt, we can obtain the following decomposition:
\begin{align*}
\phi_{\vu}(\vrho) = \underset{\valpha_{\vu}^T\vrho}{\underbrace{\sum_{m=2}^3 a_{\vu,m}Z_m(\vrho)}}
            + \underset{\varphi_{\vu}(\vrho)}{\underbrace{\sum_{m=4}^\infty a_{\vu,m}Z_m(\vrho)}}
\end{align*}
In other words, the overall phase can be written as a linear term plus a tilt-free phase term.
\end{itemize}

Once the phase distortions are defined, the resulting point spread function is
\begin{equation}
h_{\vu}(\vx) = |\mathfrak{Fourier}( e^{-j \phi_{\vu}(\vrho)}) |^2.
\end{equation}
Since the Zernike coefficients used to define the phase distortion are random, the blur and tilt operators $\calB$ and $\calT$ are both spatially varying and random.

Inspecting the decomposition of the phase, we notice that the exponential functions have their respective Fourier transform pairs
\begin{align*}
e^{-j \phi_{\vu}(\vrho)} &=
\underset{\longleftrightarrow \; \delta(\vu-\valpha_{\vu})}{\underbrace{e^{-j \valpha_{\vu}^T \vrho}}} \; \cdot \;
\underset{\longleftrightarrow \; b_{\vx}(\vu)}{\underbrace{e^{-j \varphi_{\vu}(\vrho)}}},
\end{align*}
where $\delta(\vu-\valpha_{\vu})$ is the delta function representing the tilt and $b_{\vu}(\vx)$ is the blur. Since multiplication in the Fourier space corresponds to convolution in the spatial domain, the resulting PSF is
\begin{equation}
h_{\vu}(\vx) = \delta(\vx-\valpha_{\vu}) \circledast b_{\vu}(\vx) = b_{\vu}(\vx-\valpha_{\vu}).
\end{equation}
Therefore, the PSF is a shifted version of the blur.

Now, let's take a closer look at the tilt-then-blur and blur-then-tilt operations. In terms of matrix-vector notation, the operator $\calT$ can be written as a shifting matrix $\mT \in \R^{N \times N}$ with\index{tilt-then-blur model! tilt operator}
\begin{equation}
\mT =
\begin{bmatrix}
t_{\vu_1}(\vx_1) & t_{\vu_2}(\vx_1) & \ldots & t_{\vu_N}(\vx_1)\\
t_{\vu_1}(\vx_2) & t_{\vu_2}(\vx_2) & \ldots & t_{\vu_N}(\vx_2)\\
\vdots           & \vdots           & \ddots & \vdots \\
t_{\vu_1}(\vx_N) & t_{\vu_2}(\vx_N) & \ldots & t_{\vu_N}(\vx_N)
\end{bmatrix}.
\end{equation}
The $(i,j)$th entry of $\mT$ is $t_{\vu_j}(\vx_i)$. If a pixel located at $\vu_j$ is relocated to location $\vx_i$, then $t_{\vu_j}(\vx_i) = 1$. Otherwise $t_{\vu_j}(\vx_i) = 0$. For example, if $\mT$ shifts the entire image by one pixel, then it is
\begin{equation}
\mT
=
\begin{bmatrix}
0 & 1 & 0 & 0 & \ldots & 0\\
0 & 0 & 1 & 0 & \ldots & 0\\
\vdots & \vdots & \vdots & \vdots & \ddots & \vdots \\
0 & 0 & 0 & 0 & \ldots & 0
\end{bmatrix}.
\end{equation}
Remark: If the shifting operator needs to include subpixel shifts, we can allow each row to be a local kernel (e.g. a Gaussian kernel) such that the values sum to one.

The operator $\calB$ is a collection of tilt-free but spatially varying blurs. In the matrix notation, we can define a matrix $\mB$ where $[\mB]_{ij} = b_{\vu_j}(\vx_i)$ where $b_{\vu_j}$ is the tilt-free blur located at $\vu_j$. As before, $b_{\vu_j}$ is generated by the phase distortion at $\vu_j$ using high-order Zernike coefficients. The value $b_{\vu_j}(\vx_i)$ is the tilt-free blur $b_{\vu_j}$ evaluated at pixel $\vx_i$. The overall structure of the matrix $\mB$ is\index{tilt-then-blur model! blur operator}
\begin{equation}
\mB =
\begin{bmatrix}
b_{\vu_1}(\vx_1) & b_{\vu_2}(\vx_1) & \ldots & b_{\vu_N}(\vx_1)\\
b_{\vu_1}(\vx_2) & b_{\vu_2}(\vx_2) & \ldots & b_{\vu_N}(\vx_2)\\
\vdots           & \vdots           & \ddots & \vdots \\
b_{\vu_1}(\vx_N) & b_{\vu_2}(\vx_N) & \ldots & b_{\vu_N}(\vx_N)
\end{bmatrix}.
\end{equation}
The difference between $b_{\vu_j}(\vx_i)$ and $h_{\vu_j}(\vx_i)$ is the presence of the shifting action by the tilt $\mT$. In fact, the source of tension is about whether $\calH_{\vtheta} = \calT\circ\calB$ (i.e. $\mH = \mT\mB$?) or $\calH_{\vtheta} = \calB\circ\calT$ (i.e. $\mH = \mB\mT$?).

\subsection{Tilt-then-Blur or Blur-then-Tilt?}
\begin{figure}[h]
\centering
\includegraphics[width=\linewidth]{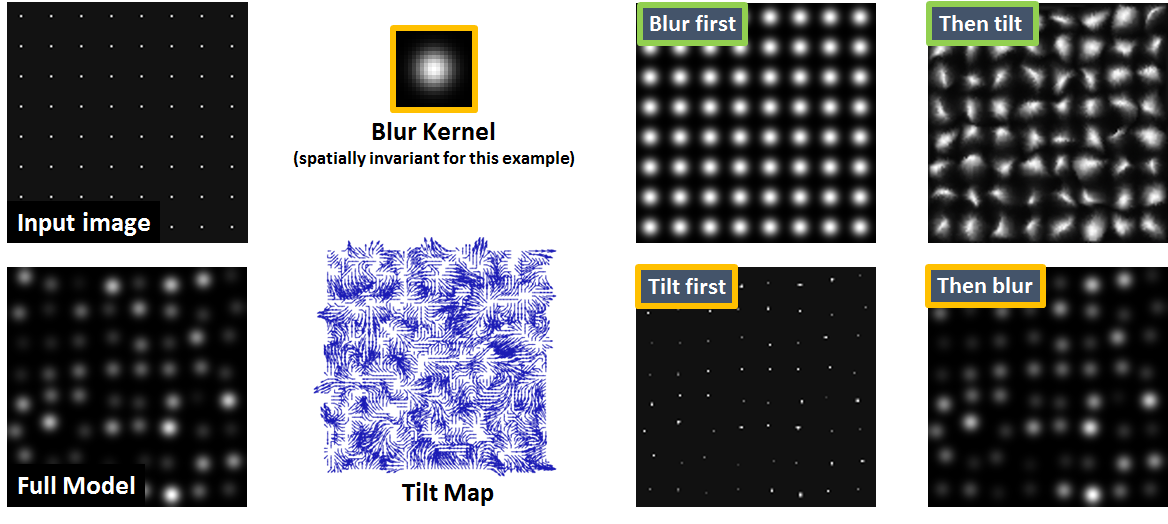}
\caption{Comparing blur-then-tilt $\calT \circ \calB$ and tilt-then-blur $\calB \circ \calT$. Given a grid of point sources, a spatially invariant blur, and a dense tilt map, the result of $\calT \circ \calB$ (the green case) shows a grid of destroyed blurs whereas the result of $\calB \circ \calT$ (the yellow case) shows a grid of shifted blurs. We remark that $\calT \circ \calB$ is incorrect whereas $\calB \circ \calT$ is correct.}
\label{fig: ch4 which first}
\end{figure}

To understand the difference between $\calT\circ\calB$ and $\calB\circ\calT$, we consider a grid of point sources as shown in \fref{fig: ch4 which first}. The tilt map is synthesized using the first two Zernike coefficients with some spatial correlation (aka the angle of arrival correlation). The blur in this example is a spatially invariant Gaussian blur so that the visualization is clearer.

Ideally, if we apply $\calH_{\vtheta}$ to this grid of points, we should expect that the output is a grid of shifted blurs. The reason is that if a single point source (a delta function) is convolved with a shifted blur (a point spread function constructed using \emph{all} Zernike coefficients), then the output must be the tilt-free blur shifted to the new location specified by the tilt.

Inspecting the case of \emph{tilt-then-blur} $\calB \circ \calT$, we see a perfect match with the full operator $\calH_{\vtheta}$. This can be understood as first starting with an unshifted delta function. When we apply the tilt to the delta function, the delta function is moved to a new position specified by the amount of the tilt. The result of this tilting operation to the grid of points is then a grid of shifted delta functions. If we apply a blur, then the blur will change the deltas to blurs. This is exactly the same as the original operator $\calH_{\vtheta}$.

Mathematically, we can show that the original operator $\calH_{\vtheta}$ is a set of shifted blurs. For example, if we assume that the tilt is globally shifting \emph{all} the tilt-free blurs $\{b_{\vx_i}(\vu) \,|\, i = 1,\ldots,N\}$ by one pixel, then each row of $\mH$ will be a shifted blur with a one-pixel shift. The $i$th blur is still located at the $i$th row, but the entries of the row are $h_{\vu_j} = [b_{\vx_i}(\vu_2), b_{\vx_i}(\vu_3), \ldots, b_{\vx_i}(\vu_{N}), 0]$. Putting all the entries together, we can show that $\mH$ is the matrix:
\begin{align}
\mH
=
\begin{bmatrix}
b_{\vu_2}(\vx_1) & \ldots & b_{\vu_N}(\vx_1) & 0\\
b_{\vu_2}(\vx_2) & \ldots & b_{\vu_N}(\vx_2) & 0\\
\vdots           & \vdots & \ddots             & \vdots \\
b_{\vu_2}(\vx_N) & \ldots & b_{\vu_N}(\vx_N) & 0
\end{bmatrix}.
\end{align}
With a simple calculation, we can show that this matrix is identical to $\mB\mT$:
\begin{align}
\mB\mT
&=
\begin{bmatrix}
b_{\vu_1}(\vx_1) & b_{\vu_2}(\vx_1) & \ldots & b_{\vu_N}(\vx_1)\\
b_{\vu_1}(\vx_2) & b_{\vu_2}(\vx_2) & \ldots & b_{\vu_N}(\vx_2)\\
\vdots           & \vdots           & \ddots & \vdots \\
b_{\vu_1}(\vx_N) & b_{\vu_2}(\vx_N) & \ldots & b_{\vu_N}(\vx_N)
\end{bmatrix}
\begin{bmatrix}
0 & 1 & 0 & 0 & \ldots & 0\\
0 & 0 & 1 & 0 & \ldots & 0\\
\vdots & \vdots & \vdots & \vdots & \ddots & \vdots \\
0 & 0 & 0 & 0 & \ldots & 0
\end{bmatrix} \notag \\
&=
\begin{bmatrix}
b_{\vu_2}(\vx_1) & \ldots & b_{\vu_N}(\vx_1) & 0\\
b_{\vu_2}(\vx_2) & \ldots & b_{\vu_N}(\vx_2) & 0\\
\vdots           & \vdots & \ddots             & \vdots \\
b_{\vu_2}(\vx_N) & \ldots & b_{\vu_N}(\vx_N) & 0
\end{bmatrix} = \mH.
\end{align}
Therefore, in this example, we have shown that $\mH = \mB\mT$.

If we use the \emph{blur-then-tilt} model $\calT \circ \calB$, the result in \fref{fig: ch4 which first} shows that the blurs are not shifted but destroyed. The reason is that if we first blur the grid of points, the result is a grid of blurs. When we apply the tilt map to the grid of blurs, each tilt vector will move its corresponding pixel value to a different location. In other words, since the tilt map is dense, every pixel has a tilt vector. These include pixels that are not the delta functions; we refer to the tilt vectors at these locations as the non-centered tilts. In the full model $\calH_{\vtheta}$ and the blur-after-tilt $\calB\circ\calT$ model, these off-centered tilts are \emph{not} used because the tilt vectors are only valid for the centered tilts. In the blur-then-tilt model $\calT\circ\calB$, the grid of blurred points will have non-zero values at those non-centered tilts. As a result, the non-centered tilts will move the blurred pixel values to a different location. From the image formation perspective, this is a grid of destroyed point spread functions.

In terms of matrices and vectors, we can show that
\begin{align}
\mT\mB
&=
\begin{bmatrix}
0 & 1 & 0 & 0 & \ldots & 0\\
0 & 0 & 1 & 0 & \ldots & 0\\
\vdots & \vdots & \vdots & \vdots & \ddots & \vdots \\
0 & 0 & 0 & 0 & \ldots & 0
\end{bmatrix}
\begin{bmatrix}
b_{\vu_1}(\vx_1) & b_{\vu_2}(\vx_1) & \ldots & b_{\vu_N}(\vx_1)\\
b_{\vu_1}(\vx_2) & b_{\vu_2}(\vx_2) & \ldots & b_{\vu_N}(\vx_2)\\
\vdots           & \vdots           & \ddots & \vdots \\
b_{\vu_1}(\vx_N) & b_{\vu_2}(\vx_N) & \ldots & b_{\vu_N}(\vx_N)
\end{bmatrix} \notag \\
&=
\begin{bmatrix}
b_{\vu_1}(\vx_2) & b_{\vu_2}(\vx_2) & \ldots & b_{\vu_N}(\vx_2)\\
\vdots           & \vdots           & \ddots & \vdots \\
b_{\vu_1}(\vx_N) & b_{\vu_2}(\vx_N) & \ldots & b_{\vu_N}(\vx_N)\\
0                & 0                & 0      & 0
\end{bmatrix} \not= \mH.
\end{align}

To summarize, we conclude the difference between blur-then-tilt and tilt-then-blur in the following theorem.
\boxedthm{
\begin{theorem}[\keyword{Decomposition of Tilt and Blur}]
If we decompose the turbulence operator $\calH_{\vtheta}$ as the product of the tilt $\calT$ and the blur $\calB$, then
\begin{equation}
\calH_{\vtheta} = \calB\circ\calT \quad\mbox{and}\quad \calH_{\vtheta} \not= \calT\circ\calB.
\end{equation}
\end{theorem}
}

\subsection{Geometric Analysis}
The above intuition can be formalized by a simple analysis. We consider a simplified situation where the turbulence is the combination of a dense field of tilts and a spatially \emph{invariant} blur. We want to show that under this simplified condition, the two operators $\calT\circ\calB$ and $\calB\circ\calT$ are different.

\subsubsection{Blur-then-tilt $\calT\circ\calB$}
Let's start by considering the spatially invariant blur. We define a template function $g(\vx)$. For example, we can think of this template function as a Gaussian blur kernel. For simplicity, we assume that this Gaussian blur kernel is spatially invariant. This spatially invariant blur will give us the blur $b_{\vu_j}(\vx)$ centered at pixel $\vu_j$ as a shifted version of the template function:
\begin{equation}
b_{\vu_j}(\vx) = g(\vx-\vu_j).
\end{equation}
When this tilt-free blur is applied to the clean image $J(\vx)$, we obtain the tilt-free blurred image
\begin{equation}
I_{\calB}(\vx_i) = \sum_{j=1}^N b_{\vu_j}(\vx_i)J(\vu_j) = \sum_{j=1}^N g(\vx_i-\vu_j)J(\vu_j).
\end{equation}
For the blur-then-tilt model $\calT \circ \calB$, the tilt $\vt_i$ assigns $I_{\calB}(\vx_i)$ to a new pixel location $\vx_i+\vt_i$ of the final image $I_{\calT\circ\calB}$. That is,
\begin{equation*}
I_{\calT\circ\calB}(\vx_i + \vt_i) = I_{\calB}(\vx_i).
\end{equation*}
Letting $\vv_i = \vx_i + \vt_i$, this becomes $I_{\calT\circ\calB}(\vv_i) = I_{\calB}(\vv_i-\vt_i)$. Since $\vv_i$ is a dummy variable, we can replace $\vv_i$ with $\vx_i$. Thus,
\begin{equation}
I_{\calT\circ\calB}(\vx_i) = \sum_{j=1}^N g(\vx_i-\vt_i-\vu_j)J(\vu_j).
\label{eq: I TB}
\end{equation}

\subsubsection{Tilt-then-blur $\calB\circ\calT$}
The tilt-then-blur model behaves differently. We first tilt the image by defining
\begin{equation}
I_{\calT}(\vu_j+\vt_j) = J(\vu_j).
\end{equation}
Letting $\vv_j = \vu_j+\vt_j$, it follows that $I_{\calT}(\vv_j) = J(\vv_j-\vt_j)$. Now, if we blur this shifted image, we will have
\begin{align*}
I_{\calB\circ\calT}(\vx_i)
= \sum_{j=1}^N b_{\vv_j}(\vx_i) I_{\calT}(\vv_j)
= \sum_{j=1}^N g(\vx_i-\vv_j) J(\vv_j-\vt_j).
\end{align*}
Replacing $\vu_j = \vv_j-\vt_j$, it follow that
\begin{equation}
I_{\calB\circ\calT}(\vx_i) = \sum_{j=1}^N g(\vx_i-\vt_j-\vu_j) J(\vu_j).
\label{eq: I BT}
\end{equation}

\subsubsection{The Roles of $\vt_i$ and $\vt_j$}
\begin{figure}[h]
\centering
\includegraphics[width=0.8\linewidth]{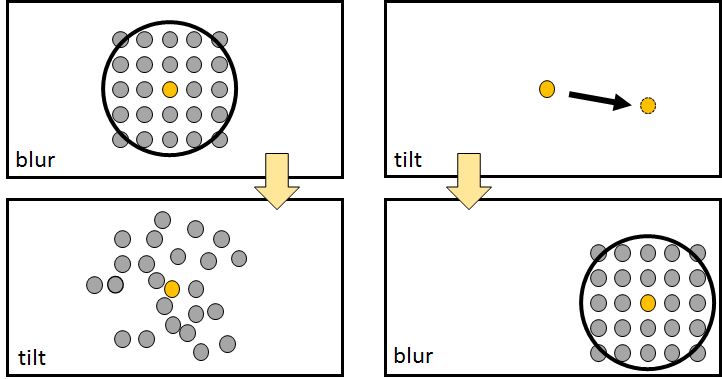}
\caption{$\calT \circ \calB$ moves the center to a new location, whereas $\calB \circ \calT$ moves the points of the blur to different locations.}
\label{fig: ch4 tilt blur equation}
\end{figure}

Comparing \eref{eq: I TB} and \eref{eq: I BT}, the difference between the two equations is that for $\calT\circ\calB$, the tilt is $\vt_i$ whereas for $\calB\circ\calT$ the tilt is $\vt_j$. In the case of $\calT \circ \calB$, the tilt $\vt_i$ is applied to the \emph{output}. That is, we blur the image (drawn as a circle in \fref{fig: ch4 tilt blur equation}) and move the output by $\vt_i$. Since each output pixel experiences a different $\vt_i$, the blur is destroyed. For $\calB \circ \calT$, although the situation is more complicated because there are $N$ tilts $\{\vt_j \;|\; j = 1\ldots,N\}$, they perturb the location of the \emph{input}. Therefore, while each $\vx_i$ sees $N$ tilts, the $N$ tilts are common for every $\vx_i$. The shape of the blur is thus preserved.

\subsection{Impact to Real Images}
Although the two operators $\calT\circ\calB$ and $\calB\circ\calT$ are theoretically different, their impacts on real images are less so. This is particularly true when the underlying image is a natural image.

To see why this is the case, rewrite \eref{eq: I TB} and \eref{eq: I BT} as
\begin{align*}
I_{\calT\circ\calB}(\vx_i)
&= \sum_{j=1}^N g(\vx_i-\vu_j)J(\vu_j-\vt_i),\\
I_{\calB\circ\calT}(\vx_i)
&= \sum_{j=1}^N g(\vx_i-\vu_j)J(\vu_j-\vt_j).
\end{align*}
Then, by approximating $J(\vu_j-\vt_i)$ and $J(\vu_j-\vt_j)$ to the first order, the pointwise difference between $I_{\calT\circ\calB}(\vx_i)$ and $I_{\calB\circ\calT}(\vx_i)$ can be evaluated as
\begin{align}
&I_{\calT\circ\calB}(\vx_i) - I_{\calB\circ\calT}(\vx_i) \notag\\
&\qquad= \sum_{j=1}^N g(\vx_i-\vu_j)\Big[J(\vu_j-\vt_i) - J(\vu_j-\vt_j)\Big] \notag\\
&\qquad\approx \sum_{j=1}^N \underset{\text{convolution}}{\underbrace{g(\vx_i-\vu_j)}} \;\;\; \underset{\text{distorted image gradient}}{\underbrace{
\underset{\text{image gradient}}{\underbrace{\nabla J(\vu_j)^T}} \;\;\;
\underset{\text{random tilt}}{\underbrace{(\vt_i-\vt_j)}}}}.
\end{align}

An intuitive argument here is that $\vt_i-\vt_j$ is the difference between two tilt vectors. Since each tilt is a zero-mean Gaussian random vector, the difference remains a zero-mean Gaussian random vector. Although they are not white Gaussian, they are nevertheless \emph{noise}. If there is a large ensemble average of these noise vectors, the result will be close to zero.

So, where does the average come from? There is a convolution by $g(\vx-\vu)$. If the support of this blur kernel is large, then many of the noise vectors will be added and this will result in a small value. However, for an image with a large field of view, the relative size of the blur kernel $g$ is usually not big (at most $30 \times 30$ for a $256\times256$ image). So, there must be another source that makes the error small.

\begin{figure}[h]
\centering
\begin{tabular}{cc}
\includegraphics[width=0.4\linewidth]{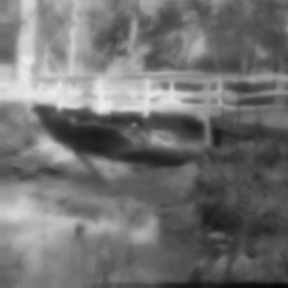}&
\includegraphics[width=0.4\linewidth]{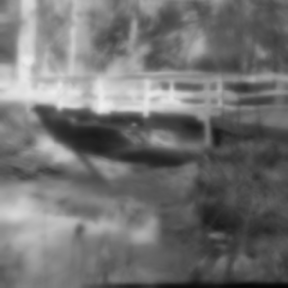}\\
(a) $\calT\circ\calB$ & (b) $\calB\circ\calT$
\end{tabular}
\caption{Simulated turbulence using $\calT\circ\calB$ and $\calB\circ\calT$. We notice that although theoretically the two are different, visually the two are nearly indistinguishable.}
\label{fig: ch4 tilt first example}
\end{figure}

The main reason why natural images tend to show a less difference between $\calT\circ\calB$ and $\calB\circ\calT$ is that the image gradient $\nabla J(\vu_j)$ is typically \emph{sparse}. For most parts, the gradient is zero except for edges. (Textures are less of a problem because they will be smoothed by the blur.) When $\nabla J(\vu_j)$ is multiplied with the noise vector $\vt_i-\vt_j$, the result is an edge map with noise multiplied by every pixel. Convolving $\nabla J(\vu_j)^T(\vt_i-\vt_j)$ with a blur kernel $g$ will further smooth out the variations. \fref{fig: ch4 tilt first example} shows a typical example with some standard optical configurations. The difference is not noticeable.

\subsection{From Forward Model to Inverse Problems}
So far we have shown that if we agree upon the phase decomposition in terms of the Zernike representation, then the correct turbulence forward model is tilt-then-blur. As a result, we need to be careful about statements such as
\begin{center}
``\emph{... The image formation of atmospheric turbulence follows the equation $\mI = \calT(\calB(\mJ))$ ... }''
\end{center}
\noindent It is more appropriate to instead state that the image formation model is $\mI = \calB(\calT(\mJ))$ and comment that for natural images it can be approximated by $\mI = \calT(\calB(\mJ))$.

From the standpoint of solving \keyword{inverse problems}, the issue becomes more interesting. The inverse problem, expressed in terms of the maximum-a-posteriori estimation, is \cite{Lou_2013_a, Lau_2017_a, Milanfar_2013_a, Gilles_2012_a, Shimizu_2008_a}
\begin{equation}
\widehat{\mJ} = \argmin{\mJ} \;\; \|\mI - \calH_{\vtheta}(\mJ)\|^2 + \lambda \; g(\mJ),
\label{eq: ch4 inverse problem}
\end{equation}
where $g(\mJ)$ is the regularization function, also known as the image prior, that confines the search space of the solution $\mJ$.

The optimization in \eref{eq: ch4 inverse problem} should look familiar to readers who know deconvolution. In the special case where $\calH_{\vtheta}$ is a spatially invariant blur, \eref{eq: ch4 inverse problem} resembles the classical deconvolution problem. Therefore, any classical/modern solution that has been invented to solve deconvolution can become a candidate for \eref{eq: ch4 inverse problem}. For turbulence image restoration, approaches following this line of thoughts are abundant, e.g., \cite{Gepshtein_2004_a, Frakes_2001_a, Xie_2016_a, He_2016_a, Lou_2013_a, Caliskan_2014_a, Hirsh_2010_a, Pellizzari_2017_a}.

In the presence of turbulence, solving \eref{eq: ch4 inverse problem} is more involved because we do not know $\calH_{\vtheta}$ which is the instantaneous turbulence operator. It is a random realization drawn from a certain probability distribution. Since we only have one observation of the distorted image $\mI$, recovering $\calH_{\vtheta}$ can be as hard as recovering $\mJ$.

There are two options that are worth discussing. Firstly, we can assume that instead of only one observation, we have $T$ observations of a static scene. In this case, the optimization objective can be written as a sum of individual turbulence terms:
\begin{equation}
\widehat{\mJ} = \argmin{\mJ} \;\; \sum_{t=1}^T \|\mI(t) - \calH_{\vtheta}^{(t)}(\mJ)\|^2 + \lambda \; g(\mJ),
\end{equation}
where $\calH_{\vtheta}^{(t)}$ is the random realization of the turbulence distortion at time $t$. Solving the new equation is slightly easier because if we knew $\calH_{\vtheta}^{(t)}$, then we will have more degrees of freedom. However, if we do not know $\calH_{\vtheta}^{(t)}$, the new problem is as difficult as the original problem.

The other option is to decouple $\calH$ into $\calT$ and $\calB$ as follows.
\begin{equation}
\widehat{\mJ} = \argmin{\mJ} \;\; \|\mI - \calB(\calT(\mJ))\|^2 + \lambda \; g(\mJ).
\end{equation}
This optimization can further be written as
\begin{align}
\widehat{\mU} &= \argmin{\mU} \;\; \|\mI - \calB(\mU)\|^2 + \lambda \; g(\mU),\\
\widehat{\mJ} &= \argmin{\mJ} \;\; \|\widehat{\mU} - \calT(\mJ)\|^2 + \lambda \; g(\mJ).
\end{align}
This is a two-step approach. After we have recovered the intermediate variable $\widehat{\mU}$, we can recover $\widehat{\mJ}$. The difficulty of this approach is that when solving for $\mU$, we need to estimate the spatially varying blur $\calB$. This task is as difficult as the original problem because we need to recover the blur kernel at every pixel.

Today's turbulence mitigation algorithms are mostly inspired by the \keyword{lucky imaging} concept which will be discussed in the next subsection. The core idea is to \emph{flip} the order of the blur and the tilt so that the optimization is
\begin{equation}
\widehat{\mJ} = \argmin{\mJ} \;\; \|\mI - \calT(\calB(\mJ))\|^2 + \lambda \; g(\mJ).
\label{eq: ch4 inverse problem 2}
\end{equation}
As discussed earlier, tilt-then-blur and blur-then-tilt are different. However, for natural images (i.e., not point sources), the visual difference may not be too obvious. The benefit of flipping the order of tilt and blur is that estimating the tilt from a stack of images is feasible. Moreover, after the lucky imaging step, a significant portion of the spatially varying blur will be removed because we will be fusing sharp regions of the images. As a result, $\calT$ can be inverted to a reasonable degree of accuracy, and $\calB$ will become easier because the spatially varying part of the blur will be suppressed. In the following Chapter, we will discuss these two steps one by one.

\boxedthm{
\vspace{2ex}
\textbf{Tilt-Blur $\calT \circ \calB$ or Blur-Tilt $\calB \circ \calT$?}
\begin{itemize}
\item The correct model is $\calB \circ  \calT$. So, for simulation, we must use $\calB \circ \calT$.
\item For solving inverse problems, it does not matter. It does not mean that we shouldn't use $\calT \circ \calB$; it's just that there are other considerations. For example, inverting the blur is often much harder than inverting the tilt.
\end{itemize}
}

\section{Lucky Imaging}
\label{sec: sec4_2}
\index{lucky imaging}After discussing the forward imaging model, we now discuss methods to solve the inverse problem. \cref{sec: sec4_2} is dedicated to the concept of lucky imaging. The goal of lucky imaging is to handle the tilt operator $\calT$ and remove some of the spatially varying blur $\calB$ in \eref{eq: ch4 inverse problem 2}.

\subsection{Lucky Probability via Zernike}
Atmospheric turbulence is a random process caused by random phase distortion. If we model the phase using the Zernike representation, the randomness is encoded by the Zernike coefficients which are sampled from a probability distribution. Among many properties, the probability distribution is a high dimensional \emph{zero-mean} Gaussian according to Tatarskii \cite{Tatarski_1967_a}. Thus, if we observe a Zernike coefficient over a long period of time, the ensemble average will be zero. What this means is that, from time to time, and/or across the spatial locations, we will see instants where the Zernike coefficients are approximately zero. The probability of getting a zero Zernike vector is not high, but in theory, it is possible. When this happens, we say that there is a \keyword{lucky event}. The resulting image is called a \keyword{lucky image} \cite{Fried_1978_a}.

\begin{figure}[h]
\centering
\includegraphics[width=\linewidth]{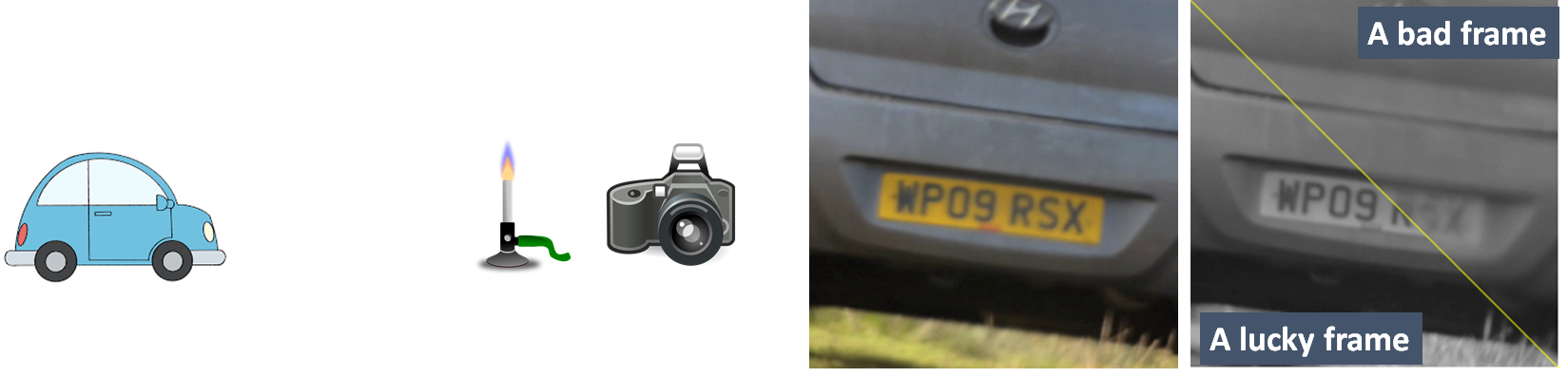}
\caption{An lucky event can be observed by placing a local heat source near the camera. This will create isoplanatic turbulence where the image is globally distorted. Over a period of observations, we will see distorted and distortion-free frames. A key for this experiment is to ensure that the exposure is short enough that motion blur is not accumulated.}
\label{fig: ch4 lucky experiment}
\end{figure}

To gain more intuition of the lucky image, we can conduct an experiment by putting a heat source in the optical path close to the camera. Since the heat is localized, the turbulence is \keyword{isoplanatic}\index{isoplanaticism}, i.e., the entire field of view will experience the same turbulence. If we use a high-speed shutter to capture the frames, we will see that occasionally the image is sharp. \fref{fig: ch4 lucky experiment} shows an example, where we see that an unlucky frame suffers from severe distortions whereas a lucky frame is distortion free.

In a truly long-range situation, the turbulent effects become \keyword{anisoplanatic}\index{anisoplanaticism}. Across different regions of the image, we can still observe the lucky effect but it appears locally. Some regions of the image will experience a lucky effect while another region of the same image may not. In this case, we say that there is a local lucky effect, and the corresponding lucky region is known as a lucky \emph{patch}.

\begin{figure}[h]
\centering
\includegraphics[width=0.55\linewidth]{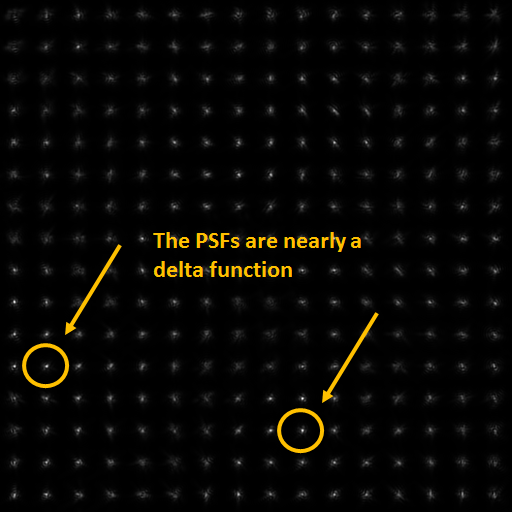}
\caption{A grid of point sources after going through turbulence simulation. From time to time, and from location to location, the PSF can be close to a delta function. We say that when this happens, there is a lucky event.}
\label{fig: ch4 lucky PSF}
\end{figure}

The lucky event can be described mathematically by considering the PSFs. \fref{fig: ch4 lucky PSF} shows a grid of PSFs resulting from passing point sources through a turbulence simulator. For this grid of PSFs, some PSFs are nearly a delta function, e.g., $h_{\vu}(\vx) \approx \delta(\vx-\vu)$.\footnote{Technically speaking, the PSF will never become a delta function because the finite aperture of the camera will at best give us a diffraction-limited blur. So, instead of having a pure delta function, we will obtain a diffraction-limited PSF. } If a PSF is a delta function, the convolution with an image will be distortion free. Such a phenomenon can happen in space across a large field of view. It can also happen in time over a long period of observation. If we use short exposures to capture a large number of frames over time, then occasionally we will see a delta function for the PSF.

We can define the lucky event via the Zernike coefficients.
\boxedthm{
\begin{definition}[\keyword{Lucky event}]
Let $\mathbf{a}_{\vu} = [a_{\vu,m} \,|\, m = 2,\ldots,M]$ be the Zernike coefficients for a pixel coordinate $\vu$, a lucky event happens when
\begin{equation}
\text{lucky event at pixel $\vu$} = \Big\{\|\mathbf{a}_{\vu}\|_2^2 \le \tau \Big\},
\end{equation}
where $\tau$ is some user-defined threshold.
\end{definition}
}

In plain words, a lucky event happens when the magnitude of the Zernike coefficients is small enough. The magnitude of the Zernike coefficients measures the amount of turbulence energy. If the scene is turbulence-free, then the phase distortion is $\phi_{\vu}(\vrho) = 0$, which is equivalent to requiring all Zernike coefficients to be zero, i.e., $\|\mathbf{a}_{\vu}\|_2 = 0$. As turbulence becomes stronger, $\|\mathbf{a}_{\vu}\|_2$ will grow.

The analysis of the lucky probability was first documented by David Fried in his 1978 paper \cite{Fried_1978_a}. Fried's approach is similar to what we described above where he uses tilt-corrected phase functions. However, instead of using the Zernike functions as the basis, Fried considered the Karhunen–Loeve (KL) expansion. The reason is that the Zernike basis functions, although they are orthogonal, don't produce statistically independent coefficients. Therefore, if we consider a pair of coefficients $a_{\vu,i}$ and $a_{\vu,j}$, the joint expectation cannot be factorized as $\E[a_{\vu,i}a_{\vu,j}] \not= \E[a_{\vu,i}]\E[a_{\vu,j}]$, and hence we cannot ensure $\E[a_{\vu,i}a_{\vu,j}]=0$ even if we know $\E[a_{\vu,i}] = 0$ for any $i$. This will cause difficulties when considering the norm $\E[\|\mathbf{a}_{\vu}\|_2^2]$. The KL representation alleviates the difficulty because the basis functions are computed via the principal component analysis and so they are statistically independent.\footnote{The caveat is that the Karhunen–Loeve (KL) expansion is not as physically interpretable as the Zernike coefficients. Mathematically, the KL decomposition is in a non-closed form determined by the phase function's covariance equation.}

The definition of the lucky event by using the KL representation is the same as that using the Zernike representation. Let $\vbeta = [\beta_1,\ldots,\beta_M]$ be the KL expansion coefficients, the lucky event can be defined as
\begin{equation}
\text{lucky event at pixel $\vu$} = \Big\{\|\vbeta_{\vu}\|_2^2 \le \tau \Big\},
\end{equation}
for some user-defined parameters $\tau$. With the KL expansion, Fried showed the probability of obtaining a lucky event.
\boxedthm{
\begin{theorem}[\keyword{Probability of a Lucky Event}]
Let $\vbeta_{\vu}$ be the Karhunen–Loeve expansion coefficients of the phase function at pixel $\vu$. The probability of obtaining a lucky event is
\begin{equation}
\underset{\text{probability of a lucky event}}{\underbrace{\Pb\Big[\|\vbeta_{\vu}\|_2^2 \le 1 \Big]}} \approx 5.6 e^{-0.1557\left(D/r_0\right)^2},
\end{equation}
for $D/r_0 \ge 3.5$, where $D$ is the aperture diameter and $r_0$ is the Fried parameter.
\end{theorem}
}

The prediction by Fried is useful in some ways but also limited in other ways. On the positive side, it does tell us the probability of obtaining a lucky event. The equation is interpretable because it is written in terms of $D/r_0$. For stronger turbulence, $D/r_0$ increases and so Fried's formula predicts that the probability will drop.

The downside of the prediction can be seen in several ways. First, $D/r_0 \ge 3.5$ is quite big for many ground-to-ground imaging systems. Typically, for ground-to-ground imaging (especially if we consider a passive imaging system that is incoherent), any $D/r_0 \ge 3.5$ is considered difficult for image restoration algorithms. For astronomical imaging, we might be able to handle a larger $D/r_0$ because we have access to more advanced tools such as adaptive optics.

Another reason why Fried's result has limited applicability for computer vision is that the lucky event Fried considered is within the isoplanatic angle. Isoplanatism means that the entire frame is distorted homogeneously by the same turbulent effect. For astronomical imaging, isoplanatic turbulence happens because the field of view is small. For ground-to-ground applications, especially those associated with computer vision tasks, the turbulence is anisoplanatic, implying that the distortions are spatially varying. Therefore, instead of seeing a lucky frame, we can only see lucky \emph{patches}. To compute the probability of getting an algorithmically fused lucky frame, we need to compute the correlation across different patches. This remains an open problem, to our knowledge.

\subsection{Optical Lucky Imaging}
\begin{figure}[h]
\centering
\includegraphics[width=0.5\linewidth]{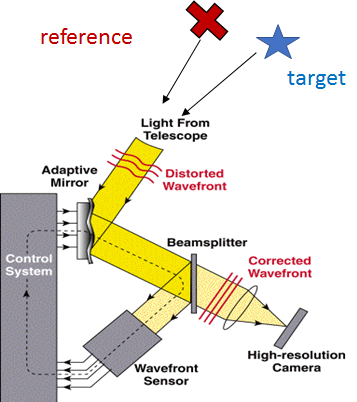}
\caption{An adaptive optics system implements the concept of lucky imaging using optical instrumentation. With the help of deformable mirrors, the adaptive optics system measures the phase of the reference star and compensates for the phase distortion by moving the deformable mirrors \cite{Ealey_1992_a}. In terms of turbulence correction, the random tilts are removed but the residue blur remains. Therefore, some deblurring algorithms are still needed to produce the result. Source: \url{https://lyot.org/background/adaptive_optics.html}.}
\label{fig: ch4 adaptive optics}
\end{figure}

Lucky imaging can be achieved optically and digitally. The technique on the optical side is known as \keyword{adaptive optics}\index{adaptive optics}. This technique is mostly used for astronomical imaging because the star does not move (as in the sense of a car moving in the scene). \fref{fig: ch4 adaptive optics} shows a schematic diagram of an adaptive optics system \cite{Fried_1982_a, Tyson_2010_a, Roddier_1999_a}.

An adaptive optics system requires a \keyword{reference star} in addition to the \keyword{target star}\index{adaptive optics! reference star}\index{adaptive optics! target star} \cite{Watnik_2018_a}. Usually, astronomers will look for a bright star near to the target star. Since we know the distance of the reference star, waves originating from the source can be measured and tracked. In adaptive optics, we use a wavefront sensor to record the incident wavefront, hence determining the instantaneous phase of the wave. If atmospheric turbulence is present, the measured phase will deviate from the theoretical prediction. A feedback signal measuring such a deviation will be sent to the control system which is connected to a set of adaptive mirrors. The adaptive mirrors rotate their angles in order to correct the phase lead and lag measured by the wavefront sensor. Once the phase lead and lag are compensated by the adaptive mirror, the optics will be used to capture the target. Since the optics have already been configured to compensate for the turbulence, the target star can be captured without any part of the phase distortions.

The mirrors of the adaptive optics system have limited precision of the phase it can correct. In most of cases, the adaptive optics system can correct for the first-order phase deviation. Using our terminologies, this is exactly the first two Zernike coefficients representing the \emph{tilt}. Mathematically, what adaptive optics seeks to find is the best linear fit so that we can decompose the phase as \cite{Fried_1965_a, Fried_1966_a}
\begin{equation}
\phi_{\vu}(\vrho) = \underset{\text{linear term: tilt}}{\underbrace{\valpha^T\vrho}} + \underset{\text{high order aberrations}}{\underbrace{\varphi_{\vu}(\vrho)}}.
\end{equation}
In this equation, the vector $\valpha = [\alpha_1, \alpha_2]$ represents the normal vector of the best fitted linear plane. An intuitive way of thinking about $\valpha$ is to treat them as the first two Zernike coefficients (which are the tilts). The adaptive optics system can compensate for this term, thus leaving us just the high order aberration term $\varphi_{\vu}(\vrho)$. However, $\valpha$ is not exactly the Zernike coefficients because the Zernike basis functions are not statistically independent. Some high-order aberration terms can also have the tilting effect. Therefore, by performing the adaptive optics, we can sometimes compensate for more than just the first two Zernike coefficients.

The other limitation of the adaptive optics system is that its spatial resolution is limited by the number of deformable mirrors that can be installed and controlled. We should remind ourselves that the deformable mirror is adjusted mechanically. While advanced MEMS can be used to reduce the power and space, precise control of the angles of the mirrors is still challenging. The speed of the deformable mirrors is also an issue. We are not able to adjust the mirrors too quickly for moving scenes. Finally, adaptive optics often require a reference star. For many ground-to-ground imaging applications, this may not be available.

\subsection{Digital Lucky Imaging: Sharpness Metric}
\index{lucky imaging! digital}Digital lucky imaging is based on capturing a stack of short-exposure images and then processing them using algorithms. The overall idea is to identify the distortion-free frames by finding which frames have the sharpest image content \cite{Roggemann_1994_a}. These frames will then be aggregated to construct a lucky image \cite{Aubailly_2009_a}.

\begin{figure}[h]
\centering
\includegraphics[width=\linewidth]{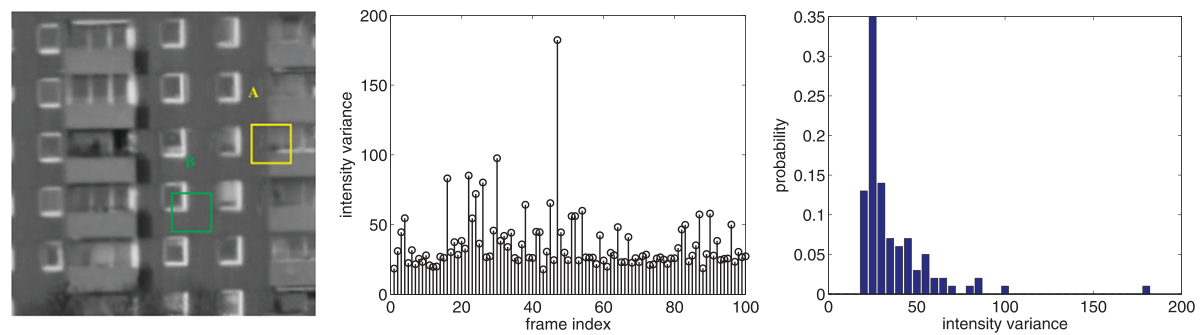}
\caption{One way to identify sharp frames from a stack of turbulence images is to compute the local intensity and evaluate its variance. [Left] An image. [Middle] The variance of the patch as a function of time. [Right] Histogram of the variance. The idea here is to identify the outliers from the histogram. Image courtesy: \cite{Milanfar_2013_a}.}
\label{fig: ch4 Milanfar Lucky}
\end{figure}

The idea of identifying the sharpest frame can be realized in multiple ways. In this Chapter, we mention a few of them:

\textbf{Intensity Variance}. In turbulence literature, the Strehl ratio is often used as a metric to predict the quality of a frame if a reference star is available. The Strehl ratio is the ratio between the peak of the intensity in the aberrated PSF and that of a diffraction-limited PSF. In \cite{Milanfar_2013_a}, it was mentioned that the Strehl ratio has a relationship to the \emph{variance} of the pixel intensity in a local neighborhood. More specifically, for a fixed coordinate $\vx$, we consider a local neighborhood $\Omega_{\vu}$ and compute the variance as
\begin{align*}
\mu_{\vx} &= \frac{1}{|\Omega_{\vu}|}\sum_{\vs \in \Omega_{\vu}} I(\vs),\\
V_{\vx}   &= \frac{1}{|\Omega_{\vu}|}\sum_{\vs \in \Omega_{\vu}} \Big( I(\vs) - \mu_{\vx} \Big)^2.
\end{align*}
If we plot the variance as a function of time, we can obtain a plot similar to \fref{fig: ch4 Milanfar Lucky}. A lucky frame is then identified if the variance is an \emph{outlier} of the plot. There are different ways to determine if a pixel is an outlier. We refer readers to \cite{Milanfar_2013_a} for a hypothesis testing approach.

\textbf{Gradient}. Another commonly used algorithmic strategy to identify a sharp image is by means of checking the gradient. The idea was reported in Joshi and Cohen \cite{Joshi_2010_ICCP} and later refined in \cite{Mao_2020_a}. To compute the gradient of a region, we calculate the horizontal and vertical finite differences:
\begin{align*}
\vg_x &= \mI \circledast [1, -1],\\
\vg_y &= \mI \circledast [1; -1].
\end{align*}
In this pair of equations, the filters $[1,-1]$ and $[1;-1]$ are the first-order horizontal and vertical difference operators, respectively. When we convolve (i.e., the $\circledast$ operation) the input image $\mI$ with these filters, we effectively extract the edges of the image and store them as $\vg_x$ and $\vg_y$, respectively. A compact notation for this operation is the gradient operator $\nabla: \R^N \rightarrow \R^{N \times 2}$:
\begin{align*}
[\vg_x, \; \vg_y] = \nabla \mI.
\end{align*}

The sharpness of the image can be computed via the norm of the gradient. For example, we can consider the \keyword{isotropic total variation}\index{total variation! isotropic} norm
\begin{align*}
\|\mI\|_{\text{TV}_2} = \sqrt{ \sum_{n=1}^N \left([\vg_x]_n^2 + [\vg_y]_n^2\right) },
\end{align*}
where $[\cdot]_n$ denotes the $n$-th element of the vector. Another approach is to consider the \keyword{anisotropic total variation norm}\index{total variation! anisotropic}, defined as
\begin{align*}
\|\mI\|_{\text{TV}_1} = \sum_{n=1}^N |[\vg_x]_n| + |[\vg_y]_n|.
\end{align*}

If we want to compute the \emph{local} gradient, we can replace the image $\mI$ by a patch $\underline{\mI}(\vx)$ located at coordinate $\vx$. If we further want to include the time axis, we can consider the patch located at $\vx$ at time $t$, i.e., $\underline{\mI}(\vx,t)$. Then, a sharpness metric $\delta^S$ can be defined as
\begin{equation}
\delta^S(\vx,t) = \|\underline{\mI}(\vx,t)\|_{\text{TV}_1}.
\end{equation}
Notice that the choice of the anisotropic or the isotropic total variation is not very important, because the two behave similarly.

Variations of the gradient are abundant. For example, instead of considering the horizontal or vertical gradient, we can consider the Laplacian of a local patch. In the age of deep learning, we can train a network to extract features; and from the features, we can identify the sharp patches.

\subsection{Digital Lucky Imaging: Fusion Step}
\index{lucky imaging! fusion}Suppose that we have a stack of $T$ frames. How should we construct the lucky image? Shall we just use one of the sharpness metrics and pick \emph{the} sharpest? The answer is no, because the so-called ``sharpest'' frame may not even be a good frame. For example, if for some reason an object in a frame is over-exposed, then there will be a surge in the gradient around the object boundary and background noise. If we simply just pick according to the gradient, then this poorly exposed frame will be picked. So in digital lucky imaging, we almost never pick the sharpest frame but a set of similarly sharp frames.

The way to accomplish the frame selection can be a hard decision (i.e., by setting a threshold) or a soft decision (which will be discussed shortly). The benefit of the hard decision is that we are more aggressive in terms of completely ignoring frames that are below the threshold, so their influence on the lucky image is eliminated. The downside, however, is that the quality of the lucky image will depend heavily on the threshold. Since there are no universal rules for choosing the threshold (besides a few statistical hypothesis techniques), it is sometimes better to consider the soft decision approach.

In a soft decision approach, we \keyword{fuse} multiple images to construct one lucky image. Using the sharpness metric as an example, we can consider a weight:
\begin{equation}
w_{\vx,t}(\Delta t) = \exp\left\{\alpha \|\underline{\mI}(\vx,t+\Delta t)\|_{\text{TV}_1}\right\}.
\end{equation}
This definition involves two parts. The first part is the exponential function. We take the exponential function to scale the total variation norm. If a patch is sharp, $\|\underline{\mI}(\vx,t+\Delta t)\|_{\text{TV}_1}$ will be large and so the exponential function further amplifies the magnitude. The second part is the coordinate offset $\Delta t$. This offset is a running index that defines the weight, which is the weight for a reference pixel located at $(\vx,t)$, and the weight is computed with respect to the temporal neighbors of $(\vx,t)$. In other words, we are computing the weight along the time axis and finding out which frame contains the sharpest content.

Given the weight, we can now fuse the lucky patch, at location $(\vx,t)$, as
\begin{equation}
\underline{\mI}_{\text{lucky}}(\vx,t) = \frac{\sum\limits_{\Delta t \in \Omega_T} w_{\vx,t}(\Delta t) \underline{\mI}(\vx,t+\Delta t)}{\sum\limits_{\Delta t \in \Omega_T} w_{\vx,t}(\Delta t)}.
\end{equation}
This equation is reminiscent of the classical non-local means \cite{Buades_2005_a} where we construct weights to form a linear combination of pixels (or patches). The slight difference here is that we are forming the linear combination purely along the temporal axis because we are looking for a lucky patch in time.

In \cite{Mao_2020_a}, it was commented that if we only do the above sharpness-based non-local averaging, we still have not overcome the limitation of the sharpness metric that it cannot differentiate a reliable sharp patch versus a poor-quality but a sharp patch. As such, it was proposed to include a \keyword{geometric consistency}\index{geometric consistency} metric:
\begin{equation}
\delta^G(\vx,t) = \|\underline{\mI}(\vx,t) - \underline{\mI}_{\text{ref}}(\vx,t)\|^2.
\end{equation}
Here, $\mI_{\text{ref}}$ is the \keyword{reference patch}\index{reference patch} constructed from the stack of frames. The reference patch is typically the temporal average of all the frames if the scene is static. As it is called, the reference patch is turbulence-jittering free so that it can serve the purpose of being a reference. Physically, this temporal average can also be considered a long-exposure image.

The meaning of the geometric consistency term is that we want to tell whether the current patch $\underline{\mI}(\vx,t)$ is a faithful sample in the stack of frames. If it is a good sample, then geometrically it should not deviate too much from the reference. The norm measuring the deviation between the two terms is therefore capturing the consistency.

When geometric consistency is added, the weight will become
\begin{align}
w_{\vx,t}(\Delta t)
&=
\underset{\text{geometric consistency weight}}{\underbrace{\exp\left\{-\alpha_1 \|\underline{\mI}(\vx,t+\Delta t) - \underline{\mI}_{\text{ref}}(\vx,t+\Delta t)\|^2 \right\}}} \notag \\
&\qquad \times
\underset{\text{sharpness weight}}{\underbrace{\exp\left\{\alpha_2 \|\underline{\mI}(\vx,t+\Delta t)\|_{\text{TV}_1}\right\}}}.
\end{align}
We can now replace the weighted sum using this new weight. Note that in this definition, the exponential function has a negative exponent with respect to the geometric consistency. So, if there is a large deviation between the current and the reference, the weight will be small so that we can skip the geometrically inconsistent patch.

\begin{figure}[h]
\centering
\includegraphics[width=0.8\linewidth]{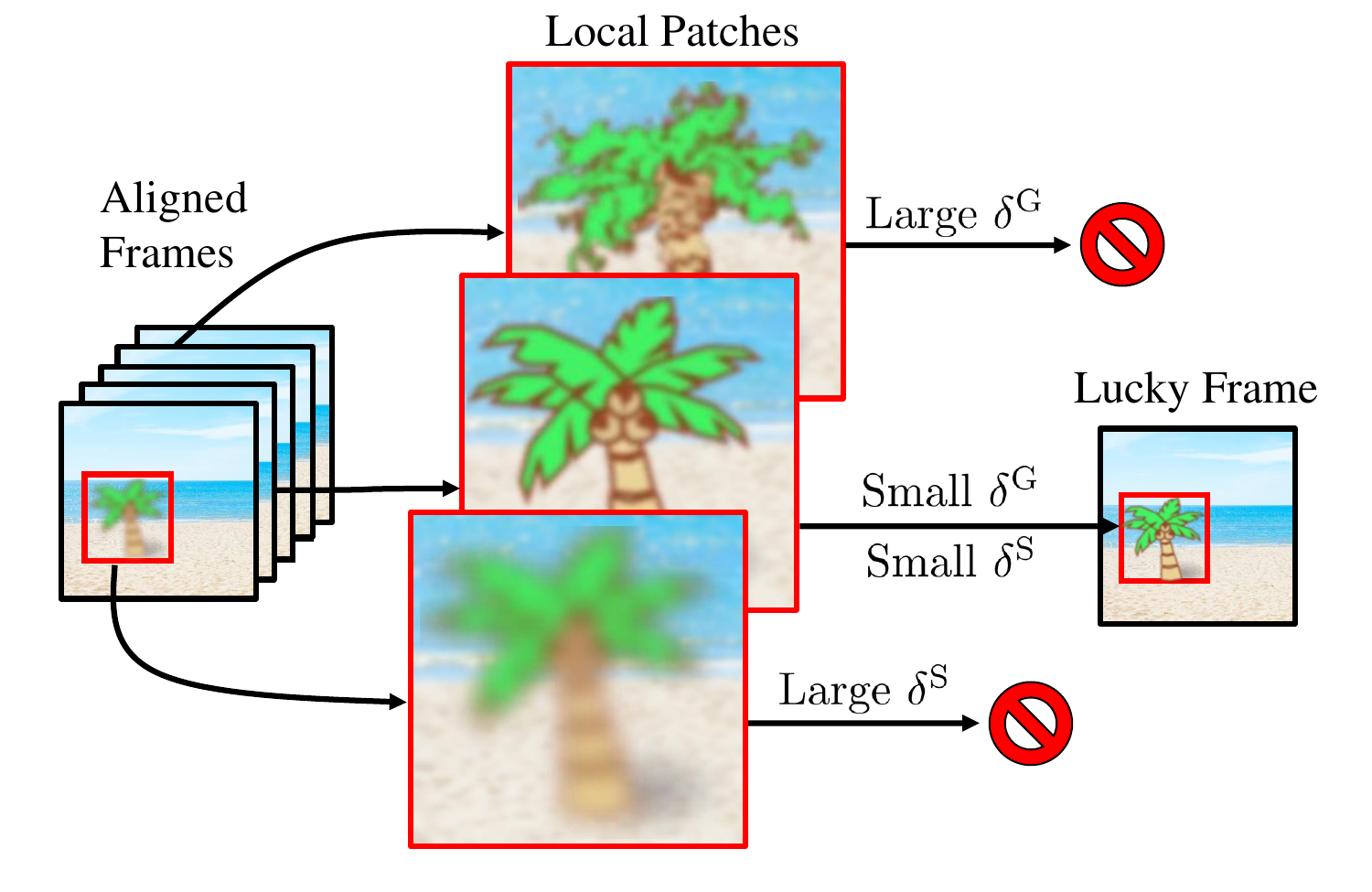}
\caption{Lucky imaging patch selection requires the patch to be geometrically consistent with the reference so that it is no longer distorted by the pixel jittering of the turbulence. The geometrical consistency score is measured by $\delta^G$. The other metric which is the sharpness metric is used to ensure that the selected patch is sufficiently sharp so that it is less affected by the atmospheric blur. Source: \cite{Mao_2020_a}.}
\label{fig: ch4 lucky}
\end{figure}

\fref{fig: ch4 lucky} is a pictorial illustration of the impact of the geometric consistency term and the sharpness term. This figure is adapted from \cite{Mao_2020_a}.

\textbf{From Lucky Patches to Lucky Image}. When we apply any of the above lucky fusion methods, the result is a patch instead of an image. To fuse an image, we can stitch multiple patches together. If there are overlaps among the patches, we take the average of the overlapped regions. The approach is similar to how BM3D performs its aggregation step. Readers interested in this can consult \cite{Dabov_2007_a}.

\textbf{Connecting to Inverse Problems}. Once the lucky image is formed, the inverse problem becomes
\begin{equation}
\widehat{\mJ} = \argmin{\mJ}  \;\; \|\mI_{\text{lucky}} - \vh \circledast \mJ \|^2 + \lambda \;\; g(\mJ).
\end{equation}
In this equation, $\vh$ represents the residual blur remaining in the lucky image $\mI_{\text{lucky}}$. We will come back to this point in the next subsection. But at the high level, we have converted \eref{eq: ch4 inverse problem 2} into a much simpler optimization problem where we only need to focus on the deconvolution task. Once the deconvolution is completed, the overall restoration problem is solved.

\textbf{Fourier Burst Accumulation}. The method we presented here broadly belongs to the family of block matching. There are alternative approaches in the literature such as Fourier burst accumulation \cite{Delbracio_2015_a, Gilles_2016_a}. \keyword{Fourier burst accumulation}\index{Fourier burst accumulation} operates in the frequency domain. If the input images are distorted by turbulence, the blur associated with the turbulence will make the Fourier spectrum of the image narrower (due to the lowpass effects). Thus, given an image stack, we look at the Fourier spectra of the individual frames. For each frequency in the spectrum, we assign a weight so that the processed spectrum is a weighted average of the input spectra. The weights are designed such that the widest spectrum (i.e., those suffering the least due to blur) will be given a higher weight. If the object does not move, a weighted average of the Fourier spectrum may be suitable to remove the blur.

\begin{figure}[h]
\centering
\includegraphics[width=\linewidth]{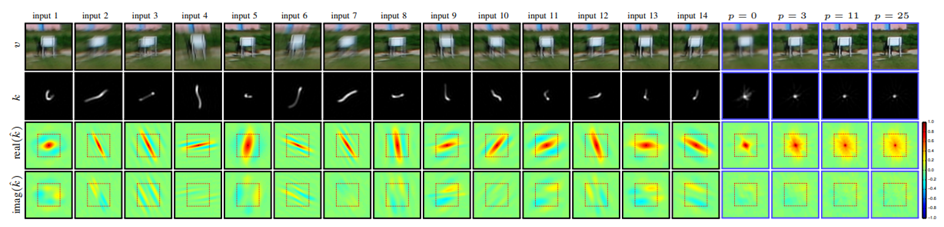}
\caption{Fourier burst photography constructs the weights by evaluating high-frequency components. Source: \cite{Delbracio_2015_a}.}
\label{fig: ch4 Fourier Burst}
\end{figure}

\section{Image Registration}
In this Chapter, we discuss another important element in turbulence mitigation: pixel registration. There are two objectives of pixel registration. Firstly, we need to compensate for the jittering effects caused by the turbulence. These effects are typically small but random. Secondly, if the scene contains moving objects, then the registration algorithm needs to compensate for the motion.

\subsection{Registration for Non-Rigid Deformation}
\index{non-rigid deformation}Consider a reference frame $I_{\text{ref}}(\vx,t)$ and a frame of interest $I(\vx,t)$. The problem of image registration is to find the displacement vector $\Delta \vx$ such that after reversing the motion, we can minimize any error between the prediction and the reference:
\begin{equation}
\widehat{\Delta \vx} = \argmin{\Delta \vx} \;\; \Big(I_{\text{ref}}(\vx,t) - I(\vx+\Delta\vx,t)\Big)^2,
\end{equation}
where the $\ell_2$ norm is just one of the many possible distance functions we can choose.

If we assume that the motion is simple and small, we can use the first-order approximation to write
\begin{align*}
I(\vx+\Delta \vx,t) = I(\vx,t) + \nabla_{\vx} I(\vx,t)^T \Delta \vx,
\end{align*}
where $\nabla_{\vx}$ denotes the spatial gradient of the image. Assuming that our goal is to set $I_{\text{ref}}(\vx,t) \approx I(\vx+\Delta \vx,t)$, we will reach the equation
\begin{equation}
\nabla_{\vx} I(\vx,t)^T \Delta \vx = I_{\text{ref}}(\vx,t)-I(\vx,t).
\end{equation}
The equation is sometimes called the \keyword{optical flow}\index{optical flow} equation. Since it is a system of linear equations, we can solve it using standard linear algebra methods. One of those is the \keyword{Lucas-Kanade}\index{Lucas-Kanade algorithm} algorithm. Because of the sheer volume of literature on this subject, we shall not repeat the implementation in this book. Generally speaking, optical flow today is very mature. There are plenty of Python / MATLAB packages one can use, including OpenCV \cite{Ranjan_2017_a, Liu_2009_a, Ilg_2016_a, Hui_2018_a}.

For atmospheric turbulence, since the turbulent effect is significantly less structured than object motion, running optical flow on a per-pixel basis is deemed impossible in the early days. Thus, during the late 00s when optimization algorithms were beginning to become popular, various motion models were proposed to improve the regularity of the problem. Among these models, the \keyword{non-rigid image registration} was the most widely adopted model for its robustness and simplicity.

\begin{figure}[h]
\centering
\begin{tabular}{cc}
\includegraphics[width=0.45\linewidth]{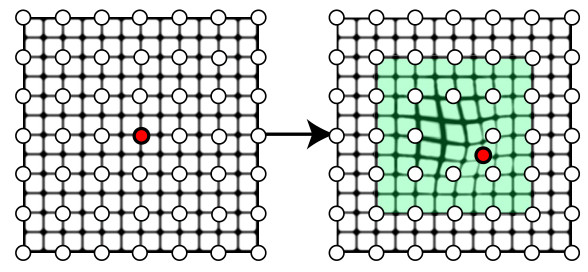}&
\hspace{-2ex}
\includegraphics[width=0.45\linewidth]{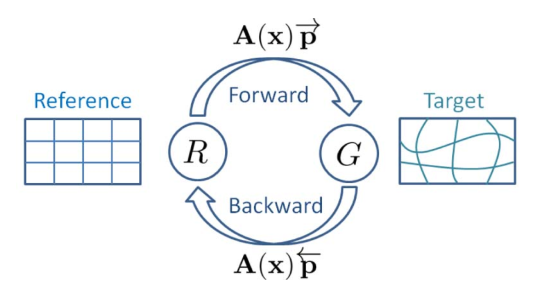}\\
(a) Source: \cite{Shimizu_2008_a} & (b) Source: \cite{Milanfar_2013_a}
\end{tabular}
\caption{Non-rigid motion estimation is commonly used to handle the turbulence image registration problem.}
\label{fig: ch4 Non-rigid}
\end{figure}

The main argument behind the non-rigid image registration methods is that in a natural image, not all pixels need to be compensated for. For example, if we have a flat region, there is no need to run any optical flow for pixels there even if we know the turbulence distortion is not static. Since there are a limited number of interest points in the image, the image registration algorithm can mainly focus on them. In addition, turbulence is correlated -- if one pixel is shifting to the right, then its adjacent pixel should also shift to the right. This allows us to use the concept of \keyword{control points} and \keyword{spline interpolation} to identify the locations of these carefully chosen control points, as shown in \fref{fig: ch4 Non-rigid}.

Non-rigid image registration has the advantage that it reduces the number of variables to a small set of control points. This makes the estimation significantly easier. The downside, however, is that the method is largely agnostic to turbulence physics. In fact, turbulence never follows a spline model. Even if some correlations can be approximated by a spline, the hyper-parameters defining the spline model need to be determined too. As algorithms evolve, we see that in the early 2020s, people have gone back to optical flow for a more precise estimation of the motion. The efforts are spent on improving the reference images and using deep neural networks to pull better features.

\subsection{Reference Frame}
For any image registration algorithm to work, we must have a \keyword{reference frame}\index{reference frame} so that the current frame has a reference point when mapping the pixels. For example, if we want to align the $(t+1)$-th frame with respect to the $t$-th frame, we would call the $t$-th frame a reference frame. In turbulence, the difficulty is that all frames in the image stack are distorted. Thus, aligning the $(t+1)$-th distorted frame to the $t$-th distorted frame does not do anything useful. As such, we need a method to first recover a reasonable reference frame.

There are multiple ways to construct a reference frame. We highlight a few methods here:

\textbf{Temporal Average}. The simplest solution is to construct a temporal average, as done in \cite{Anantrasirichai_2013_a, Milanfar_2013_a, Anantrasirichai_2018_a, Nieuwenhuizen_2019_b}. The idea is to compute
\begin{equation}
I_{\text{ref}}(\vx) = \frac{1}{T}\sum_{t=1}^T I(\vx,t),
\end{equation}
where $T$ is the number of frames we have in the stack. If the scene is static, the temporal average will give us the \keyword{long-exposure} image\index{long-exposure image} \cite{Hardie_2017_a}. A long-exposure image is tilt-free because according to Tatarskii the tilts follow a zero-mean Gaussian process \cite{Tatarski_1967_a}. However, as we presented in the turbulence physics Chapter, the long-exposure image is very blurry. Furthermore, if the scene contains a moving object, it is possible that the moving object will be washed out.

\textbf{Robust PCA}.\index{Robust principal component analysis} Another approach that can construct a reference frame is the robust principal component analysis (RPCA) technique, where we refer readers to \cite{Lau_2017_a} for more detailed discussions. The idea is to consider the observed video $\calI = [\mI(t_1),\ldots,\mI(t_T)] \in \R^{N \times T}$ that consists of $T$ frames with each frame containing $N$ pixels. RPCA solves the optimization problem
\begin{align}
\minimize{\calL, \calS} &\;\; \|\calL\|_* + \lambda \|\calS\|_0 \notag\\
\subjectto &\;\; \calI = \calL + \calS.
\end{align}
In this minimization problem, the norm $\|\cdot\|_*$ is the nuclear norm which is the sum of the eigenvalues of the matrix. The norm $\|\cdot\|_0$ is the sparsity norm that measures the number of non-zeros in the matrix. The minimization problem says that the video $\calI$ can be decomposed into a \keyword{low rank} part $\calL$ and a \keyword{sparse} part $\calS$. Using our language, we can treat the low rank part as something that does not move, i.e., the reference frame(s). The moving part is encoded by $\calS$ because we assume that the movement is mostly sparse.

RPCA is computationally heavy even though the minimization problem is convex. We need to resort to the most advanced convex optimization solvers to solve the rank minimization. Generally speaking, although many papers have been published for RPCA (beyond the context of turbulence), the actual usage in practice is not common. Besides the computational issues, the solution provided by RPCA is also not guaranteed to be good because minimizing the nuclear norm does not mean that we can extract a good reference image. If the scene contains a moving object, it is likely that RPCA will wash out it too.

\textbf{Non-local Averaging}.\index{non-local averaging} One of the more robust approaches is to perform a space-time nonlocal average instead of a simple temporal average \cite{Mao_2020_a}. The idea is to construct a reference patch $\mI_{\text{ref}}(\vx,t)$ via a linear combination of nearby patches:
\begin{equation}
\underline{\mI}_{\text{ref}}(\vx,t) = \frac{\sum\limits_{\Delta t \in \Omega_T} w_{\vx,t}(\Delta t) \underline{\mI}(\vx,t+\Delta t)}{\sum\limits_{\Delta t \in \Omega_T} w_{\vx,t}(\Delta t)}.
\end{equation}
In this equation, $w_{\vx,t}(\Delta t)$ is a weight that will be defined shortly. The weight is a function of the time stamps $\Delta t$ in a temporal neighborhood $\Omega_T = \{\Delta t_1, \ldots, \Delta t_T\}$. Thus, over the time axis, we construct the weighted average using patches $\underline{\mI}(\vx,t+\Delta t)$ that are determined to be similar to $\underline{\mI}(\vx,t)$.

Why do we ignore the spatial neighborhood? Isn't $\underline{\mI}(\vx,t)$ also a distorted patch? Both can be answered by looking at how the weight $w_{\vx,t}(\Delta t)$ is defined. We consider a standard $\ell_2$ distance between the patch $\underline{\mI}(\vx,t)$ and $\underline{\mI}(\vx+\Delta \vx, t + \Delta t)$:
\begin{equation}
\delta_{\vx,t}(\Delta \vx, \Delta t) = \|\underline{\mI}(\vx,t)-\underline{\mI}(\vx + \Delta \vx, t + \Delta t)\|^2.
\end{equation}
This is nothing but a pairwise comparison between the current patch $\underline{\mI}(\vx,t)$ and \emph{everyone} in the space-time neighborhood. Among these pairwise distances, we search along the spatial axis and pick the smallest distance, for every $\Delta t$:
\begin{equation}
\widetilde{\delta}_{\vx,t}(\Delta t) = \min_{\Delta \vx \in \Omega_x} \delta_{\vx,t}(\Delta \vx, \Delta t),
\end{equation}
where $\Omega_x = \{\Delta x_1,\ldots,\Delta x_K\}$ is the spatial neighborhood. Intuitively, what this equation does is look at each adjacent frame. If there is \emph{one} patch in the adjacent frame that is similar to $I(\vx,t)$, we will consider it a useful candidate for the weighted average. This can be an effective way of differentiating between two situations: (i) If there is object motion, then $\delta_{\vx,t}(\Delta \vx, \Delta t)$ will be large for every $\Delta t$ because we cannot find a similar patch within a spatial neighborhood as the object moves to another location. So, we can effectively preserve the moving patches by assigning a smaller weight so that the weighted average has less temporal effect. (ii) If there is turbulence, then the patches will jitter but not move away. Thus, for every $\Delta t$, it is likely that $\delta_{\vx,t}(\Delta \vx, \Delta t)$ will be small for some $\Delta \vx$. If we pick those patches, group them together, and take an average, the turbulence effect will be mitigated. Therefore, $\widetilde{\delta}_{\vx,t}(\Delta t)$ has a unique capability of handling both the object motion and the turbulence.

The weight is defined by taking the exponential function of the distance:
\begin{equation}
w_{\vx,t}(\Delta t) = \exp\{-\beta \widetilde{\delta}_{\vx,t}(\Delta t)\},
\end{equation}
where $\beta$ is a hyper-parameter that can be tuned to adjust the decay rate of the weight. Typically, $\beta$ is a function of the turbulence strength. If $D/r_0$ is large, $\beta$ will be small so that we can use more patches to form the weighted average. In contrast, if $D/r_0$ is small so that turbulence is weak, $\beta$ can be large so that we only use the least jittered patches for constructing the weighted average.

\begin{figure*}[h]
	\centering
	\begin{tabular}{ccc}
		\hspace{-2ex}
		\includegraphics[trim={3cm 0 2cm 0},clip, width=0.3\linewidth]{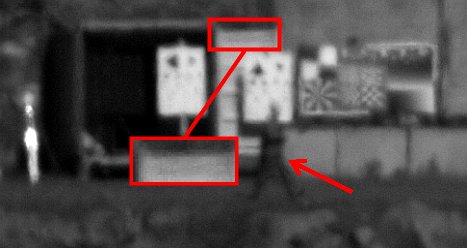} &
		\hspace{-2ex}
		\includegraphics[trim={3cm 0 2cm 0},clip, width=0.3\linewidth]{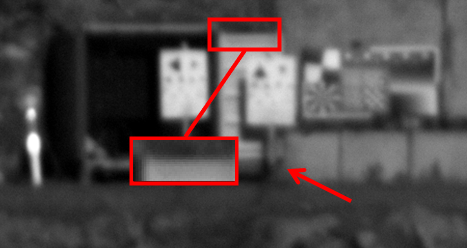} &
		\hspace{-2ex}
		\includegraphics[trim={3cm 0 2cm 0},clip, width=0.3\linewidth]{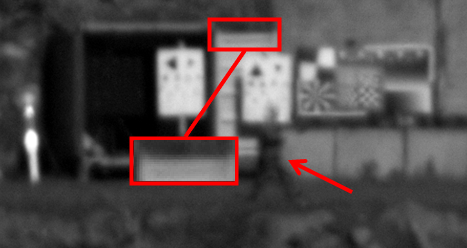} \\
		\hspace{-2ex} \small{(a) Raw input} &
		\hspace{-2ex} \small{(b) Temp. Avg} &
		\hspace{-2ex} \small{(c) Non-local}
	\end{tabular}
	\caption{Evaluating the reference frame method using real dynamic data. (a) Raw input. (b) Temporal averaging. Observe that the man is blurred over 100 frames. (c) Non-local averaging. Observe that the man remains in the image while the background is stabilized. Source: \cite{Mao_2020_a}.}
	\label{fig:ref_frame_comp 2}
\end{figure*}

A comparison of the reference frame generation method can be seen in \fref{fig:ref_frame_comp 2}, where we show the results of the temporal averaging method and the non-local method. The scene contains a static background with a moving foreground. As we can see, the temporal averaging washes out the moving foreground whereas the non-local method can preserve it.

\subsection{Other Image Registration Methods}
As far as pre-deep-learning methods are concerned, scenes containing both turbulence and moving objects are very difficult to align. People have considered approaches that extract the region of interest using segmentation methods \cite{Huebner_2012_a, Oreifej_2013_a, Nieuwenhuizen_2019_b, Halder_2015_a}. The problems with this approach are: (1) the object boundaries are often blurred due to motion. So unless we perform some kind of alpha matting algorithm to extract the transparency level, there is really no good methods to handle the boundaries. Needless to say, alpha matting itself is extremely complex and only works for a certain class of studio-level images. (2) After segmenting the object, most turbulence algorithms will proceed to mitigate the turbulence in the \emph{background}, e.g., \cite{Xue_2016_a}. While recovering the background is mathematically more well-posed, it often has little to no practical value because the object of interest is typically in the \emph{foreground}. As a result, segmentation-based methods are seldom used.

In the turbulence literature, a practically useful approach is the block matching concept \cite{Droege_2012_a, Hardie_2017_b}. The idea is similar to the non-local averaging approach we presented above. Instead of performing a soft decision as in the weighted averaging, we select patches according to some kind of similarity scores and perform frequency domain filtering. The complexity of the method is comparable to the non-local averaging. With graphics processing units (GPUs), these operations can be done in parallel and may be computed reasonably quickly.

\section{Image Deconvolution}
\label{sec: sec4_3}
\begin{figure}[h]
\centering
\includegraphics[width=\linewidth]{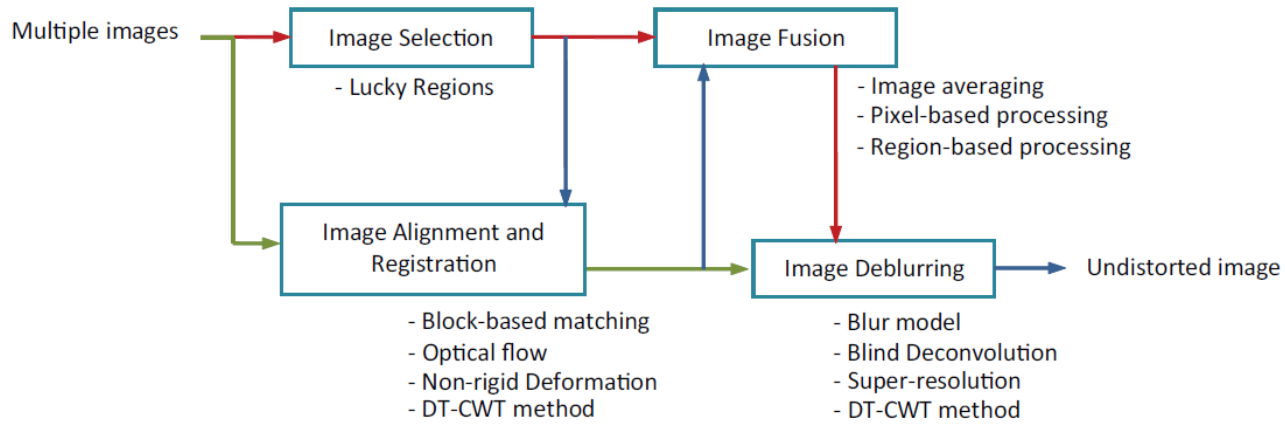}
\caption{Image restoration based on lucky imaging often requires image registration, lucky frame construction, and deblurring. Source: \cite{Vint_2020_a}.}
\label{fig: ch4 pipeline}
\end{figure}

\index{deconvolution}
Turbulence mitigation algorithms in the pre-deep-learning era are often implemented in a sequential manner where we first align the images and perform lucky fusion. Then from the lucky image, we perform another step of image deconvolution to remove the residue blur. This type of step-by-step inversion is both a reflection of the physics and a realization of the need for regularizing any optimization problem. In \fref{fig: ch4 pipeline} we show a figure taken from \cite{Vint_2020_a} that is ``representative'' of the typical procedures a turbulence mitigation algorithm takes.

\subsection{Tilt-then-Blur in Inverse Problems}
As we examine this pipeline, we recognize that the order of mitigating turbulence is often:
\begin{align*}
\widehat{\mJ} &= \texttt{Deblur}(\texttt{Lucky}(\mI)) \approx \calB^{-1}\calT^{-1}(\mI),
\end{align*}
where we recall the tilt $\calT$ and $\calB$ blur operator. The interesting observation here is that while the true forward model is $\calB(\calT(\cdot))$, i.e., tilt-then-blur, the inversion does not follow the order of deblurring-then-detilting. Instead, the inversion is detilting-deblurring.

The discrepancy between the forward model and the inverse method needs to be justified from a practical point of view. While deblurring \emph{should} be performed first, it is also \emph{much harder} to perform first. Since the blur in turbulence is spatially varying, deblurring means that we need to estimate all PSFs, one per pixel, in order to recover the image. Without further elaborating on the deblurring algorithms, this PSF estimation alone is enough to convince us that it is nearly an impossible task. Moreover, we need to make sure that the deblurred image (which is often low-passed) still contains the meaningful tilt (which is high-frequency content) for the de-tilting algorithm to work. So, although we want to deblur before de-tilt, practically it is impossible.

Tilt removal (including image registration and lucky fusion) has an immense advantage in that after these two steps, the blur in the lucky image is more or less spatially invariant. The reason is that during the lucky selection process, we only pick the sharpest patches from the image stack to construct the lucky image. Since we are selecting patches over many frames, we are effectively eliminating bad PSFs and aiming for a delta-function PSF. If all the patches are sharp, then all the corresponding PSFs will be more or less a delta function. So, we have converted the spatially varying blur problem to an invariant blur problem. The merit of the computational efficiency improvement clearly outweighs the mismatch with the forward model. Therefore, although the forward model is tilt-then-blur, the inverse algorithm is often de-tilting and then deblurring.

\subsection{Why Image Deconvolution?}
As the final step of the turbulence mitigation pipeline, image deconvolution aims to boost the image resolution by removing any residue blur in the image. But where does the residue blur come from?

\begin{figure}[h]
\centering
\includegraphics[width=\linewidth]{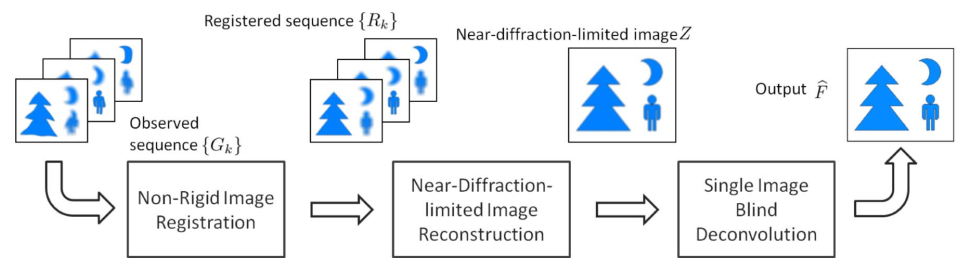}
\caption{A visualization of the deconvolution process after registration. Source: \cite{Milanfar_2013_a}.}
\label{fig: ch4 Diffraction Limit}
\end{figure}

\fref{fig: ch4 Diffraction Limit} is an excerpt from Zhu and Milanfar \cite{Milanfar_2013_a}. In this figure, the authors commented that after the lucky fusion step, the images are not completely sharp because the optical system is limited by diffraction. Therefore, even in the absence of turbulence, the acquired image will still experience some amount of blur due to the Airy disc of a finite aperture. An image deconvolution is thus needed to remove the diffraction-limited blur.

Readers at this point may ask: if the blur is purely caused by the optical system, then it should be fixed. It seems a fairly straightforward problem to measure the point spread function (PSF) of the optical system and perform a standard deconvolution step using any off-the-shelf algorithm such as Lucy-Richardson. The problem, however, is that although the lucky image is constructed from the sharpest patches from the image stack, there is no guarantee that the sharpest patches are blur-free. Since the digital lucky imaging step is not perfect, there is always some residual blur. These residual blurs are induced by the algorithm and are data-dependent, we cannot say let's measure the PSF and perform Lucy-Richardson. A critical component here is to \emph{estimate} the blur. This is why our deconvolution must be a \keyword{blind deconvolution}.

\subsection{Alternating Minimization}
\index{alternating minimization}
In what follows, we discuss a framework for handling the blur in turbulence mitigation. We emphasize that this framework is not the only way of removing the blur. It is based on the classical alternating minimization, which is relatively easy to understand the intuition behind. We shall focus on \emph{spatially invariant} blur, as we explained at the beginning of this deblurring Chapter.

Simultaneously estimating the image and the blur kernel is known as blind deconvolution \cite{Levin_2006_a, Fergus_2006_a}. In digital image restoration, the associated inverse problem is often written in terms of matrices and vectors. Denoting $\mI_{\text{lucky}} \in \R^N$ as the input blurry image (i.e., the whole lucky fused image instead of a lucky patch), $\vh \in \R^K$ as the blur kernel, $\mJ \in \R^N$ as the latent clean image, the optimization for the blind deconvolution is
\begin{equation}
(\widehat{\mJ}, \widehat{\vh}) = \argmin{\mJ,\vh} \; \|\mI_{\text{lucky}} - \mathbf{h} \circledast \mJ\|^2 + \lambda g(\mJ) + \gamma r(\vh).
\end{equation}
In this equation, the operator $\circledast$ denotes the spatially invariant convolution. The two functions $g(\mJ)$ and $r(\vh)$ are the regularization functions that constrain the solution space of the optimization. The two scalars $\lambda$ and $\gamma$ are used to control the relative emphasis between the forward model and the regularizations.

Since simultaneously optimizing for $\mJ$ and $\vh$ is a non-convex problem, the standard approach to solve the problem is to consider an alternating minimization strategy by fixing one variable and minimizing over the other variable \cite{Levin_2011_a, Michaeli_2014_a, Cai_2009_a, Krishnan_2011_a}:
\begin{align*}
\mJ^{k+1}           &= \argmin{\mJ} \; \|\mI_{\text{lucky}} - \mathbf{h}^k \circledast \mJ\|^2       + \lambda g(\mJ),\\
\mathbf{h}^{k+1}    &= \argmin{\vh} \; \|\mI_{\text{lucky}} - \mathbf{h}   \circledast \mJ^{k+1}\|^2 + \gamma r(\vh).
\end{align*}
In this pair of equations, the superscript $k = 1,2,\ldots$ denotes the iteration number. As the algorithm makes progress, we expect the estimate of the blur kernel and the image to improve.

The success/failure of the alternating minimization is largely driven by the choice of the regularization functions $g(\mJ)$ and $r(\vh)$. The regularization function $g(\mJ)$ encapsulates the image prior which is usually learned through data. One relatively mature procedure to solve the $\mJ$-subproblem is the \keyword{plug-and-play ADMM}\index{plug-and-play ADMM} (PnP) algorithm \cite{Venkatakrishnan_2013_a, Zhang_2017_a, Chan_2017_a, Romano_2017_a}. The idea is to split the $\mJ$-subproblem into two steps where one step inverts the forward imaging model using the current blur kernel $\vh^k$ and the other step uses a pre-trained deep neural network image denoiser to handle the noise. The neural network denoiser provides an implicit modeling of the image prior $g(\mJ)$. We skip the details of the PnP algorithm and refer the readers to monographs such as \cite{Ahman_2020_a, Chan_2019_a}.

For the other regularization $r(\vh)$, since we are working on turbulence-related blur, we should leverage this piece of information. Generic blur models such as those for motion blur or out-of-focus blur are less relevant although they are popular in the literature. The approach we present here is derived from the phase-to-space (P2S) transform. In some sense, we are re-using the outcomes of the P2S framework for the deblurring purpose.

\subsection{Modeling the Blur}
Recall that when we train the P2S transform, we have constructed a dataset of PSFs. We denote these PSFs as $\vh_1,\ldots,\vh_M$, where $M$ is a large number presenting the number of example PSFs we have simulated. The PSFs are tilt-free, meaning that they are generated from the turbulence simulator with the tilts removed. The PSFs are sent to a principal component analysis (PCA) to extract the principal components. Denote these principal components as $\boldsymbol{\varphi}_1,\ldots,\boldsymbol{\varphi}_L$, which are constructed from
\begin{equation}
\{\boldsymbol{\varphi}_\ell\}_{\ell=1}^L = \texttt{PCA}(\vh_1,\ldots,\vh_M),
\end{equation}
where \texttt{PCA} stands for the principal component analysis.

Assuming that the principal components are determined, the PSFs can then be represented via
\begin{equation}
\vh = \sum_{\ell=1}^L w_\ell \boldsymbol{\varphi}_{\ell}.
\end{equation}
for some representation coefficients $\{w_{\ell} \;|\; \ell = 1,\ldots,L\}$. Substituting this into the alternating minimization, we obtain a different optimization from an optimization in $\vh$
\begin{align*}
\mathbf{h}^{k+1}
&= \argmin{\vh} \; \|\mI_{\text{lucky}} - \mathbf{h}   \circledast \mJ^{k+1}\|^2 + \gamma r(\vh),
\end{align*}
to an optimization in $\vw = \{w_{\ell} \;|\; \ell = 1,\ldots,L\}$:
\begin{align}
\mathbf{w}^{k+1}
&= \argmin{\vw} \; \left\|\mI_{\text{lucky}} - \left(\sum_{\ell=1}^L w_\ell \boldsymbol{\varphi}_{\ell} \right) \circledast \mJ^{k+1} \right\|^2 + \gamma r(\vw) \label{eq: w update}\\
\mathbf{h}^{k+1}
&= \sum_{\ell=1}^L w_\ell^{k+1} \boldsymbol{\varphi}_{\ell}. \notag
\end{align}
The difference between the first and the second equation is the change of optimization variable from $\vh$ to $\vw$. In the first equation, we directly estimate the blur kernel $\vh$ without any constraint. In the second equation, we use a parametric model to confine the search space of the blur kernel. All our kernels are generated from the linear combination of the basis functions $\boldsymbol{\varphi}_{\ell}$. As long as we can specify the basis coefficients $w_{\ell}$, we can construct the blur kernel. Here we have written the PSF as spatially invariant (an approximation that may be justified by the use of lucky fusion beforehand), though works such as Novak et al. \cite{Novak_2021_a} have performed similar decomposition and a spatially varying deconvolution or spatially varying pixel-shift estimation as in the case of water distortions \cite{Yuandong_2009_a}.

The advantage of the optimization of $\vw$ over the optimization of $\vh$ is that the regularization function $r(\vh)$ is difficult to formulate whereas $r(\vw)$ can be statistically determined. In particular, by analyzing the distribution of the coefficients for a large number of example PSFs, it was found that a reasonable choice of the regularization function is
\begin{equation}
r(\vw) = \sum_{\ell=1}^L \frac{|w_\ell|}{\sigma_{\ell}},
\label{eq: ch4 model r(w)}
\end{equation}
where $\sigma_{\ell}$ is a learnable parameter, in this case the standard deviation of $w_\ell$ learned from data. This $r(\vw)$ is a mixture of weighted $\ell_1$ regularization. With combined with a standard $\ell_2$ loss in \eref{eq: w update}, the problem can be solved with convex solvers such as CVX or the ADMM algorithm.

\begin{figure}[h]
\centering
\includegraphics[width=0.75\linewidth]{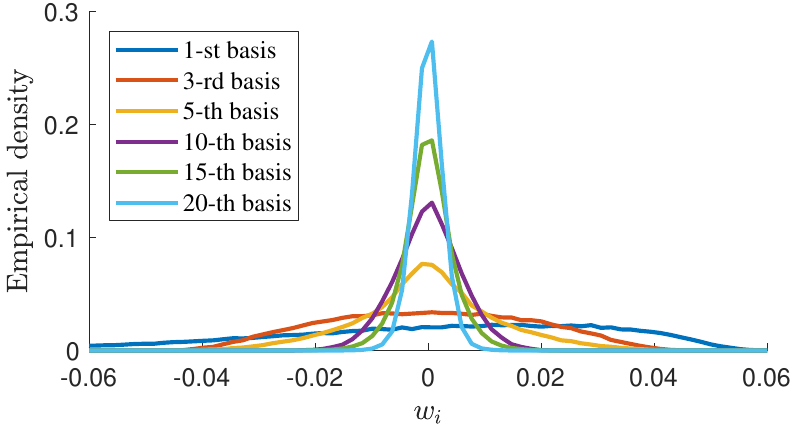}
\caption{Statistical distribution of the basis coefficients $w_{\ell}$. Except for the first few bases, the other basis coefficients demonstrate an exponentially decaying distribution. We can empirically fit this distribution with a parametric distribution. It was found that this is the exponential of the weighted sum of absolute values. Source: \cite{Mao_2020_a}.}
\label{fig: ch4 weight}
\end{figure}

Empirically, we can plot the distribution of the actual coefficients (by numerically calculating them from the raw PSFs). This will tell us how good \eref{eq: ch4 model r(w)} is and how to estimate the parameters $\sigma_{\ell}$. As shown in \fref{fig: ch4 weight}, we observe a Laplacian type of distribution for the weights $w_{\ell}$ using a fixed PCA basis. The parameters $\sigma_{\ell}$ are estimated from this plot to maximize the fit of \eref{eq: ch4 model r(w)} to the data. This also explains why if we naively assume that $r(\vw) = \|\vw\|_1$, we will not be able to capture the weights $\sigma_\ell$ as demonstrated in \fref{fig: ch4 weight}.

\begin{figure*}[t]
    \centering
    \begin{tabular}{c c c c c c}
    \hspace{\nspacetwo ex}\includegraphics[width=\nwidth\linewidth]{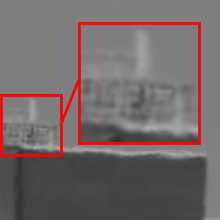} &
	\hspace{\nspace ex}\includegraphics[width=\nwidth\linewidth]{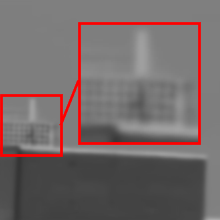} &
    \hspace{\nspace ex}\includegraphics[width=\nwidth\linewidth]{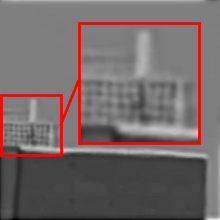} &
    \hspace{\nspace ex}\includegraphics[width=\nwidth\linewidth]{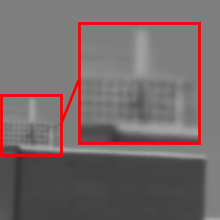} &
    \hspace{\nspace ex}\includegraphics[width=\nwidth\linewidth]{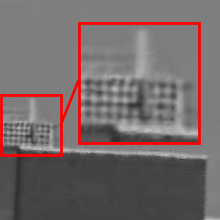} &
    \hspace{\nspace ex}\includegraphics[width=\nwidth\linewidth]{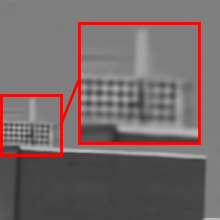} \\
	\hspace{\nspacetwo ex}\includegraphics[width=\nwidth\linewidth]{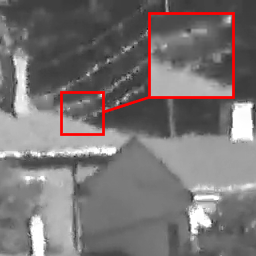} &
	\hspace{\nspace ex}\includegraphics[width=\nwidth\linewidth]{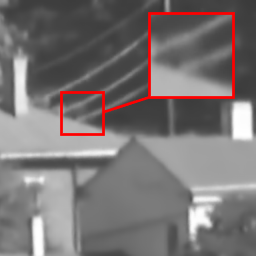} &
    \hspace{\nspace ex}\includegraphics[width=\nwidth\linewidth]{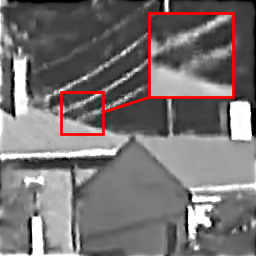} &
    \hspace{\nspace ex}\includegraphics[width=\nwidth\linewidth]{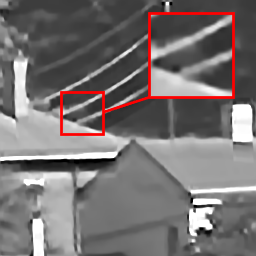} &
    \hspace{\nspace ex}\includegraphics[width=\nwidth\linewidth]{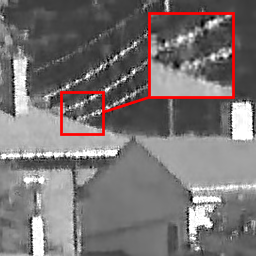} &
    \hspace{\nspace ex}\includegraphics[width=\nwidth\linewidth]{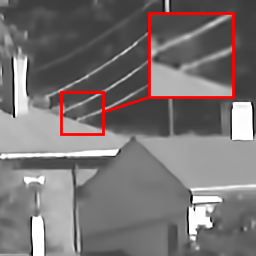} \\
	\hspace{\nspacetwo ex}\includegraphics[width=\nwidth\linewidth]{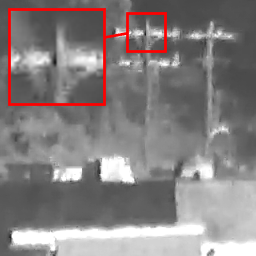} &
	\hspace{\nspace ex}\includegraphics[width=\nwidth\linewidth]{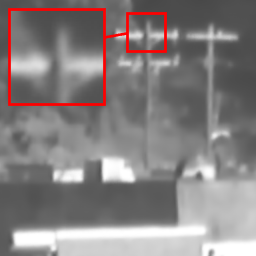} &
    \hspace{\nspace ex}\includegraphics[width=\nwidth\linewidth]{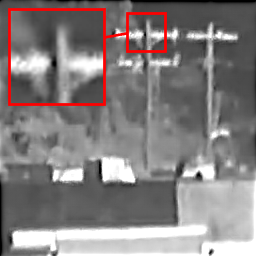} &
    \hspace{\nspace ex}\includegraphics[width=\nwidth\linewidth]{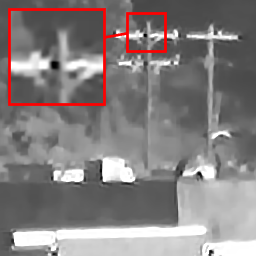} &
    \hspace{\nspace ex}\includegraphics[width=\nwidth\linewidth]{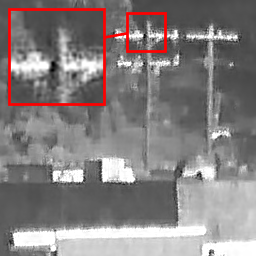} &
    \hspace{\nspace ex}\includegraphics[width=\nwidth\linewidth]{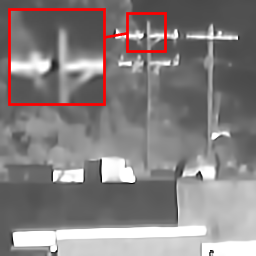} \\
    \footnotesize{(a) Input} & \hspace{-2ex} \footnotesize{(b) Avg. }& \hspace{-2ex} \footnotesize{(c) SG} & \hspace{-2ex} \footnotesize{(d) NDL} & \hspace{-2ex} \footnotesize{(e) CLEAR}    & \hspace{-2ex} \footnotesize{(f) Mao} \\
                  &                        & \cite{Lou_2013_a}       & \cite{Milanfar_2013_a}    & \cite{Anantrasirichai_2013_a} & \cite{Mao_2020_a}
    \end{tabular}
    \vspace{-1ex}
    \caption{Overall comparisons using real long-range static sequences. The first sequence was obtained from Youtube. The rest were captured by Panasonic Full HD Camcorder HC-V180K (aperture diameter 24mm and focal length 103mm), at a distance of 4km and temperature of 30$^\circ$C.}
    \label{fig:realexp}
\end{figure*}

In \fref{fig:realexp} we compared a few state-of-the-art turbulence mitigation methods (in the pre-deep learning era). The methods we study include an optimization-based approach by Lou et al. \cite{Lou_2013_a}, a classical lucky-imaging-based approach by Zhu and Milanfar \cite{Milanfar_2013_a}, a complex wavelet fusion technique CLEAR by Anantrasirichai et al \cite{Anantrasirichai_2013_a}, and an improved lucky imaging technique using all the concepts mentioned above \cite{Mao_2020_a}. As we can see in these examples, the overall deblurring result of \cite{Mao_2020_a} is convincingly better than the competitors. Especially for small objects such as those line patterns in the image, the deblurring algorithm performs well.

\subsection{Beyond Deconvolution}
Turbulence mitigation does not always stop at image deconvolution. As early as Shimizu's work \cite{Shimizu_2008_a}, it was already mentioned that one can try to super-resolve the image. In a nutshell, we can assume that in addition to the blur and tilt problem, there is also a downsampling operation happening to the image. Thus, when solving the inverse problem, we can attempt to solve
\begin{equation}
\widehat{\mJ} = \argmin{\mJ} \;\; \|\mI - \mD \calH_{\vtheta}(\mJ)\|^2 + \lambda g(\mJ),
\end{equation}
where $\mD$ is a down-sampling operator, basically the identity matrix with alternating rows skipped. Our experience working on super-resolution is that deblurring is already hard enough. To further recover the lost high-frequency content of the image, adding a down-sampling operator is not going to help. In fact, the down-sampling operator makes the inverse problem even more ill-posed. If not regularized properly, it would be very difficult to solve the minimization. But if we regularize too much, then there is a danger of hallucinating the content instead of recovering it.

\section{Deep Learning Methods}
\label{sec: sec4_4}
\index{deep learning}
Significant progresses in image restoration algorithms to mitigate atmospheric turbulence are made since the early 2020s. One of the biggest contributing factors is the promise of the deep learning methods demonstrated in a large plethora of restoration tasks such as image/video denoising/deblurring. We will not dive into the development of generic image restoration tasks because it simply evolves too rapidly. Instead, the goal of \cref{sec: sec4_4} is to highlight a few relevant concepts in the context of atmospheric turbulence.

\subsection{The Importance of Data}
Deep learning methods, as the name suggests, are a collection of algorithmic (software) solutions that use deep neural networks to \emph{learn} how to restore an image from \emph{data}. The three key components of a deep learning method are

\begin{enumerate}
\setlength\itemsep{0ex}
\item[(i)] \textbf{Model}: Whether we have a carefully designed neural network to achieve the restoration goals.
\item[(ii)] \textbf{Computing resources}: Whether we have a powerful computer with a graphics processing unit (GPU) computer accomplish the training.
\item[(iii)] \textbf{Data}: Whether we have enough data to train the model.
\end{enumerate}

Among the three factors, the data is the most important one for a few reasons. Firstly, since the deep neural network is trained based on the data, feeding the model with a poorly designed dataset will lead to poor performance of the network. For example, when handling atmospheric turbulence, if the training dataset contains only one scenario captured on a specific day in a specific place (also at just one temperature and one optical path), then no matter how good the architecture is, it will only learn how to handle that particular imaging condition. When it is presented with a different imaging condition, the network will fail. Secondly, network architectures today are more or less based on similar concepts, e.g., convolutional layers, attention layers, multi-scale, etc. While some configurations are better fit for atmospheric turbulence, their differences are not game changers if the dataset is poorly chosen.

While datasets are important, collecting one is unfortunately extremely challenging. Unlike standard problems such as image deblurring where we can easily synthesize the blur or shake the camera to collect the blurred image, in atmospheric turbulence we cannot easily ``turn off'' the turbulence as in the case of camera motion. We can capture the turbulence-distorted image on a hot day, but we will not be able to collect the ground truth on the same day. Some readers may think that if the target pattern is known a priori, we can use that as the ground truth. However, since the spectral response of the object to the camera changes from site to site (and day to day), the real ground truth is never known.

\textbf{Datasets Today (Overview)}: As of today, data collection for atmospheric turbulence is mainly divided into two categories:
\begin{itemize}
\setlength\itemsep{0ex}
\item Characterizing the turbulence, typically done by the defense/physics community, and
\item Evaluating algorithms, typically done by the image processing community.
\end{itemize}

\begin{table}[h]
\begin{tabular}{llllll}
\hline
Dataset & Publication & Approach        &  videos & range & Public? \\
\hline
NATO  & \cite{Tofsted_2007_a}, 2007  & Real                & n/a  &        & $\times$\\
OTIS  & \cite{Gilles_2017_a}, 2016       & Real           & 16   &        & $\checkmark$\\
CLEAR & \cite{Anantrasirichai_2013_a}, 2012       & Hot air        & 3    & 1km    & $\checkmark$\\
      &        & Synthetic      & 8    & --     & $\checkmark$\\
AFRL  & \cite{Velluet_2019_a}, 2019       & Real           & n/a  &  2.5km & $\times$\\
Nature& \cite{Jin_2021_a}, 2021       & Synthetic      &      &     & $\checkmark$\\
Mao   & \cite{Mao_2020_a}, 2020    & Real           & 5    & 4km & $\checkmark$\\
Army  & \cite{Yasarla_2022_a}, 2021      & Real, face     &      &     & $\times$ \\
UG2  & \cite{Mao_2022_a} \cite{UG2}, 2022   & Real, text     & 500  & 1km & $\checkmark$\\
TMT & \cite{Zhang_2022_a}, 2022           & Synthetic     &      &     & $\checkmark$\\
\hline
\end{tabular}
\caption{A partial list of atmospheric turbulence datasets reported in the literature. Most of these datasets are developed for testing/evaluating algorithms. Their volumes are usually very small compared to training data used to train deep neural networks.}
\label{table: ch4 dataset}
\end{table}

The focus and scale of the datasets vary significantly as one can see from Table~\ref{table: ch4 dataset}. For most military efforts, the data usually comes with $C_n^2$ measurements via a scintillometer. This can be useful for high-energy physics and optical communication through the atmosphere where we need to carefully characterize the received laser beam. However, because of the sensitivity of the data (and the nature of the acquisition process), the data is often classified. As such, for image processing algorithm developments, the community often collects their own datasets with a larger field of view and a wider range of image content.

\begin{figure}[h]
\centering
\includegraphics[width=\linewidth]{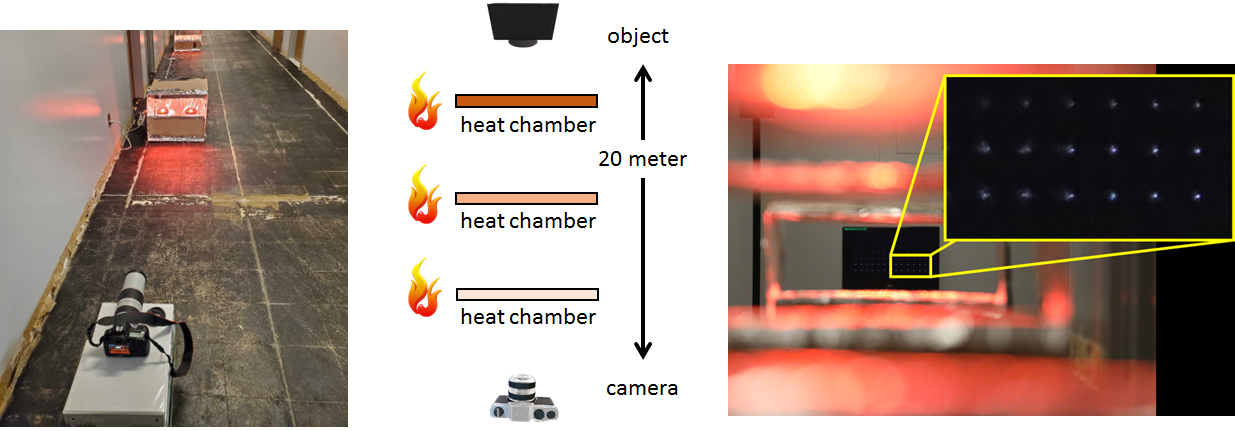}
\caption{A heat chamber is a good optical instrument to record a controllable level of turbulence. Shown here is a setup developed by Mao et al. \cite{Mao_2022_a} at Purdue University. It is a small scale, low-cost, and easier version of \cite{Velluet_2019_a}.}
\label{fig: ch4 chamber}
\end{figure}

\textbf{Testing Datasets}: Small academic research labs typically do not have access to high-end equipment. One workaround solution reported in the literature is to build a hot-air chamber \cite{Mao_2022_a}. This idea is similar to the heat tunnel reported in \cite{Velluet_2019_a} but on a much smaller scale. As shown in \fref{fig: ch4 chamber}, the heat chamber consists of multiple heat lamps located along the optical path. At the end of the optical path, a monitor is used to display the target pattern, and at the other end of the optical path, a camera with a long-range lens is used to capture the scene. By turning on and off the heat chambers, we can collect the turbulence and the ground truths, respectively. Some limitations of the heat chamber should be mentioned. For example, the heat lamps introduce a strong infrared signal that can alter the spectral profile of the scene. The color is thus distorted. Also, unless the total optical path is sufficiently long, there is no guarantee that one can accurately replicate the true long-range effect. The heat is localized and so the turbulence could be isoplanatic.\index{turbulence datasets! testing}

\begin{figure}[h]
\centering
\includegraphics[width=\linewidth]{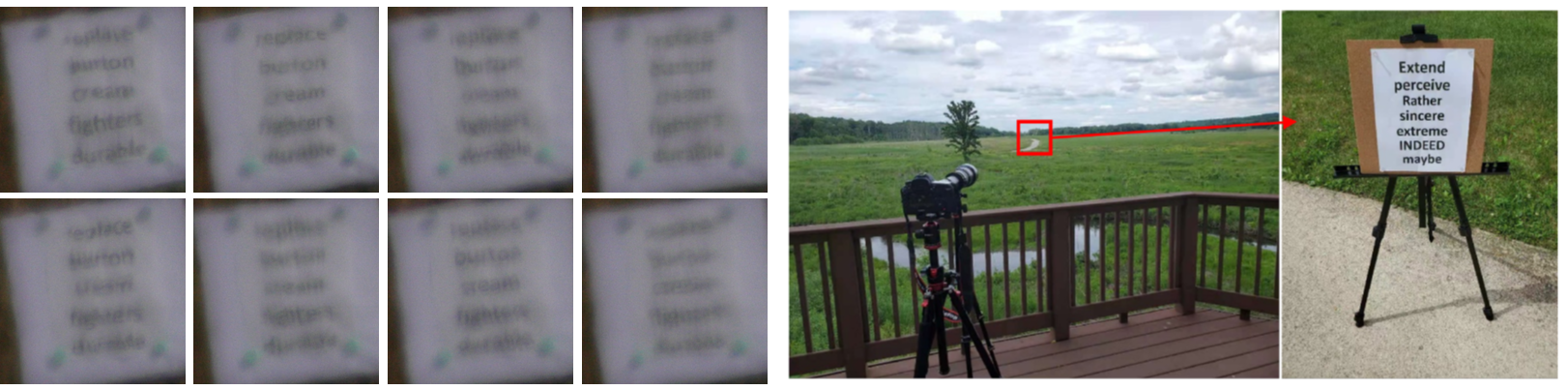}
\caption{Real text data collected by an off-the-shelf camera with an optical zoom. Shown on the left are example distorted images, whereas the subfigures on the right are the actual camera setup.}
\label{fig: ch4 text}
\end{figure}

For long-range data acquisition, the question one should ask is the purpose of the dataset. If the dataset is used to evaluate restoration algorithms, then the data must have some level of differentiating power to rank the methods. Generic images such as flowers and trees can be difficult because, in the absence of ground truth, typical metrics such as the peak signal-to-noise ratio (PSNR) cannot be used, with some analysis of metrics offered by Groff et al. \cite{Groff_2021_a}. An alternative solution here is to consider a \keyword{joint restoration-recognition} task where the restored images will be fed to a downstream object recognition algorithm to report the detected classes. For example, we can ask whether the restoration algorithm has successfully recovered the face of a subject by checking the predicted identity. \fref{fig: ch4 text} shows an example of \keyword{text recognition}. Texts are useful here because it directly tests whether the restoration can recover the desired resolution. We can prepare a list of texts of different fonts, and see to what scale the restoration can perform.

One ongoing debate in adopting deep learning-based algorithms is how much class specificity we can assume. For example, in CVPR 2022's UG2+ challenge, winning teams exploited the fact that the evaluation metric is based on optical character recognition. As such, the teams trained the neural networks using text data. While this is a legitimate approach for solving the specific task, it fails to demonstrate the other side of the developed neural networks which is to handle unseen situations. One can argue that class-specific training is valid because the users shall know the type of content and so the best model should be picked. Or, it is possible to pre-train a few models and use one model during inference. The opposite side would then argue that in many military missions, storing a suite of class-specific models can be difficult resource-wise.

\textbf{Training Dataset}: As far as training data is concerned, synthetic data is likely the path to move forward because this is the only way we can generate thousands to millions of examples images to train a deep neural network. As of today, a consistent finding in the literature is that a better turbulence simulator tends to give a better image restoration result, assuming that the network architecture is fixed. Table~\ref{tab:recon_psnr} shows a comparison between two turbulence simulators \cite{Chimitt_2020_a} and \cite{Lau_2021_a}. \cite{Chimitt_2020_a} is based on the phase-to-space transform and has a guarantee of the turbulence statistics. \cite{Lau_2021_a} assumes a non-rigid deformation with less consideration of turbulence physics. Table~\ref{tab:recon_psnr} indicates that when \cite{Chimitt_2020_a} is used to prepare a training dataset which is then used to train a U-Net, the performance is substantially better than that using \cite{Lau_2021_a}.\index{turbulence datasets! training}

\begin{table}[th]
    \centering
    \begin{tabular}{c c c c}
        \hline\hline
         $D/r_0$ & \cite{Mao_2020_a} & \cite{Chimitt_2020_a}+U-Net & \cite{Lau_2021_a}+U-Net\\
        \hline
        1.5 & 27.33dB & 27.18dB & 26.59dB \\
        3.0 & 27.04dB & 26.98dB & 26.11dB \\
        4.5 & 25.85dB & 26.01dB & 25.40dB \\
        \hline
    \end{tabular}
    \vspace{1ex}
    \caption{PSNR values of the reconstruction results, averaged over 30 testing sequences.}
    \vspace{-2ex}
    \label{tab:recon_psnr}
\end{table}

At the time of writing this book, the largest training dataset is the TMT dataset generated by Zhang et al. \cite{Zhang_2022_a}, summarized in Table~\ref{tab: TMT dataset}. The TMT dataset consists of static and dynamic parts. TMT uses the place dataset \cite{zhou2017places} for synthesizing static scenes. It randomly selects 9,017 images in the original dataset as input for the simulator. TMT generates 50 turbulence images and their associated distortion-free images for every single input, resulting in 9,017 pairs of image sequences of static scenes. TMT splits them into 7,499 pairs and 1,518 pairs for training and testing, respectively.

\begin{table}[h]
\caption{Specification of the TMT dataset, where each sequence for the static scene data has 50 frames.}
\begin{tabular}{p{1.5cm}p{3cm}p{5cm}}
\hline
& Static & Dynamic \\
\hline
Source & Place \cite{zhou2017places} & Sports \cite{safdarnejad2015sports} and TSRWGAN \cite{Jin2021NatureMI}\\
Amount & 9,017 sequences & 4,684 videos (1,979,564 frames)\\
Training & 7,499 sequences & 3,500 videos \\
Testing  & 1,518 sequences & 1,184 videos \\
\hline
\end{tabular}
\label{tab: TMT dataset}
\end{table}

For dynamic scenes, the TMT dataset contains many videos. The source datasets for our dynamic scene data are the Sports Video in the Wild (SVW) dataset \cite{safdarnejad2015sports} and all ground truth videos used in TSRWGAN \cite{Jin2021NatureMI}. These videos are mixed, generating 4,684 samples with a total number of frames of 1,979,564. TMT generates 4,684 pairs of full turbulence and distortion-free videos, then randomly split them into 3,500 videos for training and 1,184 for testing, keeping at most 120 frames per testing video.

\begin{figure}[h]
    \captionsetup[subfloat]{farskip=2pt, font=scriptsize}
    \centering
  \subfloat[Input]{%
    \includegraphics[width=0.45\linewidth]{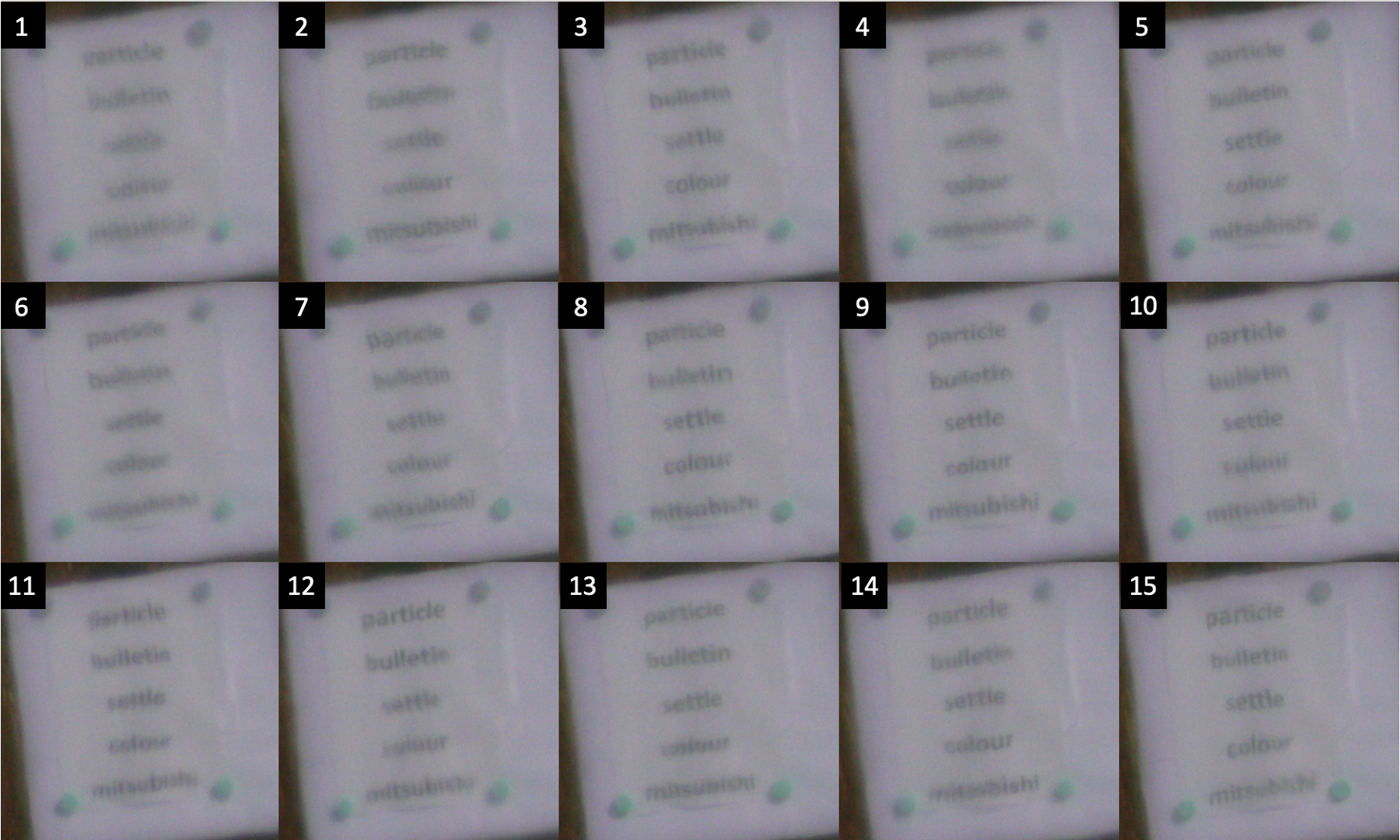}}
    \hfill
  \subfloat[Trained with TSRWGAN's data]{%
    \includegraphics[width=0.27\linewidth]{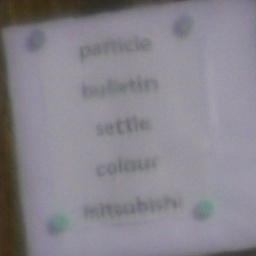}}
   \hfill
  \subfloat[Trained with TMT's data]{%
    \includegraphics[width=0.27\linewidth]{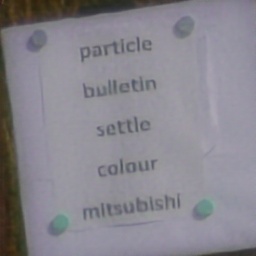}}
    \hfill
    \\
  \subfloat[Input]{%
      \includegraphics[width=0.33\linewidth, height=0.35\linewidth]{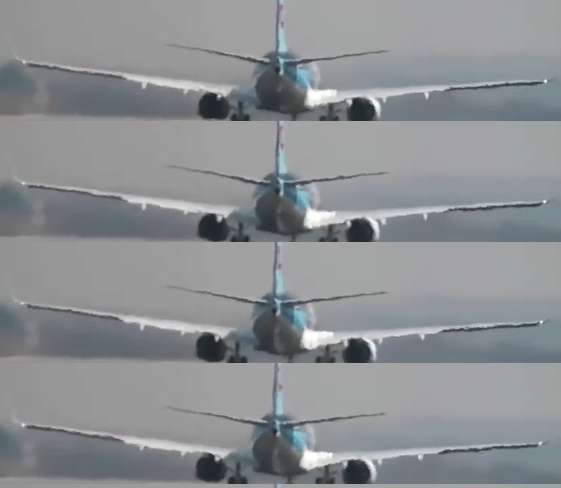}}
    \hfill
  \subfloat[Trained with TSRWGAN's data]{%
    \includegraphics[width=0.33\linewidth, height=0.35\linewidth]{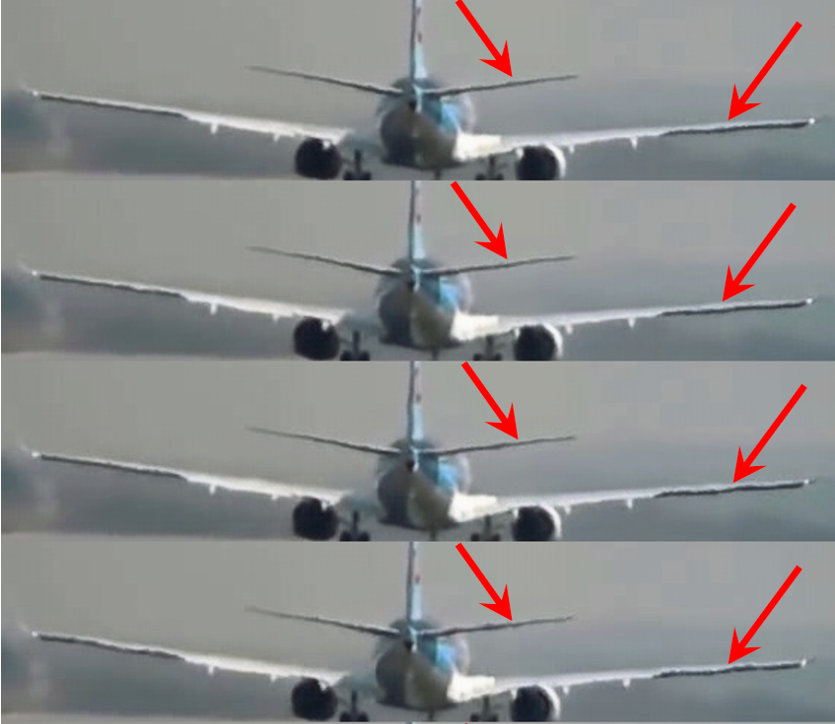}}
   \hfill
  \subfloat[Trained with TMT's data]{%
    \includegraphics[width=0.33\linewidth, height=0.35\linewidth]{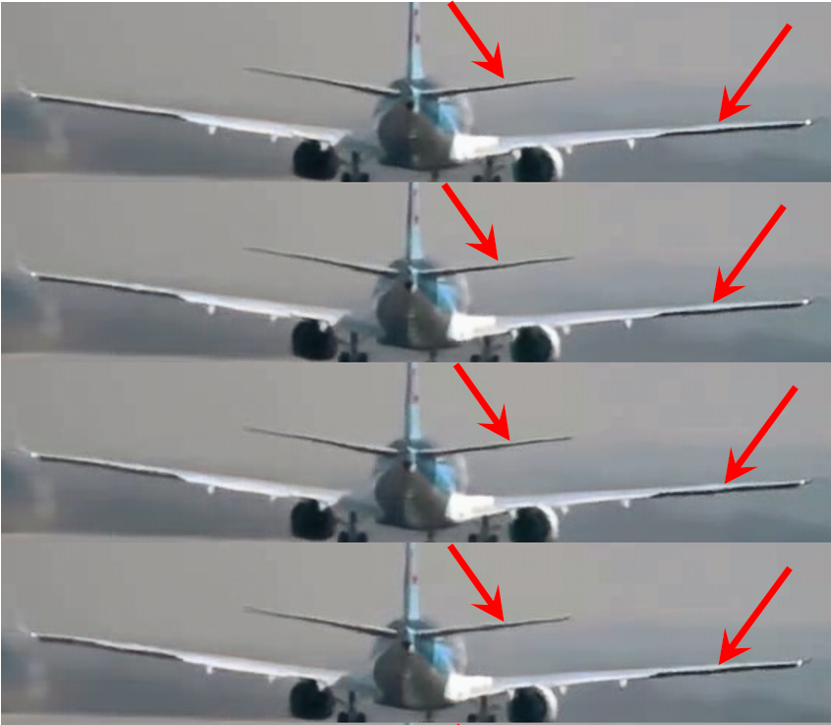}}
    \hfill
  \caption{Comparing the significance of data.}
  \label{fig:WGAN_comparison}
\end{figure}

The significance of the data can be seen from the comparison shown in \fref{fig:WGAN_comparison} where we compare the TMT dataset and the TSRWGAN dataset. The simulation tool that TSRWGAN \cite{Jin2021NatureMI} used is \cite{repasi2011computer}, which generates physically valid tilts and spatially invariant blur kernels, but higher-order aberrations are not simulated. \cite{Jin2021NatureMI} also produces synthetic data by artificial heat sources to create turbulence effects in a short distance. However, this approach tends to generate highly correlated degradation with a weak blur. We observed that their released model does not generalize well on CLEAR's real-world dataset \cite{{Anantrasirichai_2013_a}}, OTIS \cite{Gilles_2017_a} and the text dataset \cite{UG2} where the turbulence effect is stronger than that in the synthetic data used by TSRWGAN. Fine-tuning the TSRWGAN model with the TMT's data shows a significant improvement of the TSRWGAN model on those out-of-distribution datasets in Fig. \ref{fig:WGAN_comparison}.

\subsection{Network Designs}
The designs of deep neural networks for mitigating atmospheric turbulence can vary significantly from one to the other. In this subsection, we mention a few of the approaches documented in the recent literature.

\begin{figure}[h]
\centering
\includegraphics[width=\linewidth]{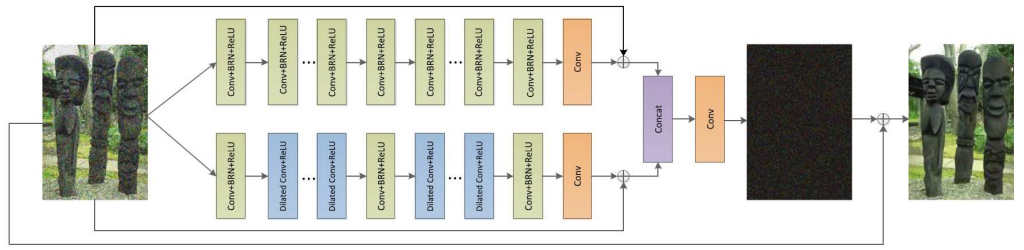}
\caption{Network architecture of BRDNet. This network is one of the older networks that use dilated convolutions and batch normalization. The performance of the network is not great for turbulence mitigation tasks as reported in \cite{Vint_2020_a}. Source: \cite{Tian_2020_a}.}
\label{fig: ch4_brdnet}
\end{figure}

In the simplest way, we can use an existing network architecture such as a UNet but trained with the synthetic turbulence data, e.g., the BRDNet shown in \fref{fig: ch4_brdnet}. A comprehensive report of the comparisons of several commonly used neural network architectures has been documented by Vint et al \cite{Vint_2020_a}. In that report, the authors compared BRDNet \cite{Tian_2020_a}, RDN \cite{Zhang_2021_a}, CAE-Unet \cite{Chen_2019_a}, SuperSR \cite{Feng_2019_a}, DnCNN \cite{Zhang_2017_a}, RCAN \cite{Zhang_2018_a}. The conclusion of the report was profoundly interesting, that none of these networks are able to produce a substantial improvement in image quality. For several of the testing scenarios, the peak signal-to-noise ratio (PSNR) of the restored image is just marginally better than the raw distorted image. Worse are some cases where the restored image introduces a large number of artifacts that make the image unrecognizable. We should mention that the networks are all trained using synthetic turbulence data containing one million images. However, to fit into the memory constraints, most network sizes are significantly reduced. Whether the shrinkage of the network capacity has an impact to the restoration is unknown.

\begin{figure}[t]
\centering
\includegraphics[width=0.75\linewidth]{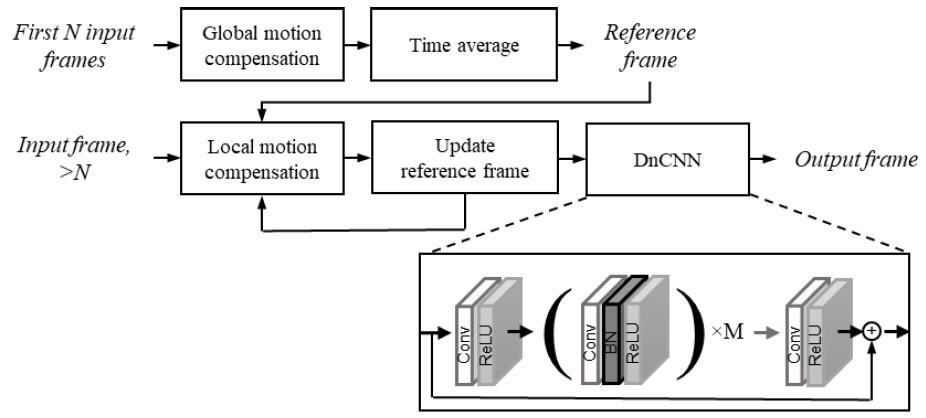}
\caption{Neural networks can be used as a module of the overall turbulence mitigation pipeline. In this work, the idea is to use DnCNN as a deblurring module after performing standard operations such as lucky imaging, registration, and fusion. The performance of such a design is limited by the first half of the pipeline which is a collection of classical techniques. Source: \cite{Nieuwenhuizen_2019_a}.}
\label{fig: ch4 simple}
\end{figure}

A natural improvement of the network architecture is to use the network as one module of the overall restoration pipeline, for example, in \fref{fig: ch4 simple}. Here, the network is used to replace a traditional image deblurring step which is usually a transform-domain method or an optimization method. Replacing a traditional module with a neural network ensures interpretability because all the other steps are based on classical approaches. It also allows users to debug if anything goes wrong.

The design shown in \fref{fig: ch4 simple} has several limitations. If we only use the deep neural network in the last stage of the pipeline, its functionality is only deblurring. The bigger challenge of aligning frames, picking the lucky patches, and handling the moving objects are all done by traditional methods. The rigidity of the framework does not fully leverage the power of deep learning that can be trained end-to-end. As we see in many of our experiments, end-to-end training is particularly important because it ensures that the intermediate features are relevant to the final output.

\begin{figure}[h]
\centering
\includegraphics[width=0.75\linewidth]{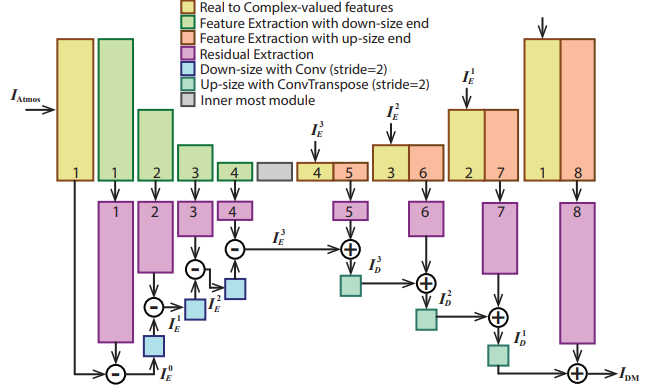}
\caption{Complex valued convolutional neural network proposed by Anantrasirichai \cite{Anantrasirichai_2022_a}. In this network, the idea is to use an encoder-decoder to extract the features and reconstruct the image. Residual connections and pyramid structures are introduced to improve the restoration. Source: \cite{Anantrasirichai_2022_a}.}
\label{fig: ch4 unet}
\end{figure}

\begin{figure}[!]
\centering
\includegraphics[width=0.75\linewidth]{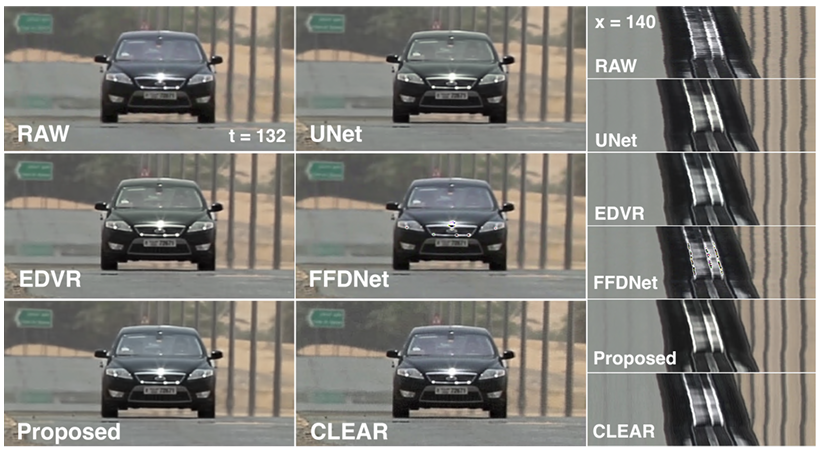}
\caption{Image reconstructions produced by \cite{Anantrasirichai_2022_a}. Source: \cite{Anantrasirichai_2022_a}.}
\label{fig: ch4 Anantrasirichai_network}
\end{figure}

If a simple re-use of an existing architecture and a simple plug-in to a classical framework both don't work, then a natural next step is to consider a fully end-to-end network that takes some turbulence physics into consideration. \fref{fig: ch4 unet} shows a complex UNet proposed by Anantrasirichai et al. \cite{Anantrasirichai_2022_a}. The idea is to use an encoder to extract the features of the distorted images, and to use a decoder to reconstruct the clean image from the codewords. In between the encoder-decoder, there is a pyramid residual module to pull and aggregate features at different scales (and in the residue domain.) A complex-valued network is used to handle the random jittering of turbulence. This makes the network reasonably similar to the classical methods such as CLEAR \cite{Anantrasirichai_2013_a} which also performs complex-value computations across different scales. In terms of performance, results shown \fref{fig: ch4 Anantrasirichai_network} appear to have fewer artifacts than some other networks such as UNet and FFDNet.

\subsection{Learning Uncertainty}
Moving away from a purely black box or a simple plug-in framework, there is an increasing effort to explicitly model the turbulence forward model during the restoration task. As we mentioned before, a turbulence inverse problem can generally be framed as (after flipping the order of tilt and blur for computational efficiency although we know that the correct order is tilt-then-blur):
\begin{equation}
\widehat{\mJ} = \argmin{\mJ} \;\; \|\mI - \calT (\calB(\mJ))\|^2 + \lambda \; g(\mJ),
\end{equation}
where $\calB$ denotes the blur operator, and $\calT$ denotes the tilt operator. The observed noisy image is denoted as $\mI$, and the optimization variable $\mJ$ is the latent recovered image. The regularization function $g(\mJ)$ is more of a placeholder to emphasize that we put regularizations over the latent image $\mJ$.

\begin{figure}[h]
\centering
\includegraphics[width=\linewidth]{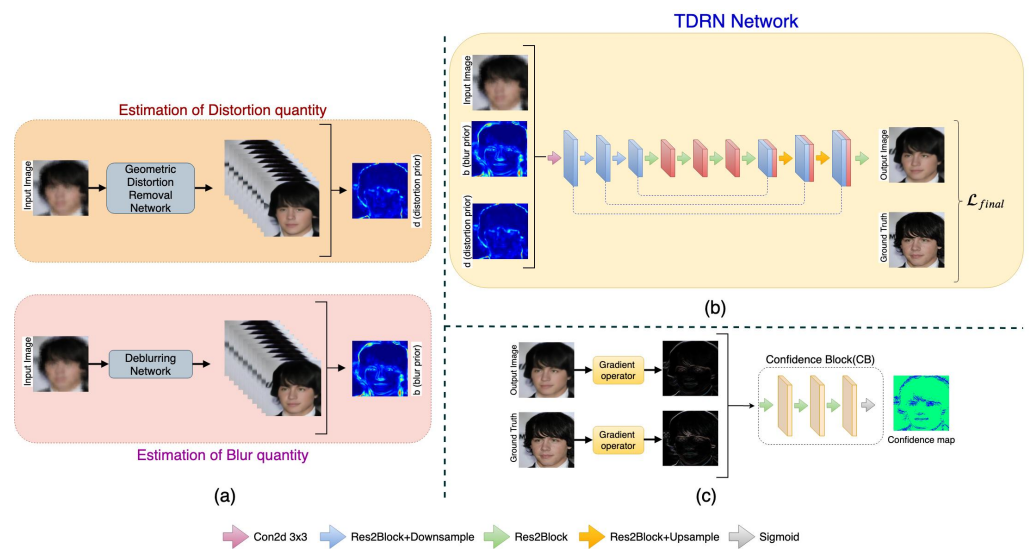}
\caption{Learning the uncertainty map alleviates the difficulty of learning the precise turbulence forward model. In this design, the network separately learns the uncertainty maps of the tilt and the blur. Then, the uncertainty maps are concatenated to the noisy input and processed by the restoration network. Source: \cite{Yasarla_2022_a}.}
\label{fig: ch4 patel}
\end{figure}

Recovering $\mJ$ would be a lot easier if we know $\calB$ and $\calT$. However, since atmospheric turbulence is random, knowing $\calB$ and $\calT$ requires us to know the instantaneous realizations of $\calB$ and $\calT$, which is not feasible in practice. The workaround solution here is to estimate the \keyword{uncertainty map}\index{uncertainty map} instead of the actual turbulence distortion. Proposed by Yasarla et al. \cite{Yasarla_2022_a}, the idea is to first generate an indicator map of the tilt and an indicator map of the blur, as shown in \fref{fig: ch4 patel}. These indicators are not the true $\calT$ and $\calB$, nor even an approximation of them. Instead, we should think of them as \emph{confidence maps} that are concatenated with the input (distorted) image to serve as an additional feature. Here, by confidence, we mean that if a pixel has a high value, then it is likely distorted by tilt (or blur). So the confidence maps tell us \emph{where} and \emph{how strong} the respective distortions are. After the estimation of the confidence maps, another network is used to perform the actual restoration.

The advantage of estimating the uncertainty map is that it bypasses the difficulty of recovering the actual turbulence distortion. However, the physical relationship between the uncertainty map and the turbulence is weak. There is no way we can take the uncertainty map, apply it to the clean image, and obtain the distorted image.

\subsection{Single-image Restoration by Re-degradation}
\begin{figure}[h]
\centering
\includegraphics[width=\linewidth]{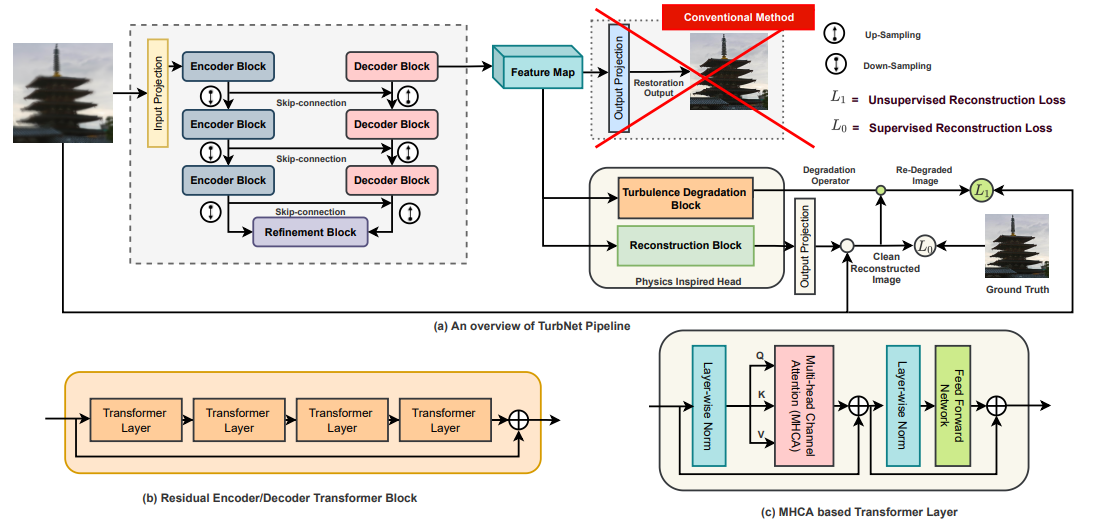}
\caption{Re-degradation and linear approximation model by Mao et al \cite{Mao_2022_a}. The core idea here is that after the features are extracted, we re-degrade the features so that the decoded image is the distorted frame. A simulated distorted frame is then compared with the re-degraded image to compute the loss. This allows us to perform self-supervised learning. Source: \cite{Mao_2022_a}.}
\label{fig: ch4 ajay}
\end{figure}

An alternative approach to the uncertainty map is the concept of \keyword{re-degradation}\index{redegradation} proposed by Mao et al \cite{Mao_2022_a}. Instead of constructing the uncertainty map and using it as an additional channel to the restoration neural network, a turbulence degradation block is introduced as a forward process to re-degrade the estimated image. The overall design is shown in \fref{fig: ch4 ajay}. Starting with the turbulence-distorted image, the method first extracts the features using any backbone feature extractor. In this case, the method uses the vision transformer. The extracted features are then sent to a pair of modules known as the turbulence degradation module and a reconstruction module. In conventional deep neural networks, the turbulence degradation module is absent and so the extracted features are directly sent to a decoder to reconstruct the image. Here, the extracted features are sent to the degradation model to re-generate the distorted image.

\begin{figure}[!]
\centering
\includegraphics[width=\linewidth]{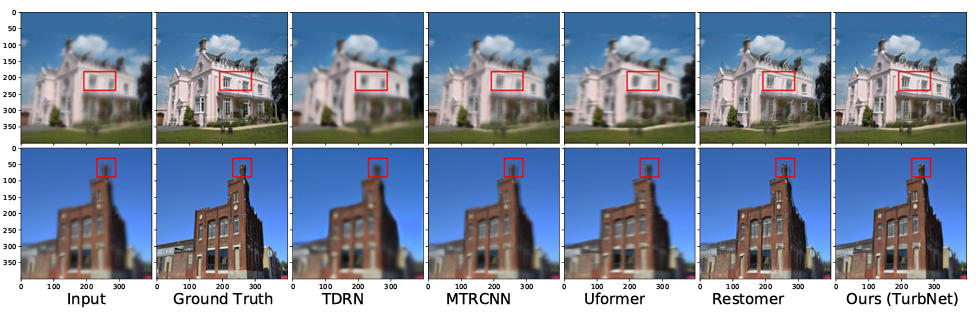}
\caption{Excerpt of the image reconstruction results reported by Mao et al. Source: \cite{Mao_2022_a}.}
\label{fig: ch4 ajay result}
\vspace{-2ex}
\end{figure}

The degradation module, in principle, should be a numerical turbulence simulator such as phase-to-space introduced in \cref{sec: sec3}. However, to use the full phase-to-space transform, it is necessary to make the simulator \keyword{differentiable}\index{differentiable simulator} so that we can back propagate the gradient to update the network parameters. Building a differentiable forward model is an ongoing effort, although some have reported initial success in doing so. In the absence of a full simulator, the degradation module in \fref{fig: ch4 ajay} can still be made by considering a \emph{linear} degradation model. As the name suggests, the linear degradation is an approximation to the full simulator. It is a trainable neural network that takes a restored image (or its features) and re-degrades it to produce the distorted image (or its features). This network can be fully convolutional or with some variations. The key is that it applies a set of linear filters to the input image (features) to produce the re-degraded image. It is a rough (but trained) approximation to the full simulator. On the positive side, the method bypasses the difficulty of recovering the exact turbulence distortion by implicitly using a network to learn how the distorted images are generated. \fref{fig: ch4 ajay result} shows some image reconstruction results.

The re-degradation concept opens the door to \keyword{self-supervised learning}\index{self-supervised learning}. By re-degrading the restored images, we avoid the difficulty of collecting a large amount of ground truth images which is often not feasible. When we start to train the network, we can use synthetic images to build a pre-trained model. Since synthetic images can be generated in whatever way the simulator allows, we can easily cover a range of turbulence conditions with ground truths. However, synthetic images always have a domain gap with real data. Being able to perform self-supervision makes the model fine-tuning  without ground truth achievable. As such, we can use self-supervision to bridge the domain gap, and hence improve generalization.

\begin{figure}[!]
\centering
\includegraphics[width=\linewidth]{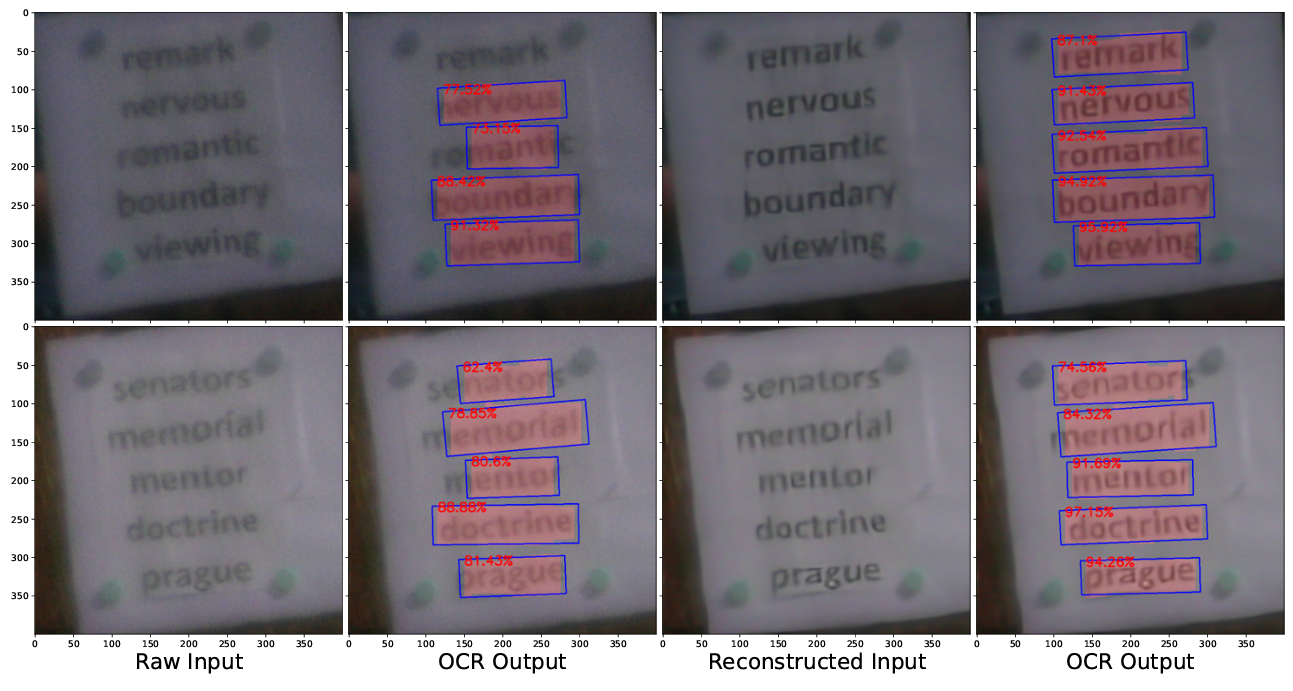}
\caption{Excerpt of the optical character recognition results reported by Mao et al. Source: \cite{Mao_2022_a}.}
\label{fig: ch4 ajay result 2}
\vspace{-2ex}
\end{figure}

One important result reported by Mao et al. \cite{Mao_2022_a} is an evaluation scheme. In turbulence mitigation, the ground truth is often very difficult to obtain because the appearance of the target pattern changes due to the weather condition. Even if there is no turbulence, the ambient light, the color, etc. will alter the ground truth. Therefore, in some applications, it makes sense to use an alternative downstream recognition task as an evaluation metric. \fref{fig: ch4 ajay result 2} reports an attempt to use optical character recognition as a way to assess how well a mitigation algorithm performs. If a mitigation algorithm performs well, we expect the optical character recognition to improve.

\subsection{Multi-frame End-to-End Transformer}
\index{transformer}
A very recent approach, when this book is written, is a direct end-to-end transformer by Zhang et al. \cite{Zhang_2022_a} as shown in \fref{fig: ch4 xingguang} known as the turbulence mitigation transformer (TMT). Building upon the highly successful video restoration transform \cite{Zamir_2021_Restormer}, the TMT presented in \cite{Zhang_2022_a} brings together two interesting ideas:

\begin{figure}[h]
\centering
\includegraphics[width=\linewidth]{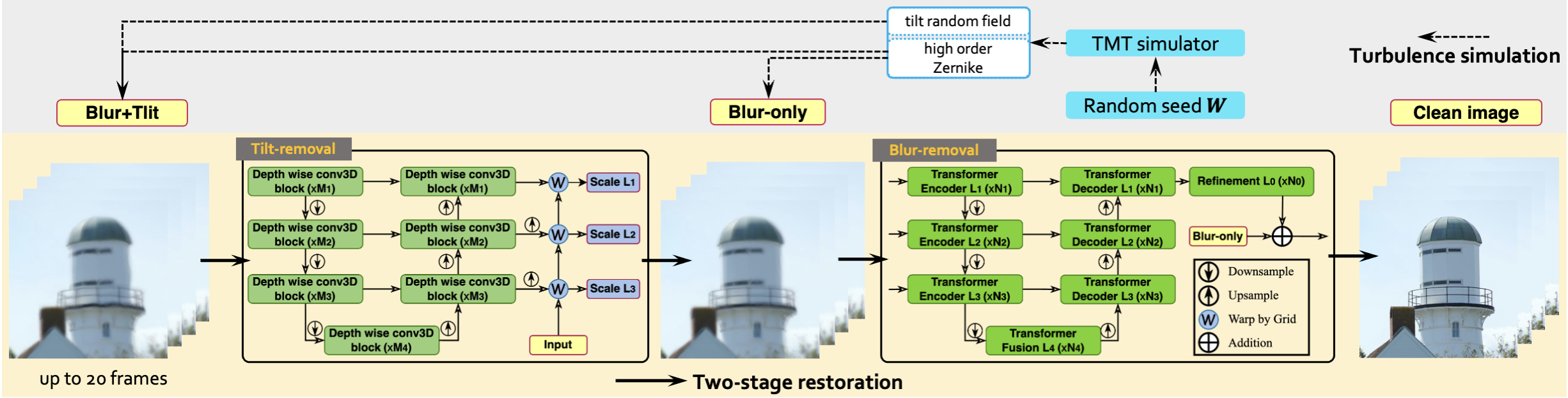}
\caption{Turbulence mitigation transformer by Zhang et al \cite{Zhang_2022_a}.}
\label{fig: ch4 xingguang}
\end{figure}

\keyword{Decoupling into tilt and blur}. Unlike the previous deep-learning turbulence mitigation methods which are mostly black boxes, TMT explicitly decouples the turbulence into tilt and blur. The method first mitigates the tilts, because tilt correction is a slightly easier task as it is evident from the success of deformable neural networks \cite{Li_2021_a}. The tilt removal sub-network shown in \fref{fig: ch4 xingguang} uses a \textbf{multi-scale} depthwise convolution to pull the features and align them. The multi-scale structure is critical here because at a lower resolution the tilt is significantly easier to be compensated than at a high resolution. The output of the sub-network is a set of three tilt-corrected images at different resolutions. The ground truth tilt-corrected images are generated by the phase-to-space simulator. In this way, the nature of turbulence is implicitly maintained.

After tilts are removed, the residual blur is still spatially varying. Convolutional neural networks are known to be a poor candidate for spatially varying degradations because the convolution kernels are global operators. Vision transformers use self-attention to extract local attention. Instead of using a global convolutional kernel, the self-attention gives weights to individual kernels so that each pixel would experience a different combination of convolutions. In other words, transformers allow us to restore the image according to the local distortion strength. This added degree of freedom plays a significant role in the restoration task.

\keyword{Temporal channel joint attention}. The challenge of designing a turbulence mitigation algorithm is how to generate global temporal attention of all the frames without suffering from a high complexity and memory requirement. In conventional transformers, generating attention has a complexity that grows quadratically with the number of pixels in the video. Thus, conventional transformers adopt a window-based strategy to compute the attention locally. In TMT's transformer, the spatial coordinates are connected purely via convolution layers. This makes the interaction among neighboring pixels smooth, which avoids the inconsistent performance around the margin of local windows. The core module of TMT is temporal-channel self-attention (TCJA). For each pixel, TMT computes the self-attention matrix on the temporal and channel axes. Although the convolution operation is spatially invariant, the combination of the features of each pixel is spatially independent, this enables spatially varying restoration. The dynamics of the blur and residue jitter largely follow a zero-mean random process. Therefore, by enabling full temporal connection over more frames, TMT can be more efficient than conventional transformers in capturing temporal dynamics.

\begin{figure}[h]
\centering
\includegraphics[width=\linewidth]{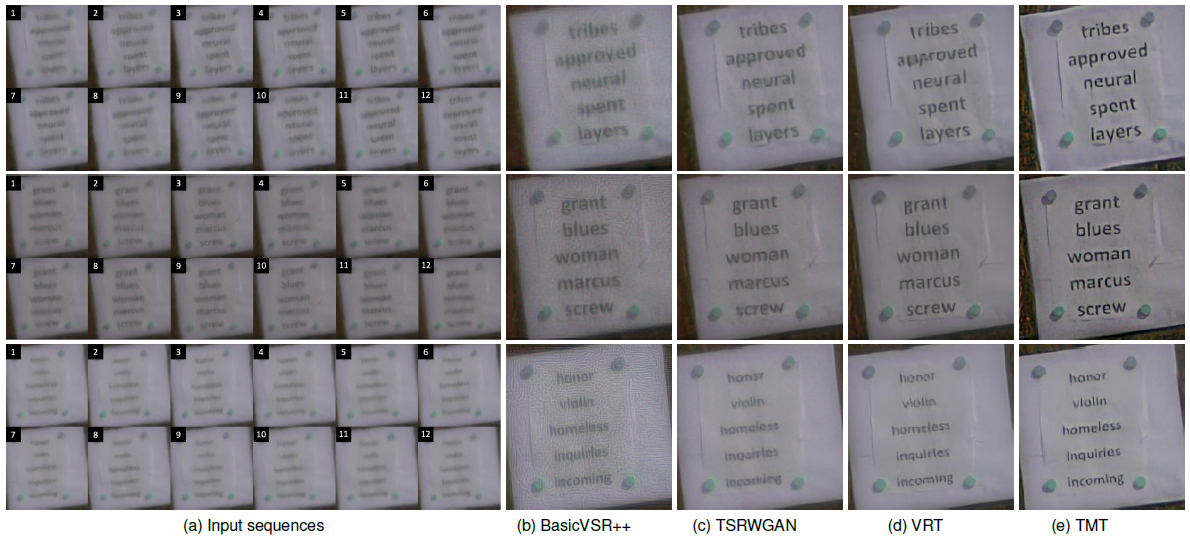}
\caption{Image reconstruction results of a \emph{real} video sequence using the turbulence mitigation transformer by Zhang et al. \cite{Zhang_2022_a}.}
\label{fig: ch4 xingguang_result}
\end{figure}

The end-to-end transformer developed by Zhang et al. \cite{Zhang_2022_a} is known as the turbulence mitigation transformer (TMT). A few snapshots of the reconstruction results of TMT, and their comparisons with other state-of-the-art restoration methods, are shown in \fref{fig: ch4 xingguang_result}. One important remark here is that the competing methods are strong baselines re-trained with the turbulence data. These baselines are significantly more powerful than the old networks (such as UNets or DnCNN) shown in the previous subsection. That TMT is able to outperform the baselines is an important message that the physics-inspired transformer networks are promising.

\subsection{Generative Methods}
\index{generative adversarial network}
Generative models have taken a significant footprint in the field of image restoration due to their realistic pixel rendering capability. The biggest strength of generative models (in particular the \keyword{generative adversarial networks}, GANs) is the way it measures the similarity of two distributions instead of using the conventional mean squared error. GANs typically consist of a generator and a discriminator. The generator is a deep neural network that takes a random code (typically a white Gaussian vector) and turns it into an image. The image will then be compared with a ground truth sample from the training dataset by a discriminator. The discriminator is another deep neural network that gives a binary decision of whether the generated image is close enough to the ground truth. If not, the generator has to be updated until the discriminator can no longer differentiate the generated image and the true image.

The advantage of a GAN is that it is not limited to the particular type of degradation caused by the physics. More interestingly, the resolution of the recovered image is not limited by the actual optical resolution limit but by the prior distribution of the training set. For example, even if one part of the image is extremely blurry, GAN can generate sharp pixels by sampling the closest samples according to the prior distribution it has learned. In some sense, GANs are a powerful tool to hallucinate image content.

\begin{figure}[h]
\centering
\includegraphics[width=\linewidth]{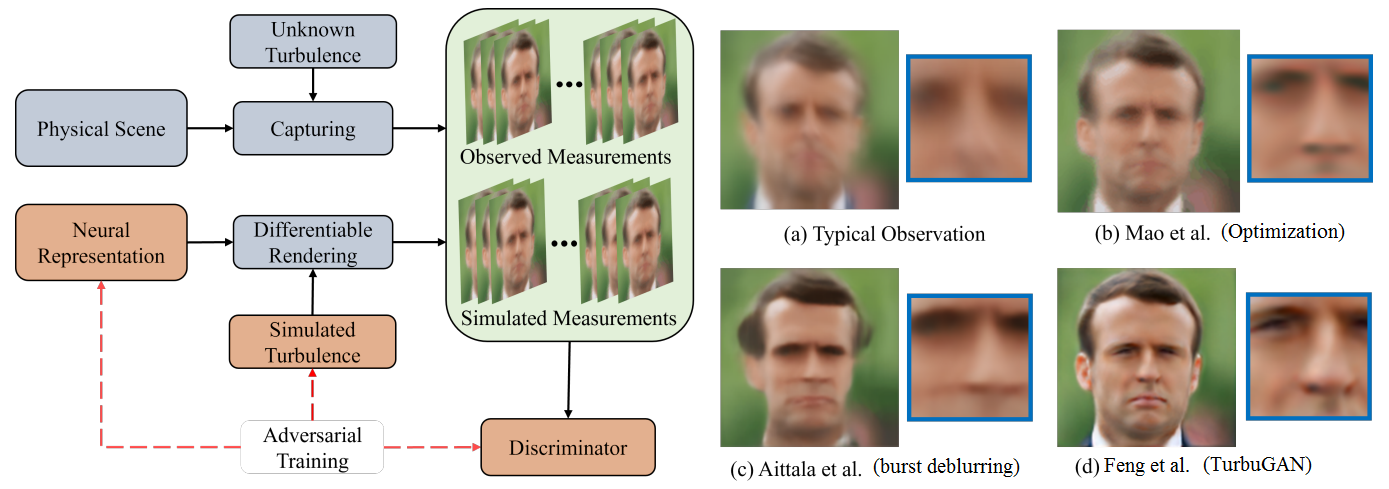}
\caption{Generative adversarial network (GAN) based algorithm by Feng et al. \cite{Feng_2022_a}.}
\label{fig: ch4 metzler}
\end{figure}

In the context of atmospheric turbulence, a recently proposed method TurbuGAN \cite{Feng_2022_a} shows how GANs can be implemented. A schematic diagram of TurbuGAN is shown in \fref{fig: ch4 metzler}. In TurbuGAN, the role of the discriminator is to differentiate between a real turbulence-distorted image and a GAN-generated image. The GAN generator takes a random code, renders a clean image, and sends it through the turbulence simulator (e.g., phase-to-space transform). The synthesized image is then compared with the real turbulence image by the discriminator.

TurbuGAN is a good demonstration that GAN is capable of handling complex distortions such as atmospheric turbulence. It also demonstrates the importance of a physics-based simulator, for without the simulator it would be impossible to render a realistic turbulence effect. In terms of image recovery, GAN generates sharper images than many traditional methods. This observation is consistent with ATFaceGAN \cite{Lau_2021_b}, another recently proposed GAN-based image restoration method. Though the nature of training a GAN is based on synthetic data, there are works such as \cite{Miller_2021_a} which seek to use a GAN for learning statistics of a scene.

Methods based on the GAN concepts generally suffer from several bottleneck challenges. For example, there is always a controversy about whether the hallucinated pixels are real or fake. For human consumption, the borderline between the real and the fake is less of an issue. However, for machine vision where we need to make a decision, the hallucinated pixels may not contain the true information. Thus, in many cases, there is no guarantee that a GAN-generated image can lead to, for example, a higher face detection performance.

From an implementation perspective, the core part of a GAN-based image restoration method is the generator. The generator can use any standard network architecture, for example, a convolutional network. If the network architecture is fixed, and if the training set is also fixed, then the only difference between a GAN-based network and a non-GAN-based network is the loss function. With complex degradations such as turbulence, the loss function alone does not make the difference. In fact, the TSRWGAN results shown in \fref{fig: ch4 xingguang_result} are based on the training of the generator of TSRWGAN \cite{Jin_2021_a}. The image quality is just on par with other networks. Therefore, GAN methods will likely have a long journey of obstacles to go through in the near future.

\begin{figure}[h]
\centering
\includegraphics[width=\linewidth]{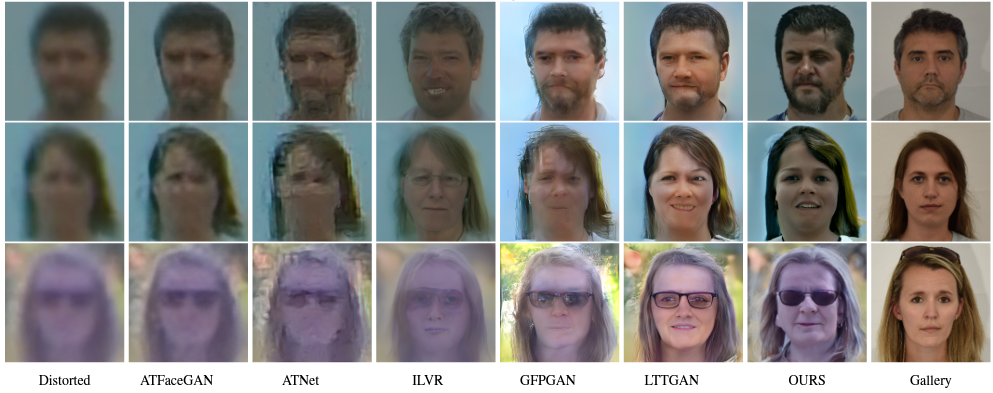}
\caption{Image restoration by denoising diffusion probabilistic model by Nair et al. \cite{Nair_2023_WACV}. Image source: \cite{Nair_2023_WACV}.}
\label{fig: ch4 diffusion}
\end{figure}

The \keyword{denoising diffusion probabilistic model}\index{denoising diffusion probabilistic model} is gaining strong momentum in the image restoration community. At a high level, the idea is to train a \emph{diffusion model} that allows us to draw samples from the posterior distribution. The sampling process can be thought of as a sequence of denoising steps where we gradually remove the noise by estimating the key parameters of the signal. Without going into the theoretical details of the method, we highlight a recent result published by Nair et al. \cite{Nair_2023_WACV}, as shown in \fref{fig: ch4 diffusion}. As we can see the restored image shows superior object details although some are noticeable artifacts. The promise offered by the model is worth noting.

\section{Summary}
\label{sec: sec4_5}
The topic of imaging through turbulence from a reconstruction perspective is presented. The effective application of deep learning to this problem is perhaps the biggest open problem today. While we have witnessed many great innovations over the past few years, we anticipate more results to come. This Chapter may be summarized as follows.

\keyword{Part 1 Atmospheric Forward Model}: We stress the importance of having an accurate and fast-forward model. This is not only for the purpose of simulating realistic training and testing data, but it also gives us the chance to integrate the forward model into the inverse problem. We have seen in this Chapter that data simulated by a more accurate forward model improves significantly the image restoration results, even if we use the exact same neural network architecture.

The problem of tilt-then-blur or blur-then-tilt has been considered. We stress that the correct order is tilt-then-blur, but we also argue that for efficient image restoration, the order can be flipped as long as the benefits to the restoration algorithm are more than the harms.

\keyword{Part 2 Lucky Imaging and Classical Modeling}: Classical methods such as lucky imaging and image registration are still highly insightful. When viewing an image through atmospheric turbulence, one may notice the image going in and out of focus, with the high-quality observations being known as lucky observations. Lucky patches, in the case of atmospheric imaging, imply that a digital lucky fusion must be performed in order to piece together a high-quality image. A good image restoration method should do its best to recover the lucky information. To this end, we argue that multi-frame restorations are significantly more beneficial than single-frame restorations.

\keyword{Part 3 Deep Learning Approaches}: Application of deep learning towards reconstruction and modeling has shown great success in many fields, though for turbulence, there still remains some gap. There are some methods that have produced convincing and promising results, and with the continued development of simulation capabilities, it is reasonable this upward trend will continue. From the perspective of the authors, integration of the true forward model is critical.

\bibliographystyle{ieeetr}
\bibliography{refs_fnt_2}

\printindex
\end{document}